\documentclass[longauth]{aa}  
\usepackage{lscape}
\usepackage{graphicx}
\usepackage{siunitx}
\usepackage{caption}
\usepackage{longfigure}
\usepackage{txfonts}

\ExplSyntaxOn

\keys_define:nn { miguel/label }
 {
  label   .tl_set:N = \l_miguel_label_tl,
  unknown .code:n   = \clist_put_right:Nx \l_miguel_label_clist
                       { \l_keys_key_tl = \exp_not:n { #1 } }
 }
\clist_new:N \l_miguel_label_clist
\box_new:N \l_miguel_label_box
\box_new:N \l_miguel_label_image_box

\NewDocumentCommand{\xincludegraphics}{O{}m}
 {
  \tl_clear:N \l_miguel_label_tl
  \clist_clear:N \l_miguel_label_clist
  \keys_set:nn { miguel/label } { #1 }
  \tl_if_empty:NTF \l_miguel_label_tl
   {
    \miguel_includegraphics:Vn \l_miguel_label_clist { #2 }
   }
   {
    \hbox_set:Nn \l_miguel_label_image_box
     {
      \miguel_includegraphics:Vn \l_miguel_label_clist { #2 }
     }
    \hbox_set:Nn \l_miguel_label_box
     {
      \skip_horizontal:n { 3pt }
      \fcolorbox{white}{white}{\footnotesize \tl_use:N \l_miguel_label_tl}
     }
    \leavevmode
    \box_use:N \l_miguel_label_image_box
    \skip_horizontal:n { -\box_wd:N \l_miguel_label_image_box }
    \hbox_overlap_right:n
     {
      \box_move_up:nn
       {
        \box_ht:N \l_miguel_label_image_box - 
        \box_ht:N \l_miguel_label_box - 3pt
       }
       { \box_use_drop:N \l_miguel_label_box }
     }
    \skip_horizontal:n { \box_wd:N \l_miguel_label_image_box }
   }
 }

\cs_new_protected:Nn \miguel_includegraphics:nn
 {
  \includegraphics[#1]{#2}
 }
\cs_generate_variant:Nn \miguel_includegraphics:nn { V }

\ExplSyntaxOff

\usepackage{soul}
\usepackage[dvipsnames]{xcolor}

\def \sini{\textrm{$\sin i$ }}
\def \msini{\textrm{$m\sin i\,$}}

\begin{document} 

   \title{The SOPHIE search for northern extrasolar planets}

   \subtitle{XVII. A wealth of new objects: Six cool Jupiters, three  brown dwarfs, and 16 low-mass binary stars}
     \titlerunning{CJS, BDS, SCS}   
   \authorrunning{S. Dalal et al.}
   \author{S.~Dalal\inst{1}\fnmsep\thanks{\email{shweta.dalal@iap.fr}},
          F.~Kiefer\inst{1,2},
          G.~H\'ebrard\inst{1,3}, 
          J.~Sahlmann\inst{4},
          S.\,G.~Sousa\inst{5,6}
          T.~Forveille\inst{7},
          X.~Delfosse\inst{7},
          L.~Arnold\inst{3},
          N.~Astudillo-Defru\inst{8},
          X.~Bonfils\inst{7},
          I.~Boisse\inst{9},
          F.~Bouchy\inst{10},
          V.~Bourrier\inst{10},
          B.~Brugger\inst{9},
          P.~Cort\'es-Zuleta\inst{9},
          M.~Deleuil\inst{9},
          O.\,D.\,S.~Demangeon\inst{5,6,21},
          R.\,F.~D\'{\i}az\inst{11},
          N.\,C.\,Hara\inst{10}, 
          N.~Heidari\inst{12,9,13},
          M.\,J.~Hobson\inst{14,15},
          T.~Lopez\inst{9},
          C.~Lovis\inst{10},
          E.~Martioli\inst{1,16},
          L.~Mignon\inst{8},
          O.~Mousis\inst{9},
          C.~Moutou\inst{17},
          J.~Rey\inst{18},
          A.~Santerne\inst{9},
          N.\,C.~Santos\inst{5,6},
          D.~S\'egransan\inst{10},
          P.~A.~Str{\o}m\inst{19,20}, \and
          S.~Udry\inst{10}
         } 

          \institute{\inst{1}Institut d'Astrophysique de Paris,   98 bis, boulevard Arago,  75014, Paris\\
            \inst{2}LESIA, Observatoire de Paris, Université PSL, CNRS, Sorbonne Université, Université de Paris, 5 place Jules Janssen, 92195, Meudon, France\\
   	          \inst{3}Observatoire de Haute-Provence, CNRS, Universit\'e d'Aix-Marseille, 04870 Saint-Michel-l'Observatoire, France\\
   	          \inst{4}RHEA Group for the European Space Agency (ESA), European Space Astronomy Centre (ESAC), Camino Bajo del Castillo s/n, 28692 Villanueva de la Ca\~nada, Madrid, Spain\\
   	           \inst{5}Instituto de Astrof{\'\i}sica e Ci\^encias do Espa\c{c}o, Universidade do Porto, CAUP, Rua das Estrelas, 4150-762 Porto, Portugal\\
                   \inst{6}Departamento\,de\,F{\'\i}sica\,e\,Astronomia,\,Faculdade\,de\,Ci\^encias,\\Universidade\,do\,Porto,\,Rua\,Campo\,Alegre,\,4169-007\,Porto,\,Portugal\\
                   \inst{7}Univ. Grenoble Alpes, CNRS, IPAG, 38000 Grenoble, France\\
                      \inst{8}Departamento de Matem\'atica y F\'isica Aplicadas, Universidad Cat\'olica de la Sant\'isima Concepci\'on, Alonso de Rivera 2850, Concepci\'on, Chile\\
                   \inst{9}Aix Marseille Univ, CNRS, CNES, LAM, Marseille, France\\
                   \inst{10}Observatoire de Gen\`eve,  Universit\'e de Gen\`eve, Chemin Pegasi, 51, 1290 Sauverny, Switzerland\\
                    \inst{11}International Center for Advanced Studies (ICAS) and ICIFI (CONICET), ECyT-UNSAM, Campus Miguelete, 25 de Mayo y Francia, (1650) Buenos Aires, Argentina.\\
                   \inst{12}Department of Physics, Shahid Beheshti University, Tehran, Iran\\
                    \inst{13}Laboratoire J.-L. Lagrange, Observatoire de la Côte d’Azur (OCA), Universite de Nice-Sophia Antipolis (UNS), CNRS, Campus Valrose, 06108 Nice Cedex 2, France\\
                    \inst{14}Instituto de Astrofésica, Pontificia Universidad Católica de Chile, Av. Vicuña Mackenna 4860, Macul, Santiago, Chile\\
                    \inst{15}Millennium Institute of Astrophysics, Av. Vicuña Mackenna 4860, 7820436 Macul, Santiago, Chile\\
                    \inst{16}Laboratório Nacional de Astrofísica, Rua Estados Unidos 154, Itajubá, MG, 37504-364, Brazil\\
                    \inst{17}Univ. de Toulouse, CNRS, IRAP, 14 Avenue Belin, 31400, Toulouse, France\\
                     \inst{18}Las Campanas Observatory, Carnegie Institution of Washington, Colina el Pino, Casilla 601 La Serena, Chile\\
                     \inst{19}Centre for Exoplanets and Habitability, University of Warwick, Gibbet Hill Road, Coventry, CV4 7AL, UK\\ 
                     \inst{20}Department of Physics, University of Warwick, Gibbet Hill Road, Coventry, CV4 7AL, UK\\
                     \inst{21}Centro de Astrof\'{\i}sica da Universidade do Porto, Rua das Estrelas, 4150-762 Porto, Portugal\\
             }

   \date{Received ; accepted }


 
\abstract{

Distinguishing classes within substellar objects and understanding their formation and evolution need larger samples of substellar companions such as exoplanets, brown dwarfs, and low-mass stars. In this paper, we look for substellar companions using radial velocity surveys of FGK stars with the SOPHIE spectrograph at the Observatoire de Haute-Provence. We assign here the radial velocity variations of 27 stars to their orbital motion induced by low-mass companions. We also constrained their plane-of-the-sky motion using HIPPARCOS and Gaia Data Release 1 measurements, which constrain the true masses of some of these companions. We report the detection and characterization of six cool Jupiters, three brown dwarf candidates, and 16 low-mass stellar companions. We additionally update the orbital parameters of the low-mass star HD~8291 B, and we conclude that the radial velocity variations of HD~204277 are likely due to stellar activity despite resembling the signal of a giant planet. One of the new giant planets, BD+631405 b, adds to the population of highly eccentric cool Jupiters, and it is presently the most massive member. Two of the cool Jupiter systems also exhibit signatures of an additional outer companion. The orbital periods of the new companions span 30 days to 11.5 years, their masses 0.72  M$_{\mathrm{J}}$ to 0.61 M$_{\odot}$, and their eccentricities 0.04 to 0.88. These discoveries probe the diversity of substellar objects and low-mass stars, which will help constrain the models of their formation and evolution.

}


\maketitle
\section{Introduction}
The first exoplanet orbiting a ``solar-type star,'' 51~Pegasi b \citep{Mayor1995}, was discovered using the radial velocity (RV) technique. This technique is currently the second-most successful planet detection method, having as of Jan 21, 2021, detected 913 exoplanets ranging from Earth-mass to more massive than Jupiter (\url{www.exoplanet.eu}). It remains much more efficient than the transit method at detecting long-period planets, and thanks to improved RV precision and increasingly long temporal baselines, RV surveys are ideal for discovering analogs of Jupiter and Saturn. The ELODIE and SOPHIE exoplanet survey at Observatoire de Haute Provence (France) is one of the longest operating RV surveys, with a total time baseline of over 25 years which extends from before the discovery of 51~Pegasi~b. 

This paper presents discoveries of new substellar and low-mass stellar companions of solar-type stars -- cool Jupiters (CJs), brown dwarfs (BDs), and M-dwarfs or low-mass stars (SCs) -- from two ongoing exoplanet detection surveys with the SOPHIE spectrograph (Section \ref{sec:surveys}).
These substellar objects are classified according to their masses. CJs are defined as massive planets with masses above 0.3 M$_{\mathrm{J}}$ and orbital periods above 100 days \citep{Wittenmyer2020}. BDs are substellar objects in the 13--75  M$_{\mathrm{J}}$ mass range and occupy the domain between massive planets and stars. SCs are low-mass stars with masses above approximately 75  M$_{\mathrm{J}}$ (0.072 M$_{\odot}$).  The mass limit that separates BDs from massive planets is conventionally the minimum mass required to fuse deuterium in the core of the substellar object, that is 13  M$_{\mathrm{J}}$ \citep{Boss2003, Chabrier2014}, and the limit between BDs and SCs comes from the minimum mass for hydrogen fusion in the core, that is 75  M$_{\mathrm{J}}$ \citep{Chabrier2000}. Both boundaries depend to a small extent on the metallicity of the object  \citep{Chabrier1997, Spiegel2011}. 

Despite the discovery of thousands of giant planets and BDs, the statistics on their occurrence and properties are still very incomplete. Most statistical studies of massive planets are for hot and warm Jupiters \citep{Howard2010, Fernandes2019}, and CJs, which are suspected to be more abundant than hot Jupiters, are poorly characterized as a population \citep{Wittenmyer2020}. Like our Solar System's Jupiter and Saturn, they are dynamically dominant in their system and influence the formation and evolution of any interior planets, including habitable worlds \citep{Raymond2006, Morbidelli2012, Raymond2017, Bryan2019}. Detecting more CJs will help to provide detailed statistics on massive planets with a range of periods from a few days to a decade.

The two leading planet formation models are core accretion \citep{Pollack1996,Mordasini2009,Guilera2010} and disk instability \citep{Cai2010,Boss2011}. Previous studies \citep{Santos2017,Narang2018,Schlaufman2018} suggest that giant planets might divide into two distinct giant planet populations. The metallicity of the host stars of giant planets with a mass above 4 M$_{\mathrm{J}}$ is, on average, lower than that of the host stars of giant planets with mass under 4 M$_{\mathrm{J}}$. This hints towards possibly two distinct planet formation scenarios for these two populations. Planet-formation models can probably form bodies up to 40  M$_{\mathrm{J}}$ \citep{Ida2004, Alibert2005, Mordasini2009}, which suggests that low-mass BDs might form like massive planets, through the disk gravitational instability scenario. High mass BDs, by contrast, are likely to form like stellar binary systems, through molecular cloud fragmentation \citep{Ma2014}.

Brown dwarfs are interesting as the transition between the formation mechanisms of giant planets and stars probably runs through their population. One interesting characteristic of the BD population is that few are detected at short orbital periods, a feature known as the BD desert \citep[e.g.,][]{Halbwachs2000, Grether2006, Sahlmann2011}. New detections have shrunk this desert in recent years\citep{Csizmadia2016, Wilson2016, Kiefer2019, Subjak2020}, but there is still a detection deficit for orbital periods under 100 days and masses between 30 and 60  M$_{\mathrm{J}}$ \citep{Ma2014, Ranc2015, Kiefer2019}. One obvious path towards a better understanding of substellar mass objects is to detect additional objects in and around the desert to better characterize its shape, which in turn will constrain the giant planet and BD formation and evolution processes.  

This paper combines the RV technique with astrometry to detect substellar objects and constrain their mass. The RV technique can, in most cases, only determine the minimum mass of the companion (\msini, where $m$ is the true mass and $i$ the orbital inclination of the planet). As mass is what distinguishes giant planets from BDs, and from SCs, the inclination ambiguity must be lifted to asserting the true nature of a substellar companion \citep[e.g.,][]{Diaz:2012fk,Curiel2020,Kiefer2021}. As just one example, HD~5388~b was first announced as a likely gas giant \citep{Santos2010} but turned out to be a BD companion when \citet{Sahlmann:2011lr} detected its astrometric signature in the HIPPARCOS measurements of HD~5388. Here as well, we combine astrometry and RV measurements to overcome the \sini ambiguity and constrain the true mass of the companions \citep{McArthur2010,Tokovinin2017}.

Section~\ref{sec:surveys} presents the two SOPHIE surveys which produced these new detections, while Sect. \ref{sec:observations} explains how the observations were performed and the data reduced. Sections~\ref{sec:hoststar}, \ref{sec:activity}, and \ref{sec:RVanalysis} respectively discuss the spectral analysis, the stellar activity, and the RV analysis of the SOPHIE observations. In Sect.~\ref{sec:astrometry} we analyze the HIPPARCOS and Gaia astrometric measurements. In Sect.~\ref{sec:results} we review the new detections of CJs, BDs and SCs. Finally, Section~\ref{sec:conclusion} discusses and summarizes our results.
 
\section{ Description of the SOPHIE surveys} \label{sec:surveys}
The SOPHIE volume-limited survey for giant planets and BDs observes a catalog of about 2300 FGK stars in the northern sky ($\delta$ $>$ $+00:00:00$). These targets are within 60~pc of the Sun and have B$-$V between 0.35 and 1.0. Around 2000 of those stars have SOPHIE observations at this point \citep{Hebrard2016,Kiefer2019}. Two ongoing programs with the SOPHIE spectrograph contribute new substellar companions to this paper. 

\subsection{Giant planets survey}
The goal of this ongoing volume-limited program is to increase the number of detections of giant planets orbiting nearby FGK stars, and to identify candidates for follow-up studies: multiplanetary systems for dynamics and transiting systems for structure characterization. This survey constrains the distributions of exoplanet parameters, which helps understand the diversity of planetary systems. The new CJs presented in this paper are the continuation of work by \citet{Boisse2010}, \citet{Moutou2014}, \citet{Diaz2016}, and \citet{Hebrard2016}. 

\subsection{Brown dwarfs survey}
The giant planet survey stops observing any star with a companion which is clearly outside the planetary mass range (i.e., $\geq$ 13 M$_{\mathrm{J}}$), and these stars are transferred to the BD survey which has looser RV precision requirements. The goal of the BD program is to obtain an unbiased inventory of companions within and about the BD mass regime for orbital periods up to 10\, yrs. This includes stellar companions with mass (or \msini) > 75  M$_{\mathrm{J}}$ (0.072 M$_{\odot}$),  because detection of these stellar companions is inevitable while aiming for completeness for BD and because they probe the connections between massive BDs and low-mass stellar companions. The new BDs and SCs presented in this paper are the continuation of work by \citet{Diaz:2012fk}, \citet{Bouchy2016}, \citet{Wilson2016}, and \citet{Kiefer2019}.

\section{Spectroscopic data} \label{sec:observations}
\subsection{Observations}
We present new observations of 27 stars with the SOPHIE spectrograph, a cross-dispersed, environmentally stabilized echelle spectrograph at the 1.93~m telescope of Observatoire de Haute Provence (OHP). SOPHIE has been in operation since 2006 and covers the 3872 to 6943 \si{\angstrom} wavelength range \citep{Perruchot2008, Bouchy2009}. The spectrograph is fed through two optical fibers, one of which is always illuminated by starlight from the telescope. Our observations illuminate the second fiber with light from the background sky to estimate its contribution to the on-star spectrum (obj AB mode), and they were carried out in the high-resolution (R = 75000) mode of the spectrograph. Wavelength calibrations and drift measurements were obtained approximately every two hours during the night, as well as at the beginning and end of each night. In 2011, the circular-section fiber was replaced with octagonal-section fiber in the fiber link to improve the stability of the spectrograph illumination \citep{Perruchot2011, Bouchy2013}, and the pre and post upgrade data have distinct characteristics.  This work, therefore, distinguishes two SOPHIE datasets, labeled SOPHIE and SOPHIE+, depending on whether the spectra were taken before or after this SOPHIE upgrade.

\subsection{Data reduction}
The SOPHIE pipeline extracts the spectra and cross-correlates them with a numerical mask \citep{Bouchy2009}. The cross-correlation functions (CCFs) are produced by considering masks corresponding to their stellar type and incorporating all of the spectral orders. The CCFs were then fitted with Gaussians to derive the radial velocities (RVs) \citep{Baranne1996, Pepe2002}. The exposure time was adjusted to reach a signal-to-noise ratio (S/N per pixel at 550 nm) of at least 50 for the giant planet survey and 30 for the BD survey (observations obtained while a BD target was still in the giant planet survey targeted S/N=50), under diverse weather conditions. Spectra that are significantly contaminated by the Moon were discarded, as were all spectra with less than half of the median S/N for a given target or a large uncertainty on the RV measurements. Spectra for the stars of the giant planet survey and the BD survey have an average S/N of 53.2 and 43.8, respectively. The  RVs were also corrected from the CCD charge transfer inefficiency following \citet{Bouchy2009b}. Parameters such as Full Width at Half Maximum (FWHM), contrast, stellar rotational velocity ($v \sin i_{\star}$, where $i_{\star}$ is the inclination of the star's rotational axis with respect to the line of sight), and the Bisector Inverse Span (BIS), were also derived from the CCF by the SOPHIE reduction software following the method of \citet{Boisse2010}. A 5 m~s$^{-1}$ systematic uncertainty is added in quadrature to the uncertainty of the RV measurements obtained before the June 2011 upgrade to account for the poor scrambling properties of the early exposures. The main characteristics of the SOPHIE and SOPHIE+ data sets are summarized in Table~\ref{tab:datadetail}. 

\section{Spectral analysis} \label{sec:hoststar}
For each star, we performed spectral analyses of an optimally weighted average of all the SOPHIE spectra unaffected by Moon pollution. We used the ARES+MOOG\footnote{We used ARES v2 \citep{Sousa2015} and MOOG vNov2019 \citep{Sneden1973} method. Details on ARES+MOOG are described in  \citet{Sousa2014}.}, following closely what was done in \citet{Santos2004} and \citet{Sousa2008} to derive the effective temperature $T_{\rm eff}$, the surface gravity $\log g$, and the metallicity ${\rm [Fe/H]}$. Using those derived spectroscopic parameters as input, stellar masses were derived from the calibration of \citet{Torres2010} with the correction of \citet{Santos2013}. Their uncertainties were computed from 10,000 random draws of the stellar parameters within their error bars and assuming Gaussian distributions. Table~\ref{tab:stellar} lists the resulting stellar parameters, as well as the $\textrm{logR'}_{\textrm{HK}}$ and $v\sin i$, which were obtained following the approach of \citet{Boisse2010}.

\section{Stellar activity analysis}\label{sec:activity}
Activity in the atmosphere of the star can alter the shape of stellar lines \citep{Queloz2001},  as can face-on binaries \citep[e.g.,][]{Diaz:2012fk,Wright2013}. This gives rise to apparent variations in RV signatures which can mimic a planetary signal. We use various indicators, such as BIS, FWHM, and $\textrm{logR'}_{\textrm{HK}}$, to probe whether the observed RV signal stems from spectral-line profile changes related to stellar activity.

We evaluate the expected activity-related RV scatter $\sigma_{a}$ (Table~\ref{tab:democor}) from our measurement of the mean $\textrm{logR'}_{\textrm{HK}}$ index (Table~\ref{tab:stellar}) using the  \citet{Santos2000} $\sigma_{a}$ vs $\textrm{logR'}_{\textrm{HK}}$ relation.  The most active star in the giant planet survey sample is HD~204277, with $\textrm{logR'}_{\textrm{HK}}=$ -4.50 $\pm$ 0.11 and  $\sigma_{a} \simeq 21.8$ m~s$^{-1}$.  The dispersion of its measured RVs is on the order of this $\sigma_{a}$ value, and Sect.~\ref{sec:notaplanet} presents a detailed discussion of the nature of its signal. We exclude it from the rest of our analysis and the next sections are dedicated to the 26 remaining stars. Their $\sigma_{a}$ range between 5 and 8 m~s$^{-1}$ for the giant planet survey targets and from 5 to 22 m~s$^{-1}$ for the  BD survey targets. The dispersion of the measured RVs of these 26 stars is significantly larger than their estimated $\sigma_{a}$ (see Table~\ref{tab:datadetail} for dispersion of the measured RVs). 

To further investigate whether the observed RV signals can be caused by stellar activity, we looked for correlations between the measured RVs and two probes of the line shape, namely the FWHM and BIS. We calculated the Pearson correlation coefficients and the significance of the correlation (p-value) (Table \ref{tab:democor}) and find that none of them is significant.   

\section{Radial velocity analysis} \label{sec:RVanalysis}

\setlength{\tabcolsep}{5pt}
\begin{table*}\centering
\small
			\caption{\label{tab:giantplanet} Keplerian solution and planetary parameters with 1-$\sigma$ uncertainties for the six CJs having 0.72  M$_{\mathrm{J}}$ $\leq$ \msini $\leq$ 3.96 M$_{\mathrm{J}}$.}
{\def\arraystretch{1.25}
\begin{tabular}{l|c|c|c|c|c|c|c}
\hline
Parameters 	&	Unit	&	BD+450564	&	BD+550362	&	BD+631405	&	HD~124330	&	HD~155193	&	HD~331093	\\
\hline
P	&	days	&	307.88   $\pm$    1.47	&	265.59   $\pm$    1.04	&	1198.48   $\pm$   60.79	&	270.66   $\pm$    1.21	&	352.65   $\pm$    2.58	&	621.62   $\pm$   16.11	\\
K 	&		m~s$^{-1}$	&	47.68   $\pm$     2.78	&	25.12   $\pm$     1.74	&	186.01   $\pm$    14.88	&	22.85   $\pm$     1.23	&	19.51   $\pm$     1.42	&	43.65   $\pm$      2.2	\\
$e$	&	-	&	0.12   $\pm$    0.06	&	0.27   $\pm$    0.06	&	0.88   $\pm$    0.02	&	0.34   $\pm$    0.05	&	0.21   $\pm$    0.08	&	0.59   $\pm$    0.03	\\
$\omega$  	& ($^\circ$)	&	70.31   $\pm$    51.85	&	-134.44   $\pm$    62.43	&	94.25   $\pm$     3.76	&	18.28   $\pm$    10.44	&	62.88   $\pm$   136.21	&	165.77   $\pm$    27.06	\\
$T_p^{\dagger}$ 	&	 BJD-2450000	&	7734.61   $\pm$      6.10	&	8194.19   $\pm$    12.29	&	9440.02   $\pm$    61.72	&	8415.57   $\pm$     3.06	&	7590.7   $\pm$     7.62	&	9022.81   $\pm$    17.77	\\
$\gamma_{S+}$  	&	km~s$^{-1}$	&	-6.817   $\pm$        0.003	&	-25.647   $\pm$        0.002	&	-15.33   $\pm$       0.01	&	-30.687   $\pm$        0.001	&	3.883  $\pm$        0.002	&	-51.29   $\pm$       0.02	\\
 $\gamma_{S}$  	&	km~s$^{-1}$	&	       --  	&	       --  	&	       -- 	&	       --  	&	       --  	&	-51.320   $\pm$       0.003	\\
 \hline
d1	&	m~s$^{-1}$yr$^{-1}$	&	         --  	&	1.76  $\pm$   0.04&	         --  	&	2.95  $\pm$   0.07	&	         --  	&	         --  	\\
d2	&	m~s$^{-1}$yr$^{-2}$	&	         --   	&	1.4365    $\pm$    0.0005	&	         --   	&	         --   	&	         --   	&	         --   	\\
$\sigma_\text{O-C, S+}$ 	&	m~s$^{-1}$	&	3.30	&	3.75	&	4.58	&	5.44	&	7.05	&	3.27	\\
$\sigma_\text{O-C, S}$ 	&	m~s$^{-1}$	&	     --        	&	     --        	&	     --        	&	     --        	&	     --        	&	3.22	\\
\hline
\msini$^{\ddagger}$	&	 M$_{\mathrm{J}}$	&	1.36   $\pm$        0.12	&	0.72   $\pm$        0.08	&	3.96   $\pm$        0.31	&	0.75   $\pm$        0.06	&	0.75   $\pm$        0.06	&	1.5   $\pm$        0.11	\\
 $a^{\ddagger}$ 	&	A.U.	&	0.83   $\pm$    0.04	&	0.78   $\pm$    0.05	&	2.06   $\pm$    0.14	&	0.86   $\pm$    0.04	&	1.04   $\pm$    0.04	&	1.44   $\pm$    0.07	\\
 \hline
\end{tabular}}
\parbox{\hsize}{\textbf{Notes:} SOPHIE measurements taken after the instrument upgrade in June 2011 are referred as S+.\\
$(^{\dagger})$ Time at periastron for BD+550362 and HD~331093; time of possible transits for BD+450564, BD+631405, HD~124330 and HD~155193.\\
$(^{\ddagger})$ The uncertainties in the host star masses are taken into account in the uncertainties on \msini and $a$.}
\end{table*}

\setlength{\tabcolsep}{7pt}		
\begin{table*}\centering
\caption{\label{tab:BDorbit} Fitted Keplerian orbital solutions with their 1-$\sigma$ uncertainties for the four BDs having 25.05  M$_{\mathrm{J}}$ $\leq$ \msini  $\leq$ 60.27  M$_{\mathrm{J}}$. }
\small
{\def\arraystretch{1.25}
\begin{tabular}{l |c |c |c| c| c}
\hline
Name 	&	Unit	&	BD-004475	&	HD~184601	&	HD~205521	&	HD~5433	\\
\hline											
P	&	days	&	723.2   $\pm$    0.74	&	849.35   $\pm$    1.45	&	2032.32   $\pm$    2.42	&	576.6   $\pm$    1.59	\\
K 	&	m~s$^{-1}$	&	711.21   $\pm$    11.51	&	1531.86   $\pm$    18.65	&	406.91   $\pm$     5.82	&	2096.86   $\pm$   164.45	\\
$e$	&	-	&	0.39   $\pm$    0.01	&	0.49   $\pm$    0.01	&	0.17   $\pm$    0.01	&	0.81   $\pm$    0.02	\\
$\omega$  	&	($^\circ$)	&	-97.55   $\pm$    2.46	&	137.71   $\pm$    0.77	&	-139.79   $\pm$    3.05	&	76.9   $\pm$    3.53	\\
$T_p^{\dagger}$	&	BJD-2450000	&	8799.64   $\pm$     3.47	&	8480.37   $\pm$     1.14	&	8468.68   $\pm$    13.29	&	8031.95   $\pm$     0.26	\\
$\gamma_{S+}$  	&	km~s$^{-1}$	&	23.58   $\pm$       0.03	&	-63.49   $\pm$       0.02	&	-7.96   $\pm$       0.01	&	30.97   $\pm$       0.19	\\
 $\gamma_{S}$  	&	km~s$^{-1}$	&	23.58   $\pm$      0.02	&	       --         	&	        --    	&	30.97   $\pm$      0.16	\\
 \hline
$\sigma_\text{O-C, S+}$ 	&	m~s$^{-1}$	&	7.41	&	7.69	&	2.57	&	16.76	\\
$\sigma_\text{O-C, S}$ 	&	m~s$^{-1}$	&	6.13	&	    --       	&	    --     	&	9.26	\\
\hline
\msini$^{\ddagger}$	&	 M$_{\mathrm{J}}$	&	25.05   $\pm$        2.23	&	60.27   $\pm$        3.65	&	26.62   $\pm$        1.64	&	49.11   $\pm$         3.4	\\
$a^{\ddagger}$	&	A.U.	&	1.48   $\pm$    0.11	&	1.76   $\pm$    0.08	&	3.26   $\pm$    0.15	&	1.37   $\pm$    0.06	\\
 \hline
\end{tabular}}
\parbox{\hsize}{\textbf{Notes:} SOPHIE measurements taken after the instrument upgrade in June 2011 are referred as S+.\\
$(^{\dagger})$ Time at periastron\\
$(^{\ddagger})$ The uncertainties in the host star masses are taken into account while obtaining the uncertainties on \msini and $a$.}
		\end{table*}

We use the \texttt{Yorbit} software \citep{Segransan2011, Bouchy2016} to fit the Keplerian RV signal induced by a companion.  \texttt{Yorbit} uses a genetic algorithm to produce starting values for a Levenberg-Marquardt optimization, which in turn provides the priors for a Markov chain Monte Carlo (MCMC) estimation of the error bars following \citet{Diaz2014}. 

The first step of the RV data analysis is to identify significant periodic signals in the data. This is done by computing the Generalized Lomb-Scargle (GLS) periodogram algorithm of the RV measurements \citep{Zechmeister2009}. In the case of giant planet survey targets, we then estimate the false-alarm probabilities (FAPs) of the tentative signals through a bootstrap permutation of the data. The GLS periodogram is however known to fail for the signals induced by companions in highly eccentricity orbits \citep{Zechmeister2009}. For two of the stars (BD+631405 and HD~331093), we, therefore, used the \textsc{PlanetPack} software to compute Keplerian periodograms \citep{Baluev2013,PP}. We ran its \textsf{kpow} command for a range of orbital periods with a frequency step of 0.01 and an upper limit on eccentricity. The latter is needed to control the computational cost, as the closer emax approaches unity, the longer is the computational time. 

When a significant period is identified, the RVs are first fit using a single Keplerian orbital model initialized at that period. The following parameters are varied while fitting the single Keplerian model: $P$ the orbital period, $K$ the RV semi-amplitude, $e$ the orbital eccentricity, $\omega$ the orbital argument of periastron, $T_{p}$ the time of passage through pericenter, $\gamma_{S}$ and $\gamma_{S+}$ the RV offsets for SOPHIE and SOPHIE+ datasets, respectively. To obtain robust confidence intervals for the free parameters, we use 1000 MCMC iterations. 

In most cases, a single companion (planet, BD, or stellar) on a Keplerian orbit is a good description of the RV measurements. For  HD~124330 and BD+550362, however, linear and quadratic drifts are considered in addition to one Keplerian. Six targets, HD~8291, HD~25603, HD~76332, HD~187057, HD~211961, and HD~352975, have too few RV measurements with either SOPHIE or SOPHIE+ to constrain both $\gamma_\text{S}$ and $\gamma_\text{S+}$. Fitting for both parameters, therefore, produced unrealistically high values for $\gamma_\text{S+}$- $\gamma_\text{S}$: they are on the order of 100 m~s$^{-1}$ when the RV offsets between SOPHIE and SOPHIE+ measurements are known to be bounded by 50 m~s$^{-1}$ \citep{Bouchy2013,Kiefer2019}. We therefore fix  $\gamma_\text{S+}$- $\gamma_\text{S}$ to 0 when deriving Keplerian solutions for those six stars. 

\begin{figure*}
	\caption{\label{fig:GiantPlanets} Keplerian orbit of the RV variations for the six CJs are plotted here. SOPHIE and SOPHIE+ RV measurements are indicated in blue squares and red circles, respectively.} 
\includegraphics[width=0.49\hsize]{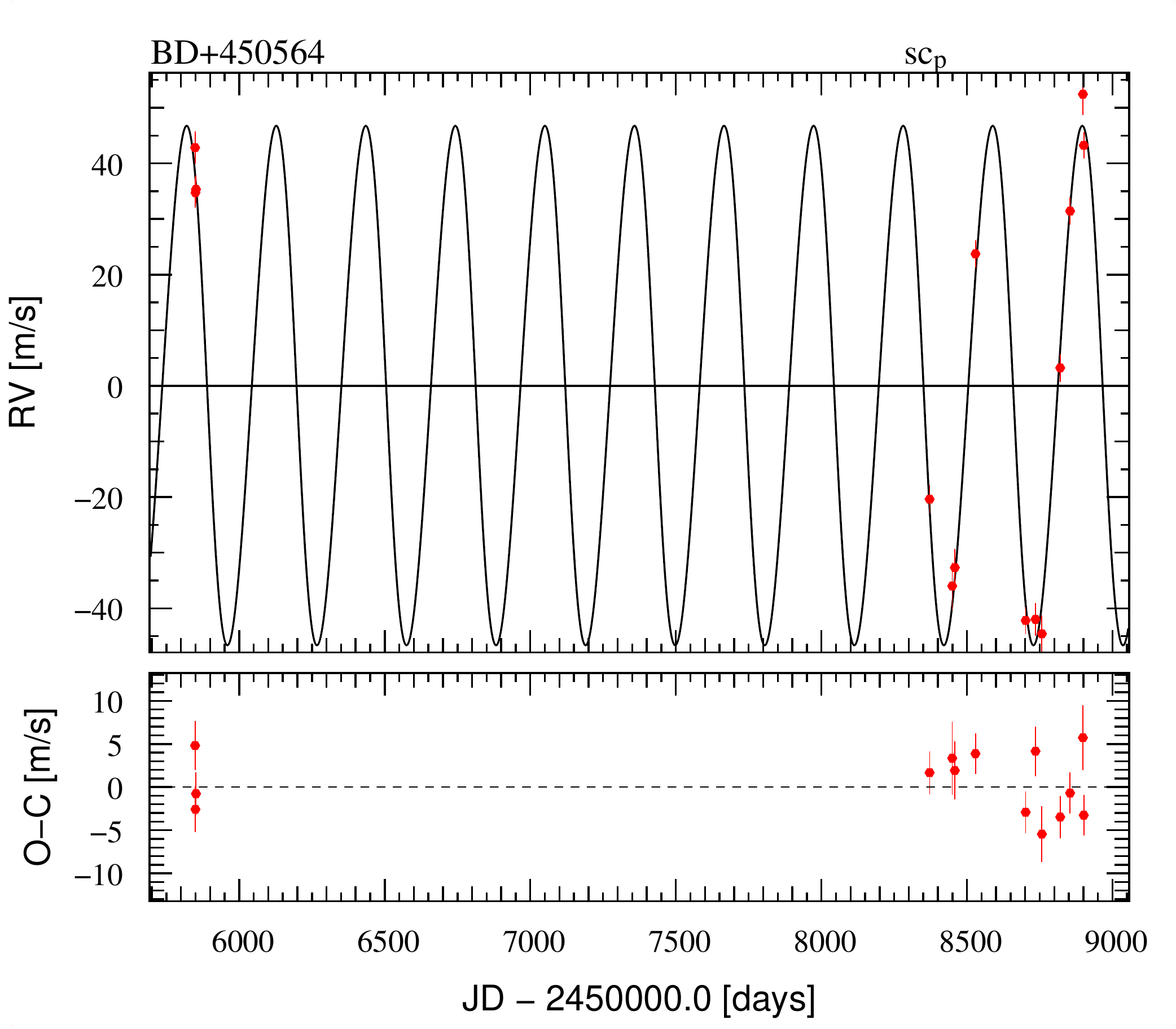} 
\includegraphics[width=0.49\hsize]{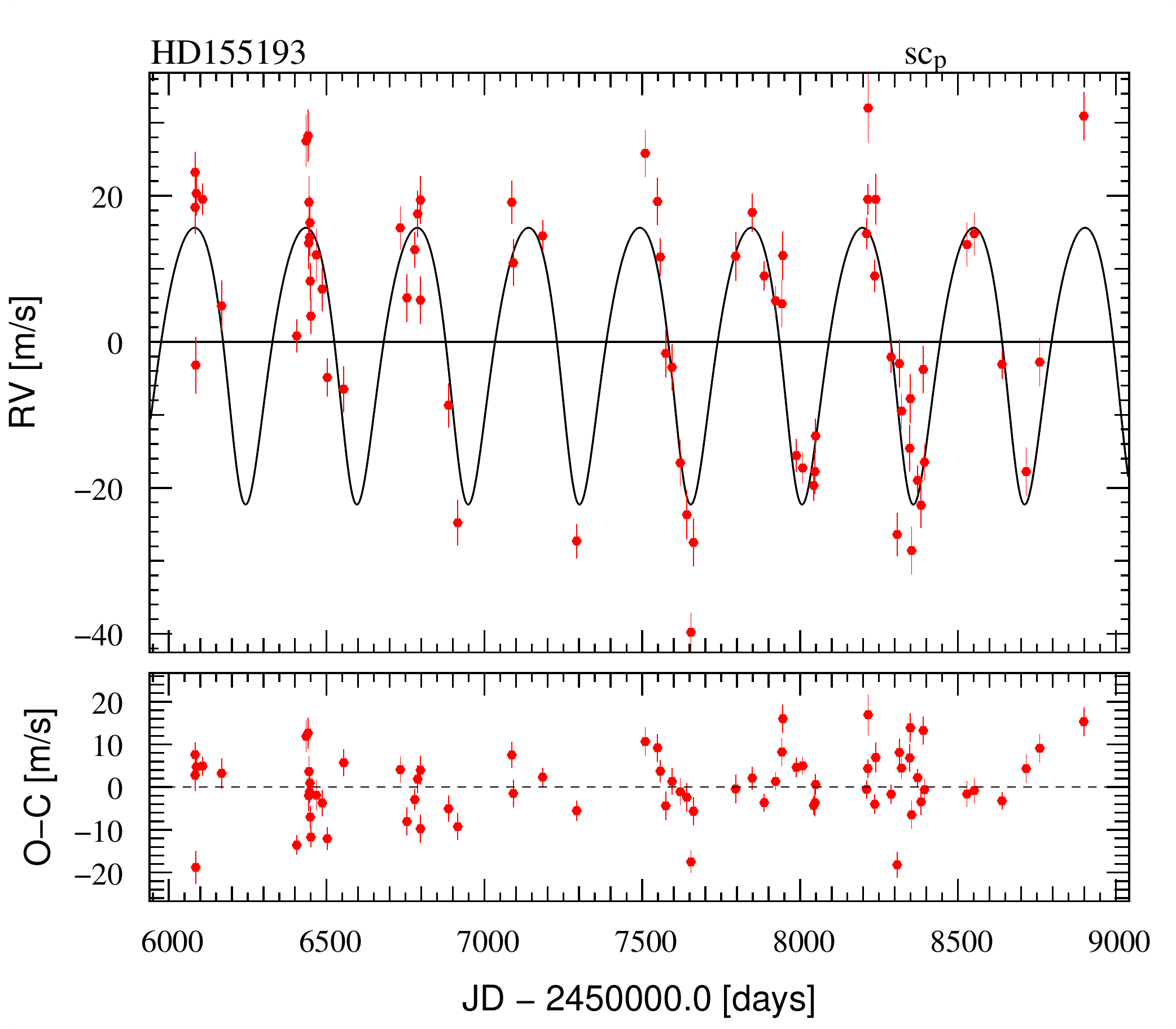}
\includegraphics[width=0.49\hsize]{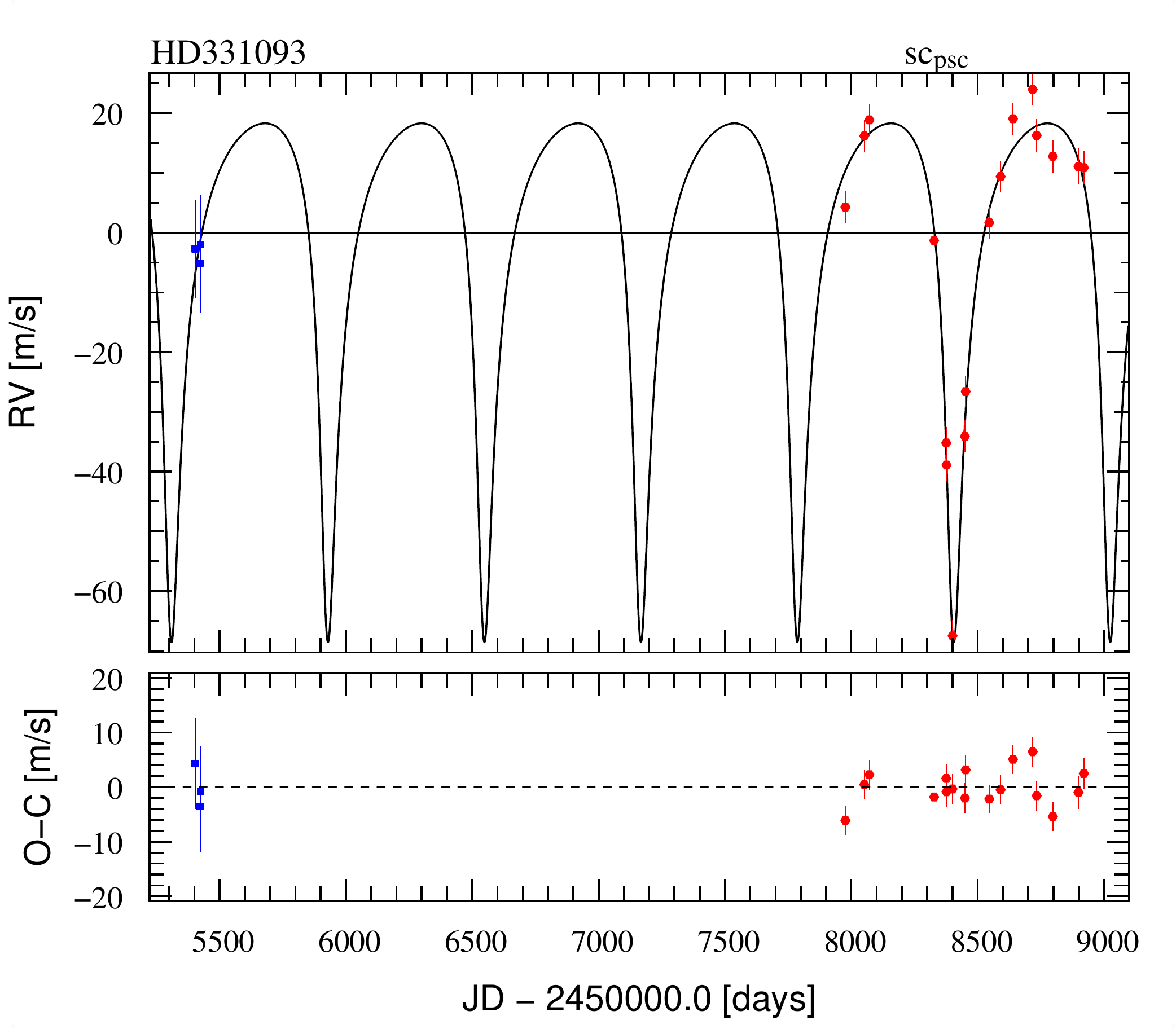}
\includegraphics[width=0.49\hsize]{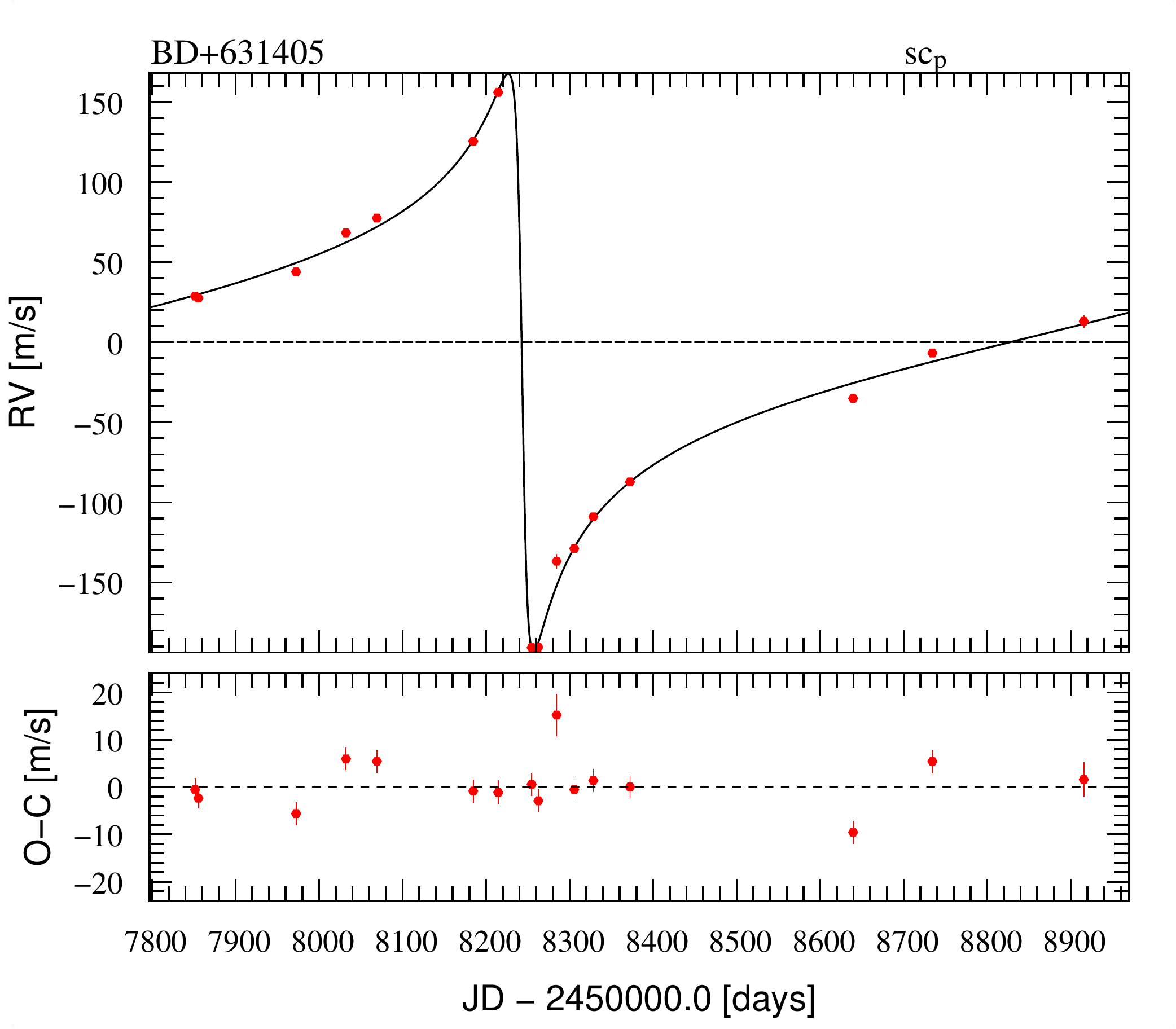}
\includegraphics[width=0.49\hsize]{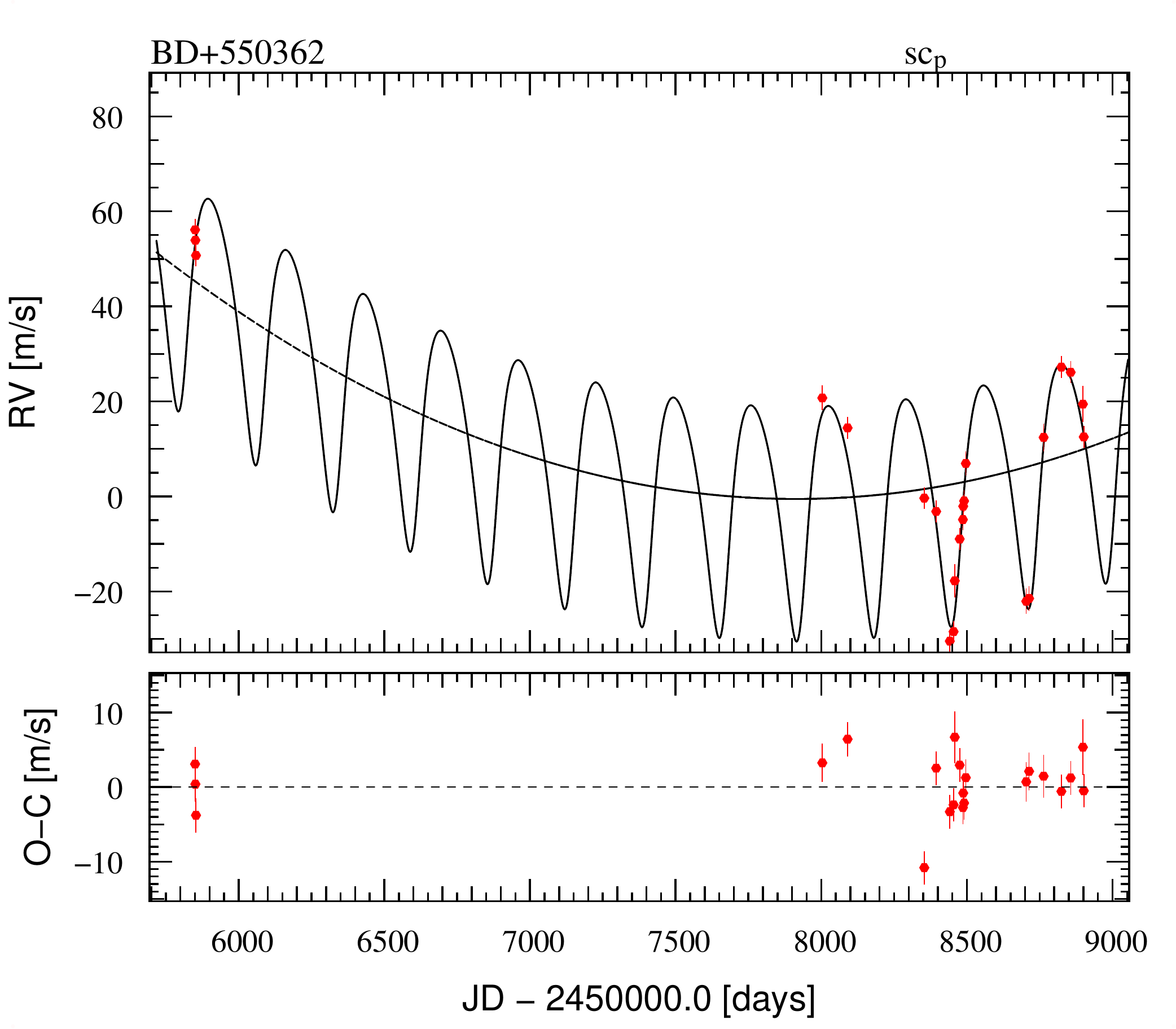}
\includegraphics[width=0.49\hsize]{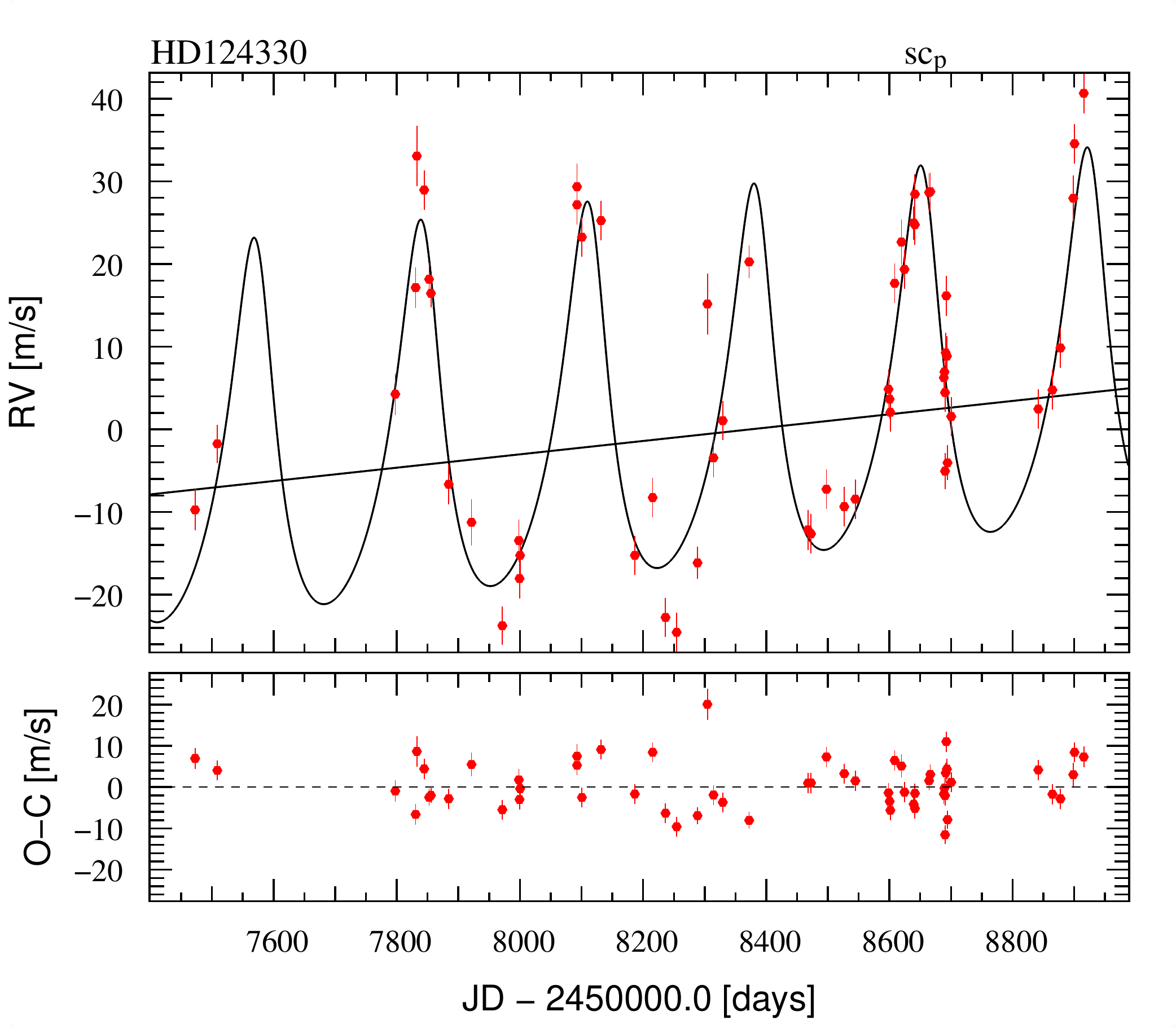}
\end{figure*}

\begin{figure*}
	\caption{\label{fig:BDorbit} Orbital solutions of the RV variations of four BDs with 25.05  M$_{\mathrm{J}}$ $\leq$ \msini  $\leq$ 60.27  M$_{\mathrm{J}}$ are depicted here. SOPHIE and SOPHIE+ RV measurements are indicated in blue squares and red circles, respectively.} 
\includegraphics[width=0.49\hsize]{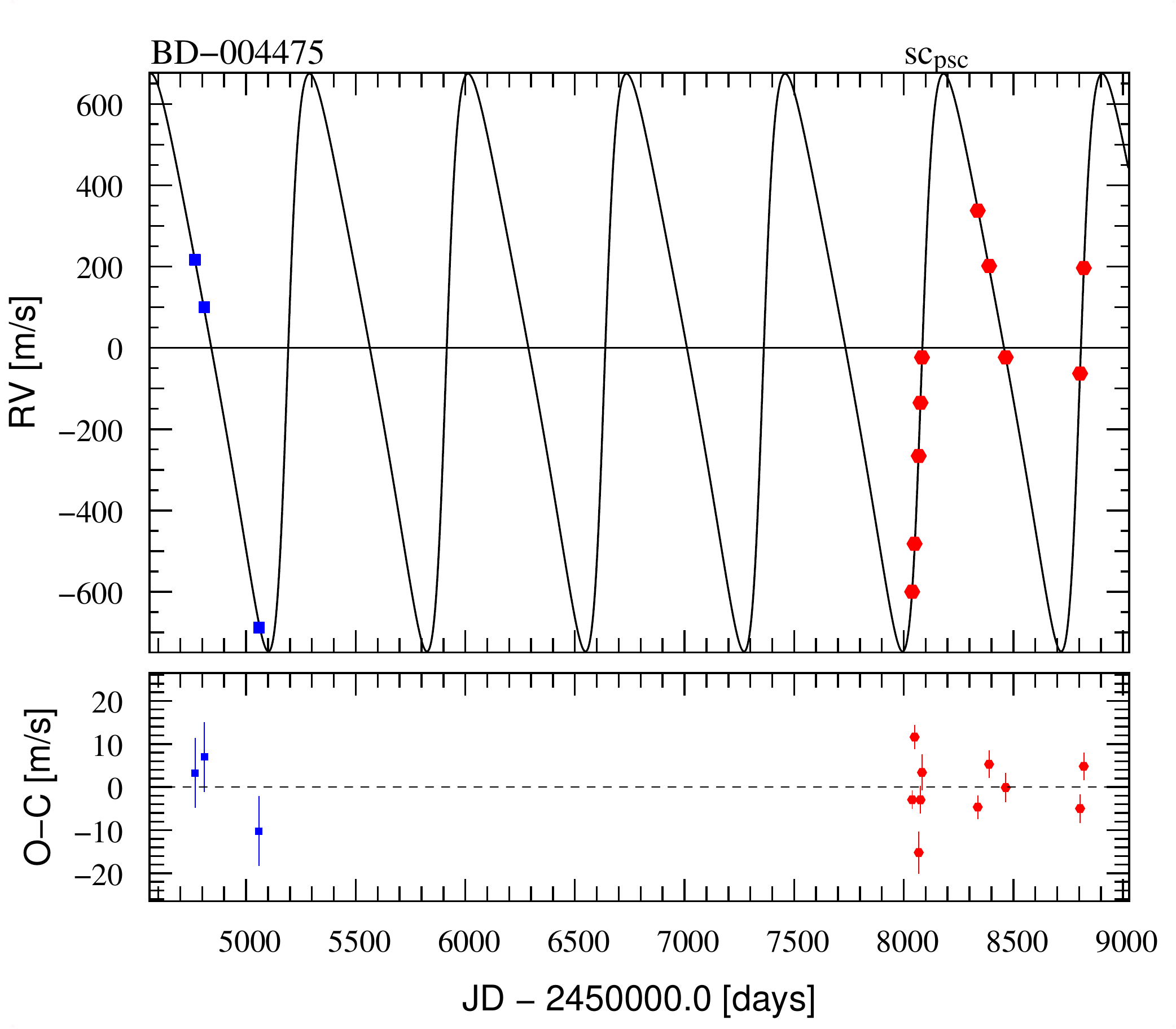}
\includegraphics[width=0.49\hsize]{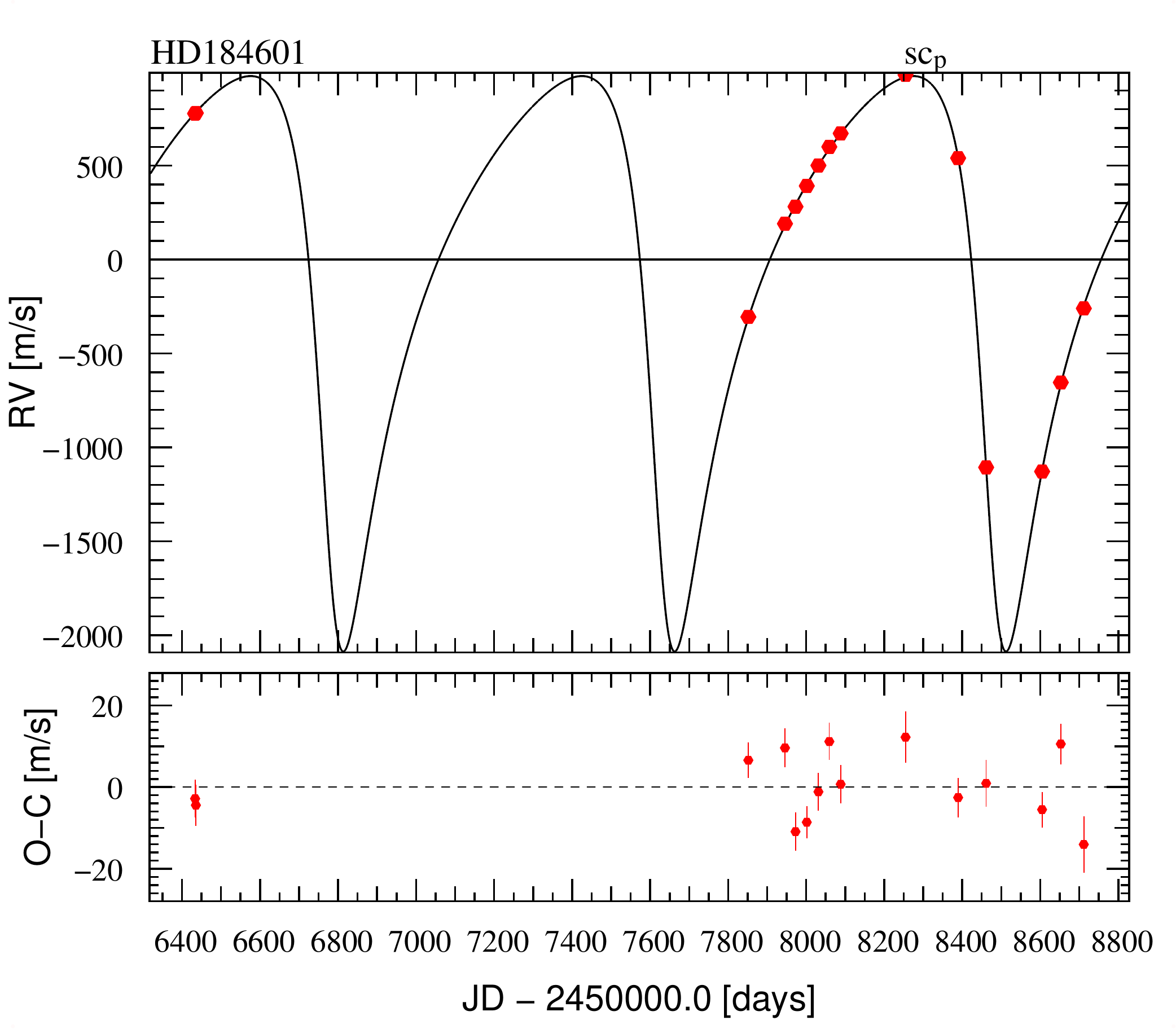}
\includegraphics[width=0.49\hsize]{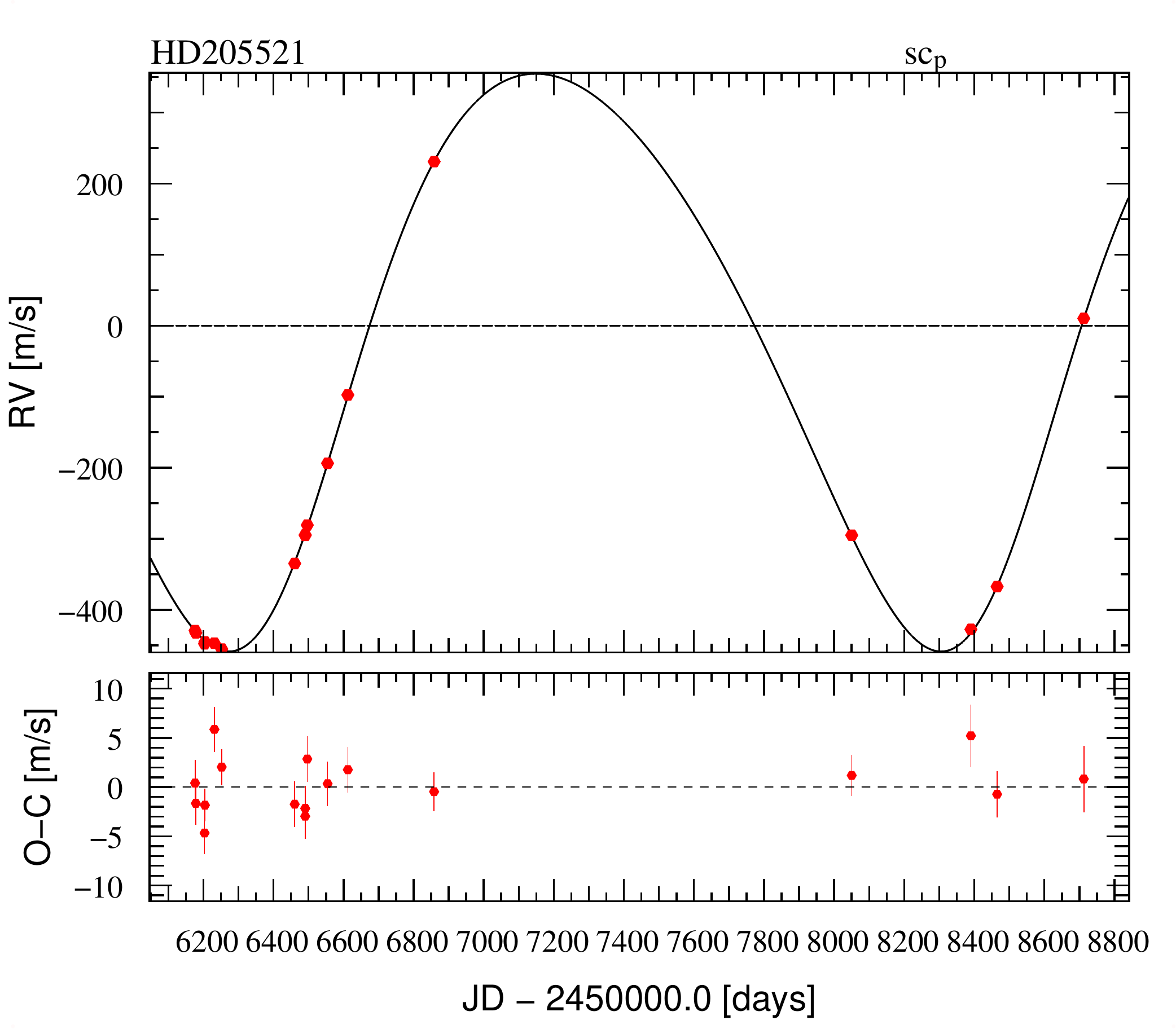}
\includegraphics[width=0.49\hsize]{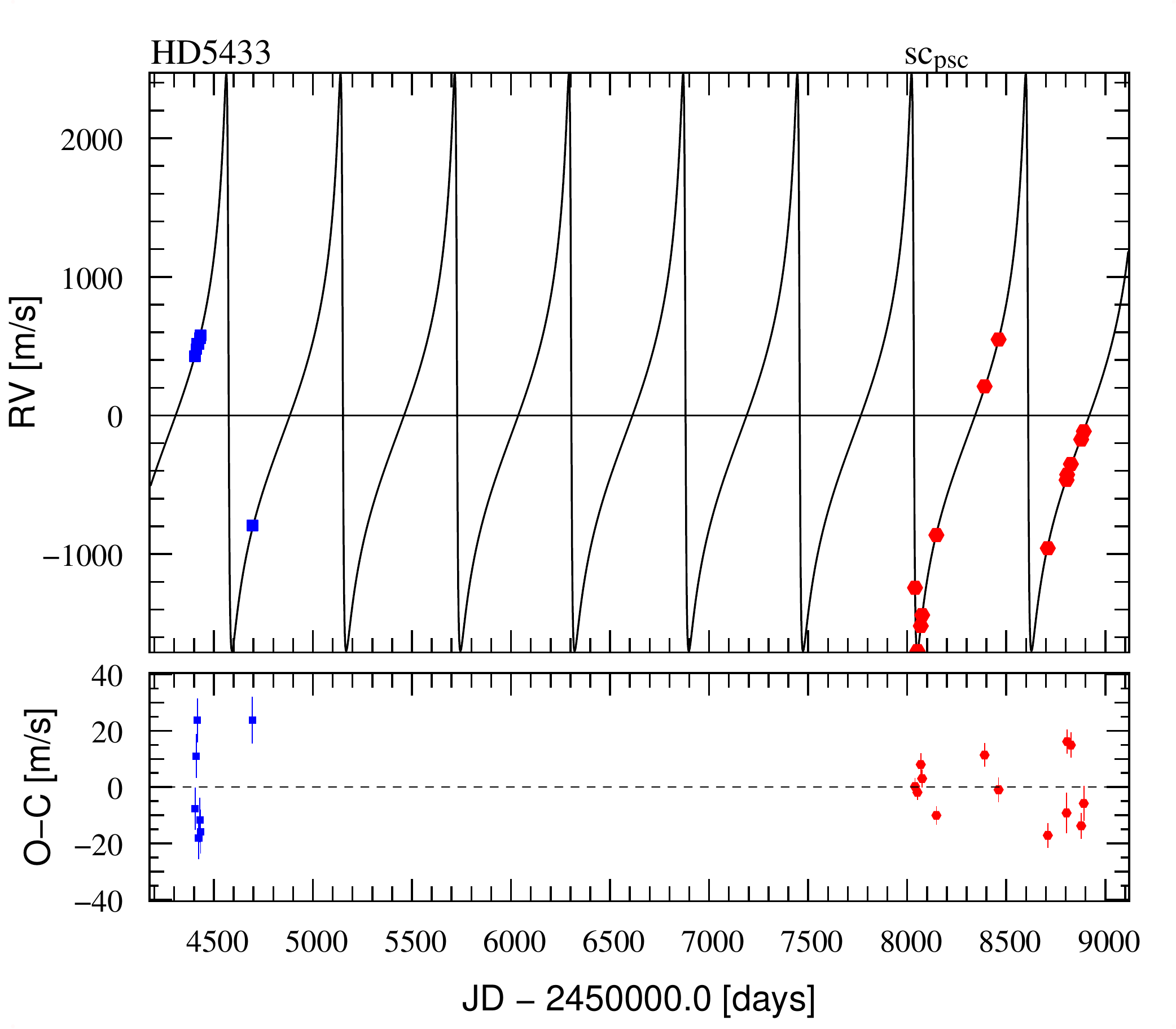}

\end{figure*}

Section~\ref{sec:results} discusses the results of the RV analysis of the newly detected objects in detail, and these objects are divided into CJs, BDs and SCs based on their minimum mass. Three tables summarize the Keplerian orbital elements as well as the \msini and $a$ derived parameters for each category. Table~\ref{tab:giantplanet} reports the orbital parameters of the six CJs, while Fig.~\ref{fig:GiantPlanets} plots their Keplerian fits as a function of time along with their RV measurements and residuals. Table~\ref{tab:BDorbit} and Fig.~\ref{fig:BDorbit} provide the same information for the four BD candidates, and Figs. \ref{fig:GiantPlanetsPC} and \ref{fig:BDorbitPC} display phase-folded plots for respectively the CJs and BDs. Finally, Table~\ref{tab:StellarC} reports the orbital parameters of the sixteen SCs and Fig.~\ref{fig:StellarC} shows their Keplerian fits and RV measurements as a function of the orbital phase. 

\section{Astrometry analysis} \label{sec:astrometry}
Astrometry can complement the RV orbital information and measure the inclination of the systems. We, therefore, used the HIPPARCOS and Gaia DR1 data to perform the astrometric analysis presented below. The true mass of the companion (either CJ, BD, or SC) is expressed by $M_\text{c}$.

\subsection{HIPPARCOS astrometry}

\subsubsection{Selecting the orbit candidates}
As the input sample for SOPHIE giant planet survey was selected from the HIPPARCOS catalog, all 26 stars listed in Table~\ref{tab:hip_params} were observed by the HIPPARCOS satellite \citep{Perryman:1997kx}. After a preliminary examination of all 26 systems, we selected 12 systems with indications of significant orbital motion \footnote{The uncertainty in the stellar mass of BD+031552 is large, which produced errors in the preliminary analysis. We, therefore, decreased its primary mass uncertainty to 0.08 $M_\sun$}. This subset contains all sources with nonstandard HIPPARCOS solution types (`1' for stochastic solutions, `5' for standard solutions, and `7' or `9'  for accelerated solutions). We list upper mass limits in Table~\ref{tab:hip_params} for those of the 26 systems where one could be derived. 

\subsubsection{Analysis of the  HIPPARCOS astrometry}
\begin{figure}
  \caption{\label{fig:orbits_3sigma_1} \emph{Top panel:} Astrometric orbit of HD~2055213$\sigma$. North is up and East is left. The solid red line shows the model orbit and the open circles mark the individual 1-dimensional HIPPARCOS measurements. \emph{Bottom panel:} O--C residuals for the normal points of the orbital solution (filled blue circles) and of the five-parameter model without companion (open squares).} 
\includegraphics[width=0.98\hsize]{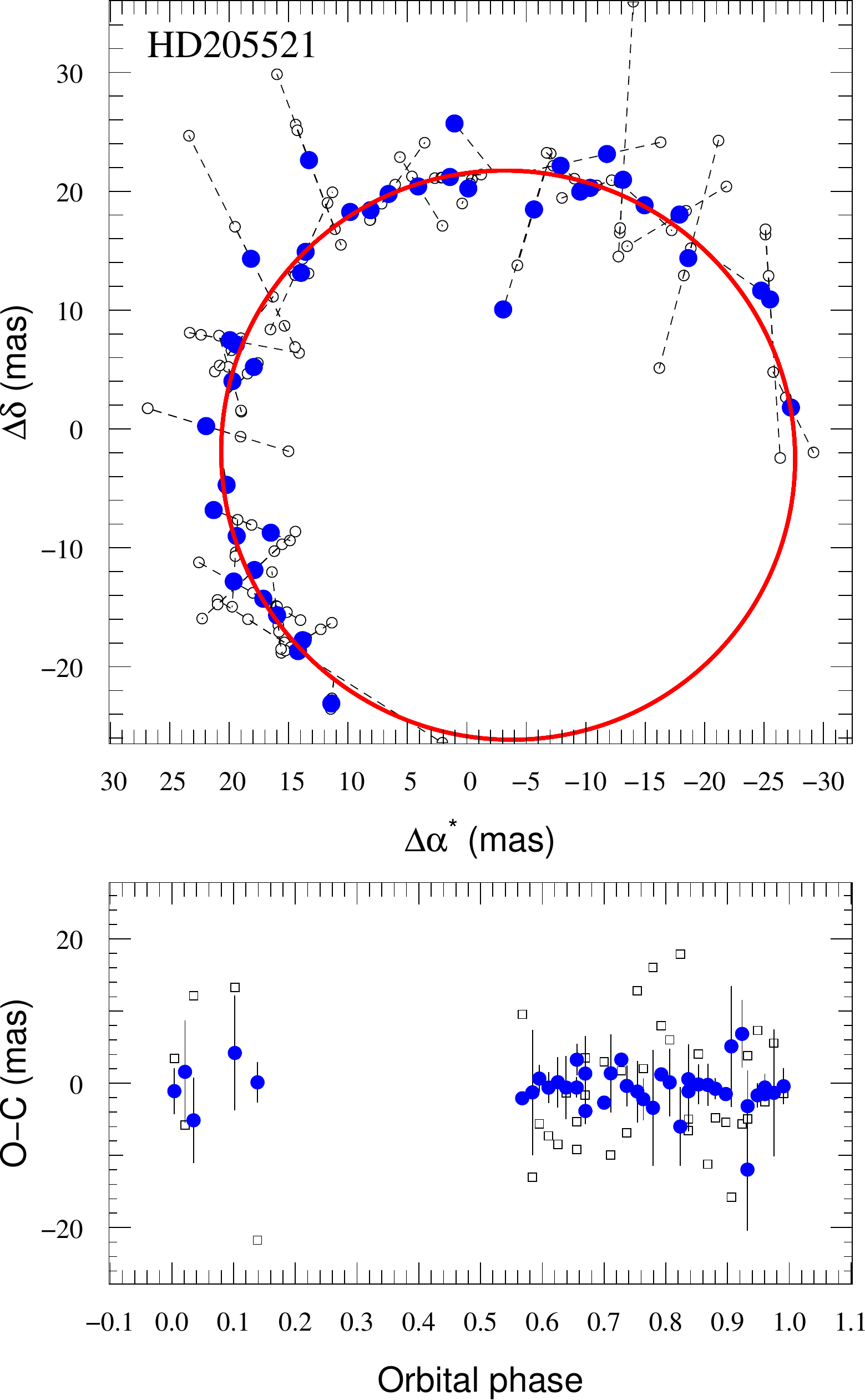} 
\end{figure}
 We analyze the Intermediate Astrometric Data (IAD) of the most recent HIPPARCOS reduction \citep{:2007kx} for signatures of orbital motion,  following \cite{Sahlmann2011} where a detailed description of the method can be found. Fixing the other orbital elements to their values from the RV orbit (Tables~\ref{tab:giantplanet}, \ref{tab:BDorbit}, and \ref{tab:StellarC}) we adjust seven free parameters to the IAD of each star: the inclination of the orbit $I_{p}$, the longitude of its ascending node $\Omega$, the parallax $\varpi$, and offsets to both coordinates ($\Delta \alpha^{\star}$, $\Delta \delta$) and both proper motion components ($\Delta \mu_{\alpha^\star}$, $\Delta \mu_{\delta}$). We search a two-dimensional grid in the two nonlinear parameters, $I_{p}$ and $\Omega$, for its global $\chi^2$-minimum. The false-alarm probability of the derived astrometric orbit was then evaluated through a permutation test employing 1000 pseudo-orbits. The uncertainties in the solution parameters were derived by Monte Carlo simulations which also propagate the uncertainties in the RV parameters. This method has a good track record in reliably detecting orbital signatures in the HIPPARCOS IAD \citep{Sahlmann2011, Sahlmann:2011lr, Diaz:2012fk, Sahlmann:2013fk3}. 
\medskip

Table~\ref{tab:hip_params} lists the target names and the basic parameters of the HIPPARCOS observations relevant for the astrometric analysis. The solution type $S_\mathrm{n}$  indicates the astrometric model adopted by the  \citep{:2007kx} reduction. The code is `5' for the standard five-parameter solution,  whereas it is `7' and `9' when the model included proper-motion derivatives of first and second order, respectively. $N_\mathrm{orb}$ represents the number of orbital periods covered by the HIPPARCOS observation time span, $N_\mathrm{Hip}$ is the number of astrometric measurements, and $\sigma_{\Lambda}$ is their median precision. Outliers in the IAD can very substantially alter the outcome of the astrometric analysis and therefore need to be clipped.

\begin{table*}
\captionsetup{width=1\textwidth}
\caption{Solution parameters determined for the significant detections in HIPPARCOS data.}
\label{tab:hip_masses} 
\centering  
\small
{\def\arraystretch{1.25}
\begin{tabular}{l r r r r r rrrrrrr} 	
\hline\hline %
Source &   $a \sin i$ & $a$ & $M_\text{c}$  & $M_\text{c}$ (3-$\sigma$)& $a_{\mathrm{rel}}$ &$O-C_5$ &$O-C_7$ & $\chi^2_{7,red}$ & Null prob. & Significance \\  
                              &(mas)     & (mas) & ($M_\sun$) & ($M_\sun$) &(mas) & (mas) & (mas) & & (\%)  & (\%) \\  
\hline 
\multicolumn{11}{c}{$>$3-$\sigma$ detections}\\
\hline
 HD~205521 &   1.77 & $ 24.2^{+ 1.4}_{-1.4}$ & $ 0.48^{+ 0.04}_{-0.04}$ & $(0.36,0.61)$ & 80.2 & 9.96 & 4.87& 0.58& 2.6e-40 &100.0  \\ 
 HD~26596 &   3.45 & $ 9.0^{+ 0.7}_{-0.8}$ & $ 0.32^{+ 0.03}_{-0.03}$ & $(0.23,0.43)$ & 38.3 & 7.05 & 4.79& 0.47& 1.3e-15 &100.0  \\ 
 HD~98451 &   2.95 & $ 9.4^{+ 1.1}_{-1.1}$ & $ 0.27^{+ 0.03}_{-0.03}$ & $(0.19,0.38)$ & 40.0 & 6.25 & 5.09& 0.92& 9.7e-08 &100.0  \\ 
 \hline 
\multicolumn{11}{c}{2--3-$\sigma$ detections}\\
\hline
HD~238135 &   16.73 & $ 18.9^{+ 1.2}_{-1.2}$ & $ 0.37^{+ 0.03}_{-0.03}$ & $(0.30,0.47)$ & 60.8 & 10.38 & 9.00& 0.91& 1.3e-13 &98.8 \vspace{1mm} \\ 
 HD~25603 &   11.33 & $ 16.1^{+ 3.0}_{-3.0}$ & $ 0.47^{+ 0.10}_{-0.10}$ & $(0.29,0.90)$ & 51.3 & 4.88 & 4.42& 0.92& 2.4e-01 &97.7 \vspace{1mm} \\ 
 HD~30311 &   16.86 & $ 37.3^{+ 3.8}_{-3.9}$ & $ 0.61^{+ 0.09}_{-0.09}$ & $(0.36,0.90)$ & 106.1 & 5.38 & 2.54& 1.61& 6.0e-19 &99.4  \\ 
 HD~76332 &   12.72 & $ 25.2^{+ 4.6}_{-4.6}$ & $ 0.47^{+ 0.10}_{-0.11}$ & $(0.24,0.84)$ & 78.0 & 4.61 & 4.11& 0.98& 6.4e-04 &99.7 \vspace{1mm} \\  
 \hline

BD+031552  &   9.30 & $ 9.8^{+ 0.6}_{-0.6}$ & $ 0.16^{+ 0.02}_{-0.01}$ & $(0.12,0.22)$ & 67.1 & 7.11 & 5.75& 1.02& 3.9e-08 &75.4  \\ 
 
 \hline
\end{tabular} }
\end{table*}

Even when the astrometric data detected no orbital signal (i.e., the derived significance is low), we can set an upper limit on the companion mass by determining the minimum astrometric signal $a_\mathrm{min}$ that would be detectable for the individual target. When the data cover at least one complete orbit, \cite{Sahlmann2011, Sahlmann:2011lr} showed that an astrometric signal-to-noise of $\mathrm{S/N} \gtrsim 6-7$ is required to obtain a detection at the $3\,\sigma$ level, where $\mathrm{S/N}=a\, \sqrt{N_\mathrm{Hip}} / \sigma_{\Lambda}$ and $a$ is the semi-major axis of the detected orbit. We conservatively use a S/N-limit of 8 to derive an upper limit on the astrometric signal of 
\begin{equation}\label{ }
a_\mathrm{min} = 8 \frac{ \sigma_{\Lambda} } { \sqrt{N_\mathrm{Hip}}} \left( 1-e^2 \right),
\end{equation}
where the factor $1-e^2$ accounts for the most unfavorable orientation of $I_{p}=90\degr$ and $\Omega=90\degr$, in which the astrometric signal is given by the semi-minor axis of the orbit. The last column in Table~\ref{tab:hip_params} lists the corresponding companion mass upper limit for systems with $N_\mathrm{orb}\gtrsim1$.

We detect the astrometric orbit of 7 sources with a significance of at least $2\,\sigma$, as determined by the permutation test. Those are listed in Table~\ref{tab:hip_masses}, where we also include BD+031552 even though its permutation significance is low. However, the null probability is small, the decrease in residual amplitude is large, the derived orbit looks visually good. Given that this source also has an accelerated HIP solution type, we indicate this solution as viable despite the failure of the permutation test. The 8 sources with orbit determinations then include all 7 sources with nonstandard HIPPARCOS solution types.

\medskip

Table~\ref{tab:hip_ppm} lists the updated parallaxes, proper motion,  inclination, and ascending node of the orbits of those 8 stars. The 8 orbits detected from HIPPARCOS astrometry include no CJ, and we astrometrically detect the orbit of one of the BD candidates, HD~205521 (Fig.~\ref{fig:orbits_3sigma_1}). Its low inclination however shifts the companion mass from the BD to the SC domain, as discussed in Sect.~\ref{sec:lowmassstars}. The remaining astrometric orbits are for SCs are plotted in  Fig.~\ref{fig:orbits_3sigma_2}.

\subsection{Gaia astrometry}\label{sec:gaia}
\setlength{\tabcolsep}{2pt}	
\begin{table*}
\captionsetup{width=1\textwidth}
 
\centering  
\small
\caption{Detected orbits with Gaia astrometric excess noise $\varepsilon_\text{DR1}$$>$$0.85$\,mas. All sources belong to the primary dataset of Gaia DR1. When the 3-$\sigma$ upper-limit on the orbital inclination is above 89.5$^\circ$, the constraint on the companion true mass is handled as an upper-limit.}
\label{tab:Gaia_results}
{\def\arraystretch{1.3}
\begin{tabular}{lcccccccccc|c}

\\
\hline\hline
Source &  $a\sin i$ & $\epsilon$ & \multicolumn{2}{c}{$a_\text{phot}$} & \multicolumn{2}{c}{$I_c$} & \multicolumn{2}{c}{$M_\text{c,true}$}  & \multicolumn{2}{c|}{$\Delta V$} & MCMC \\
			&  (mas)  & (mas)  	& (mas)  & (mas)  		& \multicolumn{2}{c}{($^\circ$)} &  \multicolumn{2}{c}{(M$_\odot$)}  & &   & Acceptance \\
	 		&				&		& 1-$\sigma$ & 3-$\sigma$ &		1-$\sigma$ & 3-$\sigma$		&	1-$\sigma$ & 3-$\sigma$  &  1-$\sigma$ & 3-$\sigma$& rate \\
\hline
\multicolumn{12}{l}{Strong constraint on inclination and mass
($I_{\text{c,max},3\sigma}$$<$$89.5^\circ$)}\\
\hline
HD~205521 &  1.42 & 2.597 & $10.23_{-1.64}^{+2.69}$ & $(6.58,24.40)$ & $7.996_{-1.702}^{+1.628}$ & $(2.97,12.47)$ &$0.20^{+0.06}_{-0.04}$ & $(0.13,0.68)$ & $8.32_{-0.86}^{+1.25}$ & $(3.27,11.13)$ & 0.2610  \\
HD~98451 & 5.68 & 2.286 & $6.92_{-0.78}^{+1.25}$ & $(5.44,15.06)$ & $55.30_{-11.48}^{+12.91}$ & $(21.08,89.50)$ &$0.16^{+0.03}_{-0.02}$ & $(0.12,0.43)$ & $8.53_{-1.22}^{+0.77}$ & $(4.57,10.11)$ & 0.2819  \\
HD~238135  & 11.15 & 3.017 & $14.00_{-1.16}^{+1.21}$ & $(11.10,17.48)$ & $53.84_{-8.30}^{+8.99}$ & $(34.73,87.58)$ &$0.45^{+0.09}_{-0.06}$ & $(0.32,0.65)$ & $4.510_{-0.949}^{+0.736}$ & $(2.513,6.421)$ & 0.2959  \\
HD~8291 &  4.691 & 1.976 & $6.287_{-0.813}^{+1.341}$ & $(4.608,16.70)$ & $48.42_{-10.72}^{+10.88}$ & $(15.43,88.39)$ &$0.14^{+0.04}_{-0.02}$ & $(0.10,0.47)$ & $9.714_{-1.314}^{+0.793}$ & $(4.699,11.43)$ & 0.2729  \\
\hline
\multicolumn{12}{l}{Lower limit on inclination and upper limit on mass ($I_{\text{c,max},3\sigma}$$>$$89.5^\circ$)} \\
\hline
HD~352975 & 2.381 & 0.9703 & $<$2.838  & $<$4.271 & $>$56.44 & $>$29.83 & $<$0.29 & $<$0.56& $>$5.529 & $>$2.769 & 0.2759  \\
HD~25603 &  9.749 & 2.225 & $<$12.43  & $<$16.67 & $>$51.23 & $>$32.04 & $<$0.42 & $<$0.70 & $>$5.689 & $>$2.654 & 0.2318  \\
HD~26596 &  2.965 & 1.158 & $<$3.389  & $<$4.352 & $>$61.19 & $>$44.52 & $<$0.14 & $<$0.19 & $>$10.10 & $>$8.442 & 0.2604  \\
HD~140208 &  3.776 & 1.454 & $<$4.186  & $<$5.605 & $>$64.11 & $>$38.07 & $<$0.42 & $<$0.68& $>$6.266 & $>$3.159 & 0.2162  \\
\hline
\end{tabular}}
\end{table*}
Recent studies have shown that information from the first Gaia Data Release (DR1) can constrain the mass of RV-detected nontransiting companions \citep{Kiefer2019, Mugrauer2020}. The \verb+GASTON+ code uses the astrometric excess noise, as published in DR1, to determine the amplitude of the astrometric motion of the host star of a known RV-detected companion \citep{Kiefer2019S,Kiefer2019, Kiefer2021}. This measures, or at least constrains from below, the inclination of the orbit, and thus resolves inclination ambiguity on the true mass. Starting from the RV-derived parameters of the companion, the parallax of the star and an estimation of its mass as priors, \verb+GASTON+ uses a Markov-Chain Monte-Carlo algorithm to explore the space of astrometric orbits compatible with the measured astrometric excess noise, hereafter $\varepsilon_\text{DR1}$, and hence to constrain the possible inclinations. It accounts for the systematics in the Gaia DR1 catalog and distinguishes between stars belonging to the TGAS (Tycho-Gaia Astrometric Solution) and the secondary subset of the DR1 catalog \citep{Lindegren2016, Kiefer2021}. For the targets in the TGAS dataset, the inputs to the 5-parameter fit include a 24-years older Tycho-2 or HIPPARCOS-2 astrometric point in addition to 14 months of Gaia measurements, leading to an improved proper motion accuracy. Targets in the secondary dataset only have Gaia observations, and the 14 months time baseline available for DR1 was insufficient to reliably decouple parallax from proper motion. DR1 only published positions (and magnitudes) for those stars and we used their Gaia DR2 \citep{Gaia2018} absolute parallax as a replacement for the missing DR1 parallax, with a conservatively allocated 10\% relative uncertainty. Lastly, \verb+GASTON+  accounts for the effect of the companion's light on the position of the photocenter, by modeling the luminosity of both components from their mass (Appendix A in Kiefer et  al. (2021)). Here we additionally impose that the secondary contributes no more than 10\% of the total luminosity since the systems we study are single-lined spectroscopic binaries.

Out of 26 targets studied in the present paper, 25 targets figure in the DR1 catalog, with HD~211961 as the sole exception. Table~\ref{tab:list_planets} summarizes for those 25 stars the DR1 inputs to \verb+GASTON+. For targets in the TGAS (respectively secondary) dataset of the DR1, the reported value of $\varepsilon_\text{DR1}$ is considered significant beyond noise, if above 0.85\,mas (resp. 1.2\,mas) \citep{Kiefer2021}. Targets with $\varepsilon_\text{DR1}$ above these thresholds can lead to, but does not guarantee, a true mass measurement from an allowed orbit inclination range that excludes 90$^\circ$. Targets with $\varepsilon_\text{DR1}$ below the thresholds cannot be distinguished from noise, so can only provide an upper-limit (lower-limit) on the mass (inclination). 

Tables~\ref{tab:Gaia_results}, \ref{tab:moreGAIA1}, and \ref{tab:moreGAIA2} summarize the results of this analysis. The results for duplicate sources should be considered with care, as the Gaia observations for a single source can be mistakenly divided between two Gaia "sources" with different IDs \citep{Gaia2016, Lindegren2016}. The astrometric solution in such cases is thus based on an incomplete astrometric data series.

Similar to what happens with the HIPPARCOS astrometry, the 8 orbits detected with the Gaia astrometry include 7 SCs, a single BD candidate, HD~205521, which Gaia DR1 likewise demonstrates is actually an SC (Section~\ref{sec:lowmassstars}), and no CJs. The remaining detected orbits are detected for the SCs. As discussed at length in Sect.~\ref{sec:results}, the RV-orbit of the HD\,76332 system is incompatible with its observed  $\varepsilon_\text{DR1}$ of 2.19\,mas: with the orbital elements of the RV orbit of for HD\,76332 as prior, all \verb+GASTON+ Monte Carlo draws produce an astrometric excess noise significantly below 2.19\,mas.  

Finally, a word of caution. \verb+GASTON+ does not take into account both companions of multiple systems at once, since this would require a totally unpractical number of MCMC steps -- and thus a ridiculous time --  to converge. We thus examine multiple companions with \verb+GASTON+, one-by-one, which means that the derived mass of each is overestimated when both contribute significantly to the astrometric signal. In the HD\,114762 system \citep{Kiefer2019} the wide orbit companion HD\,114762\,B with a$_B \,\sim$130\,au has little effect on the astrometry of the host star compared to the closer companion with P$\sim$84\,days, but a larger mass for the companion would lead to larger perturbations on the star's orbit.

\subsection{Comparison HIPPARCOS and Gaia astrometry results}
We detect astrometric orbits for eight stars in each of the HIPPARCOS and Gaia astrometric analyses, with five being common to both. The results of the two astrometric analyses agree within 3-$\sigma$ for the true mass of these five companions, with the exceptions of HD~26596. As discussed in Sect.~\ref{sec:lowmassstars}, the Gaia analysis finds a significantly lower mass for HD~26596~B.

\section{Results} \label{sec:results}
In this section, we discuss each SOPHIE RV detection in the light of the astrometric data from HIPPARCOS and Gaia, and we classify them as CJs, BDs, or SCs based on their minimum mass (or true mass for some companions). After this RV and astrometric analysis, we have six CJs, three BD candidates, and 17 low-mass stellar companions. We also present a detailed analysis of the RV signal from HD~204277, previously presented as a tentative planetary candidate and which we conclude is due to magnetic activity.

\subsection{HD~204277 : activity rather than a planet}\label{sec:notaplanet}
\begin{figure}[h]
\centering
    \includegraphics[width=0.98\hsize]{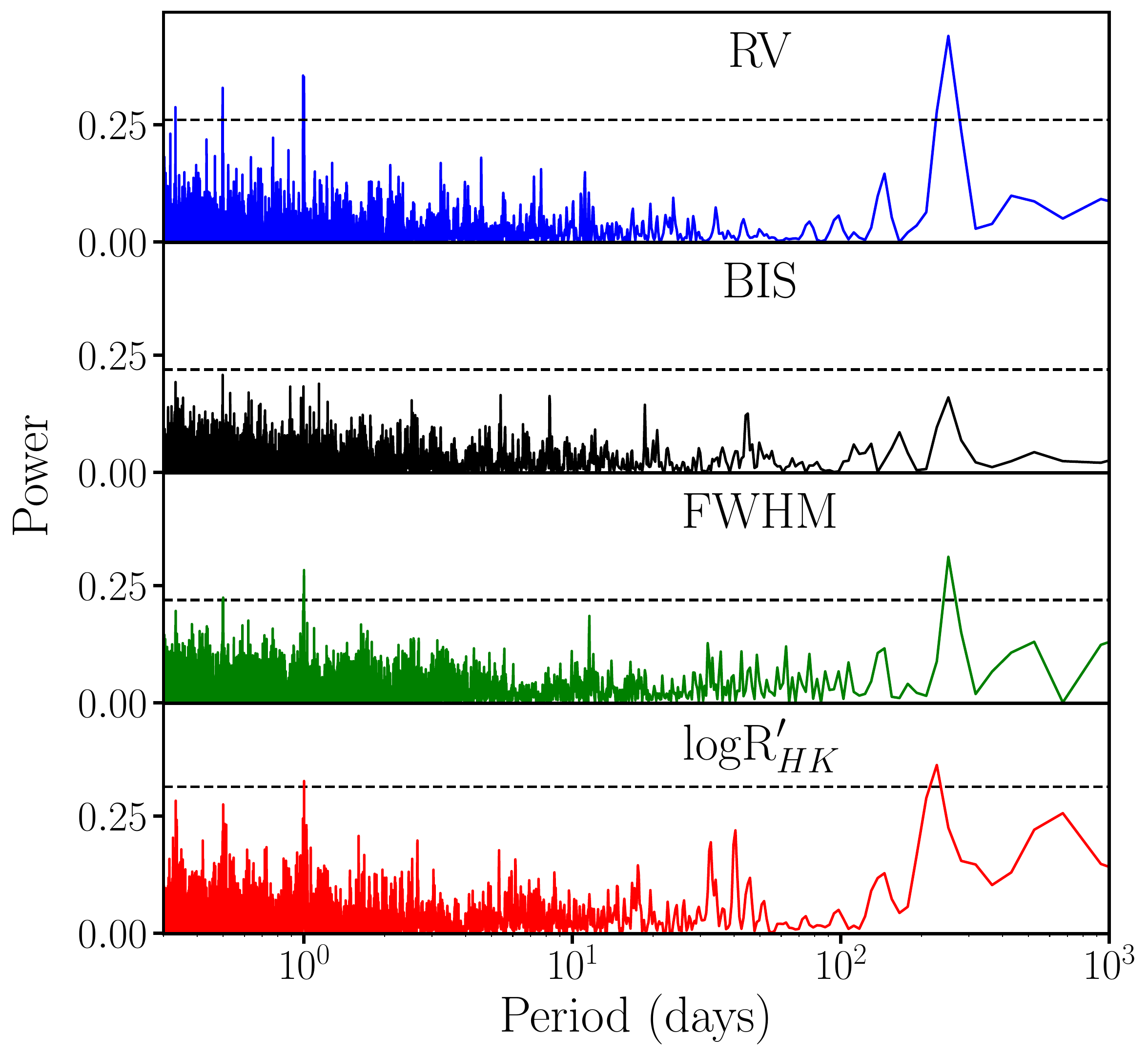}
    \caption{Lomb-Scargle periodogram of the observed radial velocities, BIS, FWHM and $\textrm{logR}'_{\textrm{HK}}$ for HD~204277. The black dashed line represents the 0.1\% False Alarm Probability.}
    \label{fig:hd204277}
\end{figure}
 HD~204277 is a V=6.7 magnitude F8V star located at a distance of 33 parsecs from the Sun and has a stellar mass of 1.14 $\pm$ 0.08 M$_{\odot}$.  It is an active star which has $\textrm{logR'}_{\textrm{HK}}$ = -4.50 $\pm$ 0.11 and therefore has $\sigma_{a} \simeq 21.8$ m~s$^{-1}$. \citet{Butler2017}, using the HIRES spectrograph, listed this system as an SRC\footnote{Signal Requiring Confirmation by additional data before rising to classification as planet candidate} with a period of 30 days. The Generalized Lomb-Scargle periodogram of the SOPHIE RV measurements for HD~204277 instead shows a strong signal around a 250~days period. However, the periodograms of both BIS and FWHM peak at the same period as the RV periodogram, and so does the periodogram of $\textrm{logR'}_{\textrm{HK}}$ (Fig.~\ref{fig:hd204277}). This suggests starspots, magnetic cycles, and other stellar activities as the likely origin of the RV signal, and there is thus a low probability that it is due to a planet. We conclude that the observed RV variations are likely due to the stellar activity, and will discuss the nature of this signal in a dedicated forthcoming paper.

\subsection{Cool Jupiters}
\subsubsection{BD+450564}
BD+450564 is a K1 type star with mass 0.81 $\pm$ 0.07 M$_{\odot}$ and $\textrm{logR'}_{\textrm{HK}}$ of -4.98 $\pm$ 0.11. It was only observed after the SOPHIE upgrade and the final dataset has 14 SOPHIE+ RV measurements. Its GLS periodogram (Fig.~\ref{fig:Gls_CJs}) has a clear peak around 300 days, with a FAP below 0.1\%. The Keplerian fit of the 14 RVs has a period of 307.8 $\pm$ 1.5 days and a semi-amplitude of 47.7 $\pm$ 2.8 m~s$^{-1}$, and it corresponds to a minimum mass of {\msini}~=~1.36 $\pm$ 0.12 M$_{\mathrm{J}}$. The orbit of BD+450564~b is quasi-circular, with $e$ = 0.12 $\pm$ 0.06. Due to the low (and possibly null) eccentricity, we fit for the time of possible transit instead of the time of periastron. The 3.3 m~s$^{-1}$ dispersion of the RV residuals is consistent with the typical uncertainty of the RVs. 

No significant astrometric orbit is detected in HIPPARCOS data. The Gaia astrometric excess noise $\varepsilon \,=\,0.25$ leads to a 3-$\sigma$ upper limit on the true mass of the giant planet companion, that is M$_{c}$ $<$ 31.4 M$_{\mathrm{J}}$.

\subsubsection{BD+550362}
BD+550362 is a K3 type star with mass of 0.91 $\pm$ 0.10 M$_{\odot}$ and $\textrm{logR'}_{\textrm{HK}}$ = -5.11 $\pm$ 0.12. Its 22 RV measurements were all obtained after the SOPHIE upgrade. The GLS periodogram (Fig.~\ref{fig:Gls_CJs}) shows two peaks around 260 and 2000 days with FAPs below 0.1\% and 1\%, respectively. We tested three models for this target: k1 (one-planet Keplerian), k1d2 (one-planet Keplerian and a quadratic drift), and k2 (two-planet Keplerian). An F-test comparison of the k1 and k1d2 models reveals that the k1d2 model gives a better fit than the k1 model with confidence of 96\% (F-value = 24.3). The F-test comparison of the k1d2 and k2 models finds no significant improvement for the k2 model (F-value=3.8, Probability= 84.5\%). The k2 model fits the shorter period planet well, but its orbital parameters for the longer period planet are very uncertain because our data do not sample its period well. We, therefore, adopt the k1d2 model for our final fit.

The k1d2 Keplerian model fit of the RVs gives a 25.1 $\pm$ 1.7 m~s$^{-1}$ semi-amplitude, and a minimum mass of \msini = 0.72 $\pm$ 0.08 M$_{\mathrm{J}}$. The orbit of BD+550362 b has a 265.6 $\pm$ 1.0 days period and a significant eccentricity of $e$ = 0.27 $\pm$ 0.06. The 3.8 m~s$^{-1}$, dispersion of its residuals is consistent with the noise of the RV measurements. The RV drift is well fitted by a quadratic trend and we obtained quadratic (d2) and linear (d1) drift coefficients of 1.4365 $\pm$ 0.0005 m~s$^{-1}$yr$^{-2}$ and 1.76 $\pm$ 0.04 m~s$^{-1}$yr$^{-1}$. If one forces a circular orbit for this additional companion, the k2 fit converges to a CJ with an orbital period of at least 3600 days and a mass of at least 2.1 M$_{\mathrm{J}}$. Other orbital solutions, with longer periods, higher eccentricity, and higher masses, are however equally consistent with the measurements. 

Additional RV data will therefore be needed to reveal the nature of this outer companion. The 3-$\sigma$ upper limit on the mass of BD+550362~b from the \verb+GASTON+ astrometric analysis of its host star, M$_{c}$ $<$ 72.45 M$_{\mathrm{J}}$, is not sufficiently tight to nail down its planetary nature. 

\subsubsection{BD+631405}
BD+631405 is a K0 type star with a mass of 0.82 $\pm$ 0.08 M$_{\odot}$ and has low magnetic activity  ($\textrm{logR'}_{\textrm{HK}}$ = -4.93 $\pm$ 0.13). Its 16 RVs were all obtained after the SOPHIE spectrograph upgrade. Figure~\ref{fig:Gls_CJse} shows both a GLS periodogram and a Keplerian periodogram of those velocities. We produced the Keplerian periodogram using the \textsf{kpow} command of the \textsc{PlanetPack} package for orbital periods between 3--1500 days with a frequency step of 0.01 days$^{-1}$ and an  emax=0.91 upper limit on the eccentricity. The GLS periodogram has no obvious peak, but its Keplerian counterpart shows a clear one around 1200 days. 

A one-planet Keplerian fit of the RVs finds a highly eccentric orbit (e = 0.88 $\pm$ 0.02) of period 1198.5 $\pm$ 60.8 days. The semi-amplitude, 186.0 $\pm$ 14.9  m~s$^{-1}$ of the Keplerian fit corresponds to a planet with minimum mass, \msini = 3.96 $\pm$ 0.31 M$_{\mathrm{J}}$. The poor sampling of the fast RV changes at periastron and the incomplete coverage of the orbital period are the main reasons for the large uncertainty on the orbital period and $T_p$ (the time of possible transit is reported in Table~\ref{tab:giantplanet} rather than the time of periastron). The  4.6 m~s$^{-1}$ dispersion of the residuals is consistent with the noise of the RV measurements. From the $\varepsilon \,=\,0.39$ in the Gaia DR1 astrometry, we set a 3-$\sigma$ upper limit on the true mass of the giant planet of M$_{c}$ < 40.23 M$_{\mathrm{J}}$. 

\subsubsection{HD~124330}
HD~124330 is a G4IV type star which has a mass of 1.15 $\pm$ 0.08 M$_{\odot}$ and $\textrm{logR'}_{\textrm{HK}}$ = -5.27 $\pm$ 0.12. Its 58 RV measurements were all secured after the SOPHIE upgrade. The GLS periodogram shows a clear peak around 270 days (Fig.~\ref{fig:Gls_CJs}). The k1d1 (one-planet Keplerian and a linear drift) model fit of the radial velocities gives a significant eccentricity, e = 0.34 $\pm$ 0.05 and an orbital period of 270.7 $\pm$ 1.2 days. This Keplerian has a semi-amplitude of 22.8 $\pm$ 1.2 m~s$^{-1}$ and corresponds to a minimum planet mass of \msini = 0.75 $\pm$ 0.06 M$_{\mathrm{J}}$ . The 5.4 m~s$^{-1}$ dispersion of the residuals is slightly larger than the noise of the RV measurements. 

The 2.95 $\pm$ 0.07 m~s$^{-1}$yr$^{-1}$ linear drift of the RVs indicates that this system contains an additional outer companion. For a circular orbit and the 3900 days minimum period defined by the extent of the measurements, its mass is at least 0.85 M$_{\mathrm{J}}$, but longer orbital periods and large eccentricities are obviously also compatible with the RV measurements. Neither HIPPARCOS nor Gaia detects any significant astrometric orbit and we set a 3-$\sigma$ upper limit on the true mass of the giant planet of M$_{c}$ < 53.76 M$_{\mathrm{J}}$. 

\subsubsection{HD~155193}
HD~155193 is a magnetically quiet F8IV type star with 
$\textrm{logR'}_{\textrm{HK}}$~=~-5.14~$\pm$~0.25 and a 1.22 $\pm$ 0.08 M$_{\odot}$ mass. Its 73 RV measurements were all acquired after the SOPHIE upgrade. 
Their GLS periodogram (Fig.~\ref{fig:Gls_CJs}) 
peaks around 350 days. The one-planet Keplerian fit to the radial velocities has a 352.6~$\pm$~2.6~days orbital period, a 19.5~$\pm$~1.4 m~s$^{-1}$ semi-amplitude, a modestly significant eccentricity of $e$ = 0.21 $\pm$ 0.08, and it corresponds to a  \msini = 0.75 $\pm$ 0.06  M$_{\mathrm{J}}$ minimum mass. Although the Keplerian fit appears robust, the nearly one year orbital period induces a large phase gap in our coverage of the planetary orbit, which limits the precision on orbital parameters such as $\omega$ and $T_p$. Table~\ref{tab:giantplanet} reports the time of possible transit instead of the time of periastron. The 7.05 m~s$^{-1}$ dispersion of the residuals is slightly above the estimated accuracy of the measurements. The $\varepsilon \,=\,0.55$ Gaia astrometric excess noise translates into a M$_{c}$ < 63.21 M$_{\mathrm{J}}$ 3-$\sigma$ \verb+GASTON+ upper limit for the true mass of the companion .

\subsubsection{HD~331093}
HD~331093 is a magnetically quiet K0 type star with $\textrm{logR'}_{\textrm{HK}}$ = -5.10 $\pm$ 0.13 and a 1.03 $\pm$0.08 M$_{\odot}$ mass. It was observed before and after the SOPHIE upgrade for a total of 20~RV measurements. Figure~\ref{fig:Gls_CJse} shows both their GLS periodogram and a Keplerian periodogram, which we produced using the \textsf{kpow} command of the \textsc{PlanetPack} package for orbital periods between 3 and 1400 days, with a  0.01 days$^{-1}$ frequency step, and an  emax=0.7 upper limit on the eccentricity. Similar to the also eccentric BD+631405 system, the GLS periodogram is featureless but the Keplerian periodogram shows a clear peak, here around 600 days. 

The Keplerian fit of the radial velocities gives a 621.6~$\pm$~16.1 days orbital period, a high eccentricity of $e$~=~0.59 $\pm$ 0.03, a 43.6~$\pm$~2.2 m~s$^{-1}$ semi-amplitude, and it corresponds to a  \msini = 1.5 $\pm$ 0.1 M$_{J}$ minimum mass. The  dispersion of the residuals,  
3.3 and 3.2 m~s$^{-1}$ for respectively SOPHIE+ and SOPHIE, is consistent with the typical uncertainty on the RV measurements. 
The Gaia astrometric excess noise only provides a very loose 3-$\sigma$ upper limit on the true mass of the companion, M$_{c}$~<~270.5~M$_{\mathrm{J}}$, and therefore contains no additional insight on the nature of HD~331093~b. 

\subsection{Brown dwarfs}
The BDs presented in this section are ``BD candidates'', as the outcome of their astrometric analysis is compatible with a substellar mass but does not demonstrate one.

\subsubsection{BD-004475}
BD-004475 is a G0 star with a 0.81 $\pm$ 0.10 M$_{\odot}$ estimated mass. It was observed both before and after the SOPHIE upgrade, for a total of 13 radial velocities. The Keplerian fit finds a 723\,days eccentric orbit ($e$=0.39) and corresponds to a \msini~=~25-M$_\text{J}$ BD candidate companion with a 1.48\,AU semi-major axis. The dispersion of the residuals of the Keplerian fit is 6.19 m/s. We detect no significant astrometric motion in the HIPPARCOS data, and the 0.72~mas astrometric excess noise measured by Gaia translates into a 125 ~M$_{\mathrm{J}}$ ($\sim$ 0.12 M$_{\odot}$) upper limit on the true mass of the companion. BD-004475~b can therefore still be either a BD or an M-dwarf star.

\subsubsection{HD~184601}
The 15 RVs of HD~184601 were all obtained after the SOPHIE upgrade. It is a G0 type star with a 0.95 $\pm$ 0.07 M$_{\odot}$ mass. HD184601 b has an orbital period of 849 days and an eccentricity of 0.49, and its minimum mass is 60.27~$\pm$ 2.15 ~M$_{\mathrm{J}}$. The dispersion of the residual of the keplerian fit to the RVs is 7.69 m~s$^{-1}$, and compatible with their measurement noise. Neither HIPPARCOS nor Gaia detect any significant astrometric motion, and the upper limit on the true mass is very loose, 276~M$_{\mathrm{J}}$ ($\sim$ 0.2 M$_{\odot}$).

\subsubsection{HD~5433}
HD~5433 is a G5 star with a mass of 0.98 $\pm$  0.07  M$_{\odot}$. Its 20 RV measurements identify a companion with an orbital period of 576 days, the highest eccentricity (e = 0.81) among the BD candidates presented in this paper, and a minimum mass of 49~$\pm$ 3.4 ~M$_{\mathrm{J}}$. The astrometric analysis of the Gaia DR1 data only sets a loose upper limit on its true mass of 236  M$_{\mathrm{J}}$ ($\sim$ 0.23 M$_{\odot}$), which leaves the true nature of HD~5433~b  undetermined.

\subsection{Low-mass stars}\label{sec:lowmassstars}
Table~\ref{tab:StellarC} and Figure~\ref{fig:StellarC} present the Keplerian orbits of the 17 stellar companions with masses (or \msini) above 75  M$_{\mathrm{J}}$ (or equivalently 0.072 M$_{\odot}$). All of these except for HD~8291 are detected for the first time. Their orbital periods range from 30 days to 4198 days and their eccentricities from 0.03 to 0.65, and we discuss a few of the more interesting ones below.
\medskip

\textbf{HD~205521} is a G5 type star and has a mass of 1.10 $\pm$ 0.082 M$_{\odot}$. The Keplerian orbit has an orbital period of 2032.32 days, an eccentricity of 0.17, and a semi-amplitude of 406.91 $\pm$ 5.82 m~s$^{-1}$. The 2.57~m/s dispersion of its residuals is compatible with the measurement uncertainties of the RVs. The minimum mass of the companion is \msini = 26.62 $\pm$ 1.64  M$_{\mathrm{J}}$ and firmly into BD candidate territory, but our analyses of the HIPPARCOS and Gaia astrometric data both find that the orbit is close to face-on and that the companion is actually a star. The astrometric orbit is firmly detected in the HIPPARCOS time series, with a $\sim$24\,mas semi-major axis of the photocenter orbit around the center of mass. The 2.6\,mas  Gaia DR1 astrometric excess noise points towards a smaller value of $\sim$10$^{+3}_{-2}$\,mas (1-$\sigma$ confidence interval) but is compatible with the HIPPARCOS estimate at the 3-$\sigma$ level. The orbit of the companion is thus within 3 to 12$^\circ$ of face-on, and its true mass is within the 0.13 to 0.7\,M$_\odot$ range. While initially classified as a BD candidate from its \msini, it is a low-mass star.
\medskip

\textbf{BD+031552} is a K5 type star with an uncertain stellar mass. Its substellar companion has a 879~days orbital period, an eccentricity of 0.47, and a minimum mass of 133.2~$\pm$~49.0 ~M$_{\mathrm{J}}$. The analysis of the HIPPARCOS time series of  BD+031552 detects its astrometric orbit with low significance but sets a 3-$\sigma$ upper limit of 0.22\,M$_\odot$ on the mass of the companion. The upper limit from the Gaia DR1 astrometric excess noise is looser and therefore not informative.
\medskip

\textbf{HD~162735} has a stellar companion with an orbital period of 4197 days,  an eccentricity of 0.65, and a minimum mass of \msini = 227.5 $\pm$ 13.1  M$_{\mathrm{J}}$. Neither HIPPARCOS nor Gaia detect any significant astrometric motion, and the 3-$\sigma$ upper limit on the true mass of the companion from the Gaia DR1 excess noise is a loose 0.73 M$_{\odot}$. HD~162375 B has the highest eccentricity and the longest orbital period of the 17 low-mass stellar companions presented in this work.

\subsubsection{Updated parameters for HD~8291b}
\textbf{HD~8291} is a well-known wide binary system with a low-mass (0.073 M$_{\odot}$) companion at a 2222~AU projected distance \citep[][and references therein]{Baron2015}.  \citet{dSantos2017} first reported the detection of a closer-in  \msini~=~124.6~$\pm$~2.1~M$_{\mathrm{J}}$  companion  on an eccentric orbit (e = 0.680 $\pm$ 0.009) of 1852.3~$\pm$~3.2~days orbital period using HARPS and SOPHIE data (Table~3 in \citet{dSantos2017}). Adding 15~new SOPHIE and SOPHIE+ RV data to the previous data, we refine this detection to an orbital period of 1862.53~$\pm$~2.9~days, an eccentricity of 0.632~$\pm$~0.001, and \msini~=~99.48~$\pm$~5.86~M$_{\mathrm{J}}$. The poorly covered periastron in \citet{dSantos2017} led to an overestimated eccentricity and therefore a generally less accurate Keplerian orbit.

This 5-year period M-dwarf companion induces a motion of the  star with semi-major axis $>$4.6\,mas, which is indeed detected in Gaia DR1 as a 1.98\,mas astrometric excess noise. To 1-$\sigma$ confidence \verb+GASTON+ finds $i_c$=49$\pm$12$^\circ$  and a companion mass of 142$^{+37}_{-21}$~M$_{\mathrm{J}}$ ($\sim$ 0.14 M$_{\odot}$), but the 3-$\sigma$ confidence region includes edge-on orbits.

\subsubsection{Incompatible orbits}
The companion of \textbf{HD~76332} has an orbital period of 2489 days, an eccentricity of 0.14, and \msini of 216.01 $\pm$ 12.46  M$_{\mathrm{J}}$. The semi-major axis of the reflex orbit is at least 13\,mas, and thus expected to be well detected by both HIPPARCOS and Gaia. They indeed both detect significant nonlinear motion even though their measurements do not span a full orbit, but cannot agree on a common true mass for this companion. The HIPPARCOS time series detect the reflex orbital motion with high significance, and find a $a$=25$\pm$5\,mas semi-major axis and a 3-$\sigma$ range of 0.24--0.84\, M$_{\odot}$ for the companion mass. Masses above $\sim$0.6\,M$_{\odot}$ would give rise to an unobserved second peak in the CCFs of this system and can therefore be excluded. The Gaia DR1 astrometric excess noise of 2.2\,mas is well below the minimum expected semi-major axis. This is qualitatively expected from Gaia DR1 only covering $\sim$1/6 of the orbit (HD\,76332 is not part of the TGAS dataset), but none of the \verb+GASTON+ simulations,  whatever the mass of their companion, could produce a Gaia DR1 astrometric excess noise above $\varepsilon$$=$$1.55$\,mas. We note that HD\,76332 has a duplicate source and that duplication may be the sign that something went wrong during the reduction for this specific source. For non-TGAS sources, the \verb+GASTON+ simulations fit the proper motion through the orbital motion observed with Gaia and we suspect that the best explanation for this discrepancy is an inaccurate proper motion fit in the DR1. If not fitting for the proper motion, \verb+GASTON+ leads indeed to $I_c$ within 40--70$^\circ$. We adopt the HIPPARCOS result as the best estimation of HD\,76332\,B's true mass.

\textbf{HD~26596} has a stellar companion in an eccentric orbit (e=0.4) at 890 days and a minimum mass of 121 $\pm$ 7  M$_{\mathrm{J}}$. To 3$\sigma$ confidence, \verb+GASTON+ derives an upper limit of $<$ 0.19 M$_{\odot}$ for the true mass and the HIPPARCOS analysis derives a range of 0.23-0.43 M$_{\odot}$. This source again has a duplicate in the DR1 database, and we suspect that its Gaia astrometric excess noise is underestimated. We therefore again prefer the HIPPARCOS confidence range for the mass of this companion to its likely underestimated \verb+GASTON+ upper limit.

\section{Discussion and conclusions} \label{sec:conclusion}
\begin{figure}[h]
\centering
    \includegraphics[width=0.99\hsize]{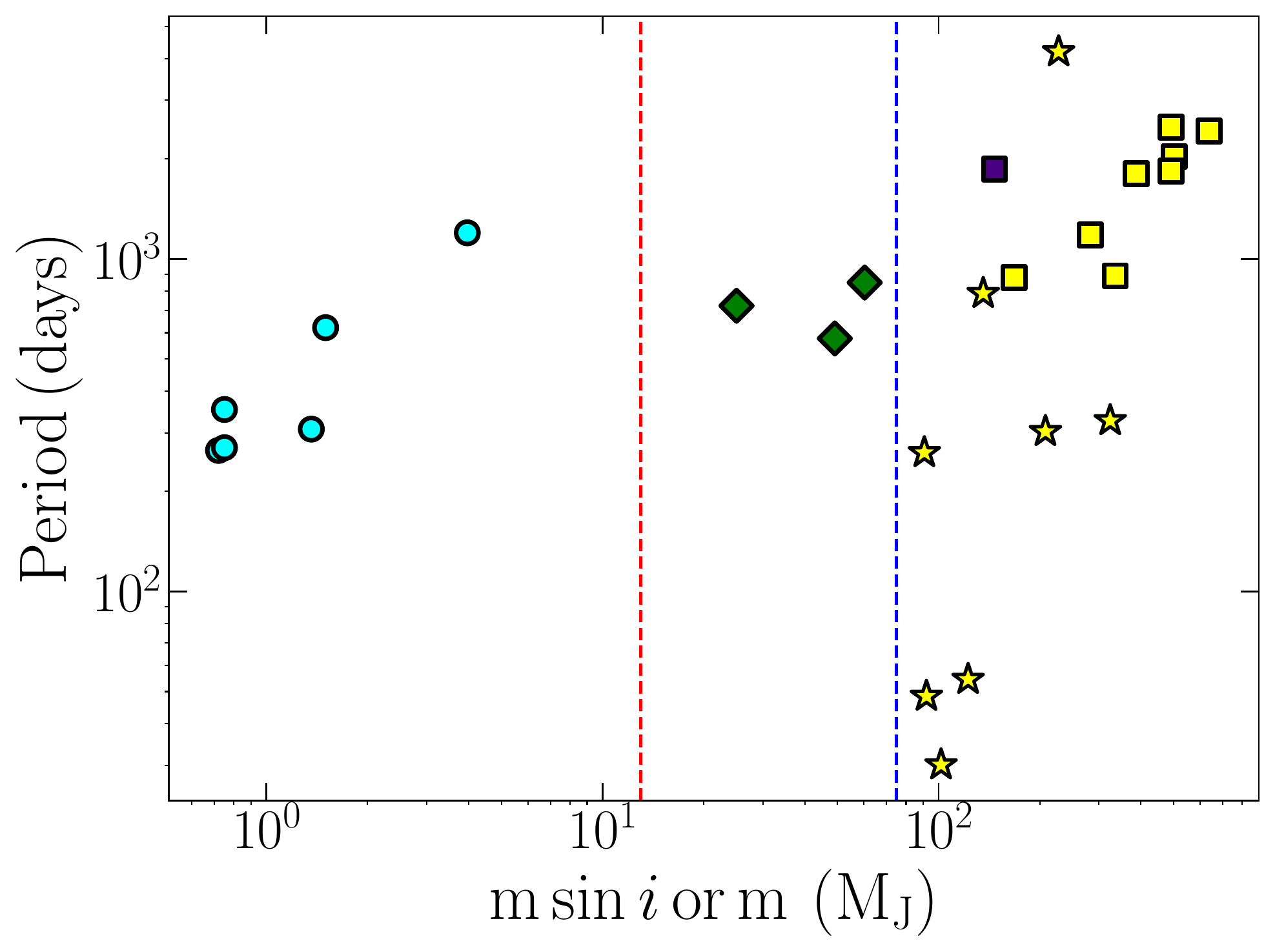}

    \caption{Period - mass distribution: The CJs and BDs candidates are shown as cyan circles and green diamonds, respectively. The SCs with \msini measurement are represented by yellow stars, while the SCs with true masses from HIPPARCOS data are represented by yellow squares. The indigo square represents HD~8291 B with its true mass inferred from the Gaia data. The red (at 13  M$_{\mathrm{J}}$) and blue (at 75  M$_{\mathrm{J}}$) dashed lines separates BDs from CJs and SCs, respectively.}
    \label{fig:periodmassall}
\end{figure}

This paper reports the discovery of 6 CJs, 3 BD candidates, and 16 SCs with the SOPHIE spectrograph at OHP. We also present updated orbital parameters for the low-mass star HD~8291 B. We analyzed the HIPPARCOS and Gaia astrometry to constrain the inclinations of the orbits and hence, the true masses of the companions.  Figure~\ref{fig:periodmassall} shows the period-mass distribution of these 26~companions.

\begin{figure}
\centering
    \includegraphics[width=0.99\hsize]{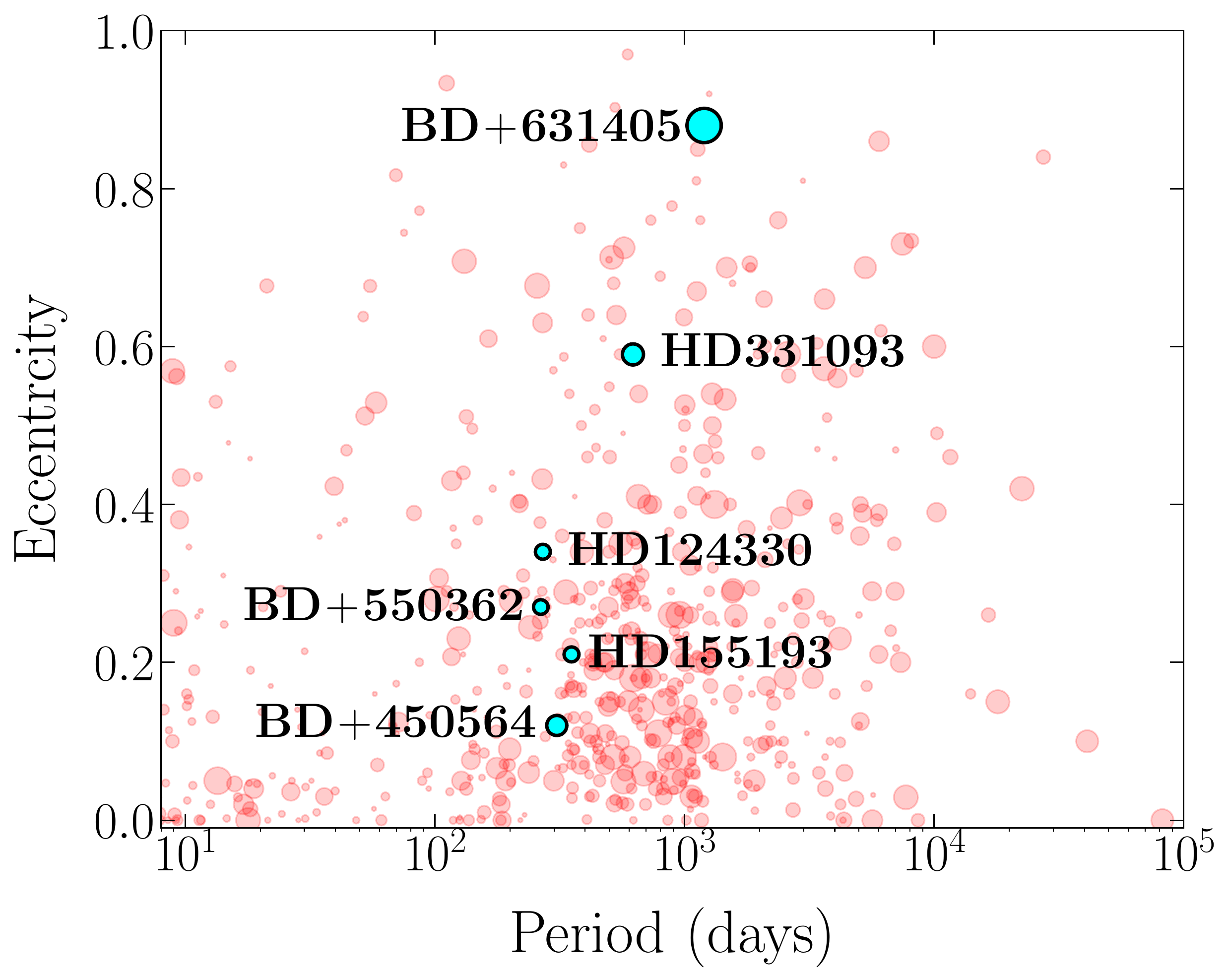}

    \caption{Period - eccentricity distribution of giant planets: The new CJs presented in this paper are shown as cyan solid circles, while the light red circles  represents all giant planets with  0.3  M$_{\mathrm{J}}$ $\leq$ $m$ (or \msini) $\leq$ 13  M$_{\mathrm{J}}$  in the Extrasolar Planets Encyclopaedia,. The size of the circles is proportional to their mass (or minimum mass), with this proportion increased by a factor of 5 for the cyan circles to emphasize the newly detected CJs.}
    \label{fig:periodecccj}
\end{figure}

The newly detected giant planets have periods that range from 266 to 1198 days, and minimum masses between 0.72 and 3.96 M$_{\mathrm{J}}$. All 6 can therefore be classified as CJs. They span a wide range in eccentricity, from the possibly circular orbit of BD+450564 b ($e$~=~0.12~$\pm$~0.06) to the highly eccentric one of BD+631405 b ($e$~=~0.88~$\pm$~0.02. BD+631405~b is the fifth most eccentric known exoplanet, after HD~20782b \citep{Jones2006,Toole2009}, HD~80606 b \citep{Naef2001,Moutou2009}, HD~7449 A b \citep{Dumusque2011}, and HD~4113 A b \citep{Tamuz2008}, and is today the most massive of these 5 highly eccentric giant planets.  Figure~\ref{fig:periodecccj} shows the period-eccentricity distribution of all known giant planets (0.3  M$_{\mathrm{J}}$ $\leq$ $m$ or \msini $\leq$ 13  M$_{\mathrm{J}}$). Its red circles are taken from The Extrasolar Planets  Encyclopaedia\footnote{\url{www.exoplanet.eu}} \citep{Schneider2011} while the cyan circles represent the 6 new discoveries. 

\begin{figure}
\centering
    \includegraphics[width=0.99\hsize]{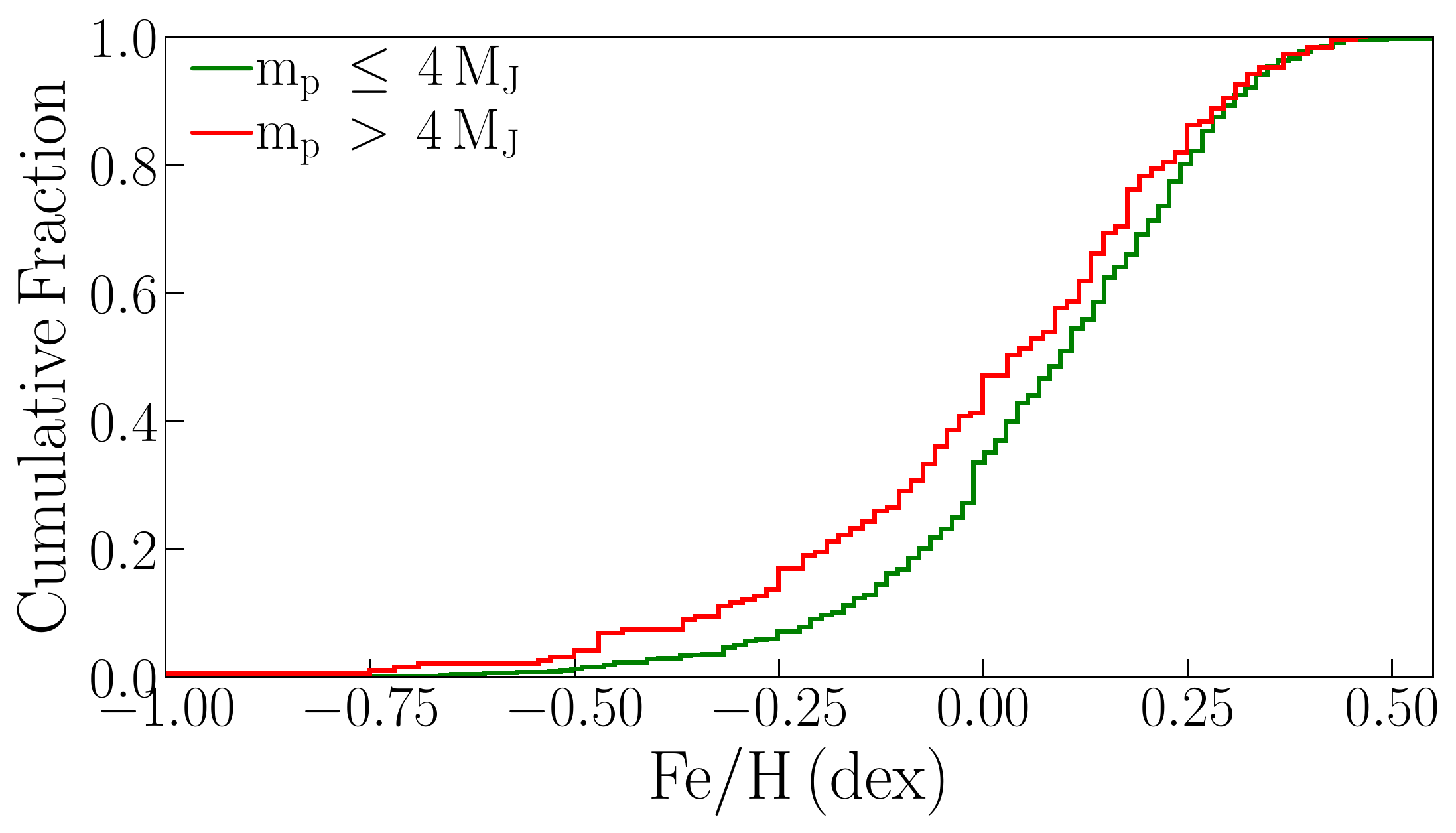}

    \caption{Cumulative metallicity distribution of the host stars of giant planets having masses above (red line) and below (green line) 4~M$_{\mathrm{J}}$ (see text for more details).}
    \label{fig:metalCJ}
\end{figure}

The radial velocities of BD+550362 and HD~124330 show additional drifts, hinting at an additional longer-period companion in both planetary systems. Additional observations of both targets will help characterize those companions. Most of the giant planet hosts in this paper are metal-rich, and their average metallicity is 0.07. We compare the metallicities of two samples of giant planet hosts (including the new detections), divided between planetary masses above and below 4~M$_{\mathrm{J}}$. Figure~\ref{fig:metalCJ} shows the cumulative metallicity distribution for host stars of these two samples, which a Kolmogorov-Smirnov (K-S) test\footnote{Python function: \textsf{scipy.stats.ks\_2samp} is used to perform the K-S test.} shows has a very low p-value of 9.53 $ \times 10^{-8}$ to be drawn from a single population. This reinforces the previous observations that the average metallicity of the host star decreases as the mass of the planet increases \citep{Santos2017,Narang2018,Swastik2021}. A possible interpretation is that the population of giant planets is bimodal, with giant planets (m$_{\mathrm{p}}$ $\leq$ 4 M$_{\mathrm{J}}$) forming via core accretion and more massive giant planets (m$_{\mathrm{p}}$ $>$ 4 M$_{\mathrm{J}}$) forming via the disk gravitational instability.

\begin{figure}
\centering
    \includegraphics[width=0.99\hsize]{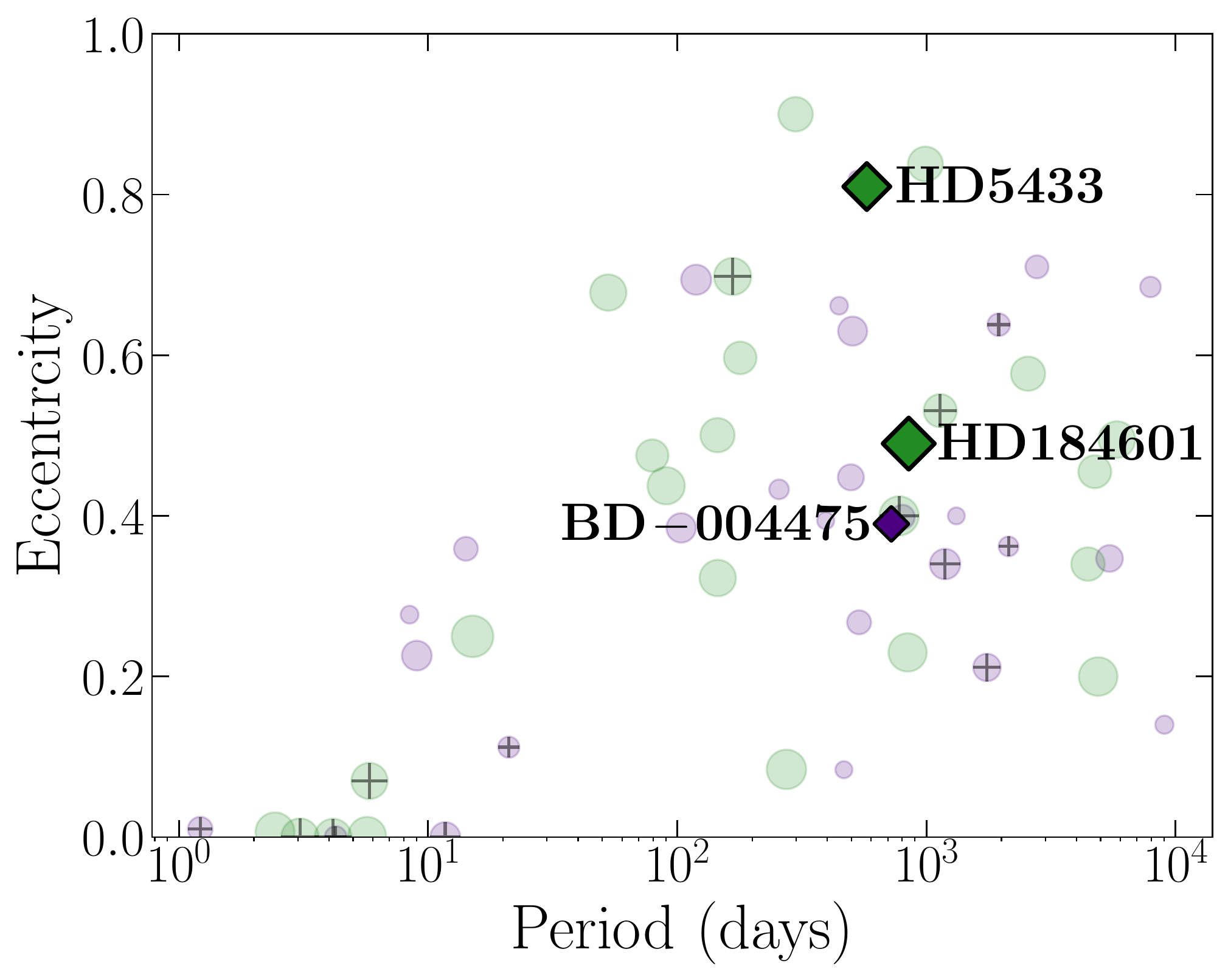}

    \caption{Period - eccentricity distribution of known BD companions. The new BD candidates from this paper are shown as solid diamonds. The circles represents the BDs from the combination of Table 1 in \citet{Ma2014} and Table A.1 in \citet{Wilson2016}. The size of all markers is proportional to the minimum mass (or mass represented by a black cross on top of the circles) of the corresponding object. The green color marks lower-mass BDs with $m$ (or \msini) < 42.5  M$_{\mathrm{J}}$, and the indigo color indicates  high mass BDs with $m$ (or \msini) > 42.5  M$_{\mathrm{J}}$.}
    \label{fig:periodeccbd}
\end{figure}

The three new BD candidates have minimum masses between 25 and 60  M$_{\mathrm{J}}$ and orbital periods between 576 and 850 days. Figure~\ref{fig:periodeccbd} shows the period–eccentricity distribution of all known BDs. All three newly detected BD candidates have significant eccentricities. The lightest of the three, BD-004475, with a minimum mass of 25 M$_{\mathrm{J}}$, has an eccentricity below 0.4, and the two BDs with masses 49 and 60 M$_{\mathrm{J}}$ have eccentricities of 0.5 and 0.8. These new detections are consistent with the observation by \citet{Ma2014} that a significant number of the BDs with (minimum) masses below 42.5 M$_{\mathrm{J}}$ and with orbital periods between 300 and 3000 days have eccentricities below 0.4. The heaviest CJ and BD companions in our small sample have higher eccentricities.

We also present RV and astrometric orbit solutions for 16 low-mass stellar companions heavier than 75  M$_{\mathrm{J}}$ or 0.072 M$_{\odot}$. These low-mass stars have orbital periods between 30 days and 11.5 years and have eccentricities between 0.04 and 0.65. These objects were followed with SOPHIE as potential BDs and turned out to be SCs. Information from the Gaia and HIPPARCOS astrometric space missions allowed us to constrain the orbital inclination and hence, the true mass of some of these objects. 

The SOPHIE giant planet and BD surveys play an important role in identifying promising targets for future direct imaging detections. Furthermore, the SOPHIE sample is volume-limited, which will help to obtain unbiased statistics for these planetary systems, which is crucial for understanding the formation and evolution of these objects.

\begin{acknowledgements}
We thank the staff of Haute-Provence Observatory for their support at the 1.93 m telescope and on SOPHIE. This work is based on observations made with SOPHIE in the context of the program “Recherche de Planétes Extra - solaire” (PI: I. Boisse, F. Bouchy) and the Programme National de Physique Stellaire (PNPS), in particular the two programs ``Recherche et caractérisation de planètes géantes avec SOPHIE'' (PI : G. Hébrard) and ``Exploring the Brown Dwarf Desert around FGK stars in the Solar neighborhood'' (PIs: F. Kiefer, IDs: 2017B\_PNPS008, 2018A\_PNPS008, 2018B\_PNPS013 ; S. Dalal, IDs: 2019A\_PNPS008, 2020A\_PNPS004, 2020B\_PNPS DALAL). F.K. acknowledges support by fellowship grants from the Centre National d’Etude Spatiale (CNES) and from the Paris-Science-Lettre (PSL) University. This work was also supported by the PNPS of CNRS/INSU co-funded by CEA and CNES. X.B., X.D, T.F. \& L.M. acknowledge funding from the French National Research Agency (ANR) in the framework of the Investissements d'Avenir program (ANR-15-IDEX-02), through the funding of the ``Origin of Life'' project of the Univ. Grenoble-Alpes. E.M. acknowledges funding from the French National Research Agency (ANR) under contract number ANR-18-CE31-0019 (SPlaSH). MH acknowledges support from ANID – Millennium Science Initiative – ICN12\_009. N. A.-D. acknowledges the support of FONDECYT project 3180063. This work was supported by FCT - Fundação para a Ciência e a Tecnologia through national funds and by FEDER through COMPETE2020 - Programa Operacional Competitividade e Internacionalização by these grants: UID/FIS/04434/2019; UIDB/04434/2020; UIDP/04434/2020; PTDC/FIS-AST/32113/2017 \& POCI-01-0145-FEDER-032113; PTDC/FISAST/28953/2017 \& POCI-01-0145-FEDER-028953. O.D.S.D. is supported in the form of work contract (DL 57/2016/CP1364/CT0004) funded by FCT.

\end{acknowledgements}

\bibliographystyle{aa}
\bibliography{aa}

\begin{thebibliography}{85}
\expandafter\ifx\csname natexlab\endcsname\relax\def\natexlab#1{#1}\fi

\bibitem[{{Alibert} {et~al.}(2005){Alibert}, {Mordasini}, {Benz}, \&
  {Winisdoerffer}}]{Alibert2005}
{Alibert}, Y., {Mordasini}, C., {Benz}, W., \& {Winisdoerffer}, C. 2005, \aap,
  434, 343

\bibitem[{{Baluev}(2013{\natexlab{a}})}]{Baluev2013}
{Baluev}, R.~V. 2013{\natexlab{a}}, Astronomy and Computing, 2, 18

\bibitem[{{Baluev}(2013{\natexlab{b}})}]{PP}
{Baluev}, R.~V. 2013{\natexlab{b}}, {PlanetPack: Radial-velocity time-series
  analysis tool}

\bibitem[{{Baranne} {et~al.}(1996){Baranne}, {Queloz}, {Mayor}, {Adrianzyk},
  {Knispel}, {Kohler}, {Lacroix}, {Meunier}, {Rimbaud}, \& {Vin}}]{Baranne1996}
{Baranne}, A., {Queloz}, D., {Mayor}, M., {et~al.} 1996, \aaps, 119, 373

\bibitem[{{Baron} {et~al.}(2015){Baron}, {Lafreni{\`e}re}, {Artigau}, {Doyon},
  {Gagn{\'e}}, {Davison}, {Malo}, {Robert}, {Nadeau}, \&
  {Reyl{\'e}}}]{Baron2015}
{Baron}, F., {Lafreni{\`e}re}, D., {Artigau}, {\'E}., {et~al.} 2015, \apj, 802,
  37

\bibitem[{{Boisse} {et~al.}(2010){Boisse}, {Eggenberger}, {Santos}, {Lovis},
  {Bouchy}, {H{\'e}brard}, {Arnold}, {Bonfils}, {Delfosse}, {Desort},
  {D{\'\i}az}, {Ehrenreich}, {Forveille}, {Gallenne}, {Lagrange}, {Moutou},
  {Udry}, {Pepe}, {Perrier}, {Perruchot}, {Pont}, {Queloz}, {Santerne},
  {S{\'e}gransan}, \& {Vidal-Madjar}}]{Boisse2010}
{Boisse}, I., {Eggenberger}, A., {Santos}, N.~C., {et~al.} 2010, \aap, 523, A88

\bibitem[{{Boss}(2011)}]{Boss2011}
{Boss}, A.~P. 2011, \apj, 731, 74

\bibitem[{{Boss} {et~al.}(2003){Boss}, {Basri}, {Kumar}, {Liebert},
  {Mart{\'\i}n}, {Reipurth}, \& {Zinnecker}}]{Boss2003}
{Boss}, A.~P., {Basri}, G., {Kumar}, S.~S., {et~al.} 2003, in Brown Dwarfs, ed.
  E.~{Mart{\'\i}n}, Vol. 211, 529

\bibitem[{{Bouchy} {et~al.}(2013){Bouchy}, {D{\'\i}az}, {H{\'e}brard},
  {Arnold}, {Boisse}, {Delfosse}, {Perruchot}, \& {Santerne}}]{Bouchy2013}
{Bouchy}, F., {D{\'\i}az}, R.~F., {H{\'e}brard}, G., {et~al.} 2013, \aap, 549,
  A49

\bibitem[{{Bouchy} {et~al.}(2009{\natexlab{a}}){Bouchy}, {H{\'e}brard}, {Udry},
  {Delfosse}, {Boisse}, {Desort}, {Bonfils}, {Eggenberger}, {Ehrenreich},
  {Forveille}, {Lagrange}, {Le Coroller}, {Lovis}, {Moutou}, {Pepe}, {Perrier},
  {Pont}, {Queloz}, {Santos}, {S{\'e}gransan}, \& {Vidal-Madjar}}]{Bouchy2009}
{Bouchy}, F., {H{\'e}brard}, G., {Udry}, S., {et~al.} 2009{\natexlab{a}}, \aap,
  505, 853

\bibitem[{{Bouchy} {et~al.}(2009{\natexlab{b}}){Bouchy}, {Isambert}, {Lovis},
  {Boisse}, {Figueira}, {H{\'e}brard}, \& {Pepe}}]{Bouchy2009b}
{Bouchy}, F., {Isambert}, J., {Lovis}, C., {et~al.} 2009{\natexlab{b}}, in EAS
  Publications Series, Vol.~37, EAS Publications Series, ed. P.~{Kern},
  247--253

\bibitem[{{Bouchy} {et~al.}(2016){Bouchy}, {S{\'e}gransan}, {D{\'\i}az},
  {Forveille}, {Boisse}, {Arnold}, {Astudillo-Defru}, {Beuzit}, {Bonfils},
  {Borgniet}, {Bourrier}, {Courcol}, {Delfosse}, {Demangeon}, {Delorme},
  {Ehrenreich}, {H{\'e}brard}, {Lagrange}, {Mayor}, {Montagnier}, {Moutou},
  {Naef}, {Pepe}, {Perrier}, {Queloz}, {Rey}, {Sahlmann}, {Santerne}, {Santos},
  {Sivan}, {Udry}, \& {Wilson}}]{Bouchy2016}
{Bouchy}, F., {S{\'e}gransan}, D., {D{\'\i}az}, R.~F., {et~al.} 2016, \aap,
  585, A46

\bibitem[{{Bryan} {et~al.}(2019){Bryan}, {Knutson}, {Lee}, {Fulton}, {Batygin},
  {Ngo}, \& {Meshkat}}]{Bryan2019}
{Bryan}, M.~L., {Knutson}, H.~A., {Lee}, E.~J., {et~al.} 2019, \aj, 157, 52

\bibitem[{{Butler} {et~al.}(2017){Butler}, {Vogt}, {Laughlin}, {Burt},
  {Rivera}, {Tuomi}, {Teske}, {Arriagada}, {Diaz}, {Holden}, \&
  {Keiser}}]{Butler2017}
{Butler}, R.~P., {Vogt}, S.~S., {Laughlin}, G., {et~al.} 2017, \aj, 153, 208

\bibitem[{{Cai} {et~al.}(2010){Cai}, {Pickett}, {Durisen}, \&
  {Milne}}]{Cai2010}
{Cai}, K., {Pickett}, M.~K., {Durisen}, R.~H., \& {Milne}, A.~M. 2010, \apjl,
  716, L176

\bibitem[{{Chabrier} \& {Baraffe}(1997)}]{Chabrier1997}
{Chabrier}, G. \& {Baraffe}, I. 1997, \aap, 327, 1039

\bibitem[{{Chabrier} {et~al.}(2000){Chabrier}, {Baraffe}, {Allard}, \&
  {Hauschildt}}]{Chabrier2000}
{Chabrier}, G., {Baraffe}, I., {Allard}, F., \& {Hauschildt}, P. 2000, \apj,
  542, 464

\bibitem[{{Chabrier} {et~al.}(2014){Chabrier}, {Johansen}, {Janson}, \&
  {Rafikov}}]{Chabrier2014}
{Chabrier}, G., {Johansen}, A., {Janson}, M., \& {Rafikov}, R. 2014, in
  Protostars and Planets VI, ed. H.~{Beuther}, R.~S. {Klessen}, C.~P.
  {Dullemond}, \& T.~{Henning}, 619

\bibitem[{{Csizmadia} \& {CoRot Team}(2016)}]{Csizmadia2016}
{Csizmadia}, S. \& {CoRot Team}. 2016, {III.6 Exploration of the brown dwarf
  regime around solar-like stars by CoRoT}, 143

\bibitem[{{Curiel} {et~al.}(2020){Curiel}, {Ortiz-Le{\'o}n}, {Mioduszewski}, \&
  {Torres}}]{Curiel2020}
{Curiel}, S., {Ortiz-Le{\'o}n}, G.~N., {Mioduszewski}, A.~J., \& {Torres},
  R.~M. 2020, \aj, 160, 97

\bibitem[{{D{\'\i}az} {et~al.}(2014){D{\'\i}az}, {Montagnier}, {Leconte},
  {Bonomo}, {Deleuil}, {Almenara}, {Barros}, {Bouchy}, {Bruno}, {Damiani},
  {H{\'e}brard}, {Moutou}, \& {Santerne}}]{Diaz2014}
{D{\'\i}az}, R.~F., {Montagnier}, G., {Leconte}, J., {et~al.} 2014, \aap, 572,
  A109

\bibitem[{{D{\'{\i}}az} {et~al.}(2012){D{\'{\i}}az}, {Santerne}, {Sahlmann},
  {H{\'e}brard}, {Eggenberger}, {Santos}, {Moutou}, {Arnold}, {Boisse},
  {Bonfils}, {Bouchy}, {Delfosse}, {Desort}, {Ehrenreich}, {Forveille},
  {Lagrange}, {Lovis}, {Pepe}, {Perrier}, {Queloz}, {S{\'e}gransan}, {Udry}, \&
  {Vidal-Madjar}}]{Diaz:2012fk}
{D{\'{\i}}az}, R.~F., {Santerne}, A., {Sahlmann}, J., {et~al.} 2012, \aap, 538,
  A113

\bibitem[{{D{\'\i}az} {et~al.}(2016){D{\'\i}az}, {S{\'e}gransan}, {Udry},
  {Lovis}, {Pepe}, {Dumusque}, {Marmier}, {Alonso}, {Benz}, {Bouchy},
  {Coffinet}, {Collier Cameron}, {Deleuil}, {Figueira}, {Gillon}, {Lo Curto},
  {Mayor}, {Mordasini}, {Motalebi}, {Moutou}, {Pollacco}, {Pompei}, {Queloz},
  {Santos}, \& {Wyttenbach}}]{Diaz2016}
{D{\'\i}az}, R.~F., {S{\'e}gransan}, D., {Udry}, S., {et~al.} 2016, \aap, 585,
  A134

\bibitem[{{dos Santos} {et~al.}(2017){dos Santos}, {Mel{\'e}ndez}, {Bedell},
  {Bean}, {Spina}, {Alves-Brito}, {Dreizler}, {Ram{\'\i}rez}, \&
  {Asplund}}]{dSantos2017}
{dos Santos}, L.~A., {Mel{\'e}ndez}, J., {Bedell}, M., {et~al.} 2017, \mnras,
  472, 3425

\bibitem[{{Dumusque} {et~al.}(2011){Dumusque}, {Lovis}, {S{\'e}gransan},
  {Mayor}, {Udry}, {Benz}, {Bouchy}, {Lo Curto}, {Mordasini}, {Pepe}, {Queloz},
  {Santos}, \& {Naef}}]{Dumusque2011}
{Dumusque}, X., {Lovis}, C., {S{\'e}gransan}, D., {et~al.} 2011, \aap, 535, A55

\bibitem[{{F.~van Leeuwen}(2007)}]{:2007kx}
{F.~van Leeuwen}, ed. 2007, Astrophysics and Space Science Library, Vol. 350,
  {Hipparcos, the New Reduction of the Raw Data}

\bibitem[{{Fernandes} {et~al.}(2019){Fernandes}, {Mulders}, {Pascucci},
  {Mordasini}, \& {Emsenhuber}}]{Fernandes2019}
{Fernandes}, R.~B., {Mulders}, G.~D., {Pascucci}, I., {Mordasini}, C., \&
  {Emsenhuber}, A. 2019, \apj, 874, 81

\bibitem[{{Gaia Collaboration} {et~al.}(2018){Gaia Collaboration}, {Brown},
  {Vallenari}, {Prusti}, {de Bruijne}, {Babusiaux}, {Bailer-Jones}, {Biermann},
  {Evans}, {Eyer}, {Jansen}, {Jordi}, {Klioner}, {Lammers}, {Lindegren},
  {Luri}, {Mignard}, {Panem}, {Pourbaix}, {Randich}, {Sartoretti}, {Siddiqui},
  {Soubiran}, {van Leeuwen}, {Walton}, {Arenou}, {Bastian}, {Cropper},
  {Drimmel}, {Katz}, {Lattanzi}, {Bakker}, {Cacciari}, {Casta{\~n}eda},
  {Chaoul}, {Cheek}, {De Angeli}, {Fabricius}, {Guerra}, {Holl}, {Masana},
  {Messineo}, {Mowlavi}, {Nienartowicz}, {Panuzzo}, {Portell}, {Riello},
  {Seabroke}, {Tanga}, {Th{\'e}venin}, {Gracia-Abril}, {Comoretto},
  {Garcia-Reinaldos}, {Teyssier}, {Altmann}, {Andrae}, {Audard},
  {Bellas-Velidis}, {Benson}, {Berthier}, {Blomme}, {Burgess}, {Busso},
  {Carry}, {Cellino}, {Clementini}, {Clotet}, {Creevey}, {Davidson}, {De
  Ridder}, {Delchambre}, {Dell'Oro}, {Ducourant},
  {Fern{\'a}ndez-Hern{\'a}ndez}, {Fouesneau}, {Fr{\'e}mat}, {Galluccio},
  {Garc{\'\i}a-Torres}, {Gonz{\'a}lez-N{\'u}{\~n}ez}, {Gonz{\'a}lez-Vidal},
  {Gosset}, {Guy}, {Halbwachs}, {Hambly}, {Harrison}, {Hern{\'a}ndez},
  {Hestroffer}, {Hodgkin}, {Hutton}, {Jasniewicz}, {Jean-Antoine-Piccolo},
  {Jordan}, {Korn}, {Krone-Martins}, {Lanzafame}, {Lebzelter}, {L{\"o}ffler},
  {Manteiga}, {Marrese}, {Mart{\'\i}n-Fleitas}, {Moitinho}, {Mora}, {Muinonen},
  {Osinde}, {Pancino}, {Pauwels}, {Petit}, {Recio-Blanco}, {Richards},
  {Rimoldini}, {Robin}, {Sarro}, {Siopis}, {Smith}, {Sozzetti}, {S{\"u}veges},
  {Torra}, {van Reeven}, {Abbas}, {Abreu Aramburu}, {Accart}, {Aerts},
  {Altavilla}, {{\'A}lvarez}, {Alvarez}, {Alves}, {Anderson}, {Andrei},
  {Anglada Varela}, {Antiche}, {Antoja}, {Arcay}, {Astraatmadja}, {Bach},
  {Baker}, {Balaguer-N{\'u}{\~n}ez}, {Balm}, {Barache}, {Barata}, {Barbato},
  {Barblan}, {Barklem}, {Barrado}, {Barros}, {Barstow}, {Bartholom{\'e}
  Mu{\~n}oz}, {Bassilana}, {Becciani}, {Bellazzini}, {Berihuete}, {Bertone},
  {Bianchi}, {Bienaym{\'e}}, {Blanco-Cuaresma}, {Boch}, {Boeche}, {Bombrun},
  {Borrachero}, {Bossini}, {Bouquillon}, {Bourda}, {Bragaglia}, {Bramante},
  {Breddels}, {Bressan}, {Brouillet}, {Br{\"u}semeister}, {Brugaletta},
  {Bucciarelli}, {Burlacu}, {Busonero}, {Butkevich}, {Buzzi}, {Caffau},
  {Cancelliere}, {Cannizzaro}, {Cantat-Gaudin}, {Carballo}, {Carlucci},
  {Carrasco}, {Casamiquela}, {Castellani}, {Castro-Ginard}, {Charlot},
  {Chemin}, {Chiavassa}, {Cocozza}, {Costigan}, {Cowell}, {Crifo}, {Crosta},
  {Crowley}, {Cuypers}, {Dafonte}, {Damerdji}, {Dapergolas}, {David}, {David},
  {de Laverny}, {De Luise}, {De March}, {de Martino}, {de Souza}, {de Torres},
  {Debosscher}, {del Pozo}, {Delbo}, {Delgado}, {Delgado}, {Di Matteo},
  {Diakite}, {Diener}, {Distefano}, {Dolding}, {Drazinos}, {Dur{\'a}n},
  {Edvardsson}, {Enke}, {Eriksson}, {Esquej}, {Eynard Bontemps}, {Fabre},
  {Fabrizio}, {Faigler}, {Falc{\~a}o}, {Farr{\`a}s Casas}, {Federici},
  {Fedorets}, {Fernique}, {Figueras}, {Filippi}, {Findeisen}, {Fonti},
  {Fraile}, {Fraser}, {Fr{\'e}zouls}, {Gai}, {Galleti}, {Garabato},
  {Garc{\'\i}a-Sedano}, {Garofalo}, {Garralda}, {Gavel}, {Gavras}, {Gerssen},
  {Geyer}, {Giacobbe}, {Gilmore}, {Girona}, {Giuffrida}, {Glass}, {Gomes},
  {Granvik}, {Gueguen}, {Guerrier}, {Guiraud}, {Guti{\'e}rrez-S{\'a}nchez},
  {Haigron}, {Hatzidimitriou}, {Hauser}, {Haywood}, {Heiter}, {Helmi}, {Heu},
  {Hilger}, {Hobbs}, {Hofmann}, {Holland}, {Huckle}, {Hypki}, {Icardi},
  {Jan{\ss}en}, {Jevardat de Fombelle}, {Jonker}, {Juh{\'a}sz}, {Julbe},
  {Karampelas}, {Kewley}, {Klar}, {Kochoska}, {Kohley}, {Kolenberg},
  {Kontizas}, {Kontizas}, {Koposov}, {Kordopatis}, {Kostrzewa-Rutkowska},
  {Koubsky}, {Lambert}, {Lanza}, {Lasne}, {Lavigne}, {Le Fustec}, {Le
  Poncin-Lafitte}, {Lebreton}, {Leccia}, {Leclerc}, {Lecoeur-Taibi},
  {Lenhardt}, {Leroux}, {Liao}, {Licata}, {Lindstr{\o}m}, {Lister}, {Livanou},
  {Lobel}, {L{\'o}pez}, {Managau}, {Mann}, {Mantelet}, {Marchal}, {Marchant},
  {Marconi}, {Marinoni}, {Marschalk{\'o}}, {Marshall}, {Martino}, {Marton},
  {Mary}, {Massari}, {Matijevi{\v{c}}}, {Mazeh}, {McMillan}, {Messina},
  {Michalik}, {Millar}, {Molina}, {Molinaro}, {Moln{\'a}r}, {Montegriffo},
  {Mor}, {Morbidelli}, {Morel}, {Morris}, {Mulone}, {Muraveva}, {Musella},
  {Nelemans}, {Nicastro}, {Noval}, {O'Mullane}, {Ord{\'e}novic},
  {Ord{\'o}{\~n}ez-Blanco}, {Osborne}, {Pagani}, {Pagano}, {Pailler},
  {Palacin}, {Palaversa}, {Panahi}, {Pawlak}, {Piersimoni}, {Pineau}, {Plachy},
  {Plum}, {Poggio}, {Poujoulet}, {Pr{\v{s}}a}, {Pulone}, {Racero}, {Ragaini},
  {Rambaux}, {Ramos-Lerate}, {Regibo}, {Reyl{\'e}}, {Riclet}, {Ripepi}, {Riva},
  {Rivard}, {Rixon}, {Roegiers}, {Roelens}, {Romero-G{\'o}mez}, {Rowell},
  {Royer}, {Ruiz-Dern}, {Sadowski}, {Sagrist{\`a} Sell{\'e}s}, {Sahlmann},
  {Salgado}, {Salguero}, {Sanna}, {Santana-Ros}, {Sarasso}, {Savietto},
  {Schultheis}, {Sciacca}, {Segol}, {Segovia}, {S{\'e}gransan}, {Shih},
  {Siltala}, {Silva}, {Smart}, {Smith}, {Solano}, {Solitro}, {Sordo}, {Soria
  Nieto}, {Souchay}, {Spagna}, {Spoto}, {Stampa}, {Steele},
  {Steidelm{\"u}ller}, {Stephenson}, {Stoev}, {Suess}, {Surdej}, {Szabados},
  {Szegedi-Elek}, {Tapiador}, {Taris}, {Tauran}, {Taylor}, {Teixeira},
  {Terrett}, {Teyssandier}, {Thuillot}, {Titarenko}, {Torra Clotet}, {Turon},
  {Ulla}, {Utrilla}, {Uzzi}, {Vaillant}, {Valentini}, {Valette}, {van Elteren},
  {Van Hemelryck}, {van Leeuwen}, {Vaschetto}, {Vecchiato}, {Veljanoski},
  {Viala}, {Vicente}, {Vogt}, {von Essen}, {Voss}, {Votruba}, {Voutsinas},
  {Walmsley}, {Weiler}, {Wertz}, {Wevers}, {Wyrzykowski}, {Yoldas},
  {{\v{Z}}erjal}, {Ziaeepour}, {Zorec}, {Zschocke}, {Zucker}, {Zurbach}, \&
  {Zwitter}}]{Gaia2018}
{Gaia Collaboration}, {Brown}, A.~G.~A., {Vallenari}, A., {et~al.} 2018, \aap,
  616, A1

\bibitem[{{Gaia Collaboration} {et~al.}(2016){Gaia Collaboration}, {Brown},
  {Vallenari}, {Prusti}, {de Bruijne}, {Mignard}, {Drimmel}, {Babusiaux},
  {Bailer-Jones}, {Bastian}, {Biermann}, {Evans}, {Eyer}, {Jansen}, {Jordi},
  {Katz}, {Klioner}, {Lammers}, {Lindegren}, {Luri}, {O'Mullane}, {Panem},
  {Pourbaix}, {Randich}, {Sartoretti}, {Siddiqui}, {Soubiran}, {Valette}, {van
  Leeuwen}, {Walton}, {Aerts}, {Arenou}, {Cropper}, {H{\o}g}, {Lattanzi},
  {Grebel}, {Holland}, {Huc}, {Passot}, {Perryman}, {Bramante}, {Cacciari},
  {Casta{\~n}eda}, {Chaoul}, {Cheek}, {De Angeli}, {Fabricius}, {Guerra},
  {Hern{\'a}ndez}, {Jean-Antoine-Piccolo}, {Masana}, {Messineo}, {Mowlavi},
  {Nienartowicz}, {Ord{\'o}{\~n}ez-Blanco}, {Panuzzo}, {Portell}, {Richards},
  {Riello}, {Seabroke}, {Tanga}, {Th{\'e}venin}, {Torra}, {Els},
  {Gracia-Abril}, {Comoretto}, {Garcia-Reinaldos}, {Lock}, {Mercier},
  {Altmann}, {Andrae}, {Astraatmadja}, {Bellas-Velidis}, {Benson}, {Berthier},
  {Blomme}, {Busso}, {Carry}, {Cellino}, {Clementini}, {Cowell}, {Creevey},
  {Cuypers}, {Davidson}, {De Ridder}, {de Torres}, {Delchambre}, {Dell'Oro},
  {Ducourant}, {Fr{\'e}mat}, {Garc{\'\i}a-Torres}, {Gosset}, {Halbwachs},
  {Hambly}, {Harrison}, {Hauser}, {Hestroffer}, {Hodgkin}, {Huckle}, {Hutton},
  {Jasniewicz}, {Jordan}, {Kontizas}, {Korn}, {Lanzafame}, {Manteiga},
  {Moitinho}, {Muinonen}, {Osinde}, {Pancino}, {Pauwels}, {Petit},
  {Recio-Blanco}, {Robin}, {Sarro}, {Siopis}, {Smith}, {Smith}, {Sozzetti},
  {Thuillot}, {van Reeven}, {Viala}, {Abbas}, {Abreu Aramburu}, {Accart},
  {Aguado}, {Allan}, {Allasia}, {Altavilla}, {{\'A}lvarez}, {Alves},
  {Anderson}, {Andrei}, {Anglada Varela}, {Antiche}, {Antoja}, {Ant{\'o}n},
  {Arcay}, {Bach}, {Baker}, {Balaguer-N{\'u}{\~n}ez}, {Barache}, {Barata},
  {Barbier}, {Barblan}, {Barrado y Navascu{\'e}s}, {Barros}, {Barstow},
  {Becciani}, {Bellazzini}, {Bello Garc{\'\i}a}, {Belokurov}, {Bendjoya},
  {Berihuete}, {Bianchi}, {Bienaym{\'e}}, {Billebaud}, {Blagorodnova},
  {Blanco-Cuaresma}, {Boch}, {Bombrun}, {Borrachero}, {Bouquillon}, {Bourda},
  {Bouy}, {Bragaglia}, {Breddels}, {Brouillet}, {Br{\"u}semeister},
  {Bucciarelli}, {Burgess}, {Burgon}, {Burlacu}, {Busonero}, {Buzzi}, {Caffau},
  {Cambras}, {Campbell}, {Cancelliere}, {Cantat-Gaudin}, {Carlucci},
  {Carrasco}, {Castellani}, {Charlot}, {Charnas}, {Chiavassa}, {Clotet},
  {Cocozza}, {Collins}, {Costigan}, {Crifo}, {Cross}, {Crosta}, {Crowley},
  {Dafonte}, {Damerdji}, {Dapergolas}, {David}, {David}, {De Cat}, {de Felice},
  {de Laverny}, {De Luise}, {De March}, {de Martino}, {de Souza}, {Debosscher},
  {del Pozo}, {Delbo}, {Delgado}, {Delgado}, {Di Matteo}, {Diakite},
  {Distefano}, {Dolding}, {Dos Anjos}, {Drazinos}, {Duran}, {Dzigan},
  {Edvardsson}, {Enke}, {Evans}, {Eynard Bontemps}, {Fabre}, {Fabrizio},
  {Faigler}, {Falc{\~a}o}, {Farr{\`a}s Casas}, {Federici}, {Fedorets},
  {Fern{\'a}ndez-Hern{\'a}ndez}, {Fernique}, {Fienga}, {Figueras}, {Filippi},
  {Findeisen}, {Fonti}, {Fouesneau}, {Fraile}, {Fraser}, {Fuchs}, {Gai},
  {Galleti}, {Galluccio}, {Garabato}, {Garc{\'\i}a-Sedano}, {Garofalo},
  {Garralda}, {Gavras}, {Gerssen}, {Geyer}, {Gilmore}, {Girona}, {Giuffrida},
  {Gomes}, {Gonz{\'a}lez-Marcos}, {Gonz{\'a}lez-N{\'u}{\~n}ez},
  {Gonz{\'a}lez-Vidal}, {Granvik}, {Guerrier}, {Guillout}, {Guiraud},
  {G{\'u}rpide}, {Guti{\'e}rrez-S{\'a}nchez}, {Guy}, {Haigron},
  {Hatzidimitriou}, {Haywood}, {Heiter}, {Helmi}, {Hobbs}, {Hofmann}, {Holl},
  {Holland}, {Hunt}, {Hypki}, {Icardi}, {Irwin}, {Jevardat de Fombelle},
  {Jofr{\'e}}, {Jonker}, {Jorissen}, {Julbe}, {Karampelas}, {Kochoska},
  {Kohley}, {Kolenberg}, {Kontizas}, {Koposov}, {Kordopatis}, {Koubsky},
  {Krone-Martins}, {Kudryashova}, {Kull}, {Bachchan}, {Lacoste-Seris}, {Lanza},
  {Lavigne}, {Le Poncin-Lafitte}, {Lebreton}, {Lebzelter}, {Leccia}, {Leclerc},
  {Lecoeur-Taibi}, {Lemaitre}, {Lenhardt}, {Leroux}, {Liao}, {Licata},
  {Lindstr{\o}m}, {Lister}, {Livanou}, {Lobel}, {L{\"o}ffler}, {L{\'o}pez},
  {Lorenz}, {MacDonald}, {Magalh{\~a}es Fernandes}, {Managau}, {Mann},
  {Mantelet}, {Marchal}, {Marchant}, {Marconi}, {Marinoni}, {Marrese},
  {Marschalk{\'o}}, {Marshall}, {Mart{\'\i}n-Fleitas}, {Martino}, {Mary},
  {Matijevi{\v{c}}}, {Mazeh}, {McMillan}, {Messina}, {Michalik}, {Millar},
  {Miranda}, {Molina}, {Molinaro}, {Molinaro}, {Moln{\'a}r}, {Moniez},
  {Montegriffo}, {Mor}, {Mora}, {Morbidelli}, {Morel}, {Morgenthaler},
  {Morris}, {Mulone}, {Muraveva}, {Musella}, {Narbonne}, {Nelemans},
  {Nicastro}, {Noval}, {Ord{\'e}novic}, {Ordieres-Mer{\'e}}, {Osborne},
  {Pagani}, {Pagano}, {Pailler}, {Palacin}, {Palaversa}, {Parsons}, {Pecoraro},
  {Pedrosa}, {Pentik{\"a}inen}, {Pichon}, {Piersimoni}, {Pineau}, {Plachy},
  {Plum}, {Poujoulet}, {Pr{\v{s}}a}, {Pulone}, {Ragaini}, {Rago}, {Rambaux},
  {Ramos-Lerate}, {Ranalli}, {Rauw}, {Read}, {Regibo}, {Reyl{\'e}}, {Ribeiro},
  {Rimoldini}, {Ripepi}, {Riva}, {Rixon}, {Roelens}, {Romero-G{\'o}mez},
  {Rowell}, {Royer}, {Ruiz-Dern}, {Sadowski}, {Sagrist{\`a} Sell{\'e}s},
  {Sahlmann}, {Salgado}, {Salguero}, {Sarasso}, {Savietto}, {Schultheis},
  {Sciacca}, {Segol}, {Segovia}, {Segransan}, {Shih}, {Smareglia}, {Smart},
  {Solano}, {Solitro}, {Sordo}, {Soria Nieto}, {Souchay}, {Spagna}, {Spoto},
  {Stampa}, {Steele}, {Steidelm{\"u}ller}, {Stephenson}, {Stoev}, {Suess},
  {S{\"u}veges}, {Surdej}, {Szabados}, {Szegedi-Elek}, {Tapiador}, {Taris},
  {Tauran}, {Taylor}, {Teixeira}, {Terrett}, {Tingley}, {Trager}, {Turon},
  {Ulla}, {Utrilla}, {Valentini}, {van Elteren}, {Van Hemelryck}, {van
  Leeuwen}, {Varadi}, {Vecchiato}, {Veljanoski}, {Via}, {Vicente}, {Vogt},
  {Voss}, {Votruba}, {Voutsinas}, {Walmsley}, {Weiler}, {Weingrill}, {Wevers},
  {Wyrzykowski}, {Yoldas}, {{\v{Z}}erjal}, {Zucker}, {Zurbach}, {Zwitter},
  {Alecu}, {Allen}, {Allende Prieto}, {Amorim}, {Anglada-Escud{\'e}},
  {Arsenijevic}, {Azaz}, {Balm}, {Beck}, {Bernstein}, {Bigot}, {Bijaoui},
  {Blasco}, {Bonfigli}, {Bono}, {Boudreault}, {Bressan}, {Brown}, {Brunet},
  {Bunclark}, {Buonanno}, {Butkevich}, {Carret}, {Carrion}, {Chemin},
  {Ch{\'e}reau}, {Corcione}, {Darmigny}, {de Boer}, {de Teodoro}, {de Zeeuw},
  {Delle Luche}, {Domingues}, {Dubath}, {Fodor}, {Fr{\'e}zouls}, {Fries},
  {Fustes}, {Fyfe}, {Gallardo}, {Gallegos}, {Gardiol}, {Gebran}, {Gomboc},
  {G{\'o}mez}, {Grux}, {Gueguen}, {Heyrovsky}, {Hoar}, {Iannicola}, {Isasi
  Parache}, {Janotto}, {Joliet}, {Jonckheere}, {Keil}, {Kim}, {Klagyivik},
  {Klar}, {Knude}, {Kochukhov}, {Kolka}, {Kos}, {Kutka}, {Lainey}, {LeBouquin},
  {Liu}, {Loreggia}, {Makarov}, {Marseille}, {Martayan}, {Martinez-Rubi},
  {Massart}, {Meynadier}, {Mignot}, {Munari}, {Nguyen}, {Nordlander}, {Ocvirk},
  {O'Flaherty}, {Olias Sanz}, {Ortiz}, {Osorio}, {Oszkiewicz}, {Ouzounis},
  {Palmer}, {Park}, {Pasquato}, {Peltzer}, {Peralta}, {P{\'e}turaud},
  {Pieniluoma}, {Pigozzi}, {Poels}, {Prat}, {Prod'homme}, {Raison}, {Rebordao},
  {Risquez}, {Rocca-Volmerange}, {Rosen}, {Ruiz-Fuertes}, {Russo}, {Sembay},
  {Serraller Vizcaino}, {Short}, {Siebert}, {Silva}, {Sinachopoulos}, {Slezak},
  {Soffel}, {Sosnowska}, {Strai{\v{z}}ys}, {ter Linden}, {Terrell}, {Theil},
  {Tiede}, {Troisi}, {Tsalmantza}, {Tur}, {Vaccari}, {Vachier}, {Valles}, {Van
  Hamme}, {Veltz}, {Virtanen}, {Wallut}, {Wichmann}, {Wilkinson}, {Ziaeepour},
  \& {Zschocke}}]{Gaia2016}
{Gaia Collaboration}, {Brown}, A.~G.~A., {Vallenari}, A., {et~al.} 2016, \aap,
  595, A2

\bibitem[{{Grether} \& {Lineweaver}(2006)}]{Grether2006}
{Grether}, D. \& {Lineweaver}, C.~H. 2006, \apj, 640, 1051

\bibitem[{{Guilera} {et~al.}(2010){Guilera}, {Brunini}, \&
  {Benvenuto}}]{Guilera2010}
{Guilera}, O.~M., {Brunini}, A., \& {Benvenuto}, O.~G. 2010, \aap, 521, A50

\bibitem[{{Halbwachs} {et~al.}(2000){Halbwachs}, {Arenou}, {Mayor}, {Udry}, \&
  {Queloz}}]{Halbwachs2000}
{Halbwachs}, J.~L., {Arenou}, F., {Mayor}, M., {Udry}, S., \& {Queloz}, D.
  2000, \aap, 355, 581

\bibitem[{{H{\'e}brard} {et~al.}(2016){H{\'e}brard}, {Arnold}, {Forveille},
  {Correia}, {Laskar}, {Bonfils}, {Boisse}, {D{\'\i}az}, {Hagelberg},
  {Sahlmann}, {Santos}, {Astudillo-Defru}, {Borgniet}, {Bouchy}, {Bourrier},
  {Courcol}, {Delfosse}, {Deleuil}, {Demangeon}, {Ehrenreich}, {Gregorio},
  {Jovanovic}, {Labrevoir}, {Lagrange}, {Lovis}, {Lozi}, {Moutou},
  {Montagnier}, {Pepe}, {Rey}, {Santerne}, {S{\'e}gransan}, {Udry},
  {Vanhuysse}, {Vigan}, \& {Wilson}}]{Hebrard2016}
{H{\'e}brard}, G., {Arnold}, L., {Forveille}, T., {et~al.} 2016, \aap, 588,
  A145

\bibitem[{{Howard} {et~al.}(2010){Howard}, {Marcy}, {Johnson}, {Fischer},
  {Wright}, {Isaacson}, {Valenti}, {Anderson}, {Lin}, \& {Ida}}]{Howard2010}
{Howard}, A.~W., {Marcy}, G.~W., {Johnson}, J.~A., {et~al.} 2010, Science, 330,
  653

\bibitem[{{Ida} \& {Lin}(2004)}]{Ida2004}
{Ida}, S. \& {Lin}, D.~N.~C. 2004, \apj, 616, 567

\bibitem[{{Jones} {et~al.}(2006){Jones}, {Butler}, {Tinney}, {Marcy}, {Carter},
  {Penny}, {McCarthy}, \& {Bailey}}]{Jones2006}
{Jones}, H. R.~A., {Butler}, R.~P., {Tinney}, C.~G., {et~al.} 2006, \mnras,
  369, 249

\bibitem[{{Kiefer}(2019)}]{Kiefer2019S}
{Kiefer}, F. 2019, \aap, 632, L9

\bibitem[{{Kiefer} {et~al.}(2021){Kiefer}, {H{\'e}brard}, {Lecavelier des
  Etangs}, {Martioli}, {Dalal}, \& {Vidal-Madjar}}]{Kiefer2021}
{Kiefer}, F., {H{\'e}brard}, G., {Lecavelier des Etangs}, A., {et~al.} 2021,
  \aap, 645, A7

\bibitem[{{Kiefer} {et~al.}(2019){Kiefer}, {H{\'e}brard}, {Sahlmann}, {Sousa},
  {Forveille}, {Santos}, {Mayor}, {Deleuil}, {Wilson}, {Dalal}, {D{\'\i}az},
  {Henry}, {Hagelberg}, {Hobson}, {Demangeon}, {Bourrier}, {Delfosse},
  {Arnold}, {Astudillo-Defru}, {Beuzit}, {Boisse}, {Bonfils}, {Borgniet},
  {Bouchy}, {Courcol}, {Ehrenreich}, {Hara}, {Lagrange}, {Lovis}, {Montagnier},
  {Moutou}, {Pepe}, {Perrier}, {Rey}, {Santerne}, {S{\'e}gransan}, {Udry}, \&
  {Vidal-Madjar}}]{Kiefer2019}
{Kiefer}, F., {H{\'e}brard}, G., {Sahlmann}, J., {et~al.} 2019, \aap, 631, A125

\bibitem[{{Lindegren} {et~al.}(2016){Lindegren}, {Lammers}, {Bastian},
  {Hern{\'a}ndez}, {Klioner}, {Hobbs}, {Bombrun}, {Michalik}, {Ramos-Lerate},
  {Butkevich}, {Comoretto}, {Joliet}, {Holl}, {Hutton}, {Parsons},
  {Steidelm{\"u}ller}, {Abbas}, {Altmann}, {Andrei}, {Anton}, {Bach},
  {Barache}, {Becciani}, {Berthier}, {Bianchi}, {Biermann}, {Bouquillon},
  {Bourda}, {Br{\"u}semeister}, {Bucciarelli}, {Busonero}, {Carlucci},
  {Casta{\~n}eda}, {Charlot}, {Clotet}, {Crosta}, {Davidson}, {de Felice},
  {Drimmel}, {Fabricius}, {Fienga}, {Figueras}, {Fraile}, {Gai}, {Garralda},
  {Geyer}, {Gonz{\'a}lez-Vidal}, {Guerra}, {Hambly}, {Hauser}, {Jordan},
  {Lattanzi}, {Lenhardt}, {Liao}, {L{\"o}ffler}, {McMillan}, {Mignard}, {Mora},
  {Morbidelli}, {Portell}, {Riva}, {Sarasso}, {Serraller}, {Siddiqui}, {Smart},
  {Spagna}, {Stampa}, {Steele}, {Taris}, {Torra}, {van Reeven}, {Vecchiato},
  {Zschocke}, {de Bruijne}, {Gracia}, {Raison}, {Lister}, {Marchant},
  {Messineo}, {Soffel}, {Osorio}, {de Torres}, \& {O'Mullane}}]{Lindegren2016}
{Lindegren}, L., {Lammers}, U., {Bastian}, U., {et~al.} 2016, \aap, 595, A4

\bibitem[{{Ma} \& {Ge}(2014)}]{Ma2014}
{Ma}, B. \& {Ge}, J. 2014, \mnras, 439, 2781

\bibitem[{{Mayor} \& {Queloz}(1995)}]{Mayor1995}
{Mayor}, M. \& {Queloz}, D. 1995, \nat, 378, 355

\bibitem[{{McArthur} {et~al.}(2010){McArthur}, {Benedict}, {Barnes},
  {Martioli}, {Korzennik}, {Nelan}, \& {Butler}}]{McArthur2010}
{McArthur}, B.~E., {Benedict}, G.~F., {Barnes}, R., {et~al.} 2010, \apj, 715,
  1203

\bibitem[{{Morbidelli} {et~al.}(2012){Morbidelli}, {Lunine}, {O'Brien},
  {Raymond}, \& {Walsh}}]{Morbidelli2012}
{Morbidelli}, A., {Lunine}, J.~I., {O'Brien}, D.~P., {Raymond}, S.~N., \&
  {Walsh}, K.~J. 2012, Annual Review of Earth and Planetary Sciences, 40, 251

\bibitem[{{Mordasini} {et~al.}(2009){Mordasini}, {Alibert}, \&
  {Benz}}]{Mordasini2009}
{Mordasini}, C., {Alibert}, Y., \& {Benz}, W. 2009, \aap, 501, 1139

\bibitem[{{Moutou} {et~al.}(2014){Moutou}, {H{\'e}brard}, {Bouchy}, {Arnold},
  {Santos}, {Astudillo-Defru}, {Boisse}, {Bonfils}, {Borgniet}, {Delfosse},
  {D{\'\i}az}, {Ehrenreich}, {Forveille}, {Gregorio}, {Labrevoir}, {Lagrange},
  {Montagnier}, {Montalto}, {Pepe}, {Sahlmann}, {Santerne}, {S{\'e}gransan},
  {Udry}, \& {Vanhuysse}}]{Moutou2014}
{Moutou}, C., {H{\'e}brard}, G., {Bouchy}, F., {et~al.} 2014, \aap, 563, A22

\bibitem[{{Moutou} {et~al.}(2009){Moutou}, {H{\'e}brard}, {Bouchy},
  {Eggenberger}, {Boisse}, {Bonfils}, {Gravallon}, {Ehrenreich}, {Forveille},
  {Delfosse}, {Desort}, {Lagrange}, {Lovis}, {Mayor}, {Pepe}, {Perrier},
  {Pont}, {Queloz}, {Santos}, {S{\'e}gransan}, {Udry}, \&
  {Vidal-Madjar}}]{Moutou2009}
{Moutou}, C., {H{\'e}brard}, G., {Bouchy}, F., {et~al.} 2009, \aap, 498, L5

\bibitem[{{Mugrauer} \& {Michel}(2020)}]{Mugrauer2020}
{Mugrauer}, M. \& {Michel}, K.-U. 2020, Astronomische Nachrichten, 341, 996

\bibitem[{{Naef} {et~al.}(2001){Naef}, {Latham}, {Mayor}, {Mazeh}, {Beuzit},
  {Drukier}, {Perrier-Bellet}, {Queloz}, {Sivan}, {Torres}, {Udry}, \&
  {Zucker}}]{Naef2001}
{Naef}, D., {Latham}, D.~W., {Mayor}, M., {et~al.} 2001, \aap, 375, L27

\bibitem[{{Narang} {et~al.}(2018){Narang}, {Manoj}, {Furlan}, {Mordasini},
  {Henning}, {Mathew}, {Banyal}, \& {Sivarani}}]{Narang2018}
{Narang}, M., {Manoj}, P., {Furlan}, E., {et~al.} 2018, \aj, 156, 221

\bibitem[{{O'Toole} {et~al.}(2009){O'Toole}, {Tinney}, {Jones}, {Butler},
  {Marcy}, {Carter}, \& {Bailey}}]{Toole2009}
{O'Toole}, S.~J., {Tinney}, C.~G., {Jones}, H.~R.~A., {et~al.} 2009, \mnras,
  392, 641

\bibitem[{{Pepe} {et~al.}(2002){Pepe}, {Mayor}, {Rupprecht}, {Avila},
  {Ballester}, {Beckers}, {Benz}, {Bertaux}, {Bouchy}, {Buzzoni}, {Cavadore},
  {Deiries}, {Dekker}, {Delabre}, {D'Odorico}, {Eckert}, {Fischer}, {Fleury},
  {George}, {Gilliotte}, {Gojak}, {Guzman}, {Koch}, {Kohler}, {Kotzlowski},
  {Lacroix}, {Le Merrer}, {Lizon}, {Lo Curto}, {Longinotti}, {Megevand},
  {Pasquini}, {Petitpas}, {Pichard}, {Queloz}, {Reyes}, {Richaud}, {Sivan},
  {Sosnowska}, {Soto}, {Udry}, {Ureta}, {van Kesteren}, {Weber}, {Weilenmann},
  {Wicenec}, {Wieland}, {Christensen-Dalsgaard}, {Dravins}, {Hatzes},
  {K{\"u}rster}, {Paresce}, \& {Penny}}]{Pepe2002}
{Pepe}, F., {Mayor}, M., {Rupprecht}, G., {et~al.} 2002, The Messenger, 110, 9

\bibitem[{{Perruchot} {et~al.}(2011){Perruchot}, {Bouchy}, {Chazelas},
  {D{\'\i}az}, {H{\'e}brard}, {Arnaud}, {Arnold}, {Avila}, {Delfosse},
  {Boisse}, {Moreaux}, {Pepe}, {Richaud}, {Santerne}, {Sottile}, \&
  {T{\'e}zier}}]{Perruchot2011}
{Perruchot}, S., {Bouchy}, F., {Chazelas}, B., {et~al.} 2011, in Society of
  Photo-Optical Instrumentation Engineers (SPIE) Conference Series, Vol. 8151,
  Techniques and Instrumentation for Detection of Exoplanets V, 815115

\bibitem[{{Perruchot} {et~al.}(2008){Perruchot}, {Kohler}, {Bouchy}, {Richaud},
  {Richaud}, {Moreaux}, {Merzougui}, {Sottile}, {Hill}, {Knispel}, {Regal},
  {Meunier}, {Ilovaisky}, {Le Coroller}, {Gillet}, {Schmitt}, {Pepe}, {Fleury},
  {Sosnowska}, {Vors}, {M{\'e}gevand}, {Blanc}, {Carol}, {Point}, {Laloge}, \&
  {Brunel}}]{Perruchot2008}
{Perruchot}, S., {Kohler}, D., {Bouchy}, F., {et~al.} 2008, in Society of
  Photo-Optical Instrumentation Engineers (SPIE) Conference Series, Vol. 7014,
  \procspie, 70140J

\bibitem[{{Perryman} {et~al.}(1997){Perryman}, {Lindegren}, {Kovalevsky},
  {Hoeg}, {Bastian}, {Bernacca}, {Cr{\'e}z{\'e}}, {Donati}, {Grenon}, {van
  Leeuwen}, {van der Marel}, {Mignard}, {Murray}, {Le Poole}, {Schrijver},
  {Turon}, {Arenou}, {Froeschl{\'e}}, \& {Petersen}}]{Perryman:1997kx}
{Perryman}, M.~A.~C., {Lindegren}, L., {Kovalevsky}, J., {et~al.} 1997, \aap,
  323, L49

\bibitem[{{Pollack} {et~al.}(1996){Pollack}, {Hubickyj}, {Bodenheimer},
  {Lissauer}, {Podolak}, \& {Greenzweig}}]{Pollack1996}
{Pollack}, J.~B., {Hubickyj}, O., {Bodenheimer}, P., {et~al.} 1996, \icarus,
  124, 62

\bibitem[{{Queloz} {et~al.}(2001){Queloz}, {Henry}, {Sivan}, {Baliunas},
  {Beuzit}, {Donahue}, {Mayor}, {Naef}, {Perrier}, \& {Udry}}]{Queloz2001}
{Queloz}, D., {Henry}, G.~W., {Sivan}, J.~P., {et~al.} 2001, \aap, 379, 279

\bibitem[{{Ranc} {et~al.}(2015){Ranc}, {Cassan}, {Albrow}, {Kubas}, {Bond},
  {Batista}, {Beaulieu}, {Bennett}, {Dominik}, {Dong}, {Fouqu{\'e}}, {Gould},
  {Greenhill}, {J{\o}rgensen}, {Kains}, {Menzies}, {Sumi}, {Bachelet},
  {Coutures}, {Dieters}, {Dominis Prester}, {Donatowicz}, {Gaudi}, {Han},
  {Hundertmark}, {Horne}, {Kane}, {Lee}, {Marquette}, {Park}, {Pollard},
  {Sahu}, {Street}, {Tsapras}, {Wambsganss}, {Williams}, {Zub}, {Abe}, {Fukui},
  {Itow}, {Masuda}, {Matsubara}, {Muraki}, {Ohnishi}, {Rattenbury}, {Saito},
  {Sullivan}, {Sweatman}, {Tristram}, {Yock}, \& {Yonehara}}]{Ranc2015}
{Ranc}, C., {Cassan}, A., {Albrow}, M.~D., {et~al.} 2015, \aap, 580, A125

\bibitem[{{Raymond} \& {Izidoro}(2017)}]{Raymond2017}
{Raymond}, S.~N. \& {Izidoro}, A. 2017, \icarus, 297, 134

\bibitem[{{Raymond} {et~al.}(2006){Raymond}, {Quinn}, \&
  {Lunine}}]{Raymond2006}
{Raymond}, S.~N., {Quinn}, T., \& {Lunine}, J.~I. 2006, \icarus, 183, 265

\bibitem[{{Sahlmann} \& {Fekel}(2013)}]{Sahlmann:2013fk3}
{Sahlmann}, J. \& {Fekel}, F.~C. 2013, \aap, 556, A145

\bibitem[{{Sahlmann} {et~al.}(2011{\natexlab{a}}){Sahlmann}, {Lovis}, {Queloz},
  \& {S\'egransan}}]{Sahlmann:2011lr}
{Sahlmann}, J., {Lovis}, C., {Queloz}, D., \& {S\'egransan}, D.
  2011{\natexlab{a}}, \aap, 528, L8+

\bibitem[{{Sahlmann} {et~al.}(2011{\natexlab{b}}){Sahlmann}, {S{\'e}gransan},
  {Queloz}, {Udry}, {Santos}, {Marmier}, {Mayor}, {Naef}, {Pepe}, \&
  {Zucker}}]{Sahlmann2011}
{Sahlmann}, J., {S{\'e}gransan}, D., {Queloz}, D., {et~al.} 2011{\natexlab{b}},
  \aap, 525, A95

\bibitem[{{Santos} {et~al.}(2017){Santos}, {Adibekyan}, {Figueira},
  {Andreasen}, {Barros}, {Delgado-Mena}, {Demangeon}, {Faria}, {Oshagh},
  {Sousa}, {Viana}, \& {Ferreira}}]{Santos2017}
{Santos}, N.~C., {Adibekyan}, V., {Figueira}, P., {et~al.} 2017, \aap, 603, A30

\bibitem[{{Santos} {et~al.}(2004){Santos}, {Israelian}, \&
  {Mayor}}]{Santos2004}
{Santos}, N.~C., {Israelian}, G., \& {Mayor}, M. 2004, \aap, 415, 1153

\bibitem[{{Santos} {et~al.}(2010){Santos}, {Mayor}, {Benz}, {Bouchy},
  {Figueira}, {Lo Curto}, {Lovis}, {Melo}, {Moutou}, {Naef}, {Pepe}, {Queloz},
  {Sousa}, \& {Udry}}]{Santos2010}
{Santos}, N.~C., {Mayor}, M., {Benz}, W., {et~al.} 2010, \aap, 512, A47

\bibitem[{{Santos} {et~al.}(2000){Santos}, {Mayor}, {Naef}, {Pepe}, {Queloz},
  {Udry}, \& {Blecha}}]{Santos2000}
{Santos}, N.~C., {Mayor}, M., {Naef}, D., {et~al.} 2000, \aap, 361, 265

\bibitem[{{Santos} {et~al.}(2013){Santos}, {Sousa}, {Mortier}, {Neves},
  {Adibekyan}, {Tsantaki}, {Delgado Mena}, {Bonfils}, {Israelian}, {Mayor}, \&
  {Udry}}]{Santos2013}
{Santos}, N.~C., {Sousa}, S.~G., {Mortier}, A., {et~al.} 2013, \aap, 556, A150

\bibitem[{{Schlaufman}(2018)}]{Schlaufman2018}
{Schlaufman}, K.~C. 2018, \apj, 853, 37

\bibitem[{{Schneider} {et~al.}(2011){Schneider}, {Dedieu}, {Le Sidaner},
  {Savalle}, \& {Zolotukhin}}]{Schneider2011}
{Schneider}, J., {Dedieu}, C., {Le Sidaner}, P., {Savalle}, R., \&
  {Zolotukhin}, I. 2011, \aap, 532, A79

\bibitem[{{S{\'e}gransan} {et~al.}(2011){S{\'e}gransan}, {Mayor}, {Udry},
  {Lovis}, {Benz}, {Bouchy}, {Lo Curto}, {Mordasini}, {Moutou}, {Naef}, {Pepe},
  {Queloz}, \& {Santos}}]{Segransan2011}
{S{\'e}gransan}, D., {Mayor}, M., {Udry}, S., {et~al.} 2011, \aap, 535, A54

\bibitem[{{Sneden}(1973)}]{Sneden1973}
{Sneden}, C.~A. 1973, PhD thesis, THE UNIVERSITY OF TEXAS AT AUSTIN.

\bibitem[{{Sousa}(2014)}]{Sousa2014}
{Sousa}, S.~G. 2014, {ARES + MOOG: A Practical Overview of an Equivalent Width
  (EW) Method to Derive Stellar Parameters}, 297--310

\bibitem[{{Sousa} {et~al.}(2015){Sousa}, {Santos}, {Adibekyan}, {Delgado-Mena},
  \& {Israelian}}]{Sousa2015}
{Sousa}, S.~G., {Santos}, N.~C., {Adibekyan}, V., {Delgado-Mena}, E., \&
  {Israelian}, G. 2015, \aap, 577, A67

\bibitem[{{Sousa} {et~al.}(2008){Sousa}, {Santos}, {Mayor}, {Udry},
  {Casagrande}, {Israelian}, {Pepe}, {Queloz}, \& {Monteiro}}]{Sousa2008}
{Sousa}, S.~G., {Santos}, N.~C., {Mayor}, M., {et~al.} 2008, \aap, 487, 373

\bibitem[{{Spiegel} {et~al.}(2011){Spiegel}, {Burrows}, \&
  {Milsom}}]{Spiegel2011}
{Spiegel}, D.~S., {Burrows}, A., \& {Milsom}, J.~A. 2011, \apj, 727, 57

\bibitem[{{Swastik} {et~al.}(2021){Swastik}, {Banyal}, {Narang}, {Manoj},
  {Sivarani}, {Reddy}, \& {Rajaguru}}]{Swastik2021}
{Swastik}, C., {Banyal}, R.~K., {Narang}, M., {et~al.} 2021, \aj, 161, 114

\bibitem[{{Tamuz} {et~al.}(2008){Tamuz}, {S{\'e}gransan}, {Udry}, {Mayor},
  {Eggenberger}, {Naef}, {Pepe}, {Queloz}, {Santos}, {Demory}, {Figuera},
  {Marmier}, \& {Montagnier}}]{Tamuz2008}
{Tamuz}, O., {S{\'e}gransan}, D., {Udry}, S., {et~al.} 2008, \aap, 480, L33

\bibitem[{{Tokovinin} \& {Latham}(2017)}]{Tokovinin2017}
{Tokovinin}, A. \& {Latham}, D.~W. 2017, \apj, 838, 54

\bibitem[{{Torres} {et~al.}(2010){Torres}, {Andersen}, \&
  {Gim{\'e}nez}}]{Torres2010}
{Torres}, G., {Andersen}, J., \& {Gim{\'e}nez}, A. 2010, \aapr, 18, 67

\bibitem[{{{\v{S}}ubjak} {et~al.}(2020){{\v{S}}ubjak}, {Sharma}, {Carmichael},
  {Johnson}, {Gonzales}, {Matthews}, {Boffin}, {Brahm}, {Chaturvedi},
  {Chakraborty}, {Ciardi}, {Collins}, {Esposito}, {Fridlund}, {Gan},
  {Gandolfi}, {Garc{\'\i}a}, {Guenther}, {Hatzes}, {Latham}, {Mathis},
  {Mathur}, {Persson}, {Relles}, {Schlieder}, {Barclay}, {Dressing},
  {Crossfield}, {Howard}, {Rodler}, {Zhou}, {Quinn}, {Esquerdo}, {Calkins},
  {Berlind}, {Stassun}, {Bla{\v{z}}ek}, {Skarka}, {{\v{S}}pokov{\'a}},
  {{\v{Z}}{\'a}k}, {Albrecht}, {Sobrino}, {Beck}, {Cabrera}, {Carleo},
  {Cochran}, {Csizmadia}, {Dai}, {Deeg}, {de Leon}, {Eigm{\"u}ller}, {Endl},
  {Erikson}, {Fukui}, {Georgieva}, {Gonz{\'a}lez-Cuesta}, {Grziwa}, {Hidalgo},
  {Hirano}, {Hjorth}, {Knudstrup}, {Korth}, {Lam}, {Livingston}, {Lund},
  {Luque}, {Rodr{\'\i}guez}, {Murgas}, {Narita}, {Nespral}, {Niraula}, {Nowak},
  {Pall{\'e}}, {P{\"a}tzold}, {Prieto-Arranz}, {Rauer}, {Redfield}, {Ribas},
  {Smith}, {Eylen}, \& {Kab{\'a}th}}]{Subjak2020}
{{\v{S}}ubjak}, J., {Sharma}, R., {Carmichael}, T.~W., {et~al.} 2020, \aj, 159,
  151

\bibitem[{{Wilson} {et~al.}(2016){Wilson}, {H{\'e}brard}, {Santos}, {Sahlmann},
  {Montagnier}, {Astudillo-Defru}, {Boisse}, {Bouchy}, {Rey}, {Arnold},
  {Bonfils}, {Bourrier}, {Courcol}, {Deleuil}, {Delfosse}, {D{\'\i}az},
  {Ehrenreich}, {Forveille}, {Moutou}, {Pepe}, {Santerne}, {S{\'e}gransan}, \&
  {Udry}}]{Wilson2016}
{Wilson}, P.~A., {H{\'e}brard}, G., {Santos}, N.~C., {et~al.} 2016, \aap, 588,
  A144

\bibitem[{{Wittenmyer} {et~al.}(2020){Wittenmyer}, {Wang}, {Horner}, {Butler},
  {Tinney}, {Carter}, {Wright}, {Jones}, {Bailey}, {O'Toole}, \&
  {Johns}}]{Wittenmyer2020}
{Wittenmyer}, R.~A., {Wang}, S., {Horner}, J., {et~al.} 2020, \mnras, 492, 377

\bibitem[{{Wright} {et~al.}(2013){Wright}, {Roy}, {Mahadevan}, {Wang}, {Ford},
  {Payne}, {Lee}, {Wang}, {Crepp}, {Gaudi}, {Eastman}, {Pepper}, {Ge},
  {Fleming}, {Ghezzi}, {Gonz{\'a}lez-Hern{\'a}ndez}, {Cargile}, {Stassun},
  {Wisniewski}, {Dutra-Ferreira}, {Porto de Mello}, {Maia}, {Nicolaci da
  Costa}, {Ogand o}, {Santiago}, {Schneider}, \& {Hearty}}]{Wright2013}
{Wright}, J.~T., {Roy}, A., {Mahadevan}, S., {et~al.} 2013, \apj, 770, 119

\bibitem[{{Zechmeister} \& {K{\"u}rster}(2009)}]{Zechmeister2009}
{Zechmeister}, M. \& {K{\"u}rster}, M. 2009, \aap, 496, 577

\end{thebibliography}
\begin{appendix}
\onecolumn

\section{Tables}\label{app:Tables}

\begin{table}[h]
\centering
 \caption{Basic characteristics of the SOPHIE (S) and SOPHIE+ (S+) observations of the 27 observed stars}
    \label{tab:datadetail}
    \begin{small}
    \begin{tabular}{l c c c c c}
    \hline\hline
    Name & Sp.type & N (S/S+)  &Time span & $\Delta$ RV &	<$\sigma_{\mathrm{RV}}$>(S/S+) \\ 
     & & &[days] &[m~s$^{-1}$]&[m~s$^{-1}$]\\
     \hline
BD-004475	&	G0	&	13	(	3	/	10	)	&	4053.95	&	315.4	&	8.00/3.19	\\
BD+031552	&	K5	&	11	(	0	/	11	)	&	2843.24	&	2031.64	&	7.76	\\
BD+450564	&	K1	&	14	(	0	/	14	)	&	3052.81	&	36.45	&	2.78	\\
BD+550362	&	K3	&	22	(	0	/	22	)	&	3050.81	&	24.67	&	2.39	\\
BD+631405	&	K0	&	16	(	0	/	16	)	&	1064.05	&	104.77	&	2.52	\\
HD~124330	&	G4IV	&	58	(	0	/	58	)	&	1443.04	&	16.96	&	2.34	\\
HD~140208	&	F5	&	11	(	4	/	7	)	&	3130.62	&	6642.73	&	10.75/12.78	\\
HD~151465	&	F5V	&	14	(	4	/	10	)	&	3228.11	&	4294.39	&	11.33/11.15	\\
HD~153915	&	F8	&	11	(	0	/	11	)	&	1204.83	&	4423.92	&	4.35	\\
HD~155193	&	F8IV	&	73	(	0	/	73	)	&	2815.13	&	16.86	&	2.88	\\
HD~162735	&	G5	&	47	(	2	/	45	)	&	3040.72	&	2011.4	&	8.17/3.10	\\
HD~166356	&	K0	&	19	(	6	/	13	)	&	4410.03	&	1353.23	&	7.64/2.91	\\
HD~184601	&	G0	&	15	(	0	/	15	)	&	2277.87	&	660.57	&	4.83	\\
HD~187057	&	G0	&	9	(	1	/	8	)	&	4761.05	&	3219.15	&	7.57/4.05	\\
HD~204277	&	F8V	&	92	(	0	/	92	)	&	1475.98	&	17.04	&	3.43	\\
HD~205521	&	G5	&	17	(	0	/	17	)	&	2536.18	&	183.99	&	2.24	\\
HD~211961	&	F8	&	10	(	3	/	7	)	&	4459.93	&	2103.23	&	9.12/6.29	\\
HD~238135	&	K0	&	8	(	0	/	8	)	&	2531.16	&	2538.24	&	3.38	\\
HD~25603	&	F5	&	21	(	6	/	15	)	&	4756.06	&	2284.57	&	8.73/5.30	\\
HD~26596	&	F9IV	&	18	(	0	/	18	)	&	3073.69	&	1560.14	&	3.89	\\
HD~30311	&	F9.5V	&	16	(	6	/	10	)	&	4144.78	&	1873.63	&	9.11/4.74	\\
HD~331093	&	K0	&	20	(	3	/	17	)	&	3516.24	&	26.02	&	8.21/2.63	\\
HD~352975	&	G5	&	9	(	3	/	6	)	&	4474.89	&	5433.19	&	8.37/3.35	\\
HD~5433	&	G5	&	20	(	7	/	13	)	&	4485.88	&	765.65	&	7.63/4.14	\\
HD~76332	&	G2V	&	11	(	4	/	7	)	&	3700.95	&	2079.7	&	8.32/4.14	\\
HD~8291	&	G5V	&	15	(	9	/	6	)	&	4416.0	&	1287.0	&	8.02/4.26	\\
HD~98451	&	K0	&	14	(	7	/	7	)	&	4308.07	&	1764.7	&	9.40/4.61	\\
        \hline
    \end{tabular}
    
    \parbox{\hsize}{\textbf{Notes:} N is the Number of RV measurements. $\Delta$ RV is the dispersion of radial velocities. <$\sigma_{\mathrm{RV}}$> is the mean of the uncertainties on radial velocities. 
  
    }
    \end{small}
   
\end{table}
\vspace{-1cm}
\begin{table}[h!]\centering

				\caption{Stellar parameters of the 27 observed targets}
    \label{tab:stellar}
				\begin{small}
				    \begin{tabular}{l c c c c c c c c c}
 
\hline
\hline
Name	&	T$_{\mathrm{eff}}$	&	log $g$	&	$v_{\mathrm{turb}}$			&	[Fe/H]		&	N$_{\mathrm{lines}}$ (FE I/	FE II)	&	 M$_\mathrm{Torres}$ & $\textrm{logR'}_{\textrm{HK}}$ & $v \sin i_{\star}$	\\
	&		[K]	&	[s.i.]	&	[km~s$^{-1}$]		&	[dex]	&			&	 [M$_{\odot}$] & [dex] & [km~s$^{-1}$]	\\
   \hline
   
BD-004475	&	5040	$\pm$ 	112	&	4.25	$\pm$ 	0.22	&	0.01	$\pm$ 	1.97	&	-0.14	$\pm$ 	0.06	&	112	/	14	&	0.810	$\pm$ 	0.097	&	-5.09	$\pm$ 	0.28	&	3.0	$\pm$ 	1.0	\\
BD+031552	&	4706	$\pm$ 	146	&	4.08	$\pm$ 	0.92	&	0.87	$\pm$ 	0.27	&	-0.26	$\pm$ 	0.06	&	89	/	2	&	0.904	$\pm$ 	0.497	&		-		&		-		\\
BD+450564	&	5004	$\pm$ 	50	&	4.22	$\pm$ 	0.13	&	0.40	$\pm$ 	0.16	&	-0.09	$\pm$ 	0.03	&	112	/	13	&	0.810	$\pm$ 	0.072	&	-4.98	$\pm$ 	0.11	&	2.0	$\pm$ 	1.0	\\
BD+550362	&	5012	$\pm$ 	78	&	4.09	$\pm$ 	0.16	&	0.42	$\pm$ 	0.22	&	0.18	$\pm$ 	0.05	&	112	/	13	&	0.909	$\pm$ 	0.099	&	-5.11	$\pm$ 	0.12	&	2.0	$\pm$ 	1.0	\\
BD+631405	&	5000	$\pm$ 	53	&	4.20	$\pm$ 	0.17	&	0.43	$\pm$ 	0.14	&	-0.09	$\pm$ 	0.03	&	94	/	9	&	0.816	$\pm$ 	0.083	&	-4.93	$\pm$ 	0.13	&	2.1	$\pm$ 	1.0	\\
HD~124330	&	5873	$\pm$ 	19	&	4.24	$\pm$ 	0.04	&	1.01	$\pm$ 	0.02	&	0.22	$\pm$ 	0.01	&	237	/	31	&	1.150	$\pm$ 	0.079	&	-5.27	$\pm$ 	0.23	&	3.2	$\pm$ 	1.0	\\
HD~140208	&	6713	$\pm$ 	85	&	4.95	$\pm$ 	0.06	&	1.76	$\pm$ 	0.14	&	-0.05	$\pm$ 	0.05	&	196	/	31	&	1.205	$\pm$ 	0.085	&	-5.09	$\pm$ 	0.32	&	9.2	$\pm$ 	1.0	\\
HD~151465	&	6622	$\pm$ 	101	&	4.88	$\pm$ 	0.07	&	1.98	$\pm$ 	0.17	&	0.00	$\pm$ 	0.06	&	196	/	29	&	1.181	$\pm$ 	0.085	&	-4.99	$\pm$ 	0.19	&	9.6	$\pm$ 	1.0	\\
HD~153915	&	6397	$\pm$ 	36	&	4.44	$\pm$ 	0.04	&	1.25	$\pm$ 	0.04	&	0.09	$\pm$ 	0.03	&	238	/	34	&	1.209	$\pm$ 	0.082	&	-5.37	$\pm$ 	0.22	&	2.3	$\pm$ 	1.0	\\
HD~155193	&	6239	$\pm$ 	20	&	4.26	$\pm$ 	0.04	&	1.36	$\pm$ 	0.02	&	0.05	$\pm$ 	0.01	&	234	/	25	&	1.221	$\pm$ 	0.083	&	-5.14	$\pm$ 	0.25	&	3.6	$\pm$ 	1.0	\\
HD~162735	&	5751	$\pm$ 	17	&	4.02	$\pm$ 	0.03	&	1.09	$\pm$ 	0.02	&	-0.01	$\pm$ 	0.01	&	242	/	34	&	1.154	$\pm$ 	0.080	&	-5.15	$\pm$ 	0.21	&	4.1	$\pm$ 	1.0	\\
HD~166356	&	5714	$\pm$ 	31	&	4.04	$\pm$ 	0.06	&	1.15	$\pm$ 	0.03	&	0.37	$\pm$ 	0.02	&	245	/	33	&	1.269	$\pm$ 	0.089	&	-5.17	$\pm$ 	0.13	&	3.3	$\pm$ 	1.0	\\
HD~184601	&	6035	$\pm$ 	50	&	4.17	$\pm$ 	0.04	&	1.11	$\pm$ 	0.09	&	-0.69	$\pm$ 	0.03	&	199	/	32	&	0.954	$\pm$ 	0.070	&	-5.07	$\pm$ 	0.13	&	1.4	$\pm$ 	1.0	\\
HD~187057	&	6170	$\pm$ 	32	&	4.13	$\pm$ 	0.04	&	1.29	$\pm$ 	0.03	&	0.09	$\pm$ 	0.02	&	243	/	33	&	1.302	$\pm$ 	0.087	&	-4.26	$\pm$ 	0.26	&	3.4	$\pm$ 	1.0	\\
HD~204277	&	6358	$\pm$ 	27	&	4.58	$\pm$ 	0.05	&	1.21	$\pm$ 	0.04	&	0.06	$\pm$ 	0.02	&	242	/	33	&	1.139	$\pm$ 	0.077	&	-4.50	$\pm$ 	0.11	&	4.6	$\pm$ 	1.0	\\
HD~205521	&	5570	$\pm$ 	36	&	4.20	$\pm$ 	0.07	&	0.90	$\pm$ 	0.05	&	0.36	$\pm$ 	0.03	&	241	/	32	&	1.100	$\pm$ 	0.082	&	-5.12	$\pm$ 	0.14	&	3.7	$\pm$ 	1.0	\\
HD~211961	&	6042	$\pm$ 	54	&	4.29	$\pm$ 	0.04	&	0.99	$\pm$ 	0.10	&	-0.49	$\pm$ 	0.04	&	205	/	33	&	0.959	$\pm$ 	0.071	&	-5.26	$\pm$ 	0.42	&	1.0	$\pm$ 	1.0	\\
HD~238135	&	5221	$\pm$ 	31	&	4.34	$\pm$ 	0.07	&	0.64	$\pm$ 	0.07	&	-0.10	$\pm$ 	0.02	&	235	/	33	&	0.828	$\pm$ 	0.061	&	-4.71	$\pm$ 	0.12	&	2.5	$\pm$ 	1.0	\\
HD~25603	&	6181	$\pm$ 	29	&	4.49	$\pm$ 	0.03	&	1.11	$\pm$ 	0.04	&	-0.22	$\pm$ 	0.02	&	189	/	24	&	1.013	$\pm$ 	0.071	&	-5.22	$\pm$ 	0.63	&	2.5	$\pm$ 	1.0	\\
HD~26596	&	6127	$\pm$ 	25	&	4.54	$\pm$ 	0.04	&	0.97	$\pm$ 	0.04	&	-0.10	$\pm$ 	0.02	&	241	/	34	&	1.023	$\pm$ 	0.071	&	-4.90	$\pm$ 	0.12	&	2.4	$\pm$ 	1.0	\\
HD~30311	&	6172	$\pm$ 	28	&	4.50	$\pm$ 	0.05	&	1.19	$\pm$ 	0.03	&	0.14	$\pm$ 	0.02	&	243	/	34	&	1.125	$\pm$ 	0.077	&	-4.51	$\pm$ 	0.12	&	6.2	$\pm$ 	1.0	\\
HD~331093	&	5544	$\pm$ 	33	&	4.20	$\pm$ 	0.08	&	0.81	$\pm$ 	0.05	&	0.17	$\pm$ 	0.03	&	246	/	33	&	1.030	$\pm$ 	0.079	&	-5.10	$\pm$ 	0.13	&	2.5	$\pm$ 	1.0	\\
HD~352975	&	5310	$\pm$ 	33	&	4.37	$\pm$ 	0.06	&	0.80	$\pm$ 	0.07	&	-0.04	$\pm$ 	0.02	&	241	/	33	&	0.854	$\pm$ 	0.063	&	-4.61	$\pm$ 	0.15	&	3.6	$\pm$ 	1.0	\\
HD~5433	&	5805	$\pm$ 	16	&	4.35	$\pm$ 	0.03	&	0.81	$\pm$ 	0.03	&	-0.08	$\pm$ 	0.01	&	251	/	33	&	0.983	$\pm$ 	0.069	&	-5.09	$\pm$ 	0.21	&	2.8	$\pm$ 	1.0	\\
HD~76332	&	5806	$\pm$ 	24	&	4.41	$\pm$ 	0.03	&	0.92	$\pm$ 	0.04	&	-0.03	$\pm$ 	0.02	&	250	/	32	&	0.980	$\pm$ 	0.069	&	-4.50	$\pm$ 	0.13	&	3.9	$\pm$ 	1.0	\\
HD~8291	&	5732	$\pm$ 	14	&	4.42	$\pm$ 	0.03	&	0.77	$\pm$ 	0.03	&	-0.10	$\pm$ 	0.01	&	238	/	26	&	0.937	$\pm$ 	0.066	&	-4.70	$\pm$ 	0.13	&	3.0	$\pm$ 	1.0	\\
HD~98451	&	5478	$\pm$ 	23	&	4.40	$\pm$ 	0.04	&	0.56	$\pm$ 	0.06	&	-0.09	$\pm$ 	0.02	&	246	/	33	&	0.878	$\pm$ 	0.063	&	-4.69	$\pm$ 	0.30	&	2.2	$\pm$ 	1.0	\\
\hline
    \end{tabular}
				\end{small}
		\end{table}

\setlength{\tabcolsep}{6pt}	
\begin{table}[h]\centering

				\caption{Stellar activity parameters of the 27 observed targets}
    \label{tab:democor}
			{\def\arraystretch{1.3}
				    \begin{tabular}{l c c c c c c c c c c}
 
\hline
\hline

Star	&	Data &	N & $\sigma_{a}$ & BIS$_{\mathrm{mean}}$	& BIS$_{\mathrm{std}}$ &	FWHM$_{\mathrm{mean}}$	& FWHM$_{\mathrm{std}}$ &	RV-BIS(p-value)	& RV-FWHM(p-value)\\
	&	&	 & m~s$^{-1}$ & km~s$^{-1}$	& m~s$^{-1}$ &	km~s$^{-1}$	& m~s$^{-1}$&		& \\
\hline

BD+450564	&	SOPHIE+	&	14	& 7.85 &			0.002	&	9.691	&	6.766	&	23.787	&	-0.012	(	0.969	)&	0.481	(	0.082   )\\					
BD+550362	&	SOPHIE+	&	22	&	7.55 &		-0.005	&	7.732	&	6.89	&	13.644	&	0.146	(	0.517	)&	0.512	(	0.015   )\\					
BD+631405	&	SOPHIE+	&	16	& 7.97 &			-0.001	&	8.207	&	6.726	&	16.154	&	0.362	(	0.168	)&	0.363	(	0.167   )\\	
HD~124330	&	SOPHIE+	&	58	&	5.61 &		-0.012	&	8.15	&	7.849	&	20.178	&	-0.001	(	0.996	)&	0.049	(	0.713   )\\	
HD~155193	&	SOPHIE+	&	73	&	7.22 &		0.046	&	12.4	&	8.668	&	44.66	&	0.337	(	0.004	)&	-0.215	(	0.068   )\\	
HD~204277	&	SOPHIE+	&	92	& 21.82 &			0.044	&	13.363	&	9.772	&	51.046	&	0.075	(	0.475	)&	0.293	(	0.005   )\\	
HD~331093	&	SOPHIE+	&	17	&	7.57 &		0.002	&	5.063	&	7.019	&	31.337	&	-0.112	(	0.669	)&	0.29	(	0.259   )\\					
	&	SOPHIE	&	3	& --	&		0.009	&	12.832	&	7.043	&	9.428	&	0.996	(	0.057	)&	0.977	(	0.138	)\\	
\hline
BD-004475	&	SOPHIE+	&	10	&	7.05 &		-0.043	&	19.212	&	7.391	&	45.266	&	-0.069	(	0.85	)&	0.125	(	0.73    )\\					
	&	SOPHIE	&	3	&--	&		-0.029	&	4.028	&	7.367	&	4.714	&	-0.515	(	0.656	)&	0.993	(	0.075   )\\		
HD~184601	&	SOPHIE+	&	15	&	7.23 &		0.025	&	22.414	&	7.615	&	23.055	&	0.167	(	0.553	)&	0.16	(	0.57    )\\		
HD~205521	&	SOPHIE+	&	17	&	6.79 &		-0.035	&	7.163	&	7.862	&	21.291	&	0.137	(	0.601	)&	0.131	(	0.617   )\\	
HD~5433	&	SOPHIE+	&	13	&	7.05 &		-0.013	&	15.494	&	7.623	&	40.265	&	-0.008	(	0.98	)&	0.308	(	0.305   )\\					
	&	SOPHIE	&	7	&--	&		-0.018	&	10.508	&	7.63	&	32.071	&	0.847	(	0.016	)&	-0.193	(	0.679   )\\	
\hline
BD+031552	&	SOPHIE+	&	11	& --&			0.004	&	35.463	&	6.162	&	41.521	&	0.315	(	0.345	)&	0.222	(	0.512   )\\					
HD~140208	&	SOPHIE+	&	7	& 7.88 &			0.029	&	31.295	&	14.867	&	51.19	&	0.458	(	0.302	)&	-0.232	(	0.617   )\\					
	&	SOPHIE	&	4	& --&			0.0		&	0.433	&	14.86	&	69.642	&	-0.997	(	0.003	)&	-0.809	(	0.191   )\\				
HD~151465	&	SOPHIE+	&	10	& 9.36 &			-0.017	&	73.547	&	15.321	&	155.013	&	0.795	(	0.006	)&	-0.42	(	0.227   )\\					
	&	SOPHIE	&	4	&	-- &		0.006	&	53.078	&	15.248	&	46.03	&	0.974	(	0.026	)&	-0.868	(	0.132   )\\					
HD~153915	&	SOPHIE+	&	11	&	4.86 &		0.043	&	18.207	&	8.4		&	45.925	&	0.095	(	0.781	)&	-0.548	(	0.081   )\\				
HD~162735	&	SOPHIE+	&	45	&	6.53 &		0.013	&	10.286	&	8.388	&	34.826	&	0.071	(	0.645	)&	0.149	(	0.328   )\\					
	&	SOPHIE	&	2	&	--&		0.01	&	5.5		&	8.37	&	10.0	&	--	&	--	\\	
HD~166356	&	SOPHIE+	&	13	&	7.41 &		0.033	&	8.391	&	7.585	&	42.538	&	-0.349	(	0.243	)&	0.056	(	0.855   )\\					
	&	SOPHIE	&	6	& --&			0.028	&	5.56	&	7.585	&	12.583	&	-0.012	(	0.982	)&	-0.099	(	0.852   )\\					
HD~187057	&	SOPHIE+	&	8	&	20.17 &		0.048	&	18.131	&	8.515	&	21.794	&	-0.331	(	0.423	)&	-0.75	(	0.032   )\\					
	&	SOPHIE	&	1	&	-- &		0.001	&	0.0		&	8.45	&	0.0		&	--	&	--	\\
HD~211961	&	SOPHIE+	&	7	& 5.87 &			0.024	&	20.041	&	7.714	&	59.966	&	-0.034	(	0.943	)&	-0.254	(	0.582   )\\					
	&	SOPHIE	&	3	&	--&		0.024	&	5.793	&	7.747	&	41.096	&	0.159	(	0.898	)&	-0.805	(	0.404   )\\					
HD~238135	&	SOPHIE+	&	8	& 8.51 &			0.011	&	13.444	&	6.936	&	32.38	&	-0.297	(	0.475	)&	0.693	(	0.057   )\\					
HD~25603	&	SOPHIE+	&	15	&	6.29 &		0.02	&	24.781	&	8.199	&	16.918	&	-0.255	(	0.36	)&	-0.176	(	0.531   )\\					
	&	SOPHIE	&	6	& --&			0.021	&	8.557	&	8.193	&	19.72	&	-0.093	(	0.86	)&	-0.214	(	0.684   )\\					
HD~26596	&	SOPHIE+	&	18	&	10.93 &		0.011	&	14.698	&	7.845	&	32.189	&	0.051	(	0.842	)&	-0.522	(	0.026   )\\					
HD~30311	&	SOPHIE+	&	10	&21.44 &			0.023	&	32.462	&	11.099	&	76.609	&	0.01	(	0.978	)&	0.729	(	0.017   )\\					
	&	SOPHIE	&	6	&	--&		0.022	&	18.991	&	11.1	&	11.547	&	-0.716	(	0.109	)&	0.004	(	0.993   )\\					
HD~352975	&	SOPHIE+	&	6	&	12.95 &		-0.016	&	21.616	&	7.838	&	55.503	&	0.361	(	0.482	)&	0.589	(	0.219   )\\					
	&	SOPHIE	&	3	&	--&		-0.02	&	5.437	&	7.793	&	54.365	&	0.726	(	0.483	)&	0.959	(	0.184   )\\					
HD~76332	&	SOPHIE+	&	7	&	14.88&		0.021	&	20.181	&	8.546	&	23.212	&	-0.445	(	0.317	)&	-0.004	(	0.994   )\\					
	&	SOPHIE	&	4	&	--&		0.038	&	19.313	&	8.602	&	10.897	&	-0.929	(	0.071	)&	0.684	(	0.316   )\\					
HD~8291	&	SOPHIE+	&	6	& 11.55&			-0.026	&	26.357	&	7.678	&	12.134	&	-0.442	(	0.38	)&	-0.175	(	0.74    )\\					
	&	SOPHIE	&	9	&	--&		-0.009	&	9.601	&	7.732	&	55.733	&	-0.272	(	0.479	)&	-0.348	(	0.359   )\\					
HD~98451	&	SOPHIE+	&	7	& 8.56 &			-0.002	&	20.767	&	6.84	&	38.545	&	-0.449	(	0.312	)&	0.19	(	0.684   )\\					
	&	SOPHIE	&	7	&	--&		0.0	&	12.938	&	6.861	&	29.966	&	-0.539	(	0.212	)&	-0.513	(	0.239	)\\				
\hline
    \end{tabular}}
    \parbox{\hsize}{\textbf{Notes:} N is the Number of RV measurements. $\sigma_{a}$ is the expected activity-related RV scatter which is estimated from the mean $\textrm{logR'}_{\textrm{HK}}$ value for each star. BIS$_{\mathrm{mean}}$ is the mean of bisector span and BIS$_{\mathrm{std}}$ is the variation in bisector span. FWHM$_{\mathrm{mean}}$ is the mean of full width at half maximum and FWHM$_{\mathrm{std}} $ is the variation in FWHM. RV-BIS and RV-FWHM is the Pearson correlation coefficient between RV-BIS and RV-FWHM respectively, and the significance of correlation is computed by p-value (For N $\ge$ 3).
  
    }

		\end{table}

	\begin{landscape}
	\setlength{\tabcolsep}{3pt}	
		\begin{table*}\centering

				\caption{\label{tab:StellarC}Table of Keplerian orbital fits for the  stellar companions with \msini $\ge$  75  M$_{\mathrm{J}}$.}
				\begin{small}
						{\def\arraystretch{2}
					\begin{tabular}{@{}lccccccccccc@{}}
					\hline\hline
  Name   &    P    &       K    &    $e$   &      $\omega$   &         $T_p$   &   $\gamma_\text{S+}$   &  $\gamma_\text{S}$      & $\sigma_\text{O-C, S+}$   &  $\sigma_\text{O-C, S}$   &   \msini  &      $a$   \\  
 
  & [days] & [m~s$^{-1}$] & & [$\deg$] & [BJD-2500000] &[km~s$^{-1}$] & [km~s$^{-1}$]& [m~s$^{-1}$] & [m~s$^{-1}$] & [ M$_{\mathrm{J}}$] & [A.U.]
  \\
 \hline
\multicolumn{12}{l}{Targets with both SOPHIE and SOPHIE+ RV measurements} \\
\hline
   HD~151465     	  	&  	54.439	  $\pm$   	0.002	  	&  	6067.05	  $\pm$   	18.56	  	& 	0.243	  $\pm$ 	0.005	  	&  	24.94	  $\pm$  	0.7	  	& 	57984.77	  $\pm$    	0.15	  	&  	-10.062	  $\pm$    	0.02	  	&   	-10.074	  $\pm$     	0.02	  	&   	11.36	  	&   	16.0	  	& 	122.43	  $\pm$     	7.15	  	& 	0.31	  $\pm$  	0.01	\\
   HD~166356     	  	&  	261.539	  $\pm$   	0.062	  	&  	2756.72	  $\pm$   	15.9	  	& 	0.452	  $\pm$ 	0.004	  	&  	95.55	  $\pm$  	0.41	  	& 	58341.61	  $\pm$    	0.16	  	&  	-6.251	  $\pm$    	0.034	  	&   	-6.28	  $\pm$     	0.012	  	&   	10.84	  	&   	11.13	  	& 	90.68	  $\pm$     	5.16	  	& 	0.89	  $\pm$  	0.04	\\
   HD~140208     	  	&  	326.422	  $\pm$   	0.02	  	&  	8807.57	  $\pm$   	13.12	  	& 	0.281	  $\pm$ 	0.003	  	&  	-176.2	  $\pm$  	0.6	  	& 	57564.83	  $\pm$    	0.37	  	&  	-15.84	  $\pm$    	0.021	  	&   	-15.835	  $\pm$     	0.015	  	&   	13.11	  	&   	10.02	  	& 	324.04	  $\pm$     	19.06	  	& 	1.06	  $\pm$  	0.05	\\
    HD~98451     	  	&  	1179.614	  $\pm$   	0.247	  	&  	2607.56	  $\pm$   	5.47	  	& 	0.189	  $\pm$ 	0.003	  	&  	-94.48	  $\pm$  	0.89	  	& 	58444.96	  $\pm$    	2.28	  	&  	-4.302	  $\pm$    	0.013	  	&   	-4.304	  $\pm$     	0.006	  	&   	12.64	  	&   	5.85	  	& 	121.91	  $\pm$     	7.01	  	& 	2.18	  $\pm$  	0.1	\\
     HD~8291     	  	&  	1862.529	  $\pm$   	2.896	  	&  	2211.55	  $\pm$   	12.56	  	& 	0.632	  $\pm$ 	0.001	  	&  	-62.51	  $\pm$  	0.56	  	& 	56932.45	  $\pm$    	3.07	  	&  	7.526	  $\pm$    	0.021	  	&   	7.503	  $\pm$     	0.019	  	&   	3.47	  	&   	8.56	  	& 	99.48	  $\pm$     	5.86	  	& 	2.99	  $\pm$  	0.13	\\
    HD~30311     	  	&  	2425.713	  $\pm$   	26.655	  	&  	3264.05	  $\pm$   	87.3	  	& 	0.383	  $\pm$ 	0.011	  	&  	-54.96	  $\pm$  	2.58	  	& 	58828.91	  $\pm$    	12.5	  	&  	42.415	  $\pm$    	0.146	  	&   	42.442	  $\pm$     	0.087	  	&   	23.63	  	&   	27.92	  	& 	215.76	  $\pm$     	14.07	  	& 	3.89	  $\pm$  	0.17	\\
   HD~162735     	  	&  	4197.884	  $\pm$   	33.269	  	&  	3427.42	  $\pm$   	4.12	  	& 	0.65	  $\pm$ 	0.002	  	&  	1.73	  $\pm$  	0.19	  	& 	58373.21	  $\pm$    	0.72	  	&  	-51.354	  $\pm$    	0.02	  	&   	-51.341	  $\pm$     	0.004	  	&   	0.22	  	&   	8.76	  	& 	227.55	  $\pm$     	13.08	  	& 	5.66	  $\pm$  	0.25	\\								
\hline
\multicolumn{12}{l}{Targets with only SOPHIE+ RV measurements} \\
\hline
  
   HD~153915     	  	&  	30.1313	  $\pm$   	0.0004  	&  	5859.78	  $\pm$   	4.17	  	& 	0.036	  $\pm$ 	0.001	  	&  	79.09	  $\pm$  	1.32	  	& 	58194.06	  $\pm$    	0.11	  	&  	-55.477	  $\pm$    	0.004	  	&   	    --     	  	&   	5.76	  	&   	  --      	  	& 	101.7	  $\pm$     	5.72	  	& 	0.21	  $\pm$  	0.01	\\
   HD~187057     	  	&  	48.3994	  $\pm$   	0.0003  	&  	4318.52	  $\pm$   	3.61	  	& 	0.098	  $\pm$ 	0.001	  	&  	-108.97	  $\pm$  	0.55	  	& 	58352.82	  $\pm$    	0.07	  	&  	21.844	  $\pm$    	0.005	  	&   	    --     	  	&   	3.72	  	&   	  --      	  	& 	92.01	  $\pm$     	5.09	  	& 	0.29	  $\pm$  	0.01	\\
   HD~352975     	  	&  	302.699	  $\pm$   	0.02	  	&  	7220.52	  $\pm$   	8.29	  	& 	0.252	  $\pm$ 	0.001	  	&  	-156.7	  $\pm$  	0.28	  	& 	58222.15	  $\pm$    	0.28	  	&  	-8.296	  $\pm$    	0.007	  	&   	    --     	  	&   	5.25	  	&   	  --      	  	& 	207.92	  $\pm$     	12.4	  	& 	0.9	  $\pm$  	0.04	\\
   HD~211961     	  	&  	787.368	  $\pm$   	0.3	  	&  	3287.9	  $\pm$   	5.12	  	& 	0.353	  $\pm$ 	0.001	  	&  	9.83	  $\pm$  	0.2	  	& 	57884.8	  $\pm$    	0.64	  	&  	23.246	  $\pm$    	0.005	  	&   	    --     	  	&   	4.29	  	&   	  --      	  	& 	135.77	  $\pm$     	8.2	  	& 	1.72	  $\pm$  	0.08	\\
BD+031552       	& 	   	879.028	$\pm$     	0.39	& 	   	3432.3	$\pm$     	8.32	& 	  	0.473	$\pm$   	0.002	& 	   	-35.44	$\pm$    	0.62	& 	  	57630.24	$\pm$      	1.3	& 	   	-15.834	$\pm$      	0.017	& 	    	 --     	& 	    	6.07	& 	    	--        	& 	  	133.21	$\pm$       	49.02	& 	  	1.81	$\pm$    	0.58	\\
    HD~26596     	  	&  	889.823	  $\pm$   	0.256	  	&  	2808.58	  $\pm$   	6.24	  	& 	0.441	  $\pm$ 	0.002	  	&  	-113.04	  $\pm$  	0.35	  	& 	57332.16	  $\pm$    	1.23	  	&  	-2.581	  $\pm$    	0.007	  	&   	    --     	  	&   	6.57	  	&   	  --      	  	& 	121.09	  $\pm$     	6.94	  	& 	1.89	  $\pm$  	0.08	\\
   HD~238135     	  	&  	1805.079	  $\pm$   	0.686	  	&  	5657.82	  $\pm$   	223.49	  	& 	0.393	  $\pm$ 	0.019	  	&  	-91.0	  $\pm$  	1.84	  	& 	57347.67	  $\pm$    	7.52	  	&  	-10.576	  $\pm$    	0.233	  	&   	    --     	  	&   	2.27	  	&   	  --      	  	& 	274.88	  $\pm$     	18.4	  	& 	2.99	  $\pm$  	0.14	\\
    HD~25603     	  	&  	1840.358	  $\pm$   	0.656	  	&  	4574.33	  $\pm$   	10.72	  	& 	0.208	  $\pm$ 	0.002	  	&  	-129.27	  $\pm$  	0.51	  	& 	58520.13	  $\pm$    	2.22	  	&  	74.276	  $\pm$    	0.012	  	&   	    --     	  	&   	10.51	  	&   	  --      	  	& 	272.31	  $\pm$     	15.66	  	& 	3.19	  $\pm$  	0.14	\\
    HD~76332     	  	&  	2489.182	  $\pm$   	7.957	  	&  	3318.31	  $\pm$   	10.97	  	& 	0.143	  $\pm$ 	0.004	  	&  	53.37	  $\pm$  	2.17	  	& 	57319.61	  $\pm$    	17.78	  	&  	13.135	  $\pm$    	0.023	  	&   	    --     	  	&   	7.16	  	&   	  --      	  	& 	216.01	  $\pm$     	12.46	  	& 	3.8	  $\pm$  	0.17	\\									
\hline
					\end{tabular}}
					
				    \parbox{\hsize}{\textbf{Notes:} The uncertainty in the host star mass is taken into account while obtaining the uncertainty on \msini and $a$.
  
    }
				\end{small}
		\end{table*}
	\end{landscape}

\onecolumn
\thispagestyle{empty}
\newpage
\begin{table}\caption{Parameters of the HIPPARCOS astrometric observations. A $\star$ in the last column indicates an orbit determination.}
\label{tab:hip_params} 
\centering  
\begin{tabular}{r r r r r r r} 	
\hline\hline %
Name & HIP & $S_\mathrm{n}$& $N_\mathrm{orb}$ & $\sigma_{\Lambda}$  & $N_\mathrm{Hip}$ & $M_{\text{c},\mathrm{max}}$\\  
       &     &  &           & (mas)                 &           &   ($M_\sun$)\\  
\hline 

 BD+450564 & 010245 & 5 & 3.6 & 5.4 & 94 & 0.27 \\ 
 BD+550362 & 007441 & 5 & 4.5 & 7.9 & 144 & 0.43 \\ 
 BD+631405 & 088617 & 5 & 1.0 & 5.5 & 154 & 0.29 \\ 
 HD~124330 & 069322 & 5 & 4.3 & 4.4 & 124 & 0.34 \\ 
 HD~155193 & 083953 & 5 & 2.6 & 2.4 & 95 & 0.15 \\ 
 HD~331093 & 096846 & 5 & 1.9 & 5.1 & 133 & (8.52) \\ 
 \hline
 BD-004475 & 114458 & 5 & 1.5 & 5.4 & 78 & 0.16 \\ 
 HD~184601 & 096049 & 5 & 1.4 & 3.7 & 106 & 0.20 \\ 
 HD~205521 & 105906 & 1 & 0.6 & 5.5 & 132 & $\cdots$ $\star$\\ 
 HD~5433 & 004387 & 5 & 2.0 & 3.5 & 70 & 0.66 \\ 
 \hline
 BD+031552 & 034341 & 7 & 1.2 & 5.6 & 88 & 0.09 $\star$ \\ 
 HD~140208 & 076748 & 5 & 3.6 & 3.3 & 117 & 0.23 \\ 
 HD~151465 & 082186 & 5 & 16.9 & 3.2 & 95 & 0.96 \\ 
 HD~153915 & 083341 & 5 & 30.7 & 2.9 & 94 & 1.72 \\ 
 HD~162735 & 087420 & 5 & 0.3 & 4.5 & 118 & $\cdots$ \\ 
 HD~166356 & 088610 & 5 & 4.4 & 3.0 & 178 & 0.21 \\ 
 HD~187057 & 097389 & 5 & 20.1 & 4.2 & 149 & (3.55) \\ 
 HD~211961 & 110330 & 5 & 1.4 & 3.7 & 83 & 0.17 \\ 
 HD~238135 & 061274 & 9 & 0.7 & 8.2 & 165 & $\cdots$ $\star$\\ 
 HD~25603 & 019134 & 5 & 0.5 & 3.9 & 84 & $\cdots$ $\star$\\ 
 HD~26596 & 019842 & 1 & 1.3 & 6.4 & 94 & 0.25 $\star$\\ 
 HD~30311 & 022221 & 7 & 0.4 & 2.1 & 67 & $\cdots$ $\star$\\ 
 HD~352975 & 102458 & 5 & 3.6 & 6.5 & 134 & 0.48 \\ 
 HD~76332 & 043882 & 1 & 0.4 & 4.1 & 87 & $\cdots$ $\star$\\ 
 HD~8291 & 006407 & 5 & 0.6 & 3.9 & 92 & $\cdots$ \\ 
 HD~98451 & 055323 & 1 & 0.9 & 4.7 & 80 & 0.20 $\star$\\ 
\hline
\end{tabular} 
 \parbox{\hsize}{\textbf{Notes:} The table is divided in three parts to distinguish CJs, BDs, and SCs, after the RV analysis (from top to bottom).}
\end{table}

\begin{table*}\caption{Updated parallax and proper motion values and the astrometric orbit parameters ($I_{p}$, $\Omega$) for 8 sources with significant orbit detections.}
\label{tab:hip_ppm} 
\centering  
\small
\begin{tabular}{r r r r r r rrrrrrr} 	
\hline\hline %

Object  & $\Delta \alpha^{\star}$ & $\Delta \delta$ & $\varpi$ &$\Delta \varpi$  & $\Delta \mu_{\alpha^\star}$ & $\Delta \mu_{\delta}$ & $I_{p}$ & $\Omega$  \\  
        &  (mas)                     & (mas)             &  (mas)    &  (mas)            &  (mas $\mathrm{yr}^{-1}$)       & (mas $\mathrm{yr}^{-1}$) & (deg)	      & (deg)	    \\  
\hline 
\hline 
\multicolumn{9}{c}{$>$3-$\sigma$ detections}\\
\hline

 HD~205521 & $-5.2^{+ 0.8}_{-0.8}$ & $-4.5^{+ 0.9}_{-0.9}$  & $21.94^{+ 0.65}_{-0.65}$ & $-1.80$ & $17.1^{+ 1.5}_{-1.5}$ & $-12.1^{+ 1.1}_{-1.1}$ & $ 176.1^{+ 0.2}_{-0.3}$ & $ 278.0^{+ 8.7}_{-8.7}$  \vspace{1mm} \\ 
 HD~26596 & $-5.5^{+ 1.0}_{-1.0}$ & $-3.7^{+ 1.0}_{-1.0}$  & $19.21^{+ 1.06}_{-1.06}$ & $2.49$ & $0.9^{+ 0.9}_{-0.9}$ & $1.7^{+ 0.7}_{-0.7}$ & $ 154.1^{+ 2.1}_{-2.5}$ & $ 136.2^{+ 9.6}_{-2.6}$  \vspace{1mm} \\ 
 HD~98451 & $0.5^{+ 1.3}_{-1.3}$ & $-8.5^{+ 1.5}_{-1.5}$  & $17.49^{+ 1.66}_{-1.67}$ & $6.88$ & $-4.7^{+ 1.5}_{-1.5}$ & $6.8^{+ 1.6}_{-1.6}$ & $ 30.9^{+ 3.3}_{-2.8}$ & $ 322.6^{+ 9.2}_{-9.2}$  \vspace{1mm} \\

\hline 
\multicolumn{9}{c}{2--3-$\sigma$ detections}\\
\hline
 
HD~238135 & $-1.7^{+ 0.8}_{-0.8}$ & $-1.7^{+ 0.9}_{-0.9}$  & $19.72^{+ 1.13}_{-1.13}$ & $0.34$ & $-13.9^{+ 1.5}_{-1.5}$ & $3.2^{+ 1.9}_{-1.9}$ & $ 116.4^{+ 4.2}_{-4.5}$ & $ 99.8^{+ 7.2}_{-7.2}$  \vspace{1mm} \\ 
 HD~25603 & $11.6^{+ 3.6}_{-3.5}$ & $-8.7^{+ 2.6}_{-2.6}$  & $15.31^{+ 0.89}_{-0.89}$ & $0.34$ & $-7.3^{+ 1.8}_{-2.1}$ & $-7.2^{+ 4.8}_{-4.9}$ & $ 48.0^{+ 13.5}_{-10.2}$ & $ 277.4^{+ 13.5}_{-14.7}$  \vspace{1mm} \\  
 HD~30311 & $9.0^{+ 4.8}_{-4.7}$ & $-36.2^{+ 5.9}_{-5.7}$  & $24.97^{+ 0.77}_{-0.76}$ & $-0.15$ & $19.2^{+ 2.9}_{-2.9}$ & $12.4^{+ 2.8}_{-2.8}$ & $ 153.2^{+ 3.0}_{-3.9}$ & $ 148.9^{+ 11.1}_{-11.1}$  \vspace{1mm} \\ 
 HD~76332 & $-3.6^{+ 5.3}_{-5.2}$ & $-22.9^{+ 4.7}_{-4.8}$  & $19.18^{+ 1.18}_{-1.18}$ & $2.25$ & $15.2^{+ 4.2}_{-4.0}$ & $-1.1^{+ 3.8}_{-3.8}$ & $ 144.5^{+ 6.3}_{-8.2}$ & $ 207.1^{+ 17.1}_{-10.4}$  \vspace{1mm} \\  
\hline
BD+031552 & $-2.6^{+ 2.0}_{-2.0}$ & $3.0^{+ 1.1}_{-1.1}$  & $36.64^{+ 1.47}_{-1.47}$ & $-1.44$ & $1.3^{+ 1.7}_{-1.7}$ & $8.2^{+ 1.2}_{-1.2}$ & $ 68.4^{+ 12.5}_{-10.7}$ & $ 357.5^{+ 10.5}_{-10.4}$  \vspace{1mm} \\

  \hline
\end{tabular} 

\end{table*}

\newpage
\setlength{\tabcolsep}{5pt}	

\begin{table}
\caption{\label{tab:list_planets} List of sources studied with GASTON. Where uncertainties are missing, we assume 10\% errors on the corresponding  parameter. The parallax with uncertainties are all taken from the Gaia DR1, while those entered without uncertainties are taken from the Gaia DR2 \citep{Gaia2018} and a relative uncertainty $\sim$10\% is assumed for them (see Sect.~\ref{sec:gaia} for explanations).} 
\centering
\begin{tabular}{@{}l|@{~~}c@{~~}c@{~~}c@{~~}c@{~~}|c@{~~}c@{~~}c@{~~}c@{~~}c@{}}  
\hline
Name & $\pi$    	& $a_\star\sin i$  & Transit  & Drift  	& $\varepsilon_\text{DR1}$ 	& $N_\text{pts}$ 	& $N_\text{FoV}$ 	& Gaia   & Duplicate \\
	  &  (mas)	&   (mas) 			& flag & flag &  (mas)          	&                      		&				&  dataset & source \\
\hline
\multicolumn{10}{l}{DETECTION SAMPLE ($\varepsilon_\text{DR1,prim}$$>$$0.85$\,mas, $\varepsilon_\text{DR1,second}$$>$$1.2$\,mas)} \\
\hline

HD~352975		& 12.40$\pm$0.34 	& 2.41$\pm$0.25 	& n 	& n 	& 0.97 	& 63 	& 8 	& 1 	& n\\
HD~26596		& 14.40$\pm$0.42 	& 2.97$\pm$0.29 	& n 	& n 	& 1.16 	& 366 	& 43 	& 1 	& y\\
HD~140208		& 15.02$\pm$0.41 	& 3.81$\pm$0.37 	& n 	& n 	& 1.45 	& 88 	& 15 	& 1 	& n\\
HD~8291	& 15.97$\pm$0.84 	& 4.63$\pm$0.50 	& n 	& n 	& 1.98 	& 79 	& 10 	& 1 	& n\\
HD~76332		& 17 	& 13 	& n 	& n 	& 2.19 	& 196 	& 30 	& 2 	& y\\
HD~25603		& 12.99$\pm$0.76 	& 10.0$\pm$1.1 	& n 	& n 	& 2.23 	& 198 	& 24 	& 1 	& y\\
HD~98451		& 20.48$\pm$0.67 	& 5.69$\pm$0.57 	& n 	& n 	& 2.29 	& 90 	& 12 	& 1 	& n\\
HD~205521	& 19.06$\pm$0.64 	& 1.42$\pm$0.15 	& n 	& n 	& 2.60 	& 106 	& 17 	& 1 	& n\\
HD~238135	& 13.99$\pm$0.66 	& 12.3$\pm$1.4 	& n 	& n 	& 3.02 	& 178 	& 23 	& 1 	& y\\
\hline
\multicolumn{10}{l}{NONDETECTION SAMPLE ($\varepsilon_\text{DR1,prim}$$<$$0.85$\,mas, $\varepsilon_\text{DR1,second}$$<$$1.2$\,mas)} \\
\hline

HD~166356		& 16 	& 0.93 	& n 	& n 	& 0.24 	& 16 	& 5 	& 2 	& n\\
BD+450564		& 18.73$\pm$0.24 	& 0.0249$\pm$0.0031 	& n 	& n 	& 0.25 	& 59 	& 11 	& 1 	& n\\
HD~5433		& 16.69$\pm$0.28 	& 1.07$\pm$0.11 	& n 	& n 	& 0.34 	& 42 	& 9 	& 1 	& n\\
BD+631405& 26.24$\pm$0.25 	& 0.250$\pm$0.034 	& n 	& n 	& 0.39 	& 36 	& 8 	& 1 	& y\\
HD~124330	& 16.50$\pm$0.22 	& 0.00880$\pm$0.00094 	& n 	& n 	& 0.49 	& 162 	& 20 	& 1 	& y\\
HD~30311		& 25 	& 17 	& n 	& n 	& 0.50 	& 133 	& 16 	& 2 	& n\\
BD+031552		& 39.18$\pm$0.31 	& 9.6$\pm$6.6 	& n 	& n 	& 0.52 	& 44 	& 7 	& 1 	& n\\
HD~184601 	& 13.38$\pm$0.24 	& 1.39$\pm$0.14 	& n 	& n 	& 0.52 	& 101 	& 15 	& 1 	& n\\
HD~331093	& 1.54$\pm$0.27 	& 0.00309$\pm$0.00064 	& n 	& n 	& 0.53 	& 62 	& 11 	& 1 	& n\\
HD~155193	& 17.54$\pm$0.56 	& 0.0107$\pm$0.0012 	& n 	& n 	& 0.55 	& 53 	& 7 	& 1 	& n\\
HD~187057	& 8.82$\pm$0.29 	& 0.169$\pm$0.016 	& n 	& n 	& 0.59 	& 105 	& 15 	& 1 	& n\\
HD~151465	& 16.85$\pm$0.31 	& 0.496$\pm$0.048 	& n 	& n 	& 0.61 	& 86 	& 13 	& 1 	& n\\
HD~162735		& 10.11$\pm$0.30 	& 10.15$\pm$0.99 	& n 	& n 	& 0.72 	& 88 	& 10 	& 1 	& n\\
BD-004475	 	& 22.86$\pm$0.37 	& 1.04$\pm$0.16 	& n 	& n 	& 0.72 	& 83 	& 11 	& 1 	& y\\
BD+550362		& 19.40$\pm$0.27 	& 0.0111$\pm$0.0017 	& n 	& n 	& 0.72 	& 284 	& 36 	& 1 	& y\\
HD~153915	& 14.10$\pm$0.44 	& 0.229$\pm$0.022 	& n 	& n 	& 0.74 	& 98 	& 14 	& 1 	& n\\
\hline
\end{tabular}
	\end{table}


\begin{table}\small\centering
\caption{\label{tab:moreGAIA1}Nondetected orbits with astrometric excess noise $\varepsilon_\text{DR1,prim}$$<$$0.85$\,mas and $\varepsilon_\text{DR1,second}$$<$$1.2$\,mas. 
The companion true mass can only be bounded from above with here given the $3\sigma$ upper-limit.}
{\def\arraystretch{1.0}
\begin{tabular}{lcccccccccc}
\\
Planet name & Period & $m\sin i$ & $a\sin i$ & $\epsilon$ & \multicolumn{1}{c}{$a_\text{phot}$} & \multicolumn{1}{c}{$I_c$} & \multicolumn{1}{c}{$M_\text{c,true}$}  & \multicolumn{1}{c}{$\Delta V$} & MCMC  \\
			& (days) & (M$_\text{J}$) & (mas)  & (mas)  	& (mas)		 & ($^\circ$) &   \multicolumn{1}{c}{(M$_\text{J}$)}  & &   Acceptance   \\
	 		&		&			&		&		 & 3-$\sigma$ &		3-$\sigma$		&	3-$\sigma$  & 3-$\sigma$& rate\\
\hline
\multicolumn{10}{l}{PRIMARY DATASET}\\
\\
BD+450564 & 307.8 & 1.36 & 0.02492 & 0.2530 & $<$0.6762 & $>$2.579 & $<$31.41 & $>$22.07 & 0.1650  \\
BD+550362 & 265.6 & 0.72 & 0.01149 & 0.7212 & $<$1.160 & $>$0.6012 & $<$72.45 & $>$17.70 & 0.02573  \\
BD+631405 & 1198.5 & 3.96 & 0.2508 & 0.3884 & $<$2.949 & $>$5.974 & $<$40.23 & $>$21.49 & 0.2425  \\
HD~124330 & 270.6 & 0.75 & 0.008802 & 0.4926 & $<$0.6918 & $>$0.8084 & $<$53.76 & $>$22.82 & 0.1360  \\
HD~155193 & 352.65 & 0.75 & 0.01073 & 0.5476 & $<$1.044 & $>$0.7112 & $<$63.21 & $>$19.92 & 0.1254  \\
HD~331093 & 621.62 & 1.5 & 0.003091 & 0.5275 & $<$0.6038 & $>$0.3704 & $<$270.5 & $>$6.451 & 0.1444  \\
\hline
BD-004475 & 723.2 & 25.05 & 1.036 & 0.7189 & $<$5.605 & $>$13.31 & $<$124.8 & $>$8.225 & 0.2210  \\
HD~5433 & 576.6 & 49.11 & 1.071 & 0.3396 & $<$5.444 & $>$13.83 & $<$236.3 & $>$6.919 & 0.2070  \\
HD~184601 & 849.35 & 60.27 & 1.394 & 0.5220 & $<$6.187 & $>$14.90 & $<$276.1 & $>$6.184 & 0.2003  \\
\hline
BD+031552 & 879.0 & 133.2 & 9.565 & 0.5194 & $<$28.01 & $>$17.44 & $<$579.6 & $>$2.614 & 0.2318  \\
HD~151465 & 54.44 & 122.4 & 0.4957 & 0.6112 & $<$1.819 & $>$16.02 & $<$571.6 & $>$4.125 & 0.1904  \\
HD~153915 & 30.13 & 101.7 & 0.2286 & 0.7443 & $<$1.123 & $>$10.11 & $<$808.2 & $>$2.539 & 0.1029  \\
HD~162735 & 4197.8 & 227.5 & 10.10 & 0.7168 & $<$21.52 & $>$24.31 & $<$764.7 & $>$2.658 & 0.2735  \\
HD~187057 & 48.40 & 92.01 & 0.1687 & 0.5872 & $<$0.9450 & $>$8.777 & $<$823.3 & $>$2.614 & 0.1341  \\
\multicolumn{10}{l}{SECONDARY DATASET} \\
\\
HD~30311 & 2424 & 215.4 & 16.48 & 0.5023 & $<$40.54 & $>$22.60 & $<$791.5 & $>$2.563 & 0.1801  \\
HD~166356 & 261.5 & 90.68 & 0.9290 & 0.2424 & $<$4.879 & $>$11.32 & $<$588.9 & $>$4.161 & 0.3022  \\
\hline
\end{tabular}}
 \parbox{\hsize}{\textbf{Notes:} The primary dataset in the table is divided in three parts to distinguish CJs, BDs, and SCs, after the RV analysis (from top to bottom).}
\end{table}

\clearpage
\begin{table}\centering
\caption{\label{tab:moreGAIA2} Companion HD~76332\,B, whose RV orbit is incompatible with Gaia astrometric excess noise. The 
minimum and maximum $\varepsilon$ that we were able to simulate for this secondary dataset source is given as $\varepsilon_\text{simu,min}$ and 
$\varepsilon_\text{simu,max}$.}
{\def\arraystretch{1.0}
\begin{tabular}{l@{~~~}c@{~~~}c@{~~~}}
\hline
Parameter & Unit &  HD~76332\,B\\
\hline
Period	& (day) & 2489.182 \\
$m\sin i$ & (M$_\text{J}$) & 216.01 \\
$a\sin i$ & (mas) &  13 \\
$\varepsilon_\text{DR1}$ & (mas) &  2.19 \\
$\varepsilon_\text{simu,min}$ & (mas) &  0.74 \\
$\varepsilon_\text{simu,max}$ & (mas) & 1.55 \\
\hline
\end{tabular}}
\end{table}

\thispagestyle{empty}
\section{Figures}

\newdimen\LFcapwidth \LFcapwidth=\linewidth
\begin{longfigure}{c}
	\caption{\label{fig:Gls_CJs} Generalized Lomb-Scargle (GLS) periodogram of the RVs for the CJs BD+450564 b (top left), BD+550362 b (top right), HD~124330 b (bottom left), and HD~155193 b (bottom right). The three dashed black lines represent the 10\%, 1\%, and 0.1\% false alarm probability in ascending order.} 
	\endLFfirsthead
	\caption{Continued.}
	\endLFhead
	
\includegraphics[width=0.49\hsize]{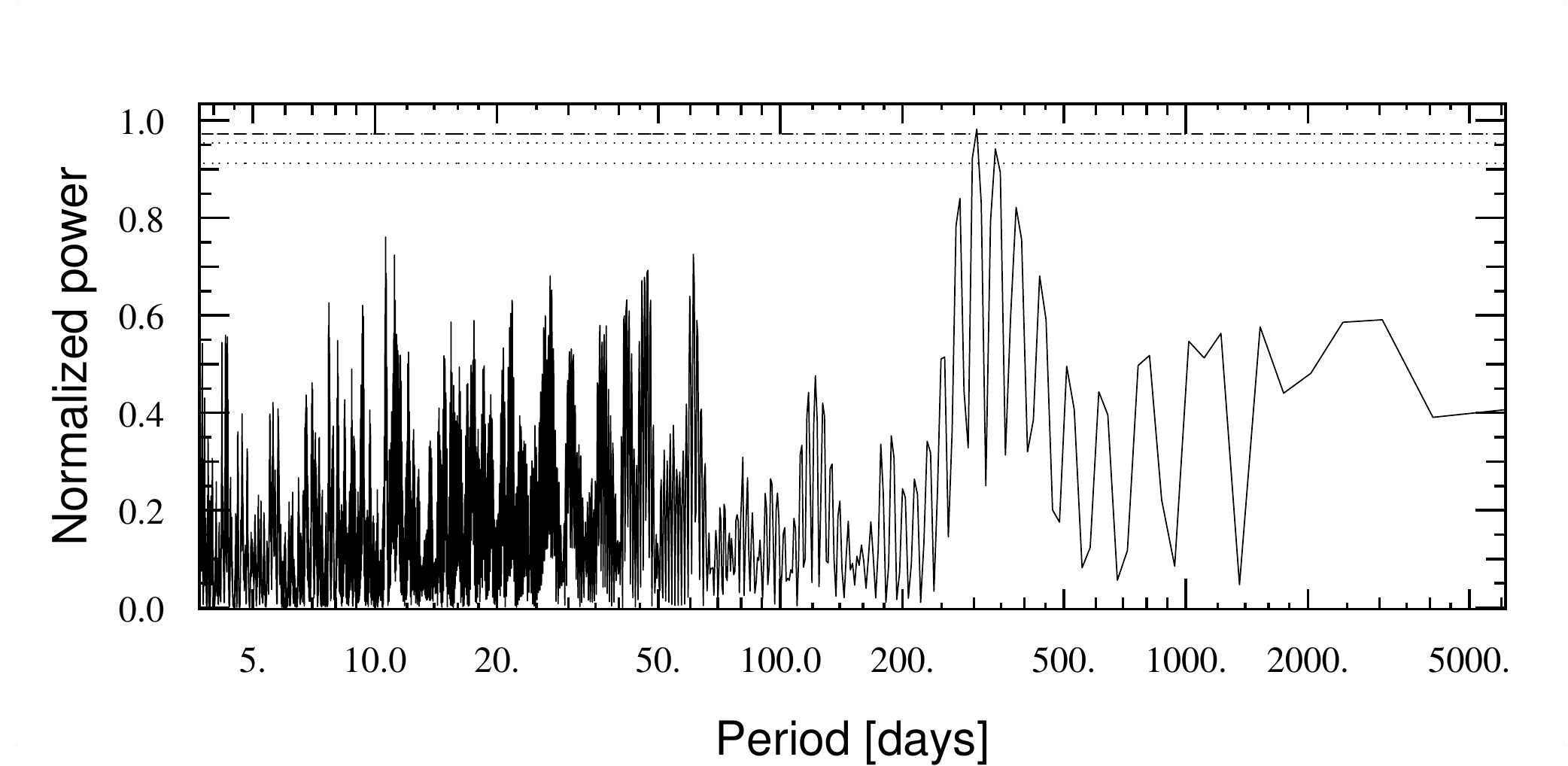}
\includegraphics[width=0.49\hsize]{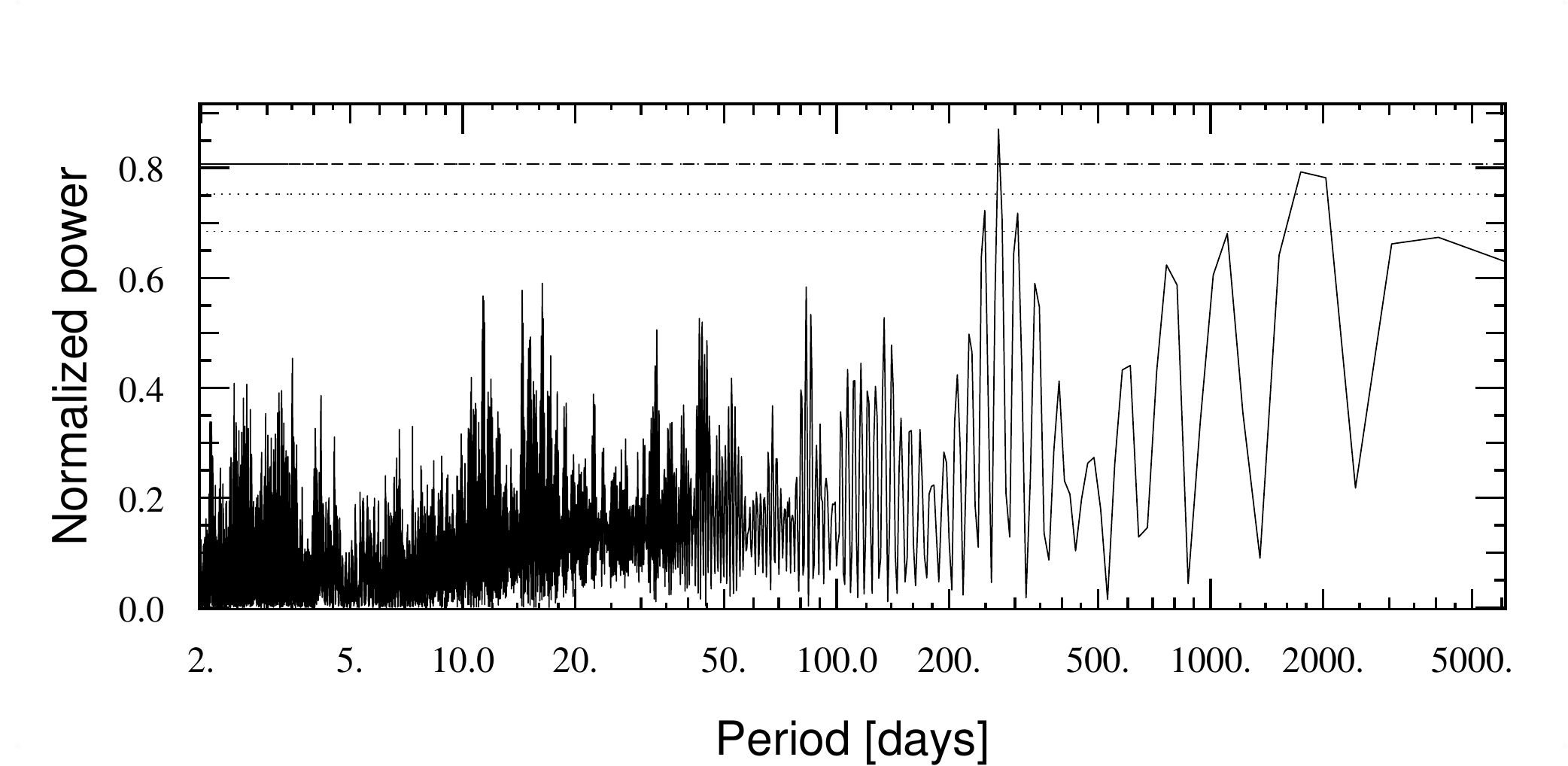}\\
\includegraphics[width=0.49\hsize]{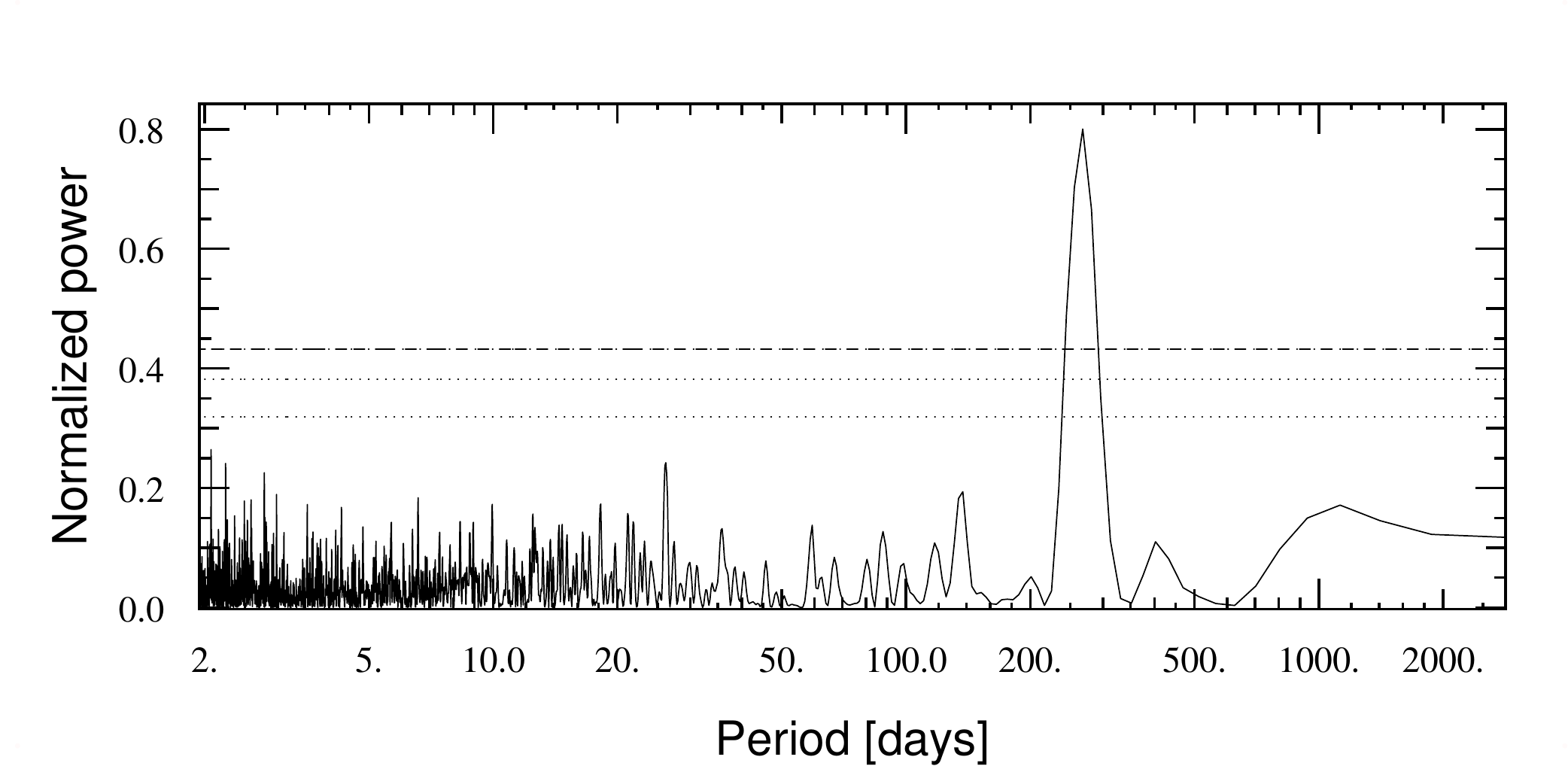}
\includegraphics[width=0.49\hsize]{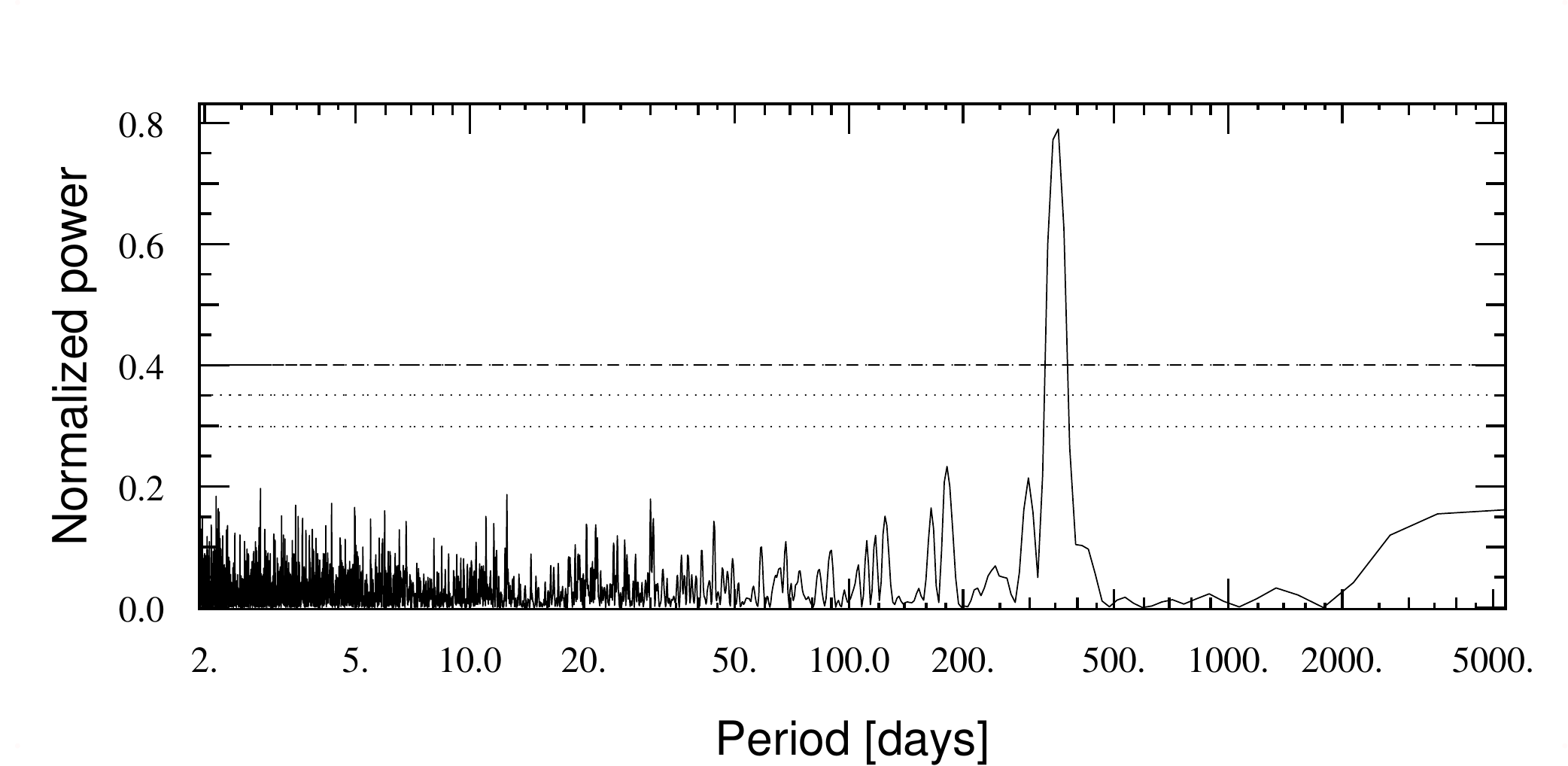}\\
\end{longfigure}
\newdimen\LFcapwidth \LFcapwidth=\linewidth

\newpage
\begin{longfigure}{c}
	\caption{\label{fig:Gls_CJse} Generalized Lomb-Scargle (GLS) periodogram (on left) and Keplerian periodogram (on right) of the RVs for the eccentric CJs, BD+631405 b and HD~331093 b. The two dashed black lines in GLS periodogram represent the 10\% and 1\% false alarm probability in ascending order.} 
	\endLFfirsthead
	\caption{Continued.}
	\endLFhead
\xincludegraphics[width=0.49\hsize,label=\textbf{BD+631405}]{{PG_planets/BD+631405_sc_p_gls_obs_16.099800_1765.470000_fap_0.010000_vs_period}.pdf}
\xincludegraphics[width=0.47\hsize]{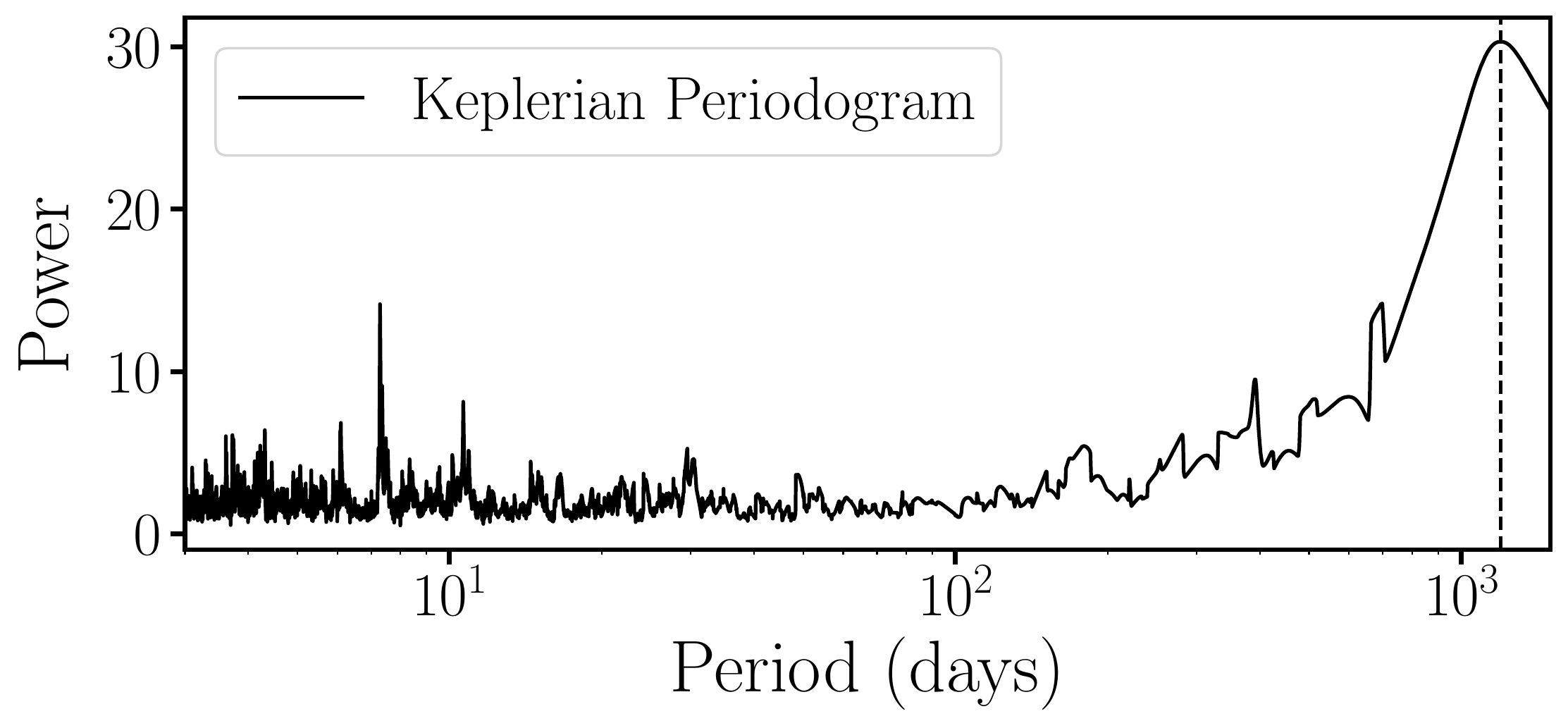}\\
\xincludegraphics[width=0.49\hsize,label=\textbf{HD331093}]{{PG_planets/HD331093_sc_psc_gls_obs_1.892800_6992.416600_fap_0.010000_vs_period}.pdf}
\xincludegraphics[width=0.47\hsize]{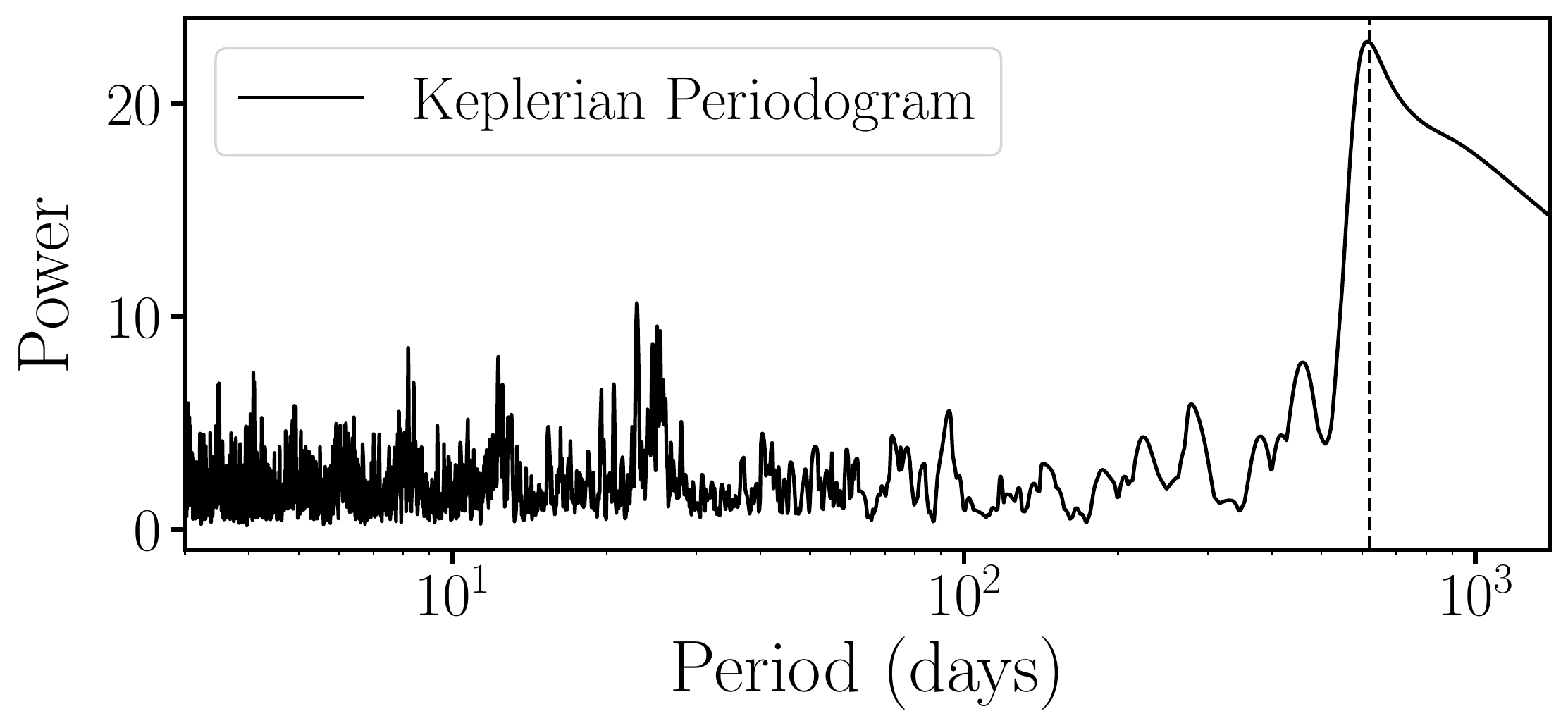}\\
\end{longfigure}

\newdimen\LFcapwidth \LFcapwidth=\linewidth
\begin{longfigure}{c}
	\caption{\label{fig:GiantPlanetsPC} Phase folded RV curve for the six CJs are plotted here. SOPHIE and SOPHIE+ RV measurements are indicated in blue squares and red circles respectively.} 
	\endLFfirsthead
	\caption{Continued.}
	\endLFhead
\includegraphics[width=0.49\hsize,  clip=true, trim=0 -50 0 0]{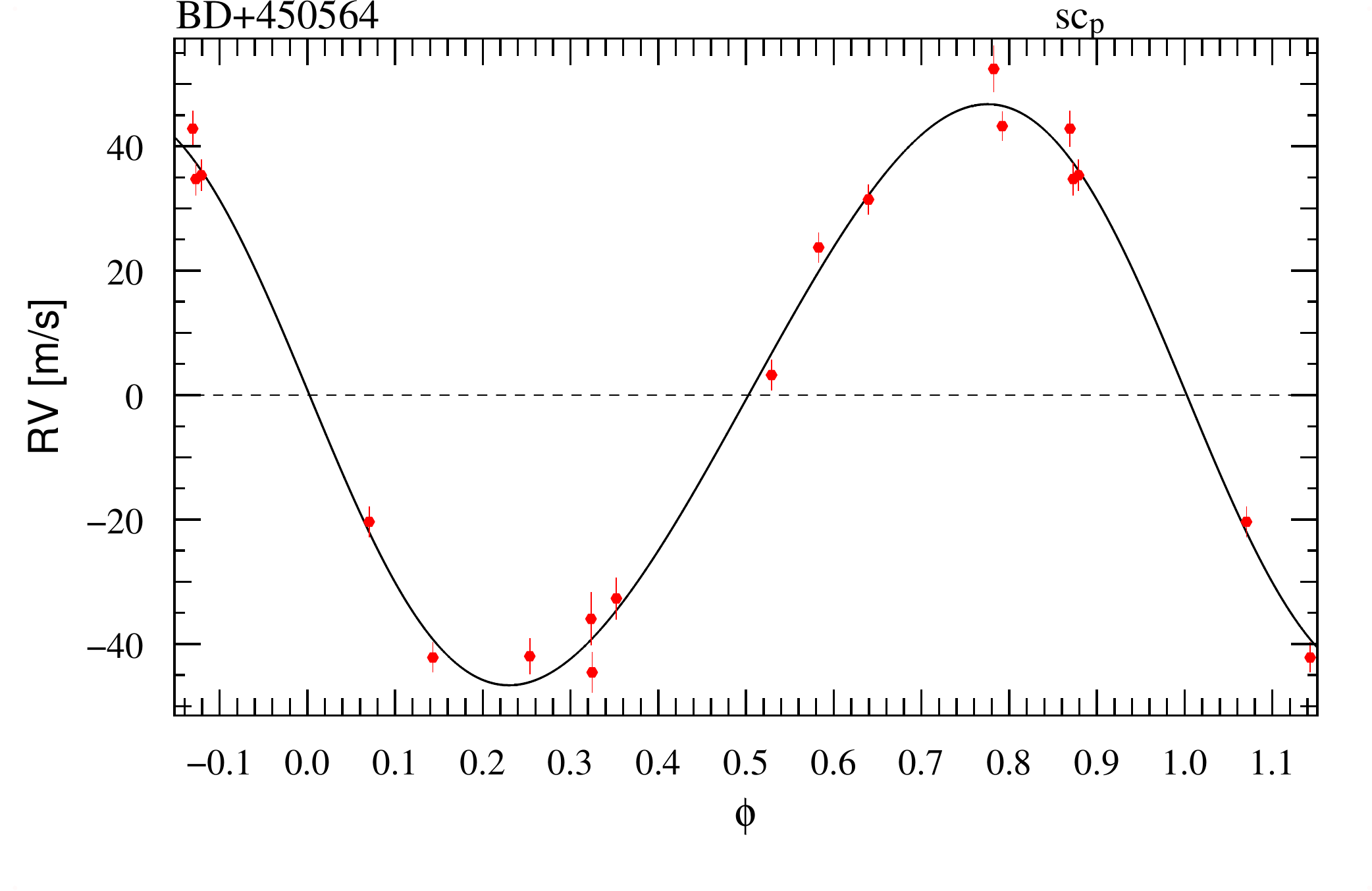}
\includegraphics[width=0.49\hsize,  clip=true, trim=0 -50 0 0]{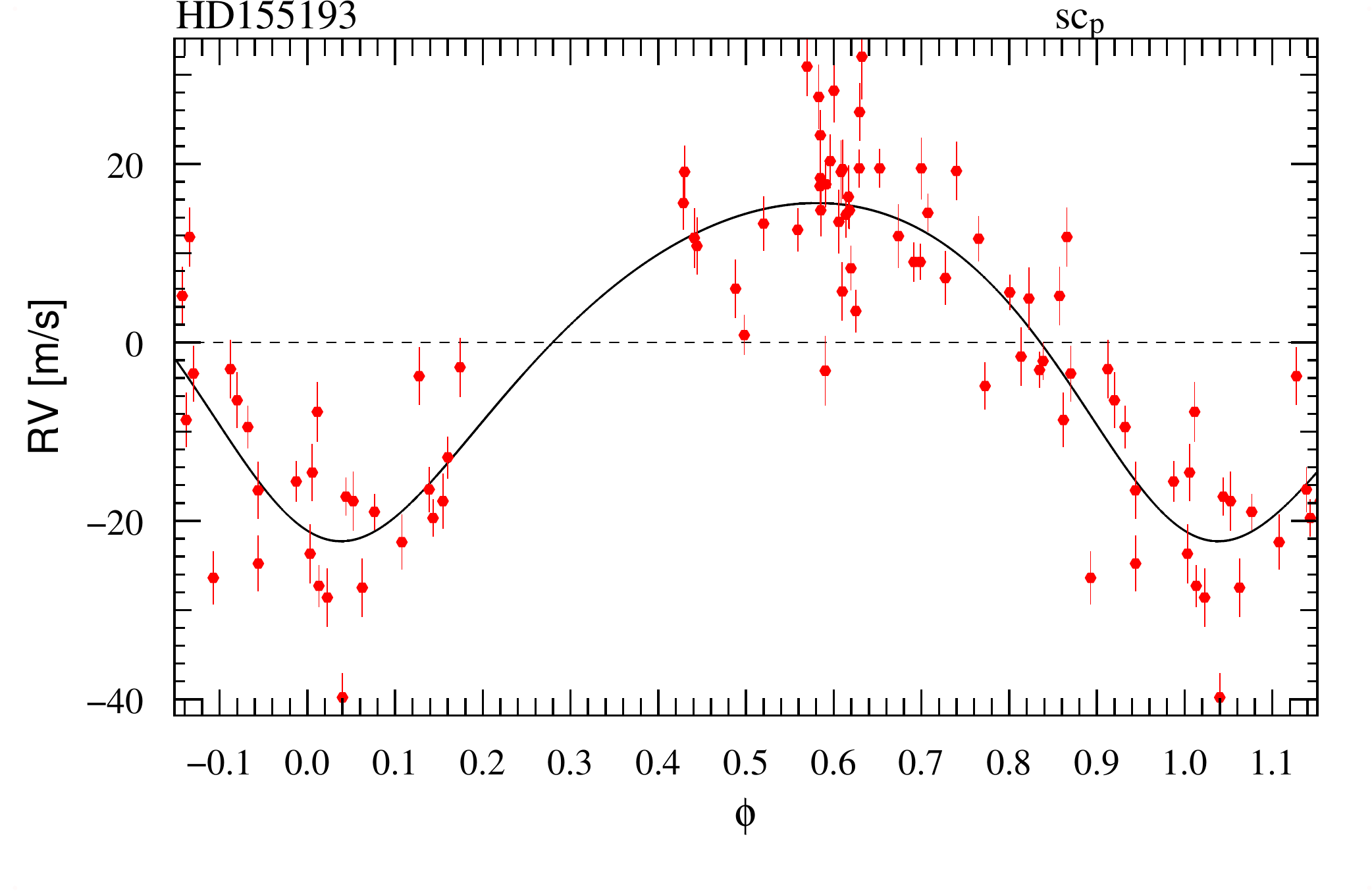}\\
\includegraphics[width=0.49\hsize,   clip=true, trim=0 -50 0 0]{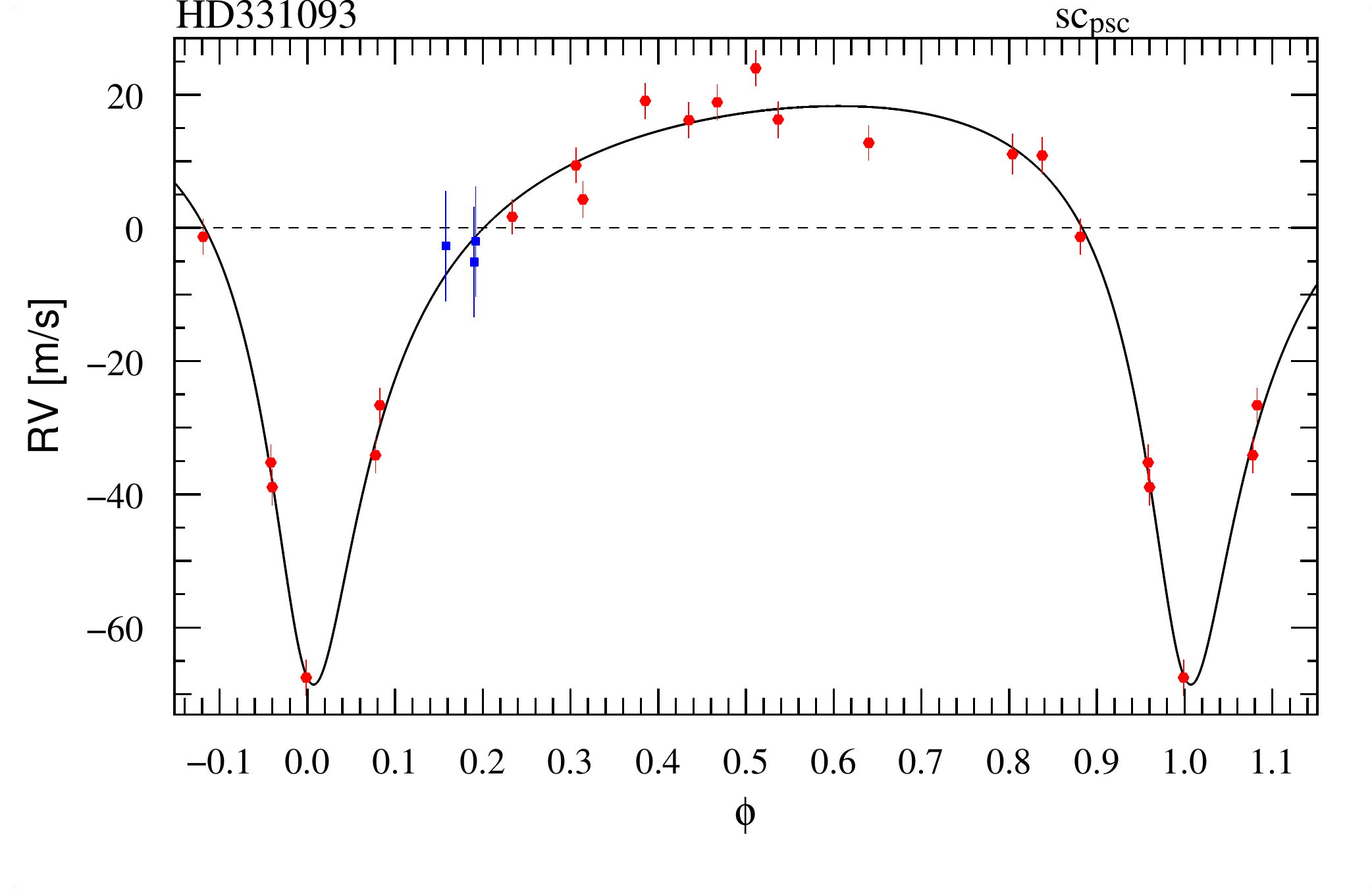}
\includegraphics[width=0.49\hsize,   clip=true, trim=0 -50 0 0]{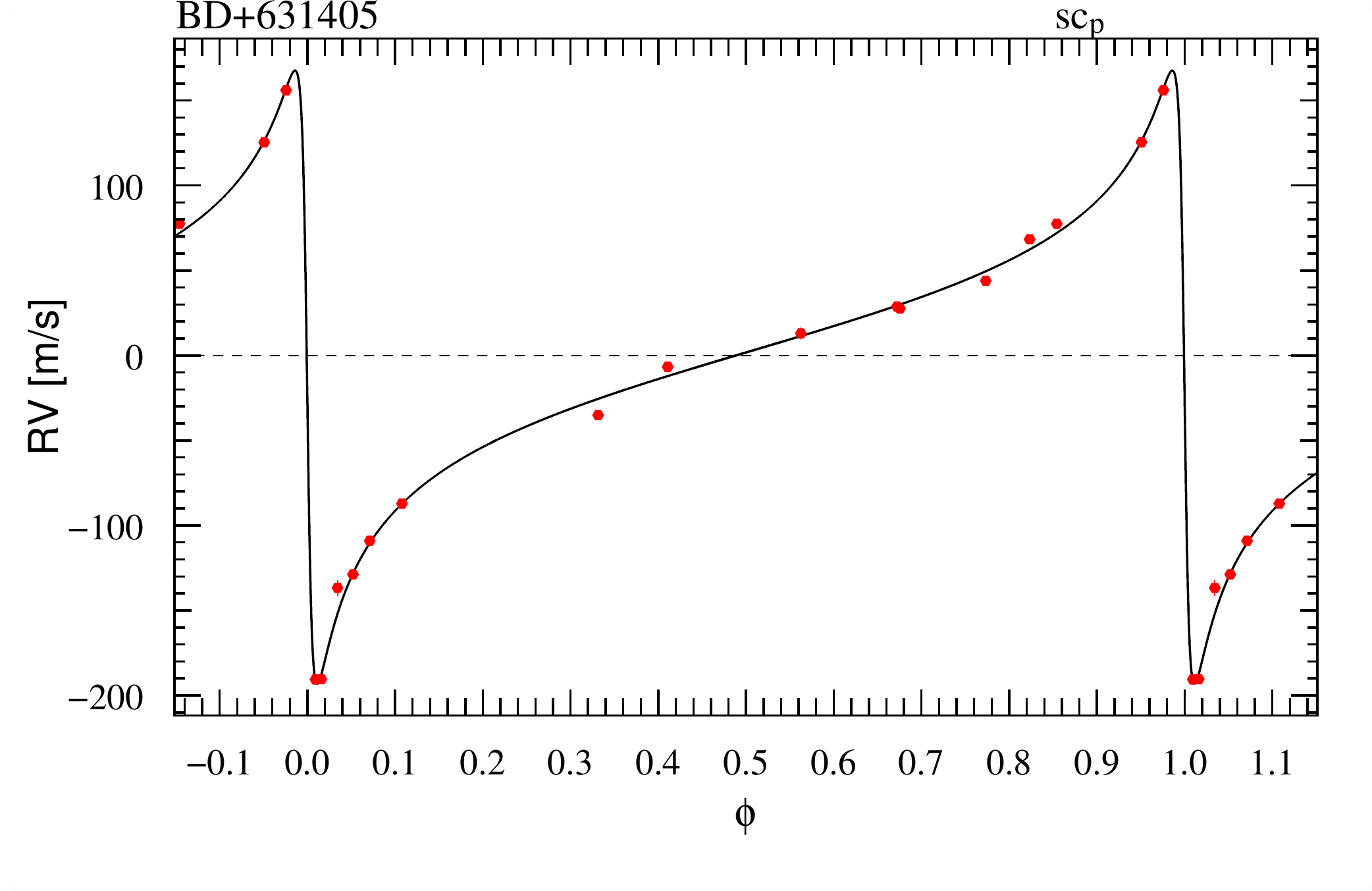}\\
\includegraphics[width=0.49\hsize,  clip=true, trim=0 -50 0 0]{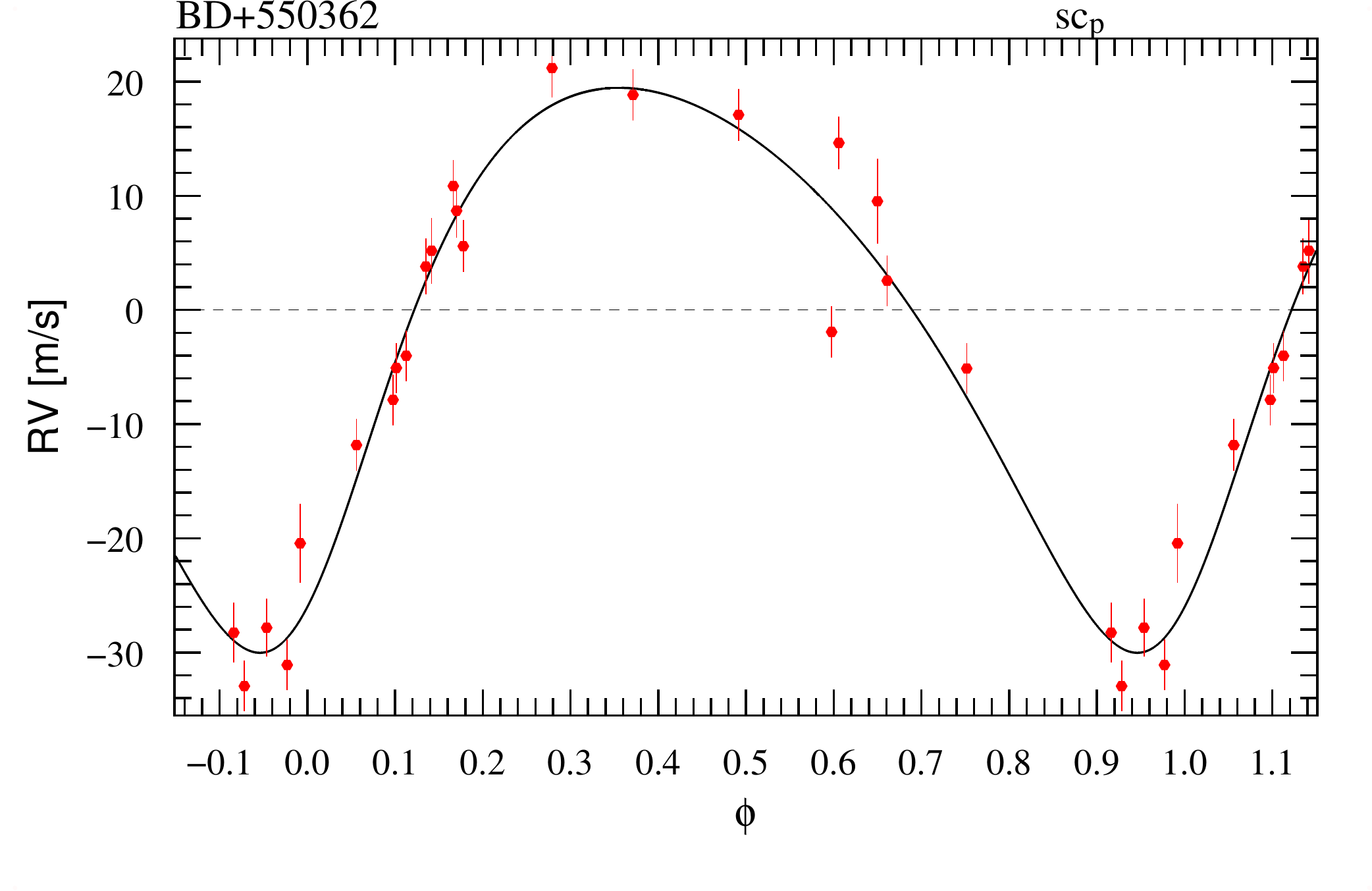}
\includegraphics[width=0.49\hsize,  clip=true, trim=0 -50 0 0]{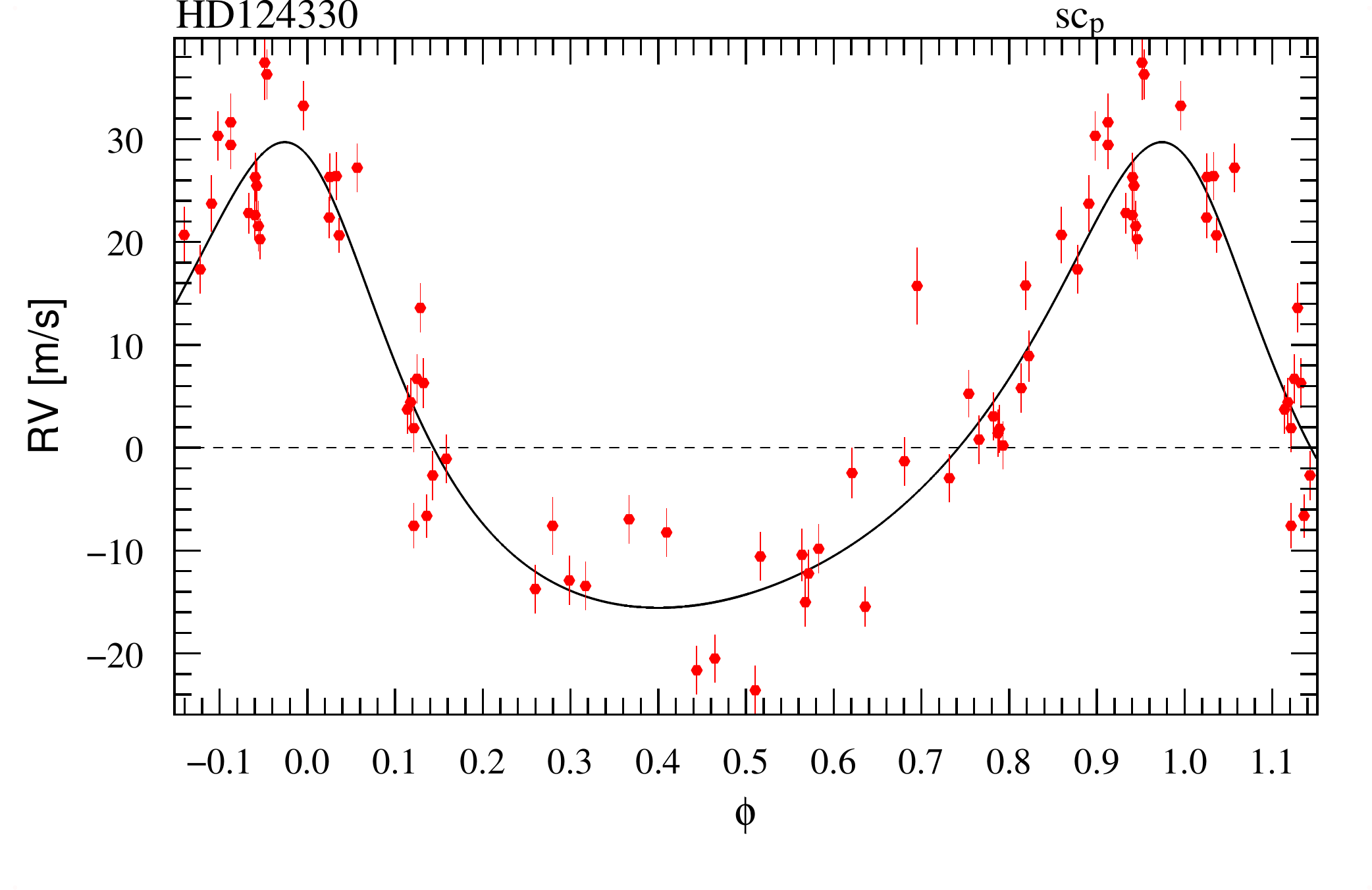}\\
\end{longfigure}

\begin{longfigure}{c}
	\caption{\label{fig:BDorbitPC} Phase folded RV curve for the four BDs with \msini in the range of 13-75  M$_{\mathrm{J}}$ are plotted here. SOPHIE and SOPHIE+ RV measurements are indicated in blue squares and red circles respectively.} 
	\endLFfirsthead
	\caption{Continued.} 
	\endLFhead
\includegraphics[width=0.49\hsize,  clip=true, trim=0 -50 0 0]{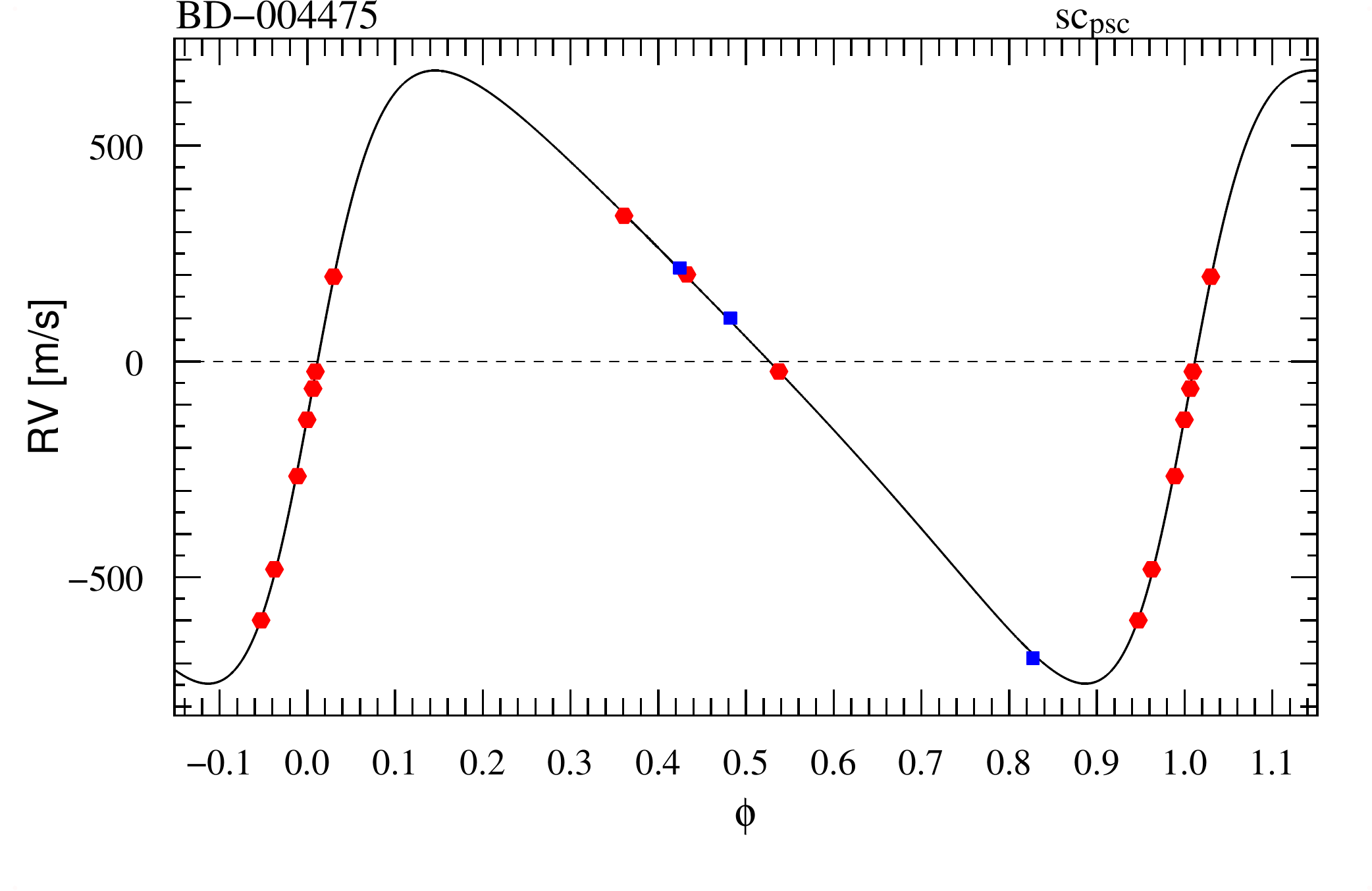}
\includegraphics[width=0.49\hsize,  clip=true, trim=0 -50 0 0]{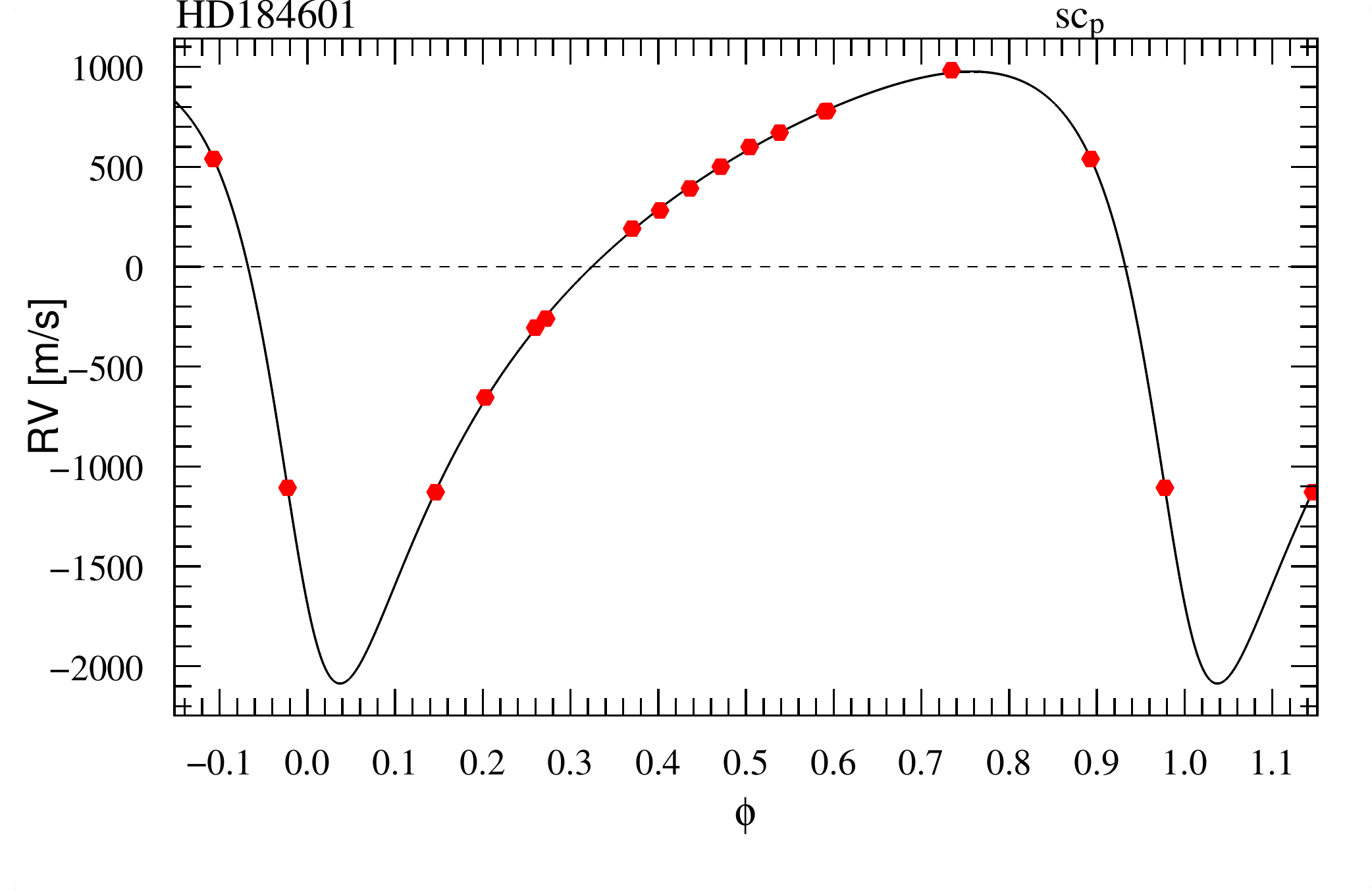}\\
\includegraphics[width=0.49\hsize,  clip=true, trim=0 -50 0 0]{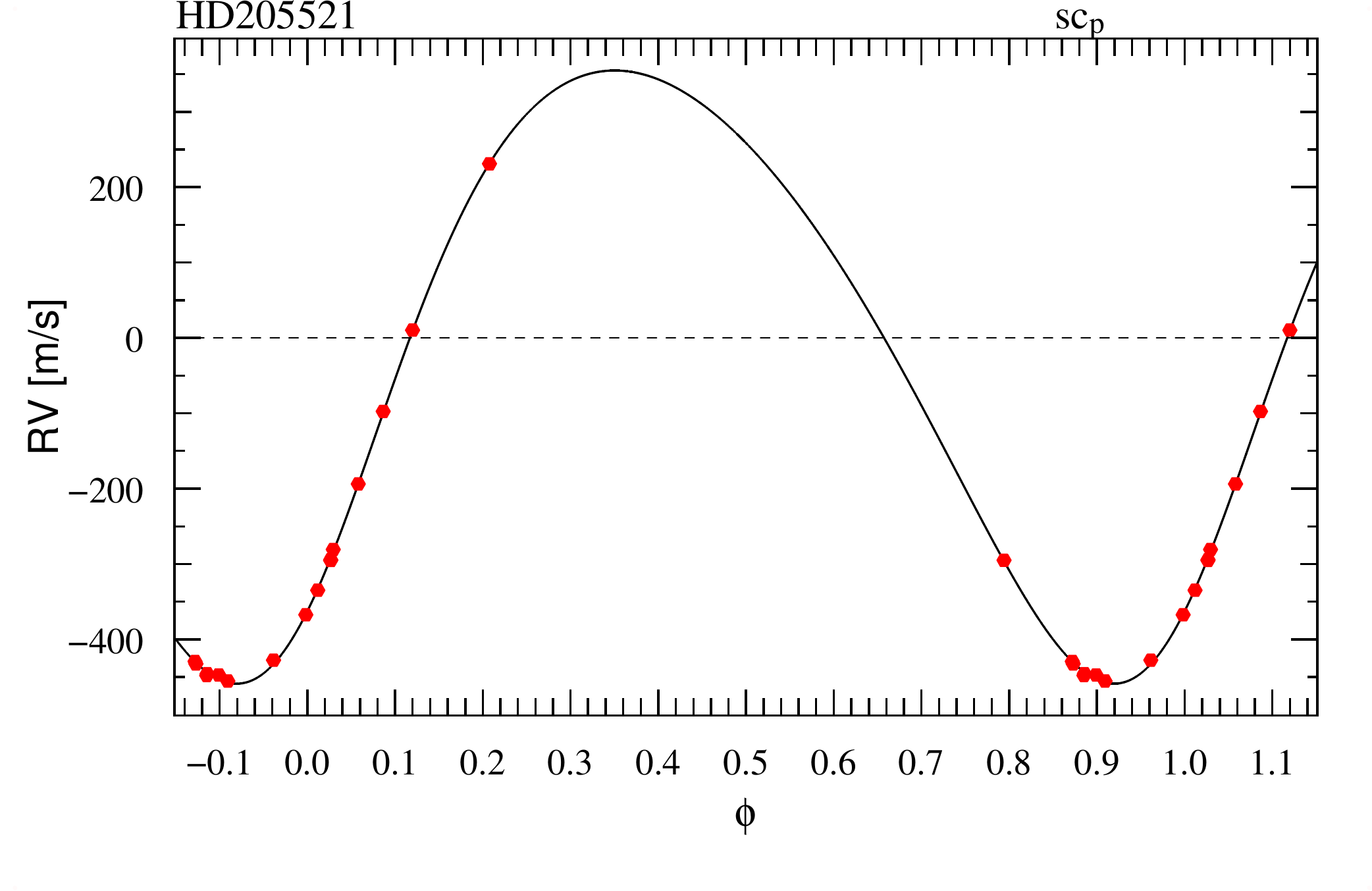}
\includegraphics[width=0.49\hsize,  clip=true, trim=0 -50 0 0]{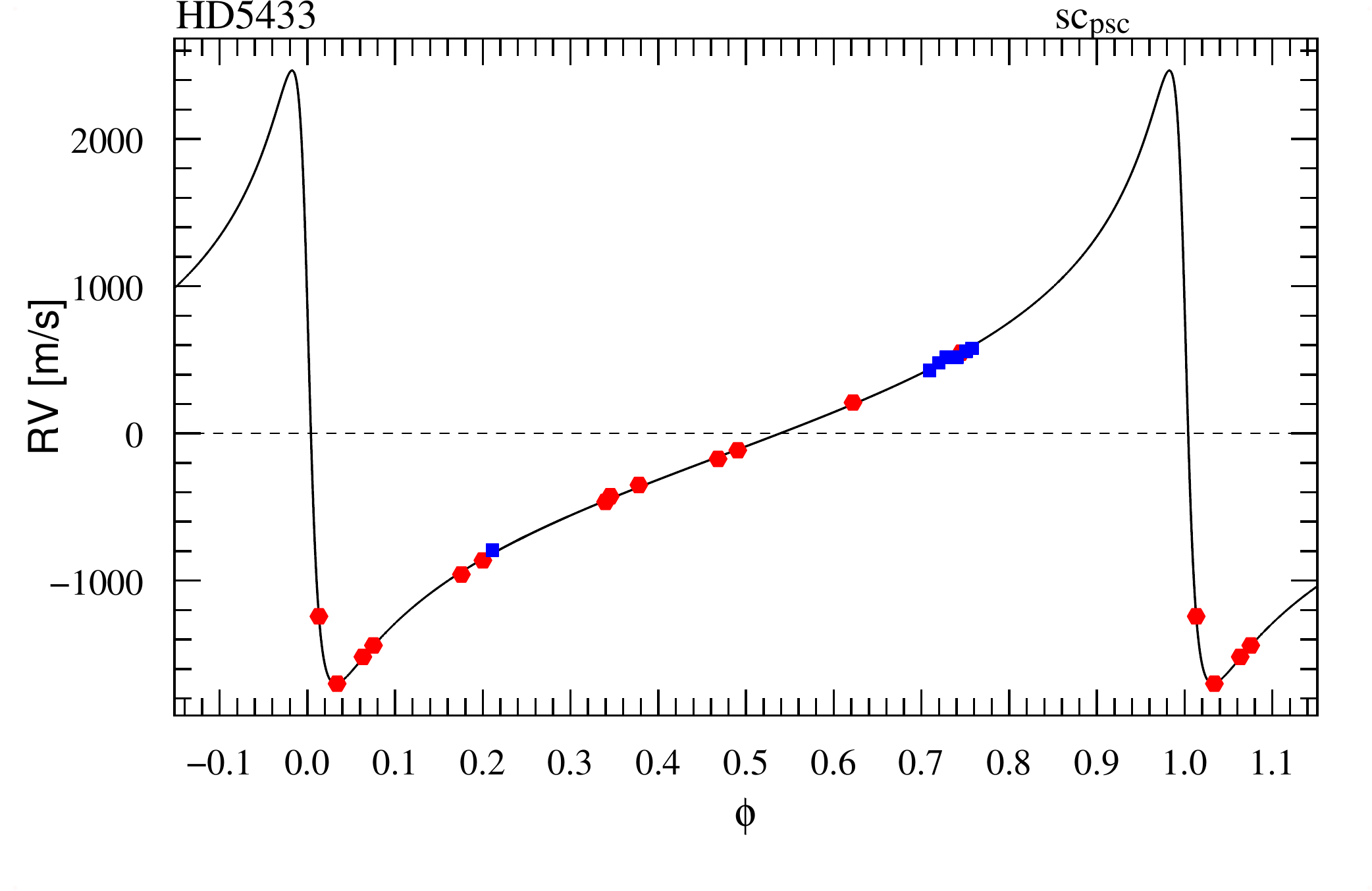}\\

\end{longfigure}

\newdimen\LFcapwidth \LFcapwidth=\linewidth
\begin{longfigure}{c}
\caption{\label{fig:StellarC} Keplerian fit in the radial velocities (RV vs time with O-C residuals and RV vs phase) for the targets with 16 SCs having \msini $\ge$ 75  M$_{\mathrm{J}}$ (0.072 M$_{\odot}$) obtained with SOPHIE spectrograph are plotted here. SOPHIE and SOPHIE+ RV measurements are indicated in blue squares and red circles respectively, except for HD~8291 where the blue squares correspond to the HARPS RV data. For targets, HD~8291, HD~25603, HD~76332, HD~187057, HD~211961, and HD~352975, SOPHIE and SOPHIE+ RV data are considered as one dataset and labeled ``sophie'' in top right corner of RV vs time plot.} 
\endLFfirsthead
\caption{Continued.}
\endLFhead
\vspace{-1mm}
\includegraphics[width=0.33\hsize,height=60mm]{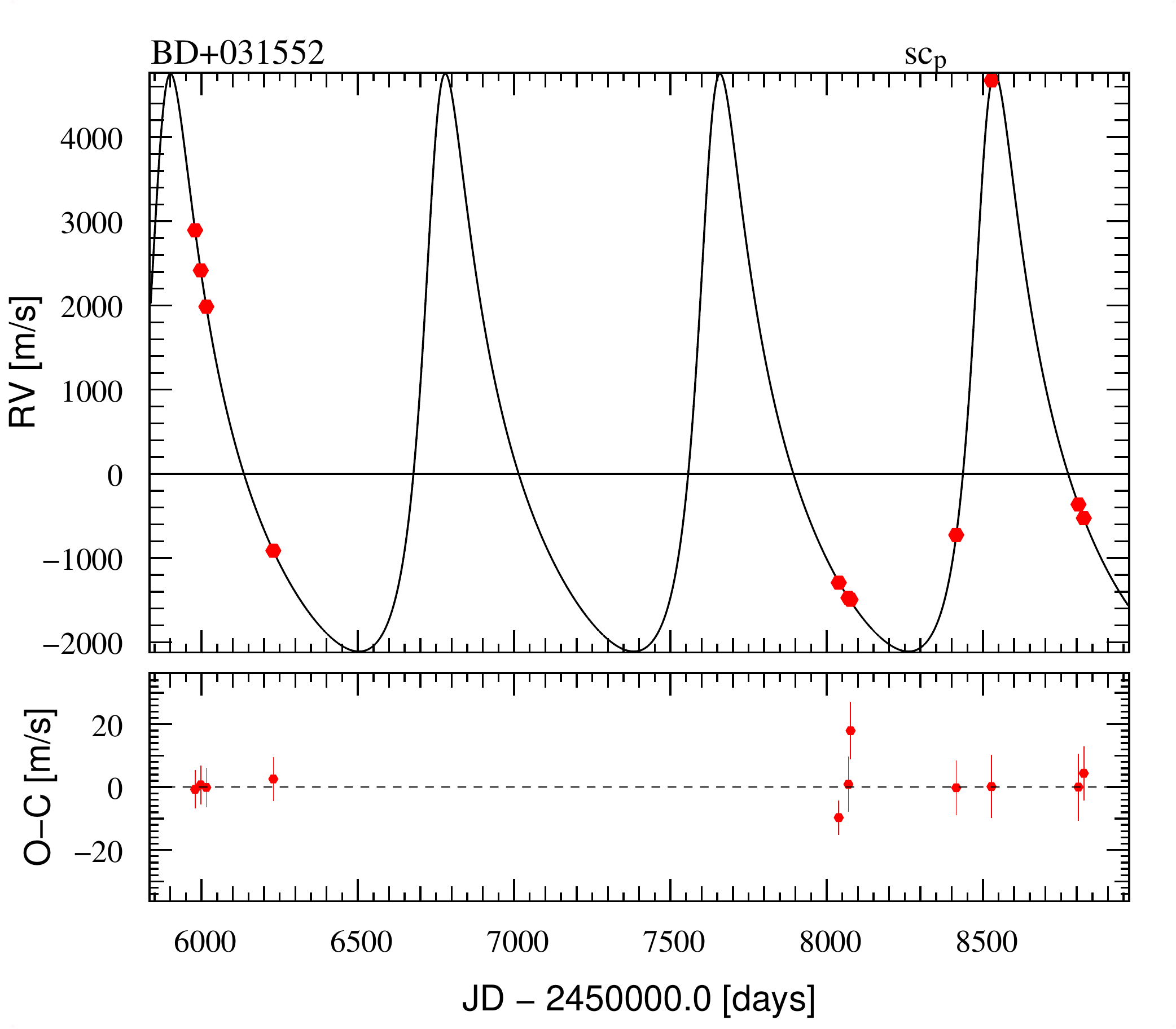}
\includegraphics[width=0.33\hsize,height=60mm]{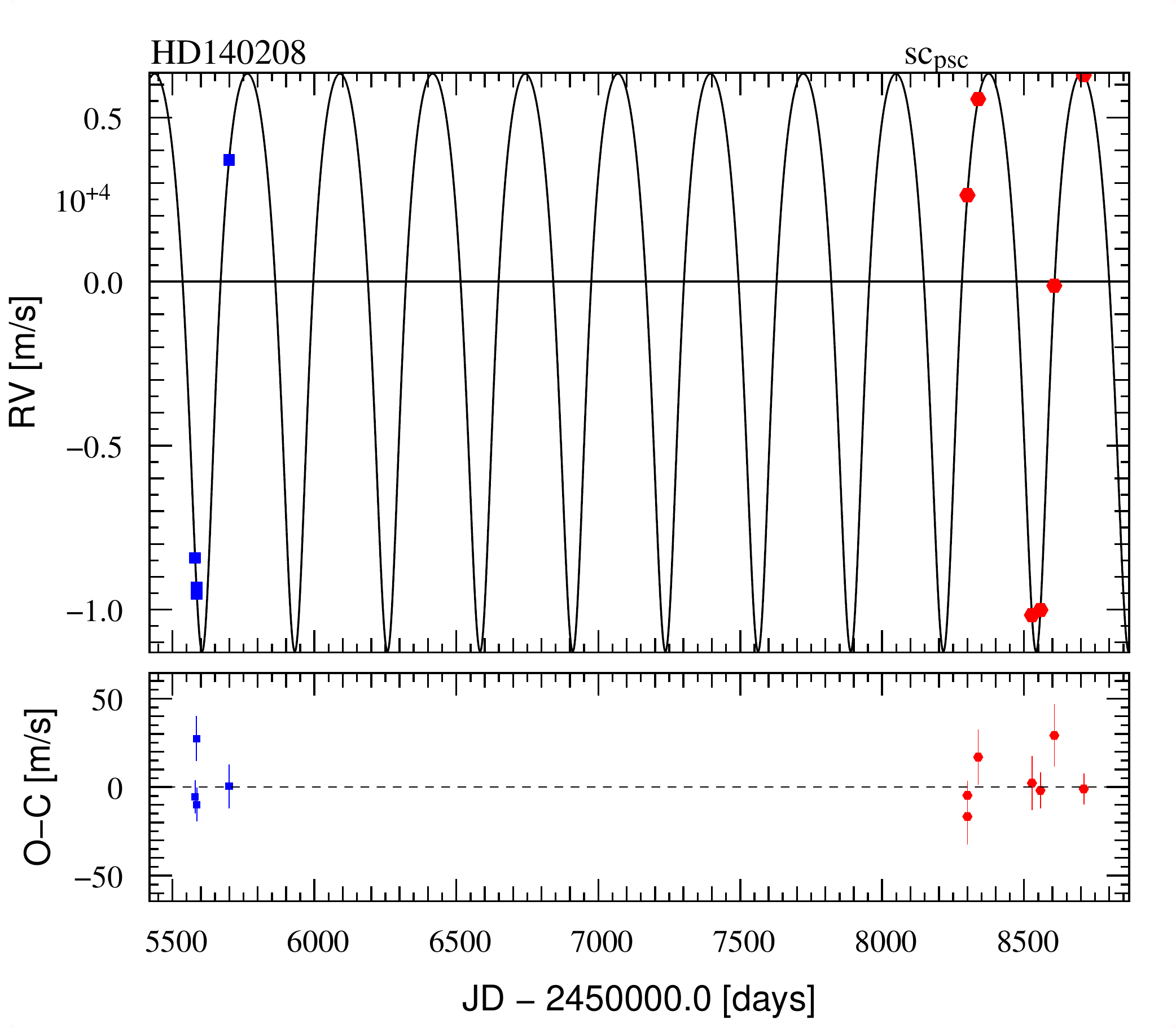}
\includegraphics[width=0.33\hsize,height=60mm]{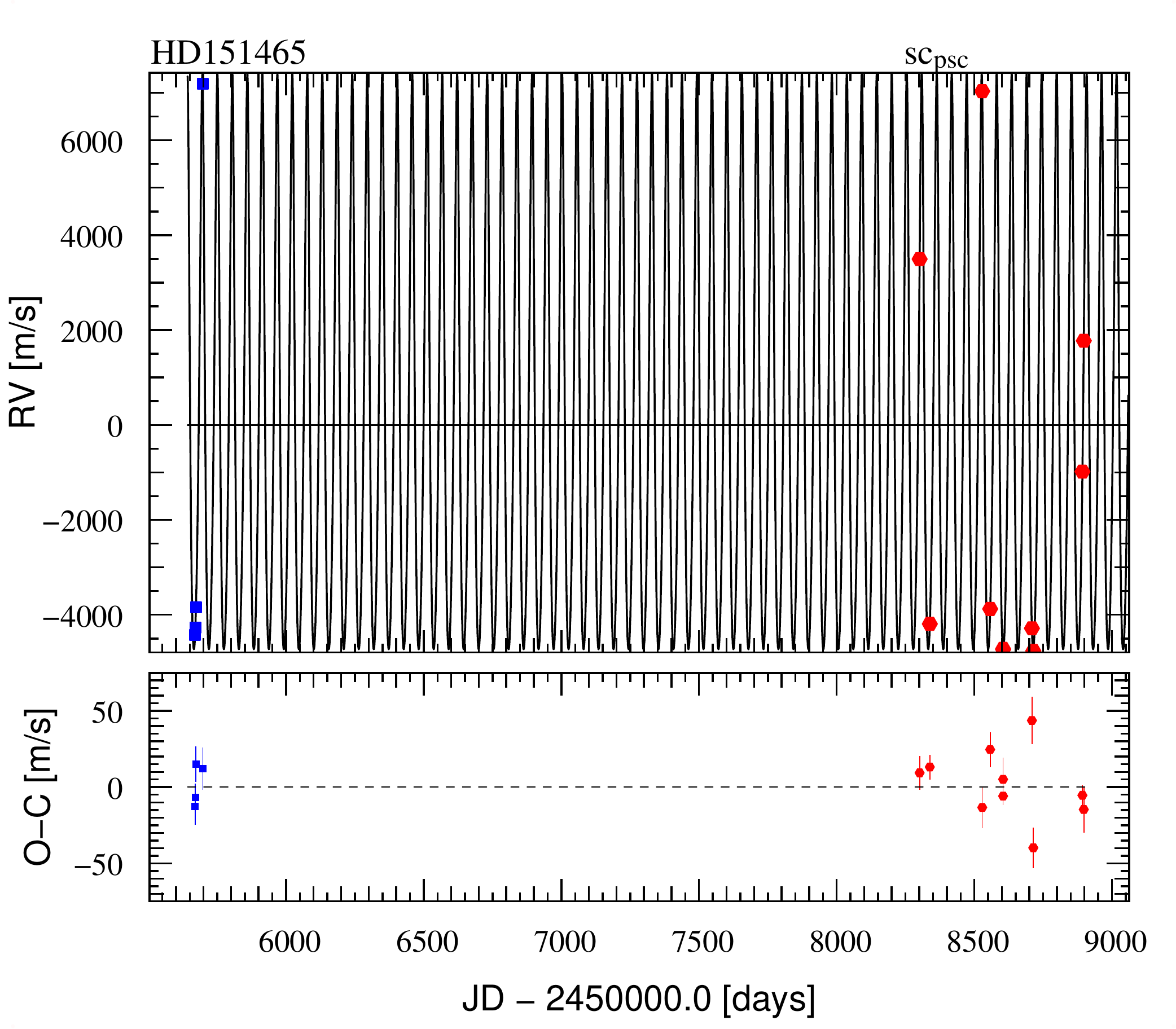}\\
\vspace{10mm}
\includegraphics[width=0.33\hsize, clip=true, trim=0 -25 0 13]{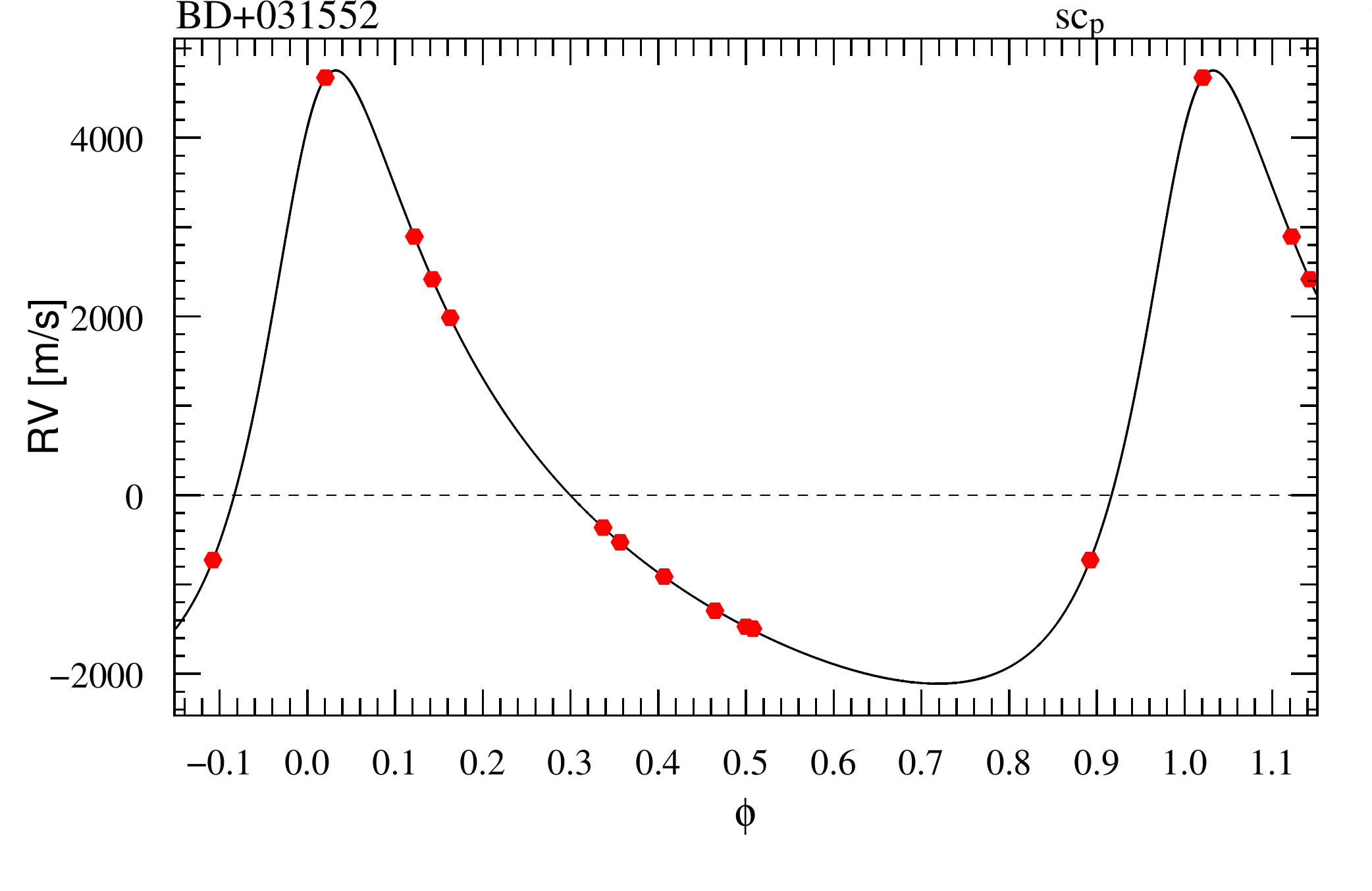}
\includegraphics[width=0.33\hsize, clip=true, trim=0 -25 0 13]{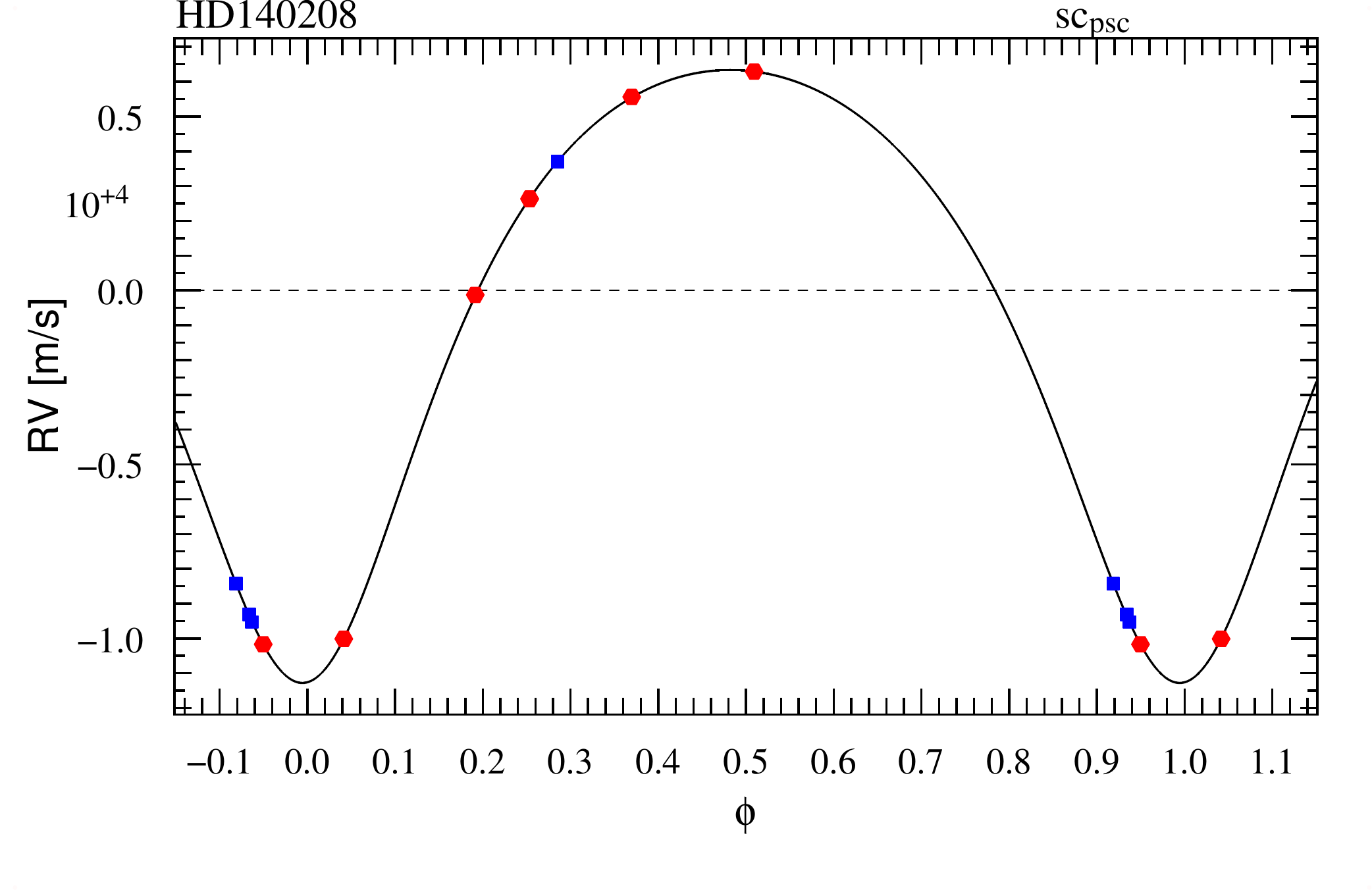}
\includegraphics[width=0.33\hsize, clip=true, trim=0 -25 0 13]{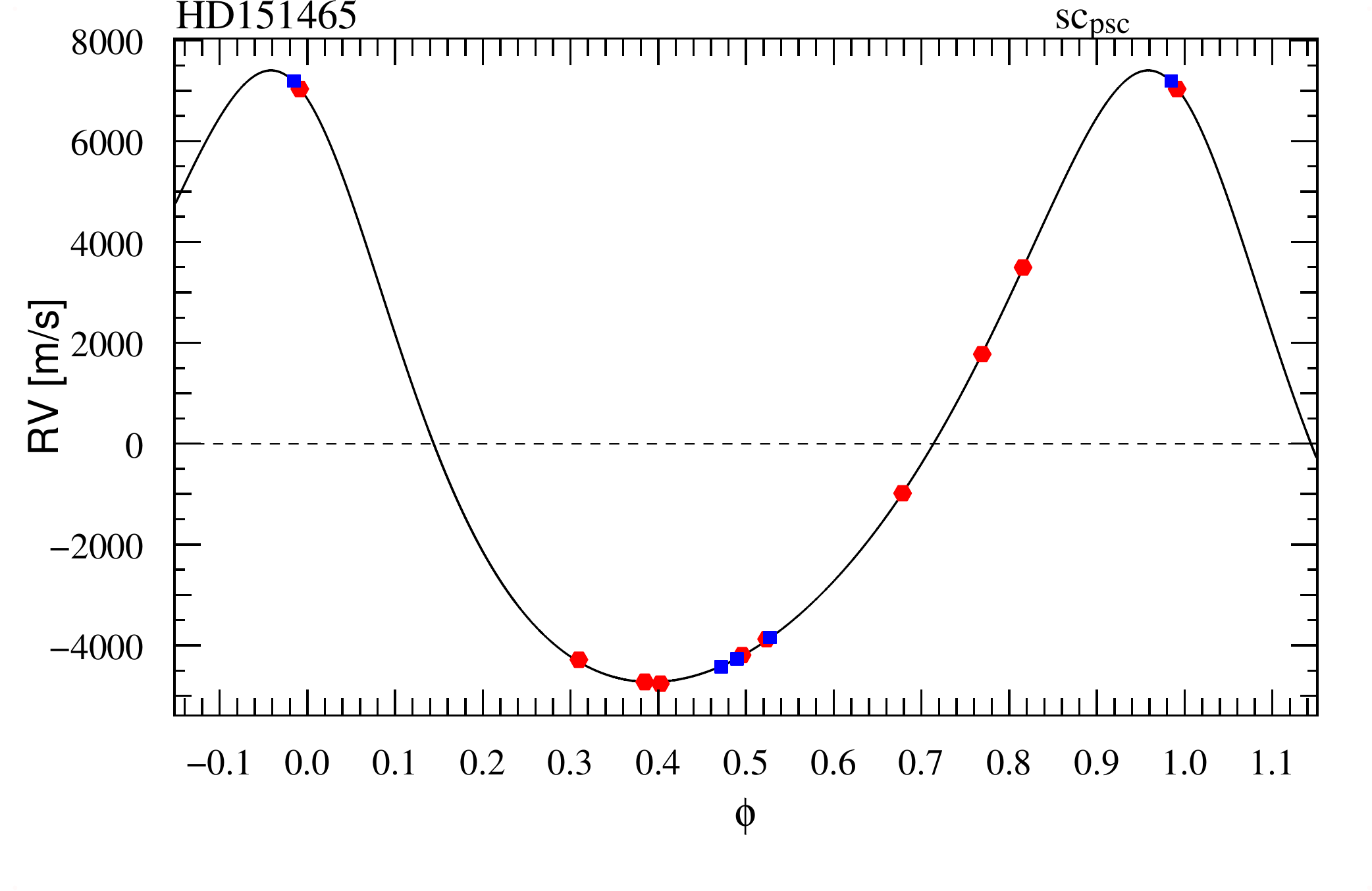}\\

\vspace{-1mm}
\includegraphics[width=0.33\hsize,height=60mm]{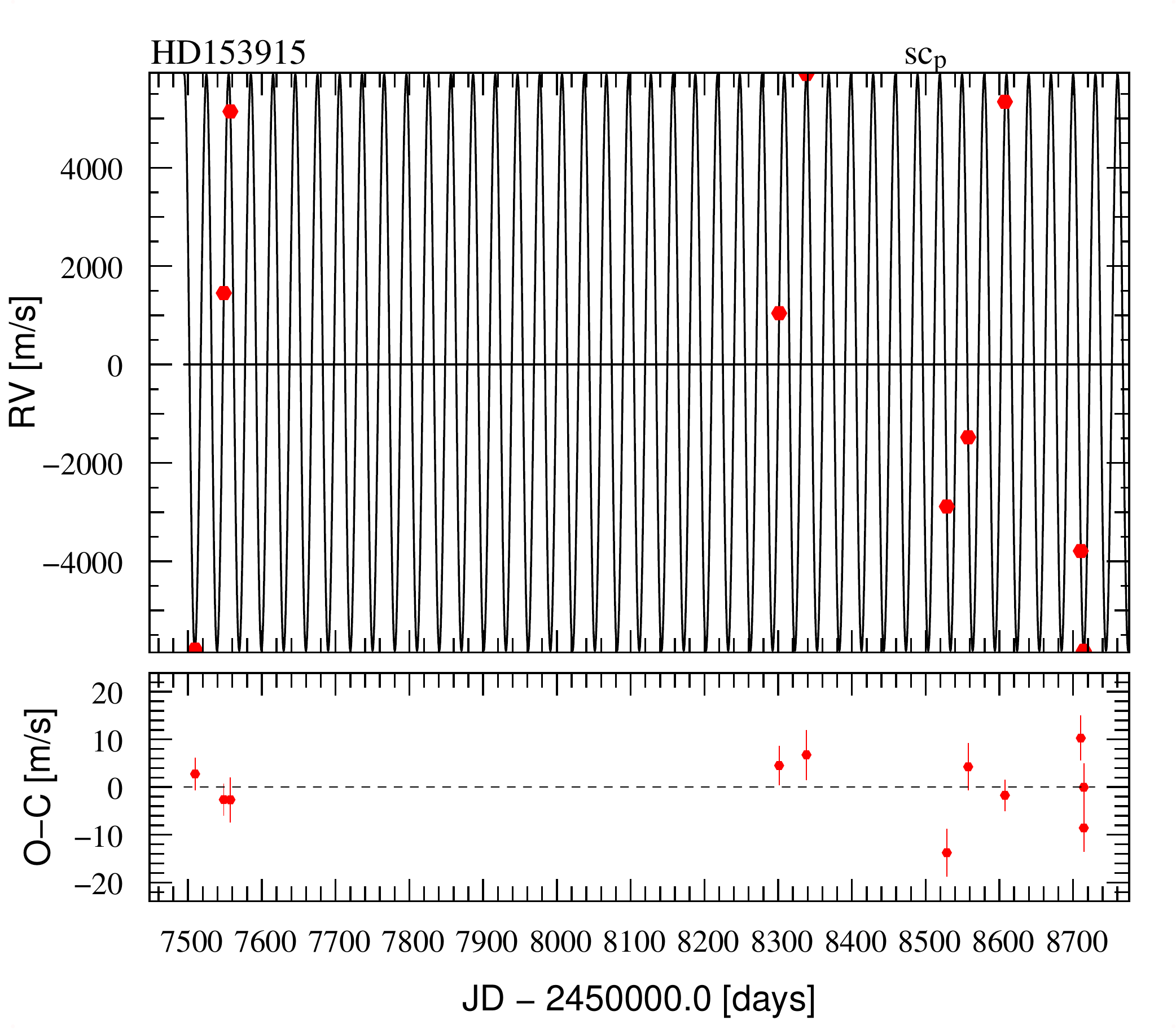}
\includegraphics[width=0.33\hsize,height=60mm]{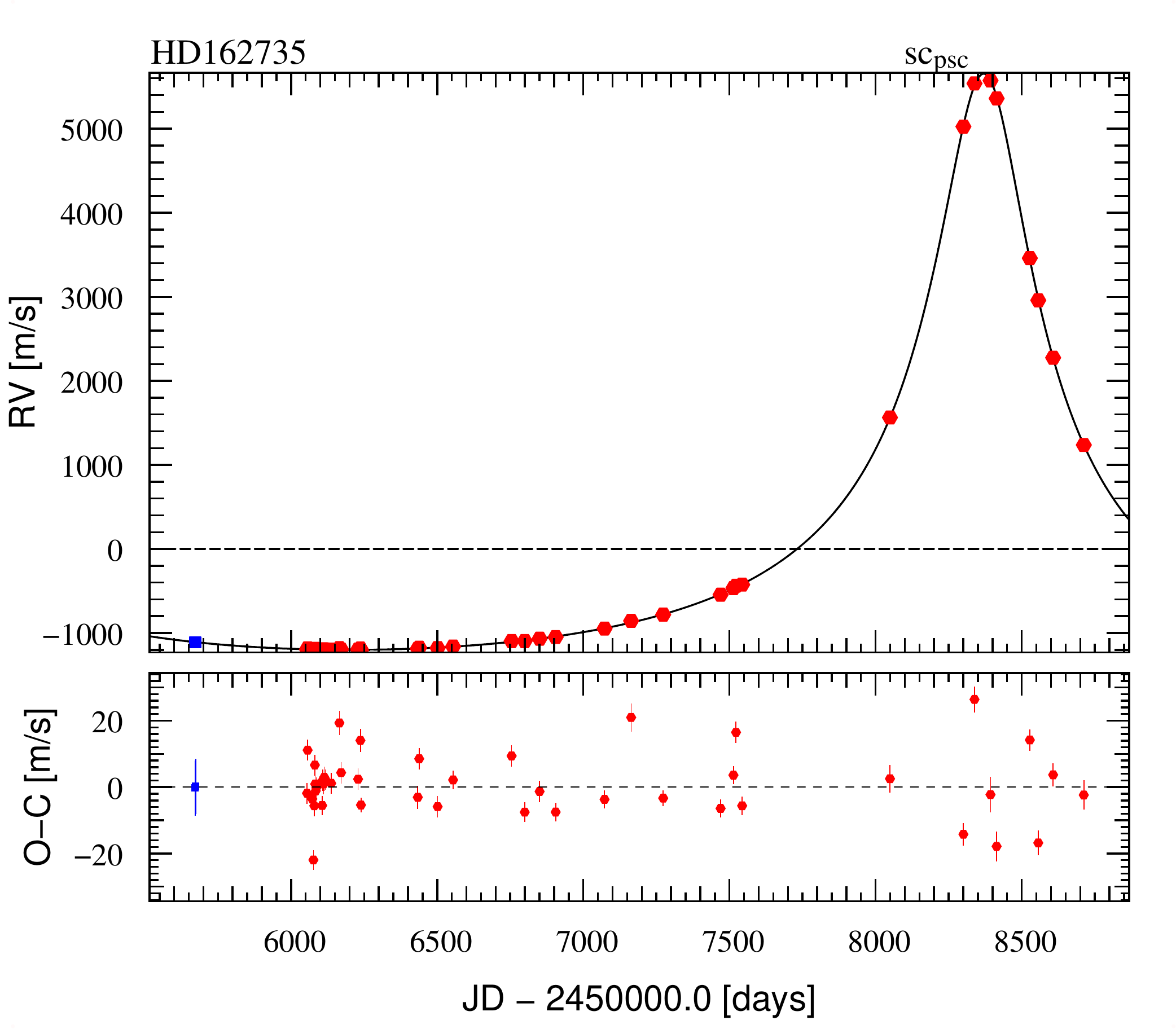}
\includegraphics[width=0.33\hsize,height=60mm]{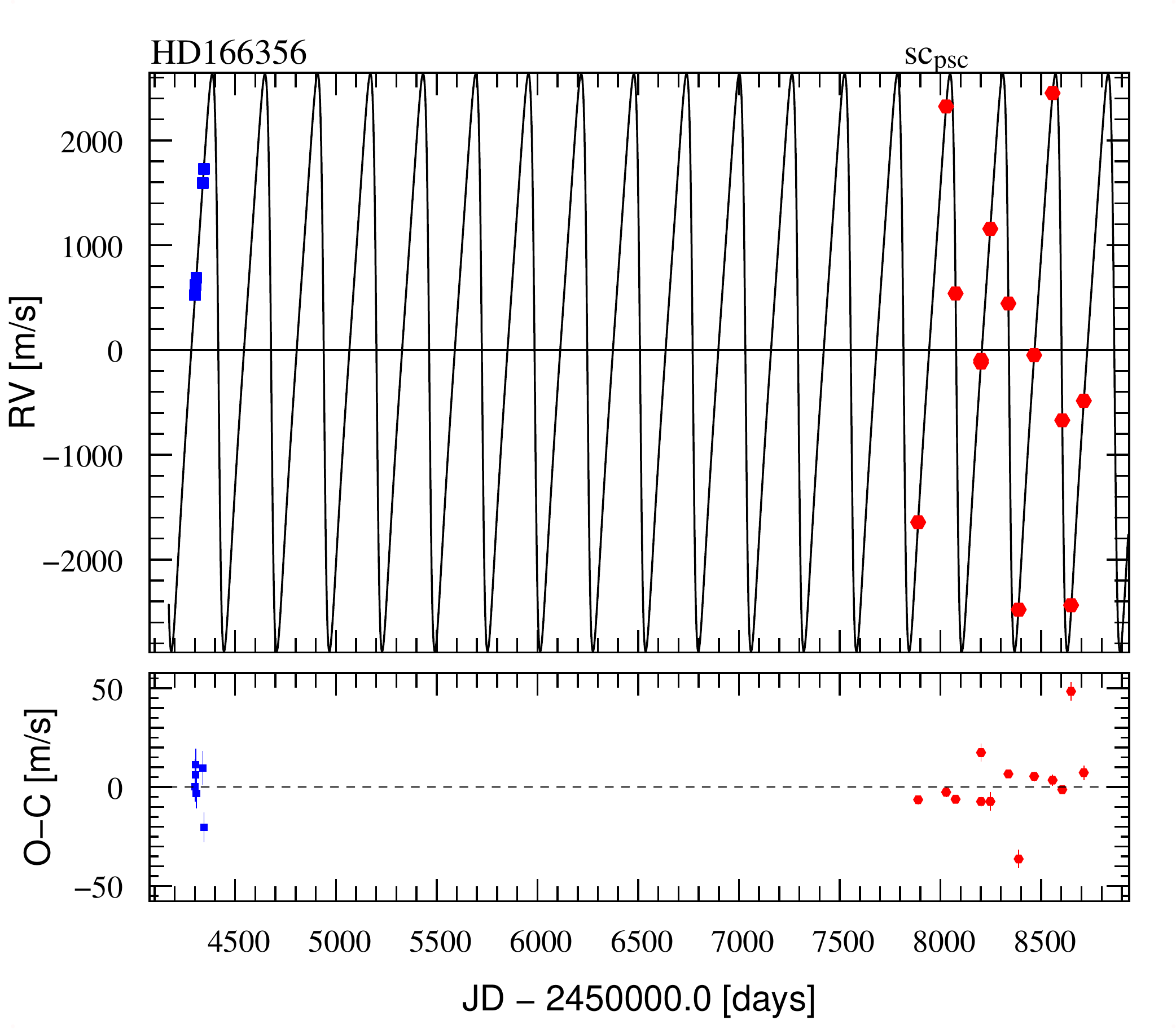}\\
\vspace{10mm}
\includegraphics[width=0.33\hsize, clip=true, trim=0 -25 0 13]{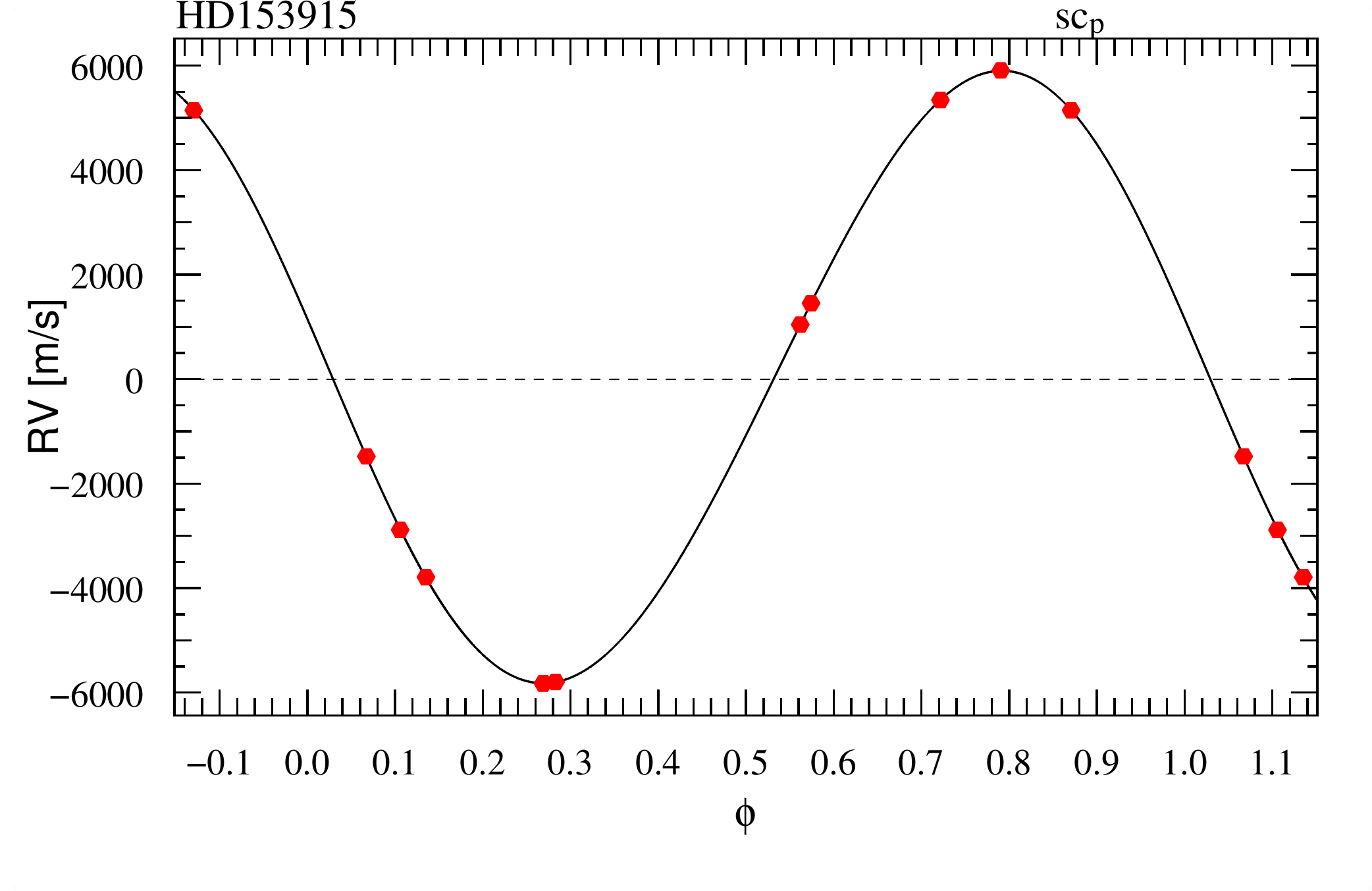}
\includegraphics[width=0.33\hsize, clip=true, trim=0 -25 0 13]{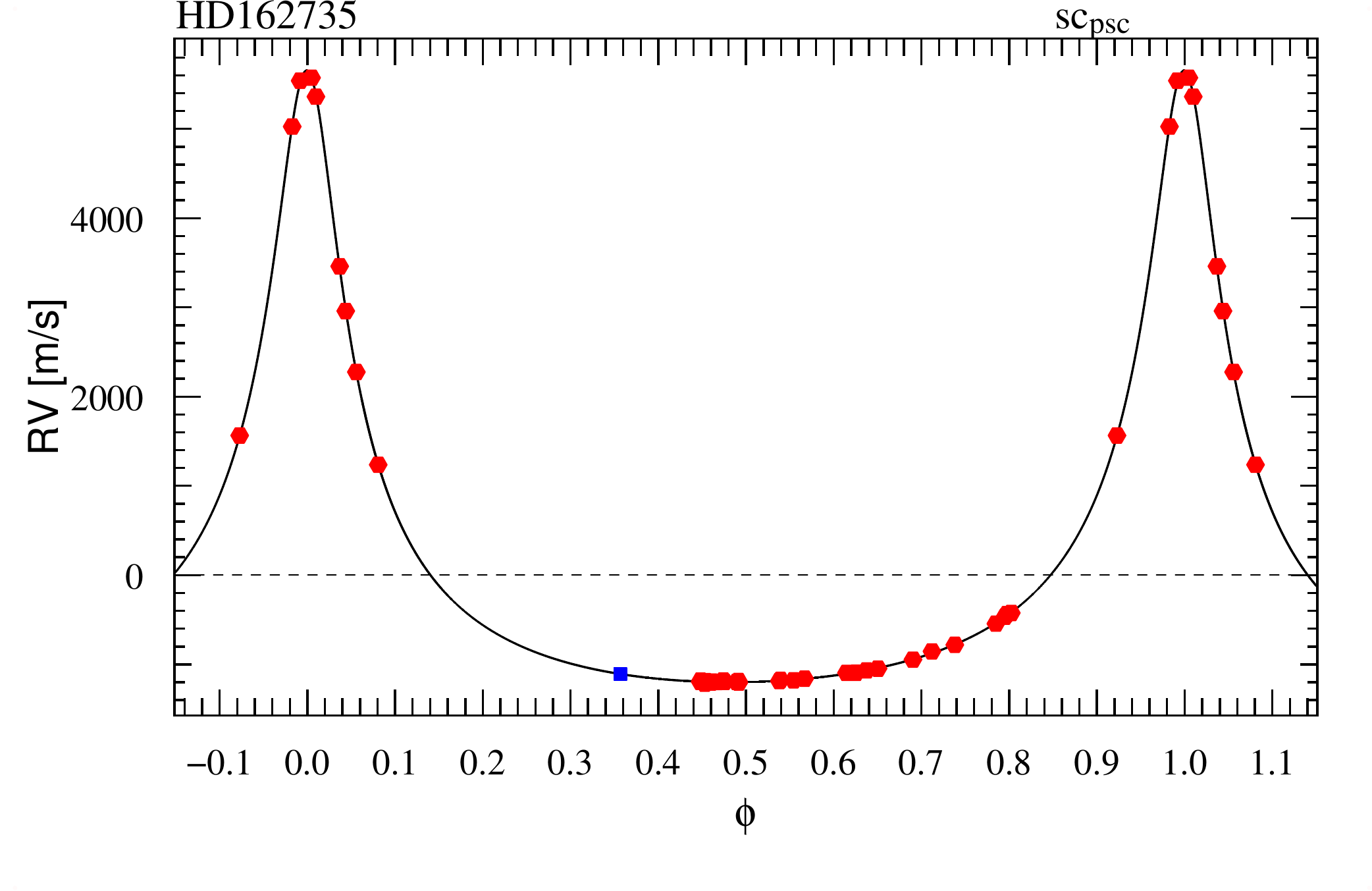}
\includegraphics[width=0.33\hsize, clip=true, trim=0 -25 0 13]{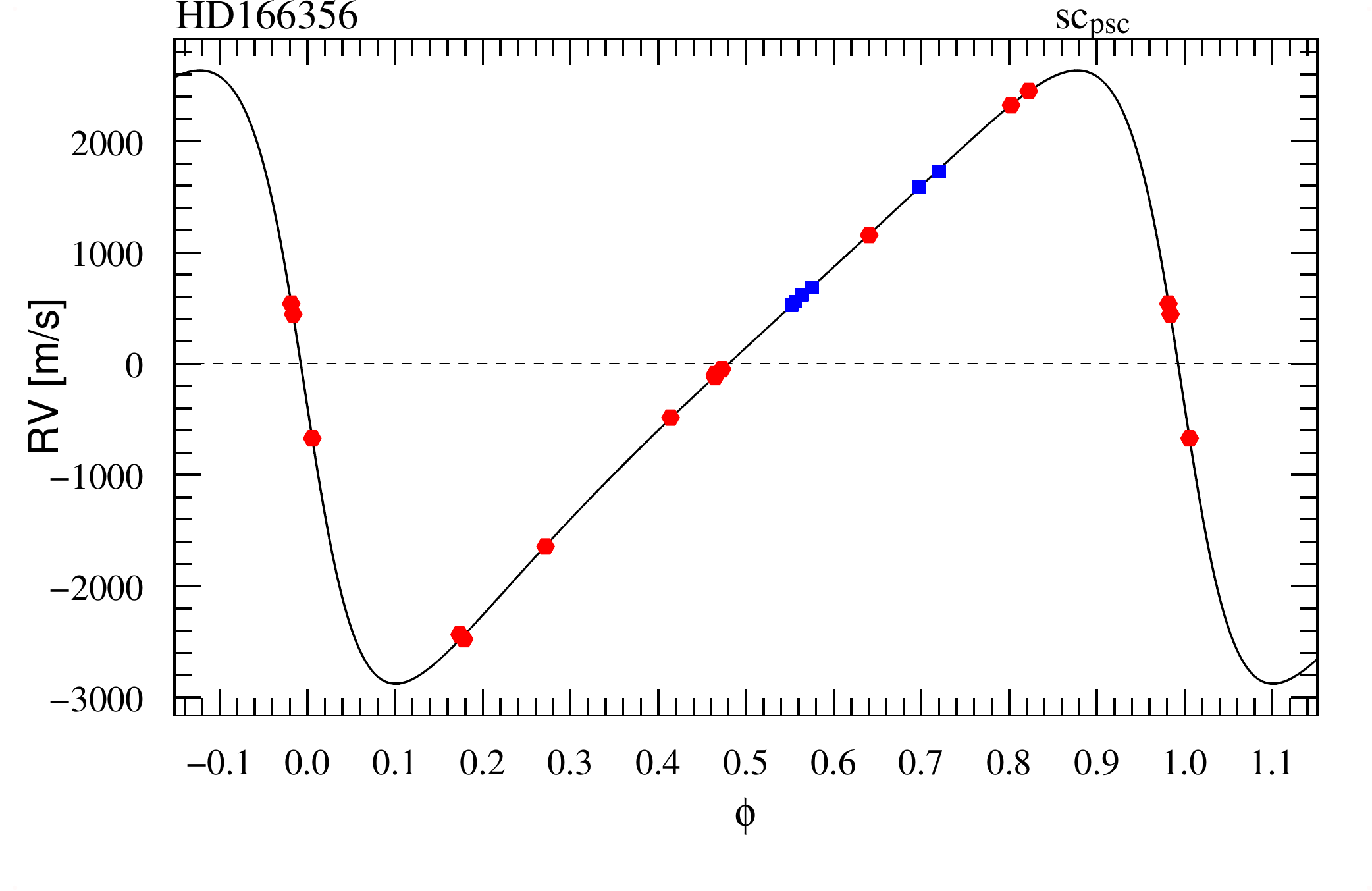}\\

\vspace{-1mm}
\includegraphics[width=0.33\hsize,height=60mm]{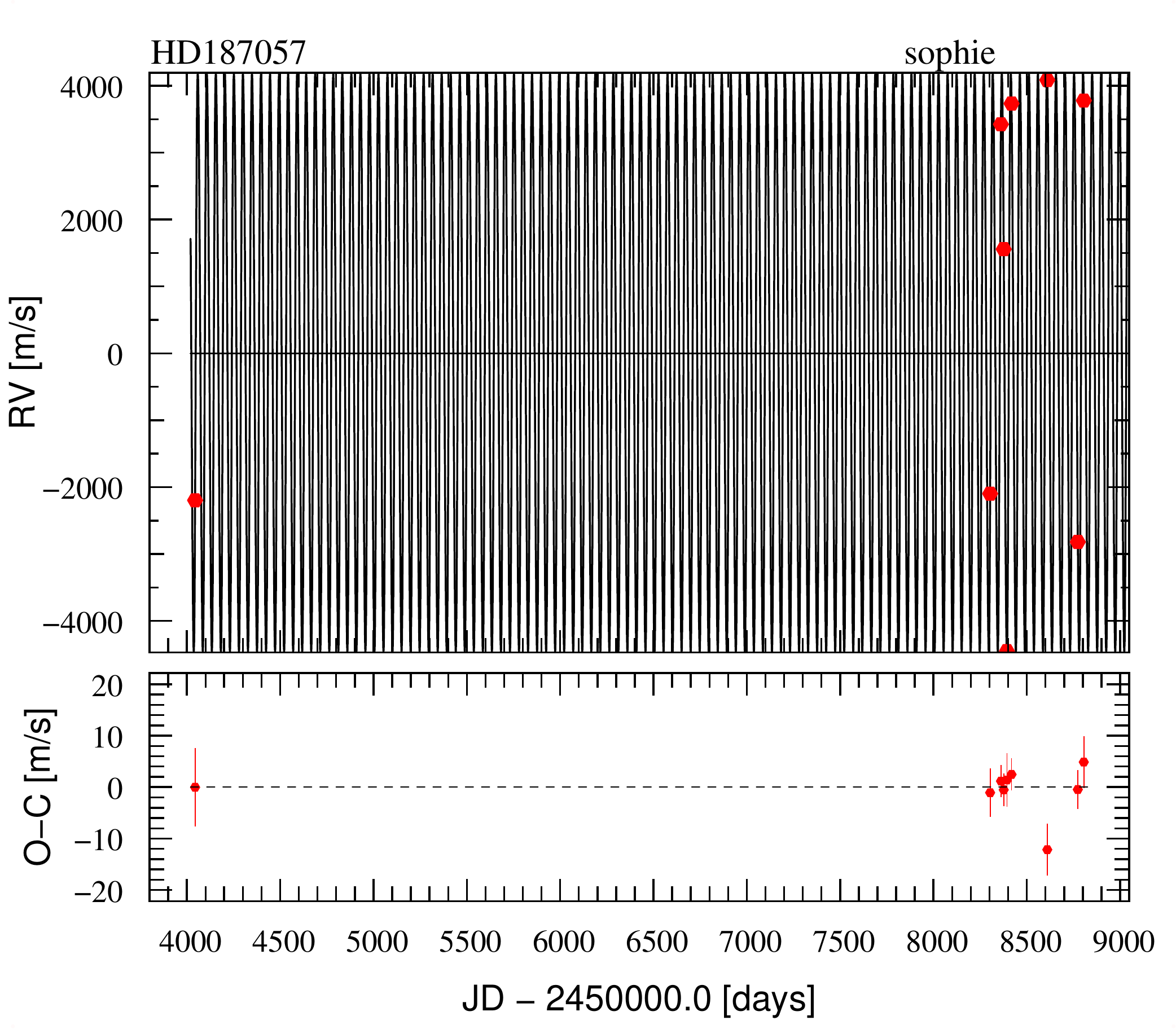}
\includegraphics[width=0.33\hsize,height=60mm]{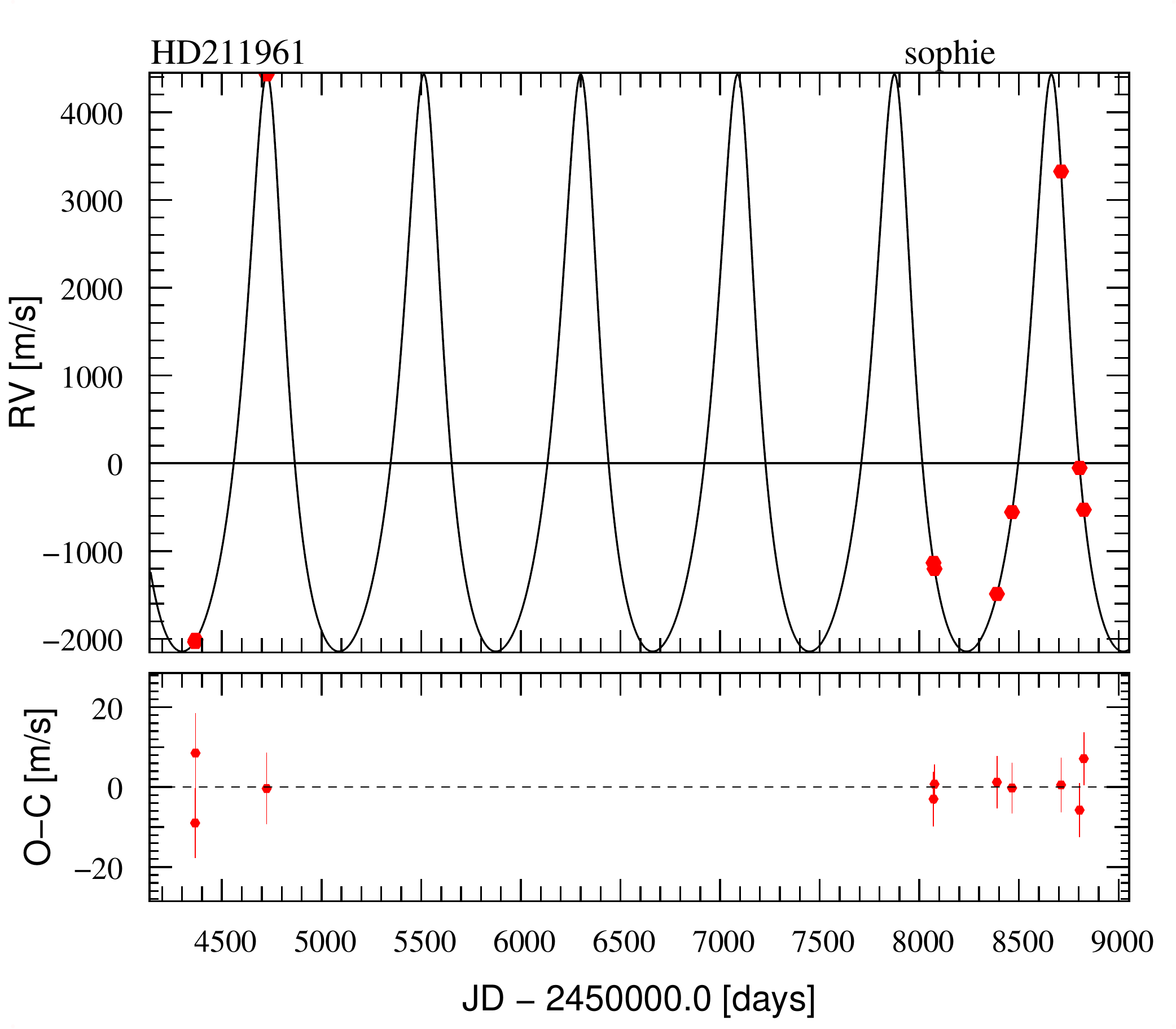}
\includegraphics[width=0.33\hsize,height=60mm]{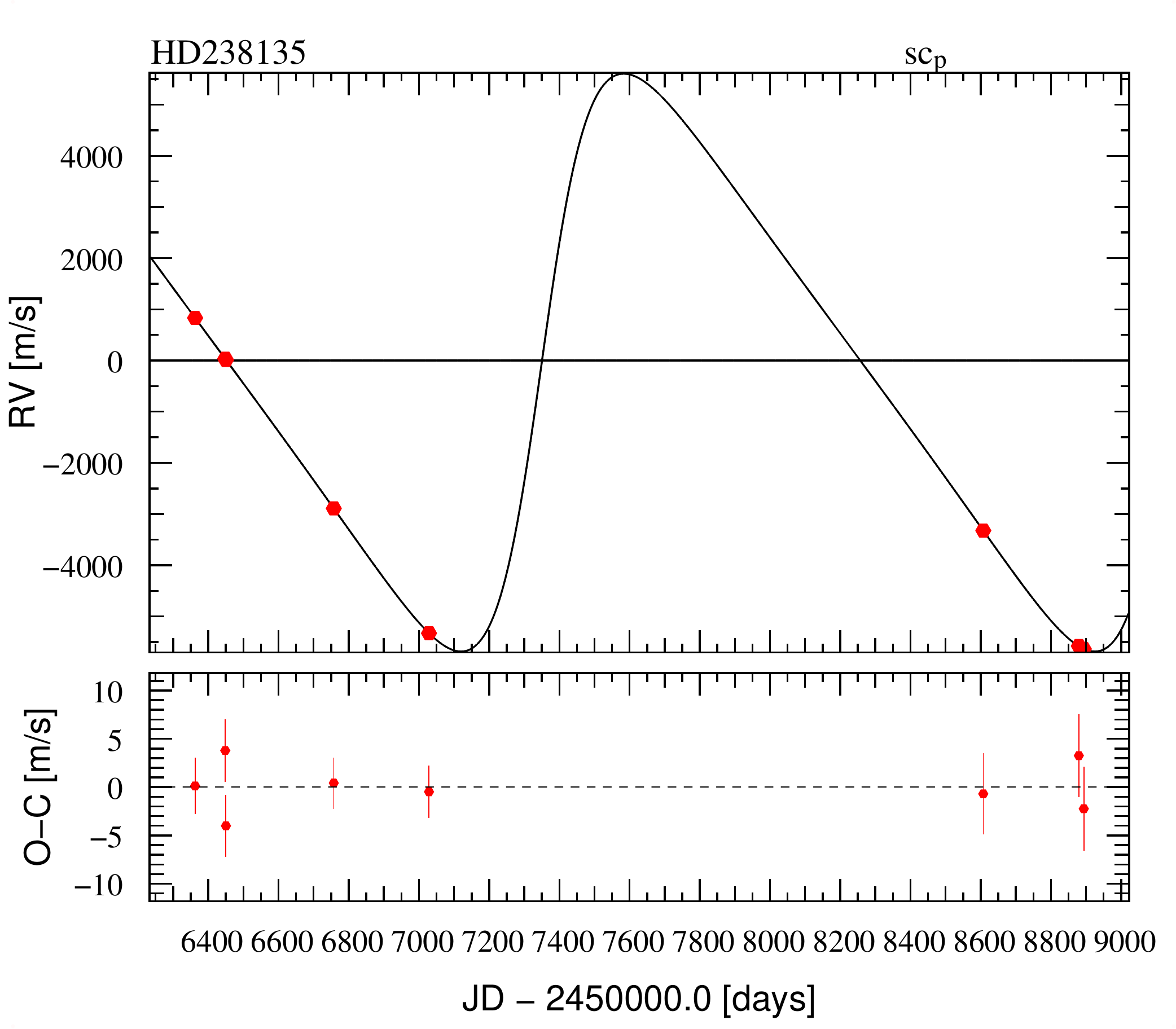}\\
\vspace{10mm}
\includegraphics[width=0.33\hsize, clip=true, trim=0 -25 0 13]{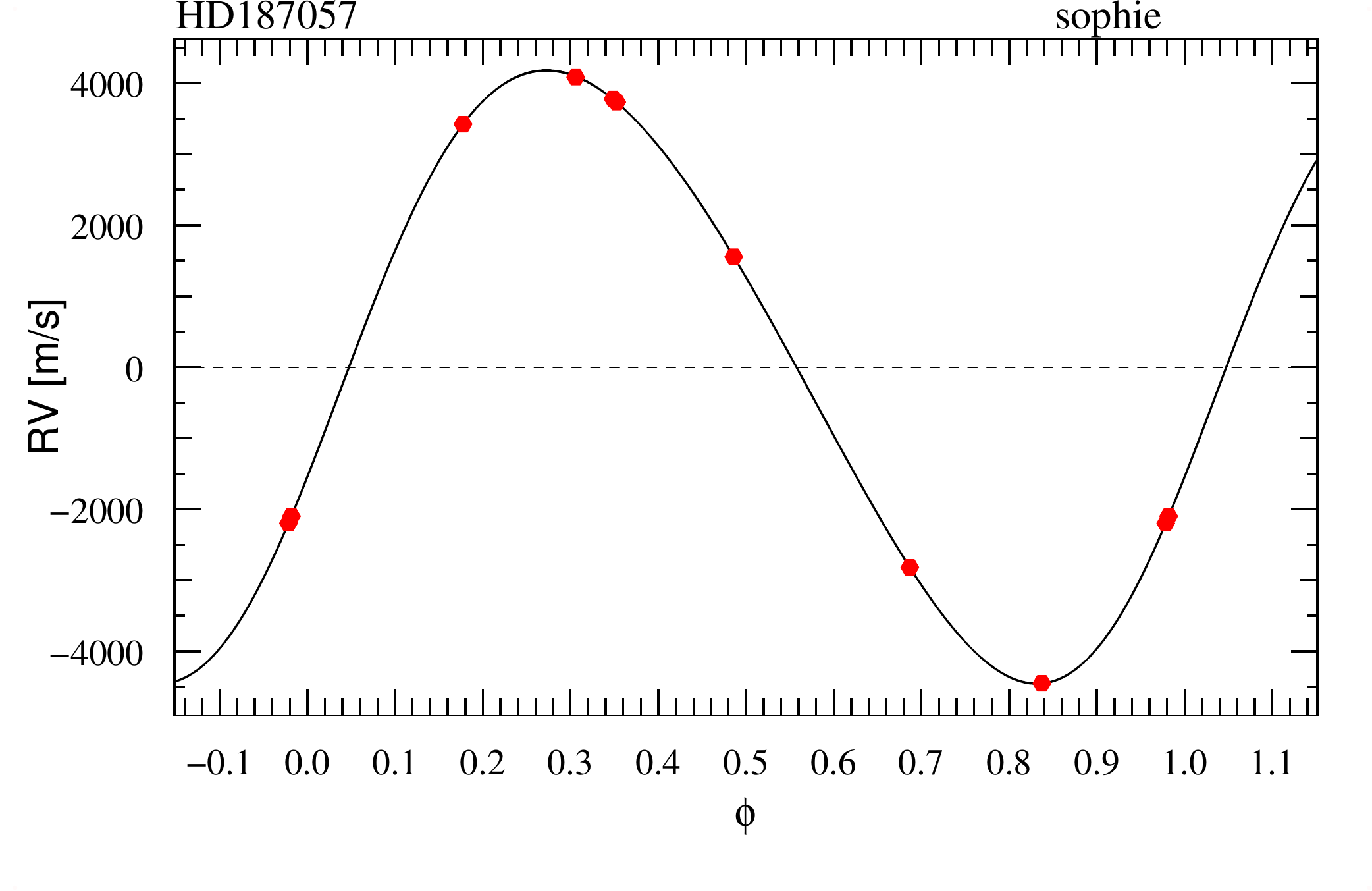}
\includegraphics[width=0.33\hsize, clip=true, trim=0 -25 0 13]{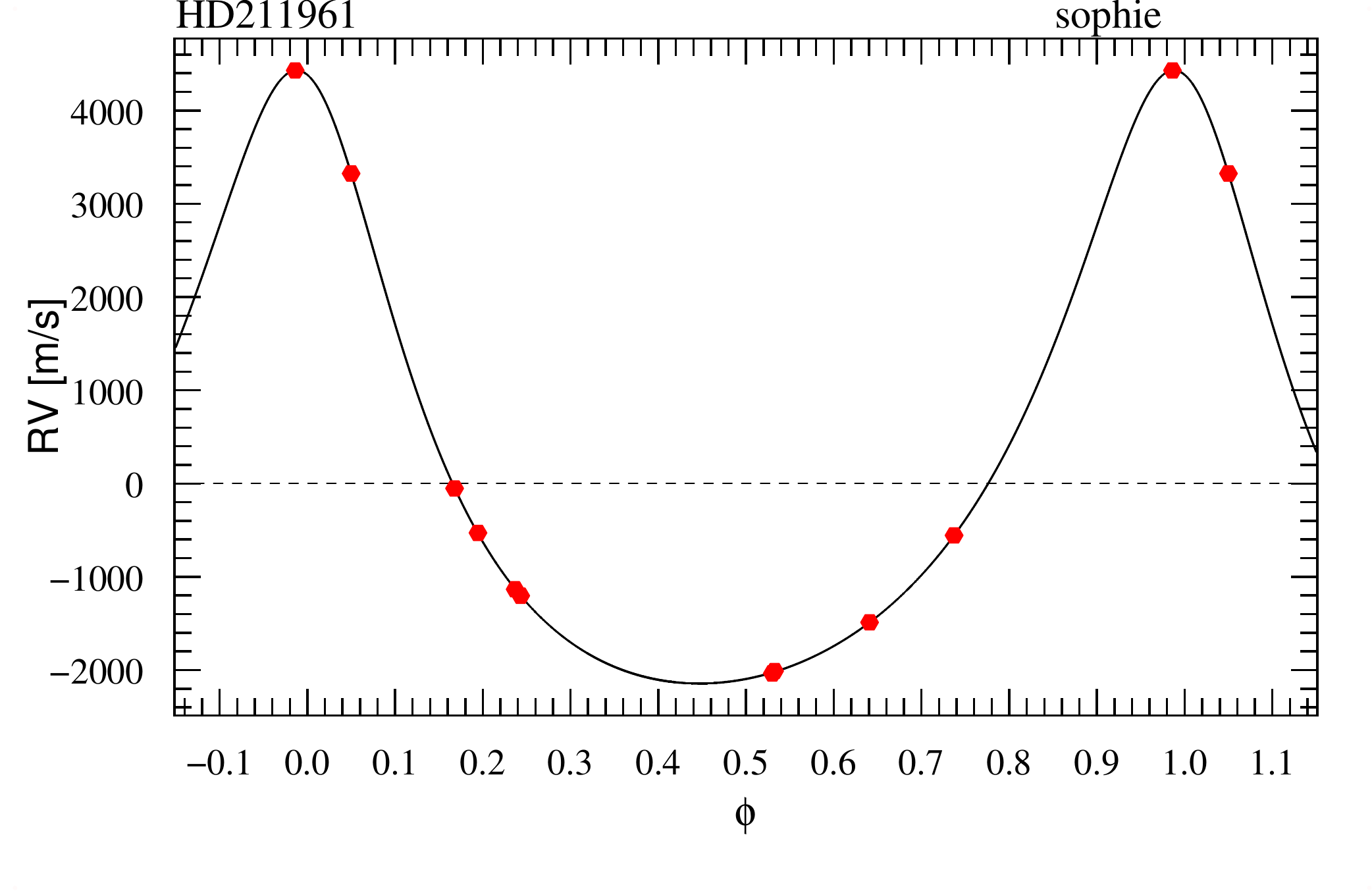}
\includegraphics[width=0.33\hsize, clip=true, trim=0 -25 0 13]{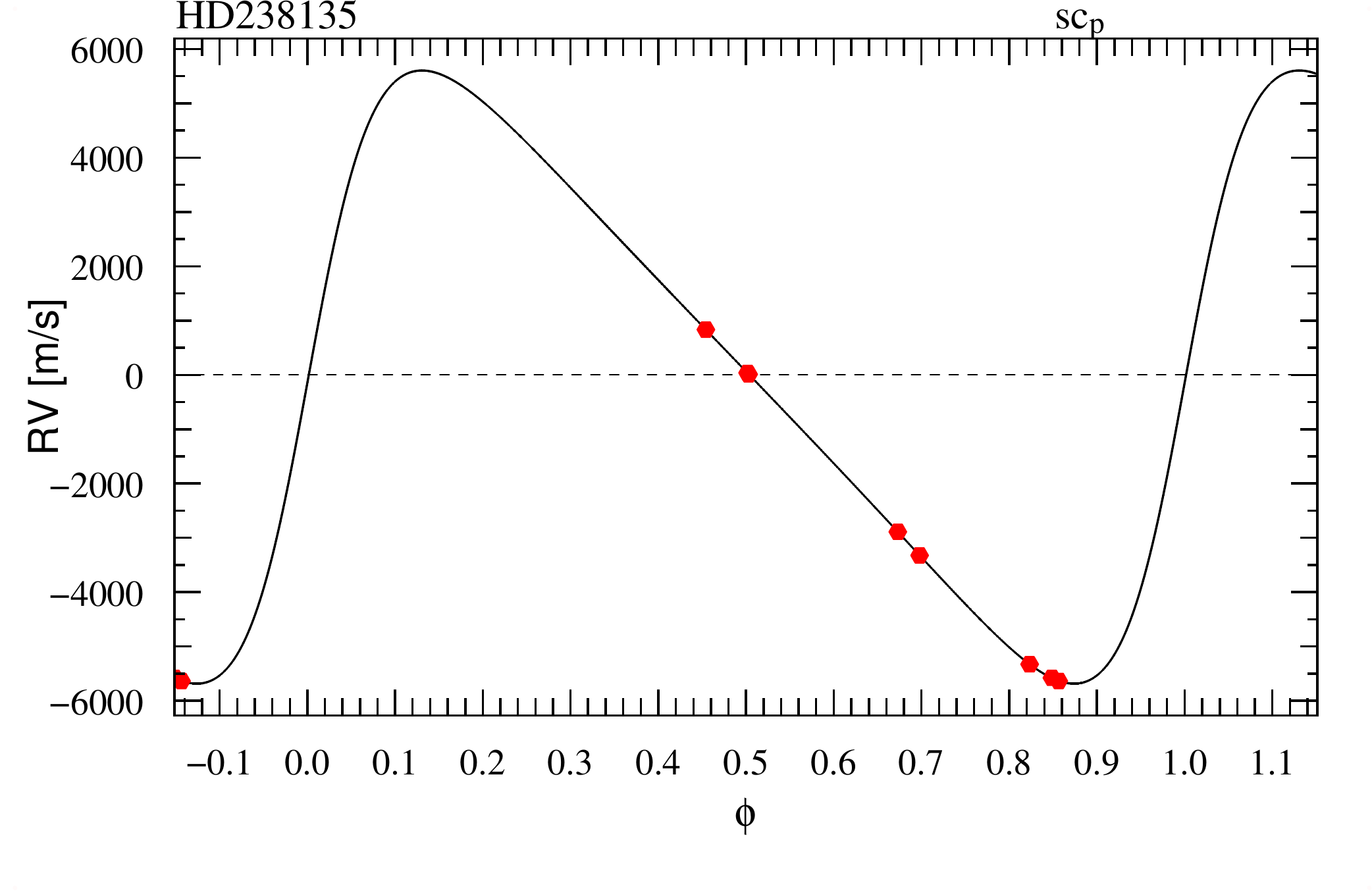}\\

\vspace{-1mm}

\includegraphics[width=0.33\hsize,height=60mm]{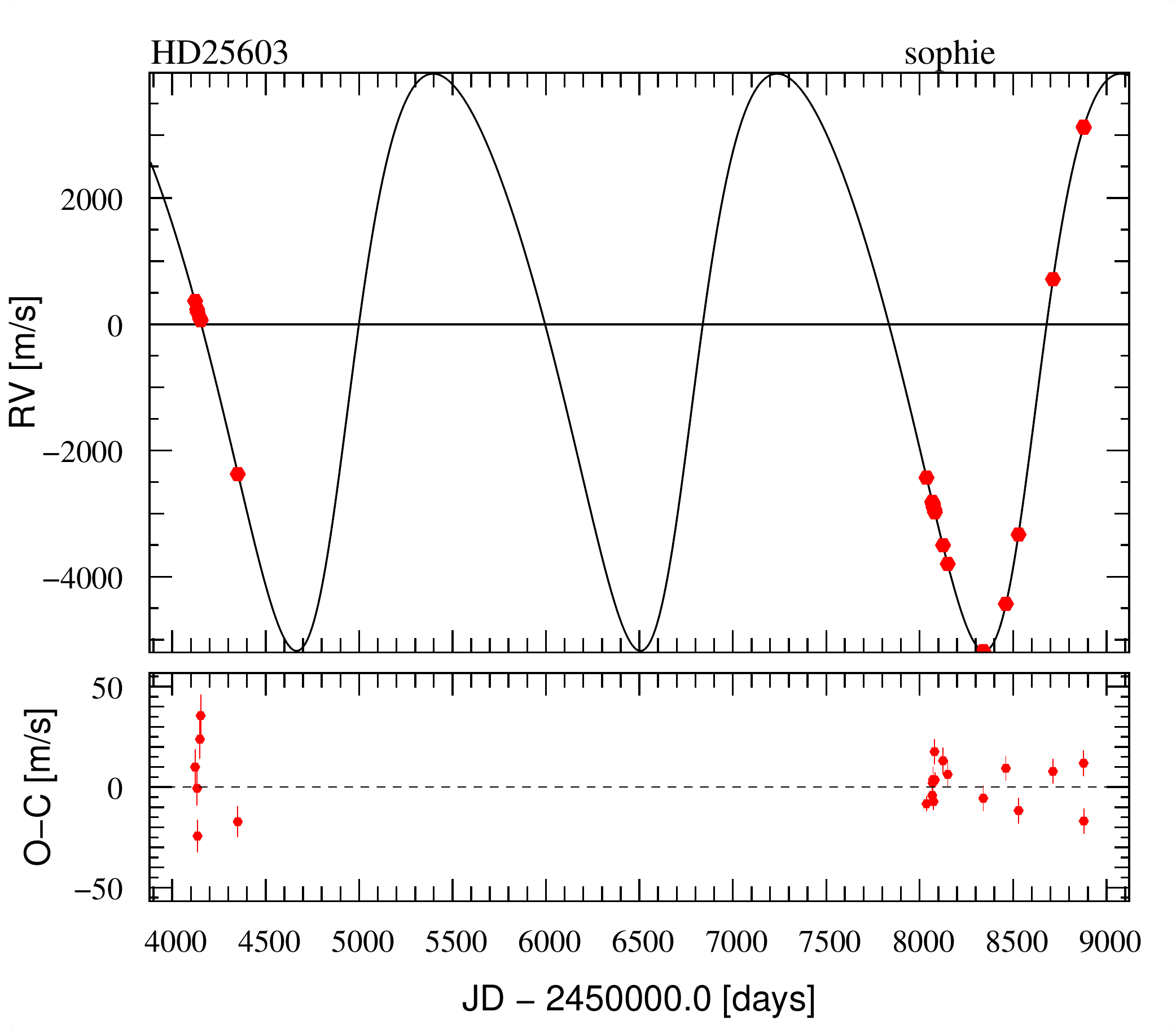}
\includegraphics[width=0.33\hsize,height=60mm]{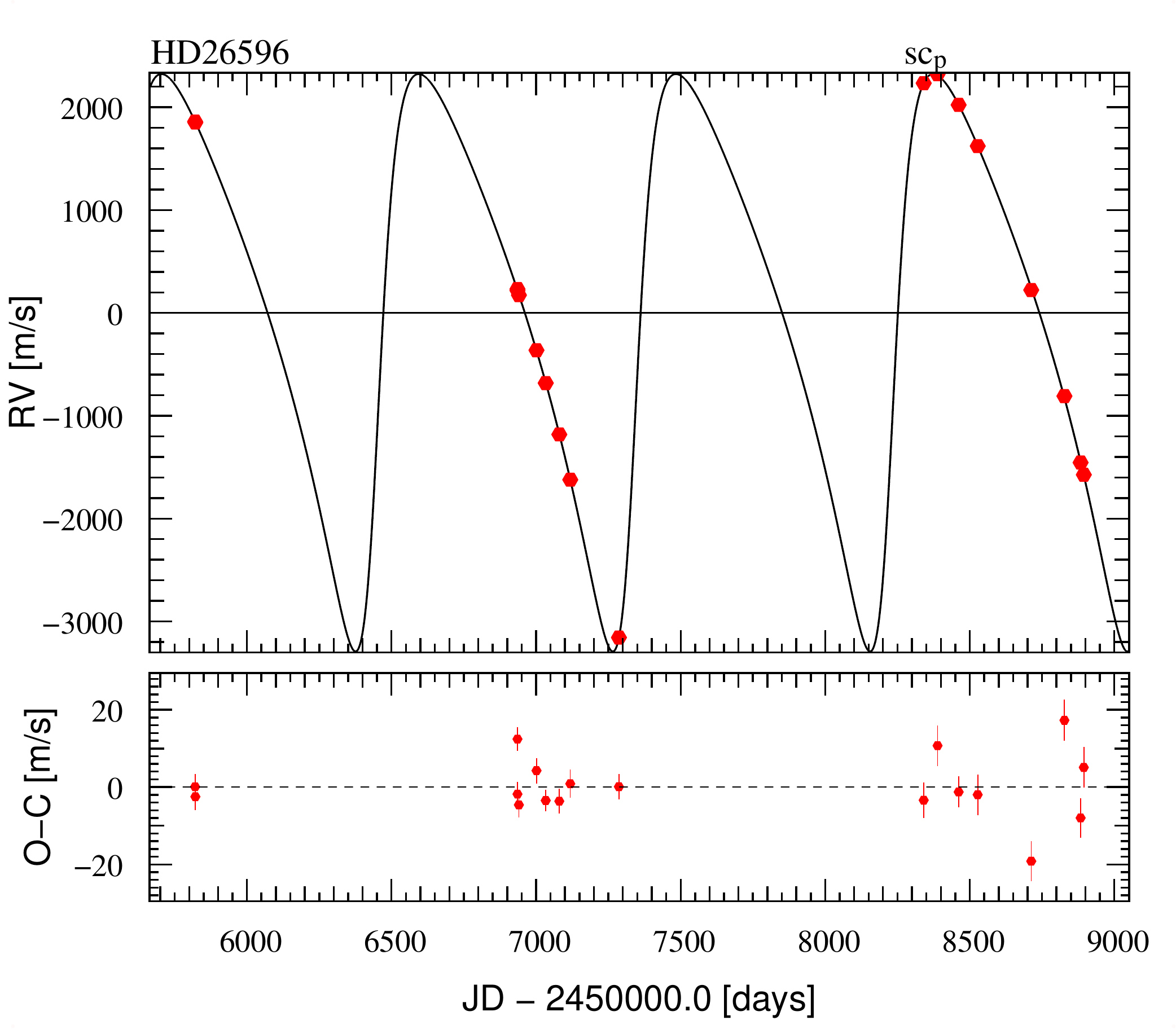}
\includegraphics[width=0.33\hsize,height=60mm]{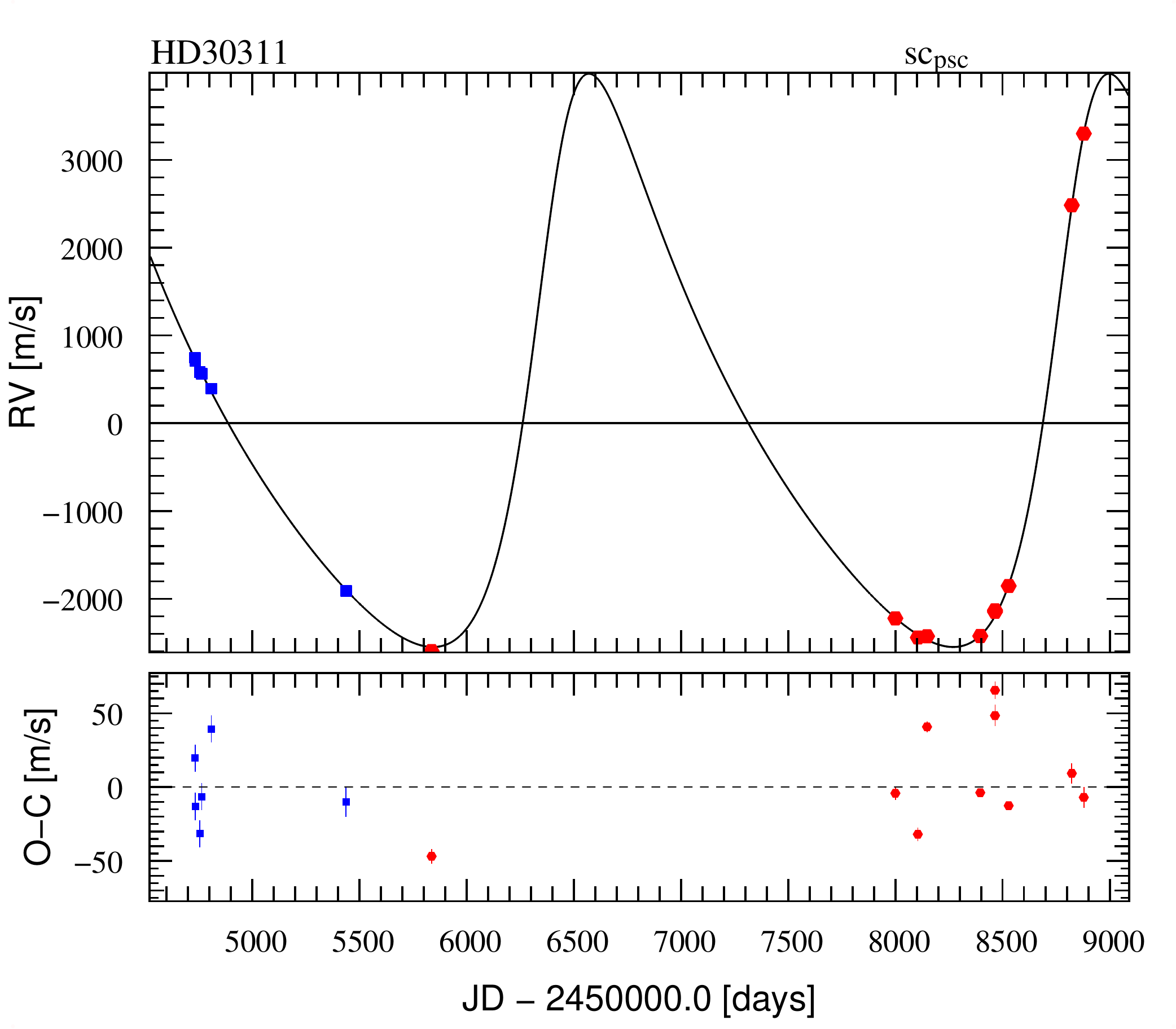}\\
\vspace{10mm}

\includegraphics[width=0.33\hsize, clip=true, trim=0 -25 0 13]{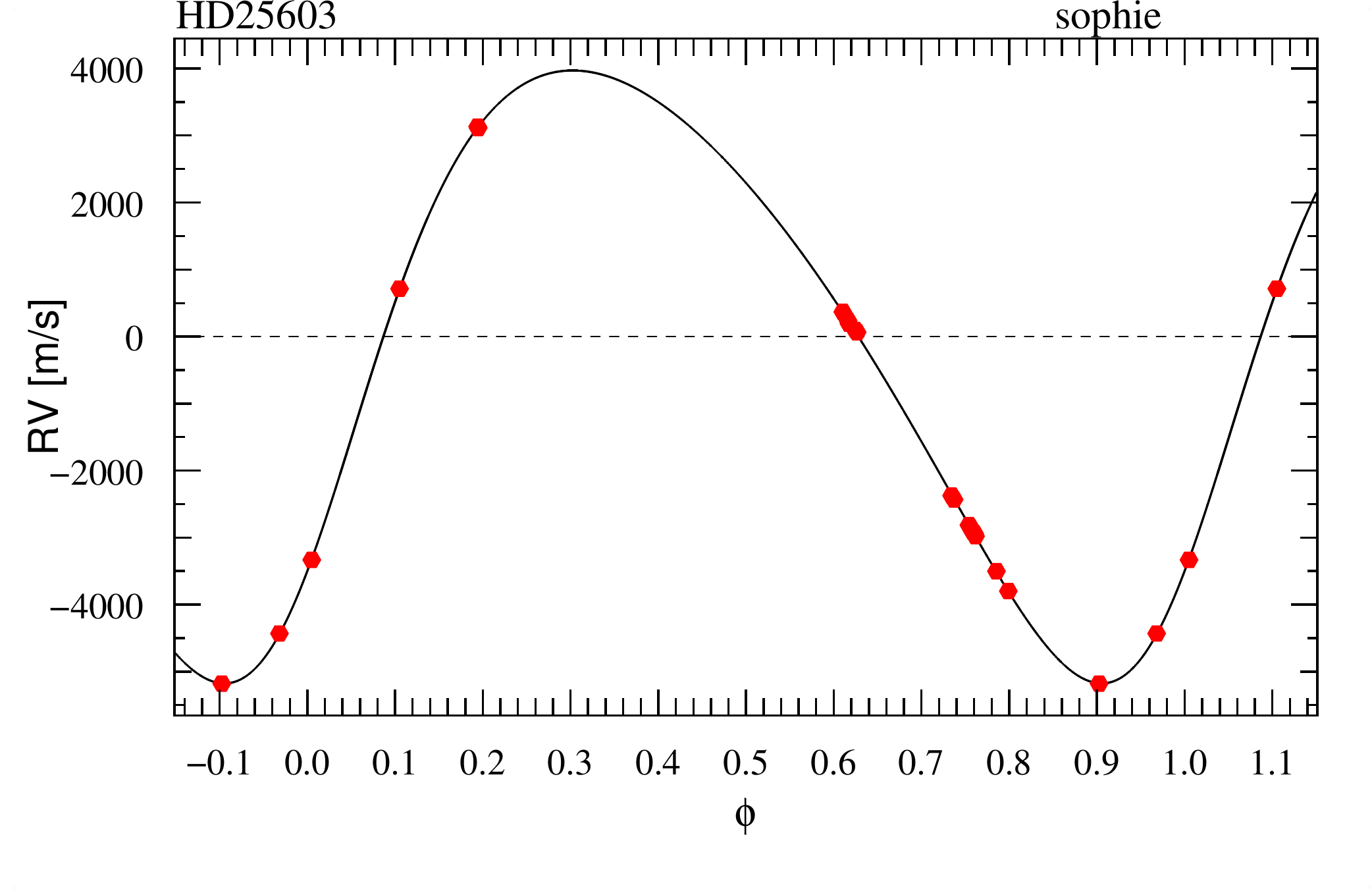}
\includegraphics[width=0.33\hsize, clip=true, trim=0 -25 0 13]{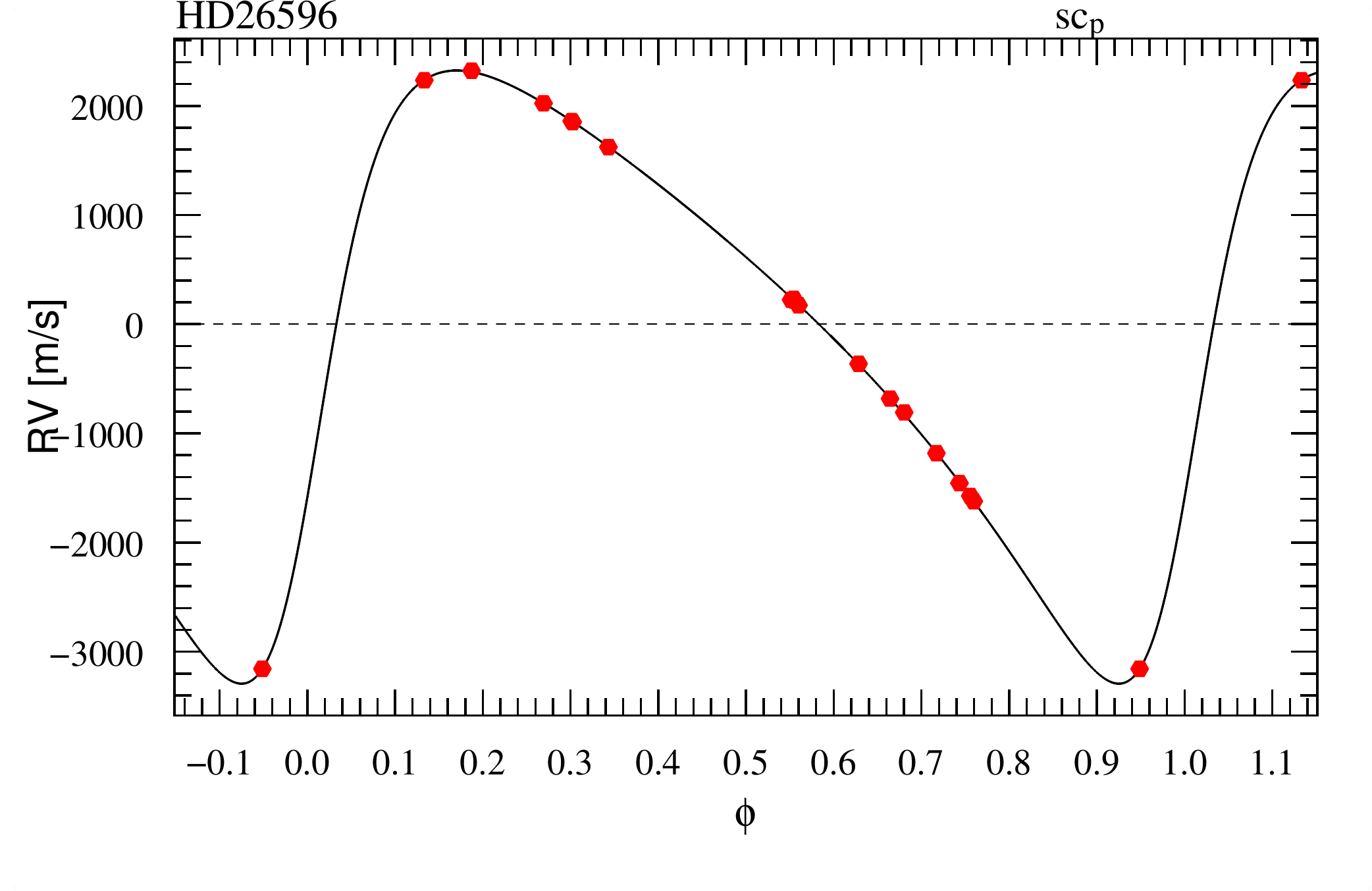}
\includegraphics[width=0.33\hsize, clip=true, trim=0 -25 0 13]{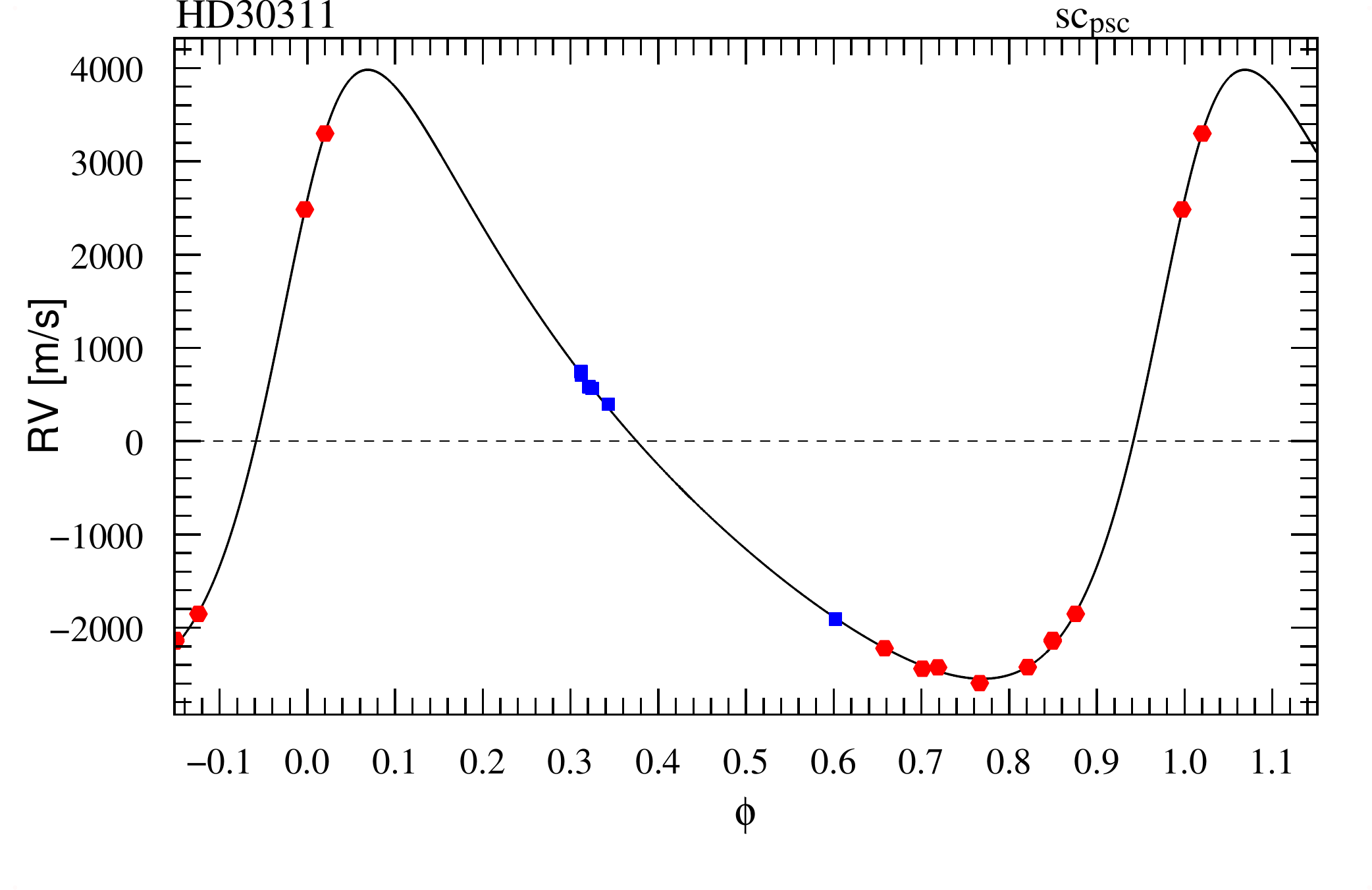}\\

\vspace{-1mm}

\includegraphics[width=0.33\hsize,height=60mm]{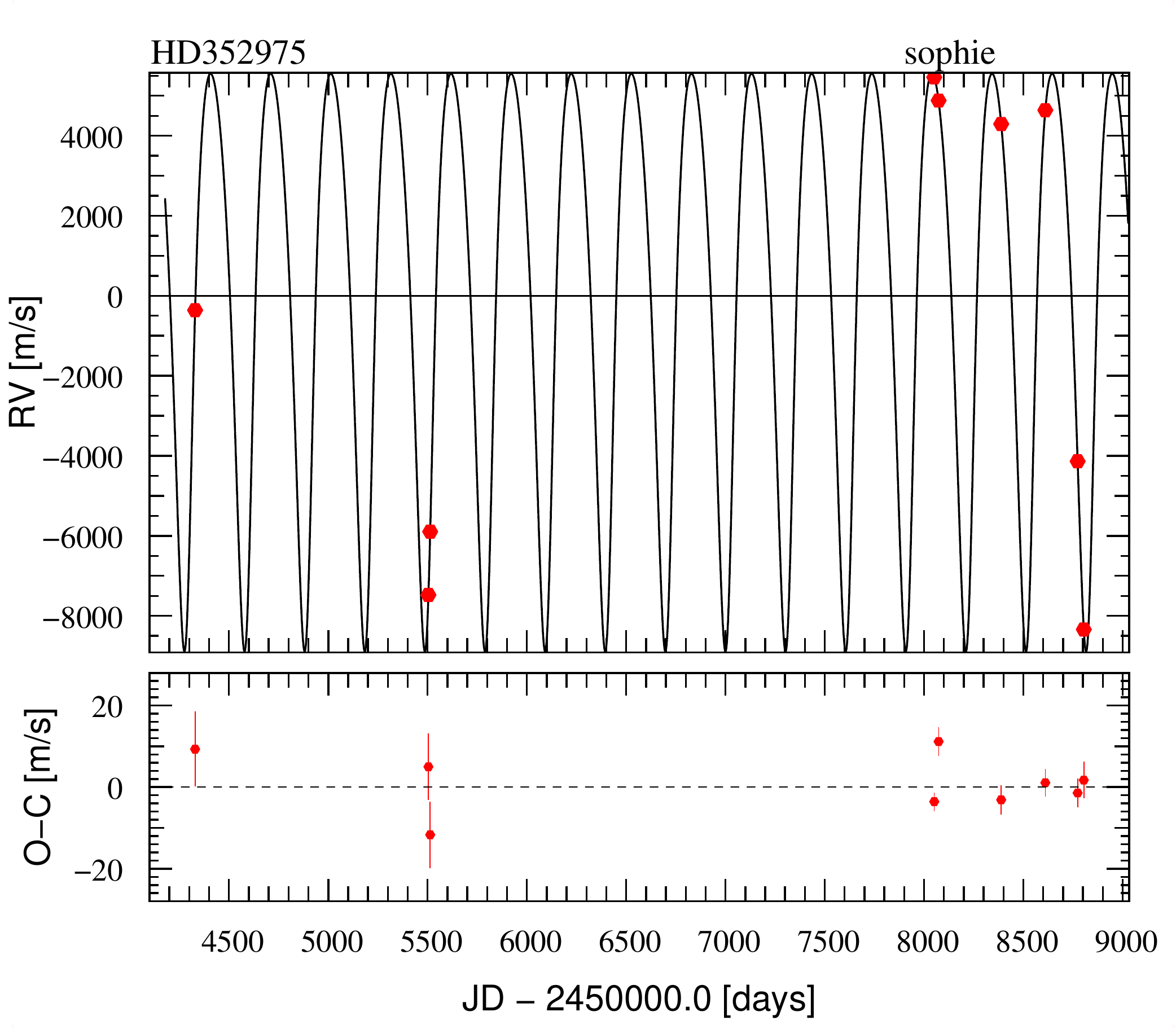}
\includegraphics[width=0.33\hsize,height=60mm]{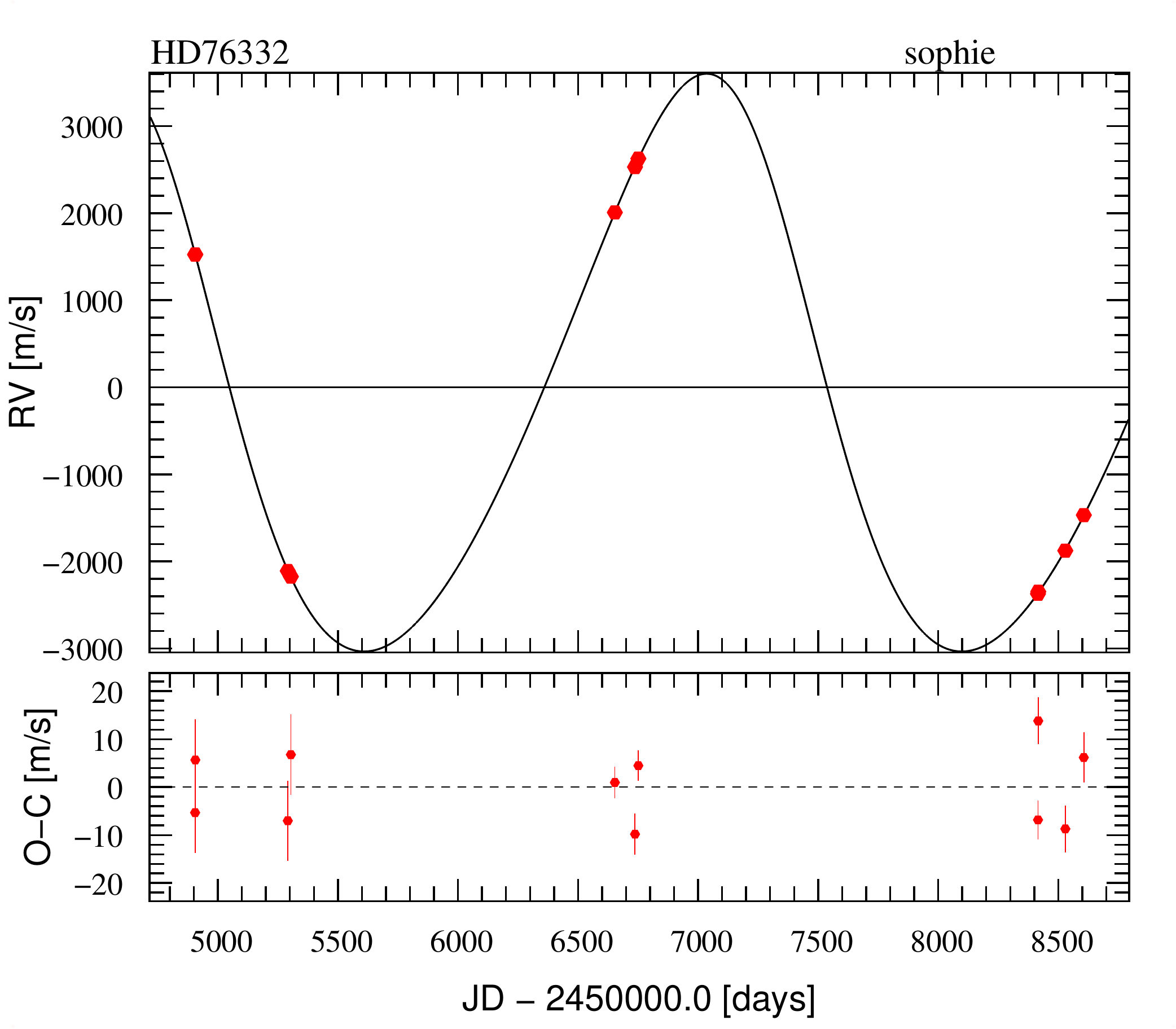}
\includegraphics[width=0.33\hsize,height=60mm]{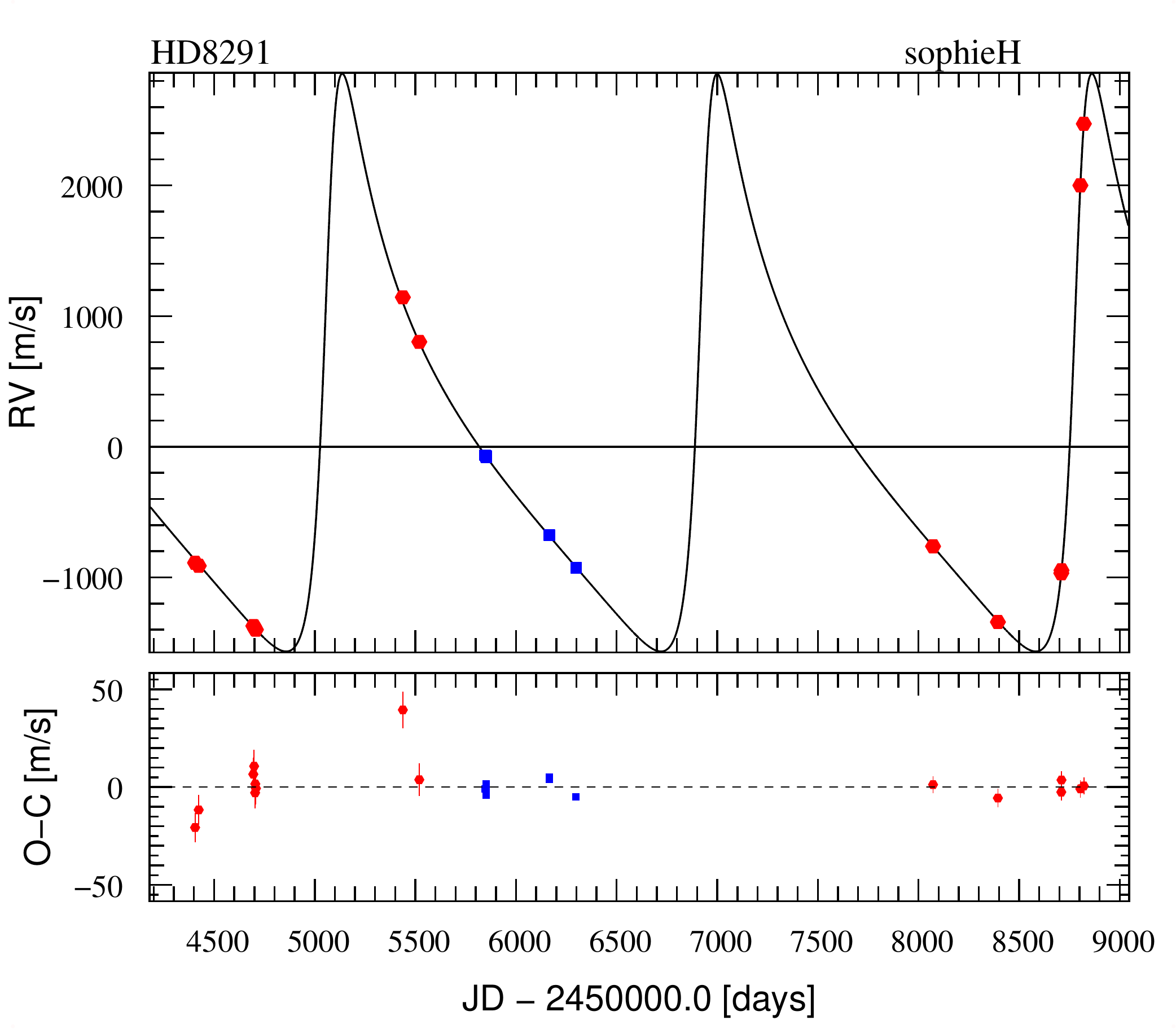}\\
\vspace{10mm}

\includegraphics[width=0.33\hsize, clip=true, trim=0 -25 0 13]{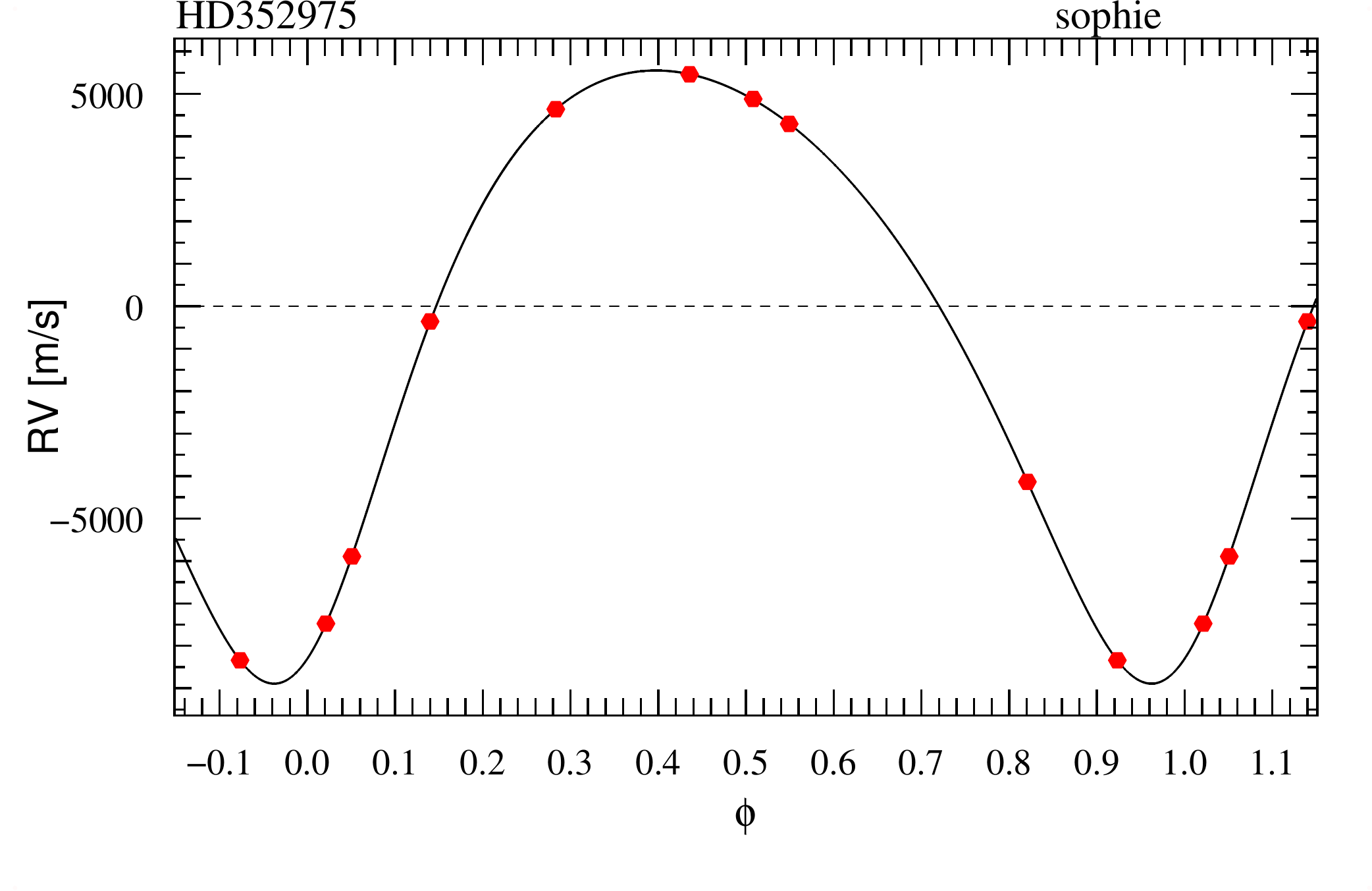}
\includegraphics[width=0.33\hsize, clip=true, trim=0 -25 0 13]{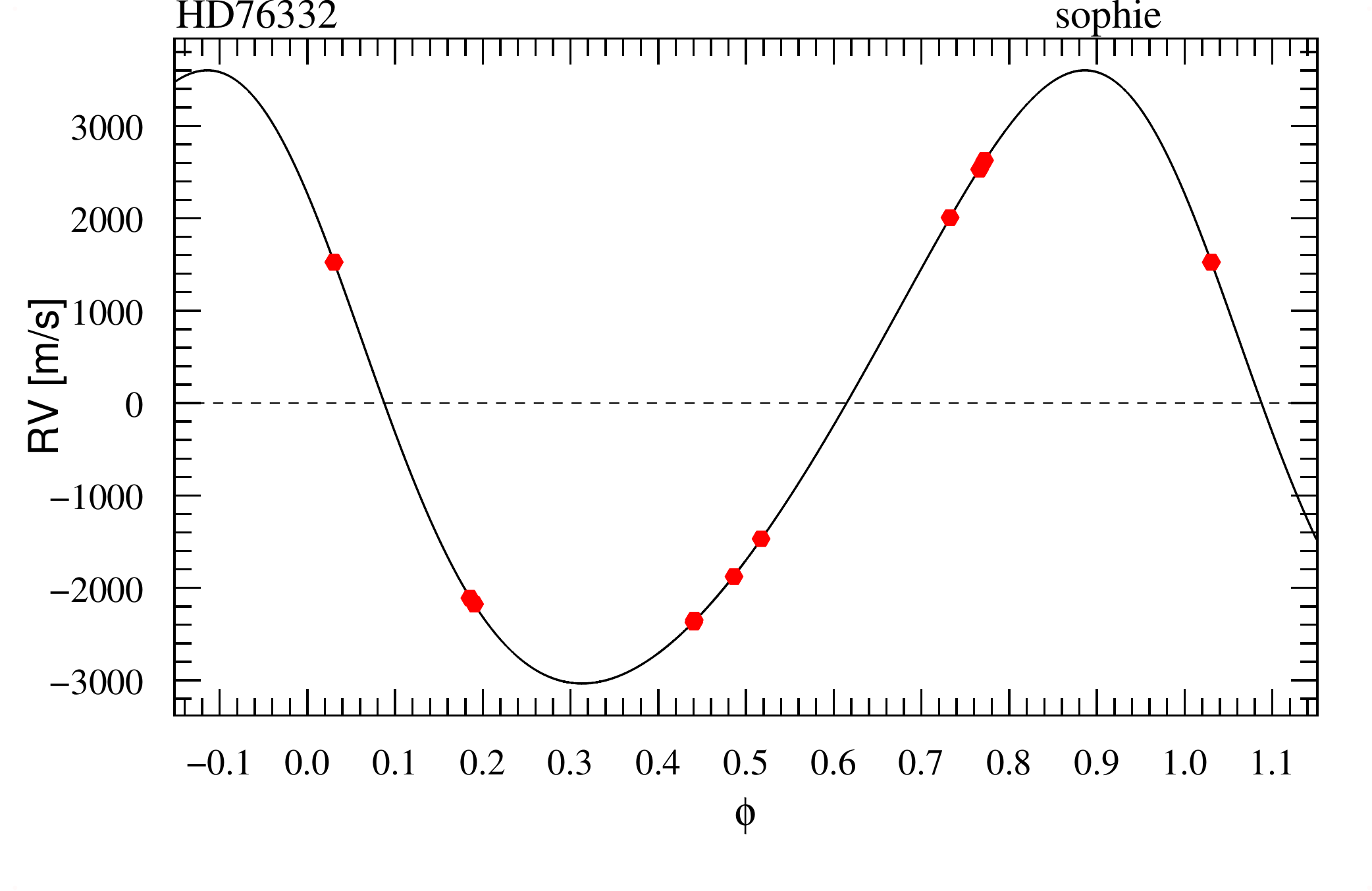}
\includegraphics[width=0.33\hsize, clip=true, trim=0 -25 0 13]{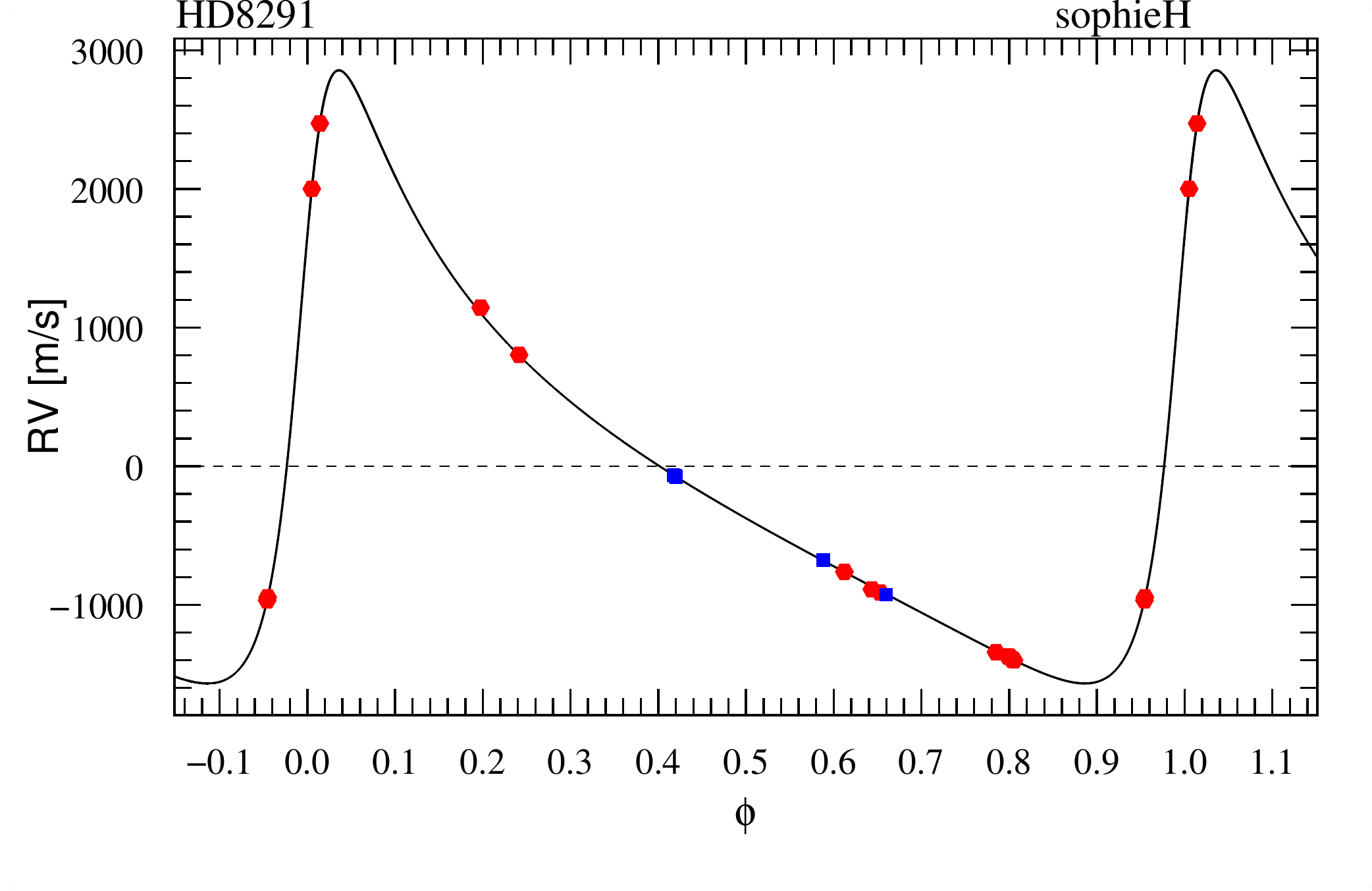}\\

\vspace{-1mm}

\includegraphics[width=0.33\hsize,height=60mm]{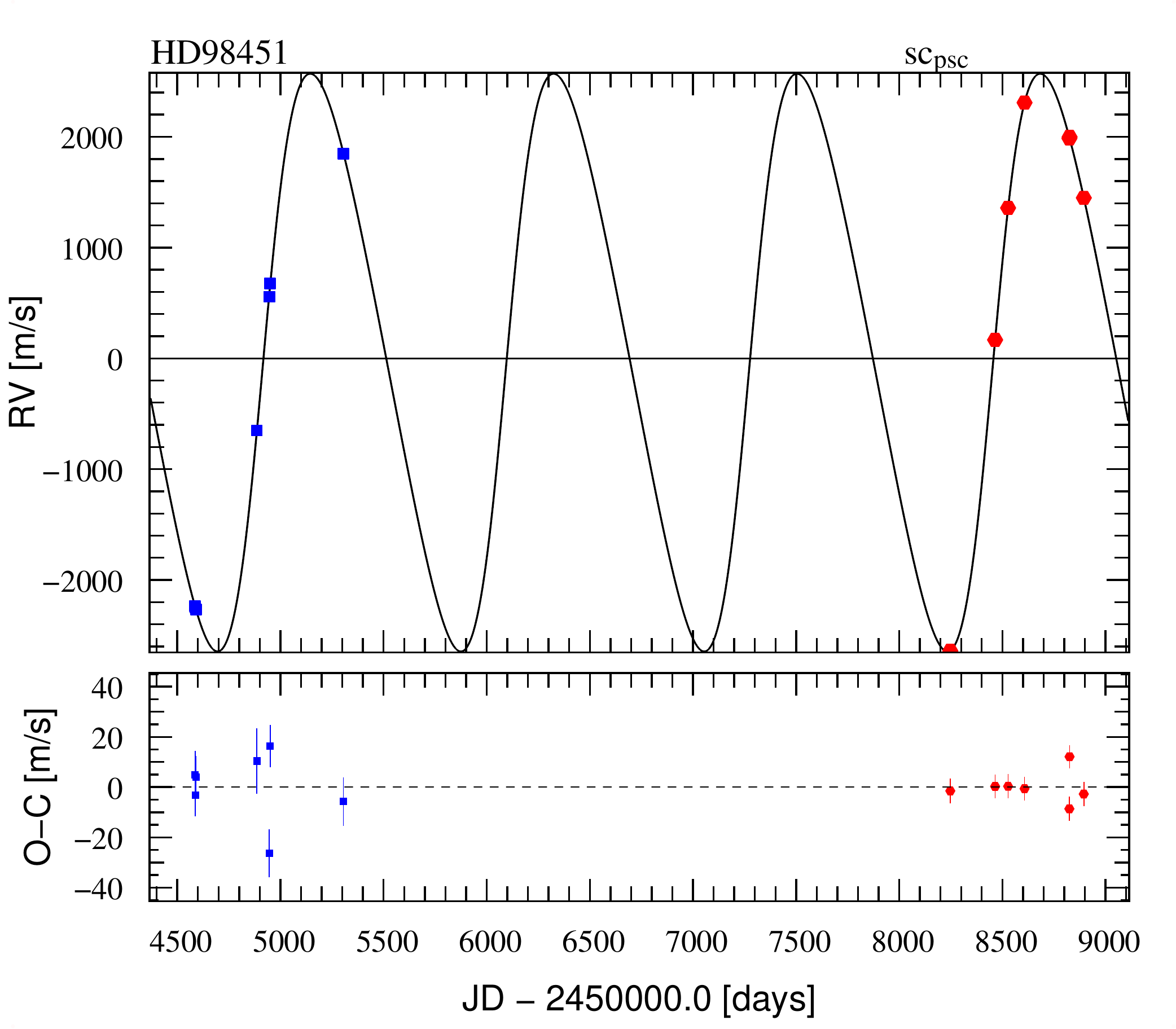}\\
\vspace{10mm}

\includegraphics[width=0.33\hsize, clip=true, trim=0 -25 0 13]{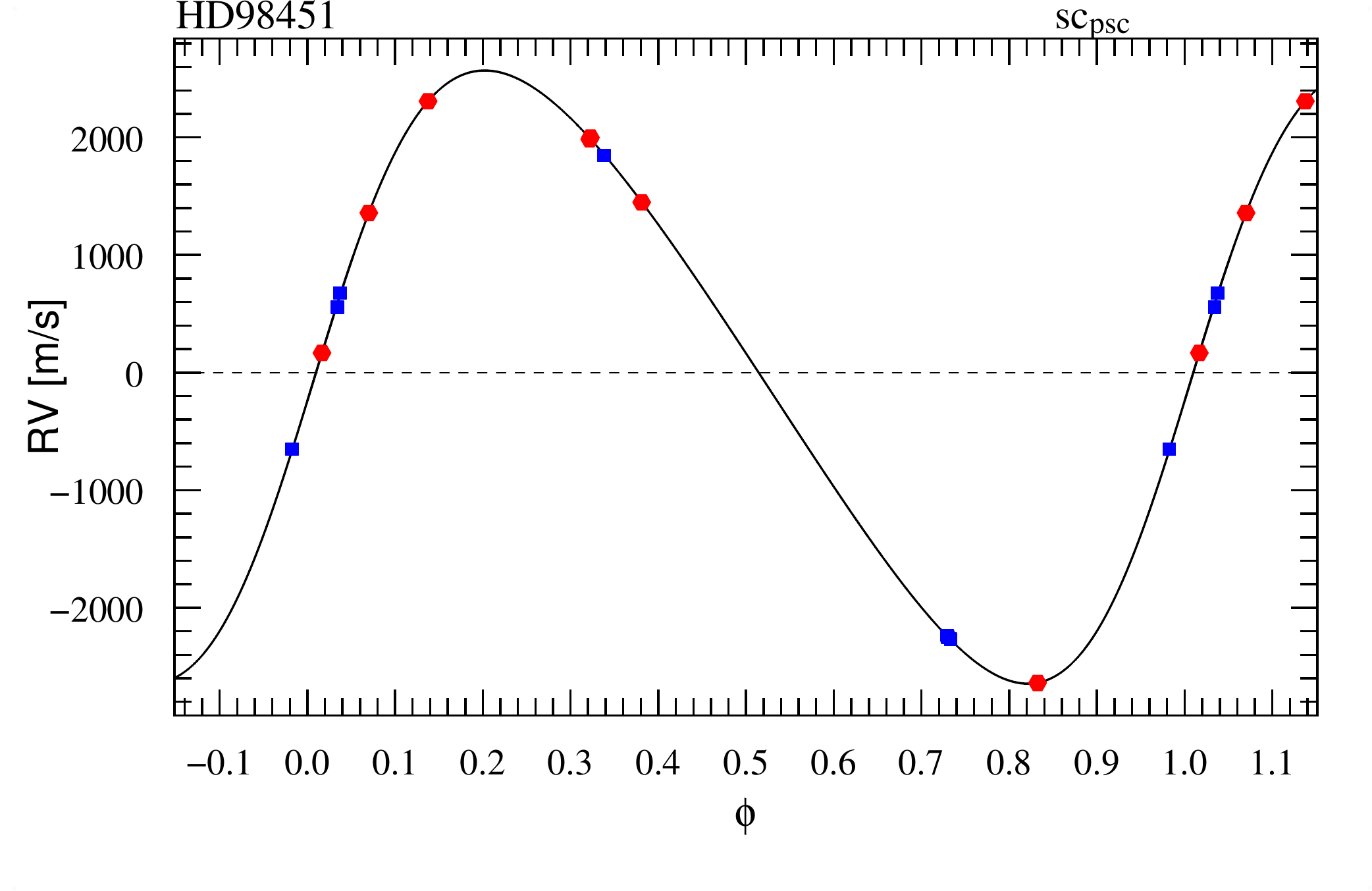}\\

\end{longfigure}

\newdimen\LFcapwidth \LFcapwidth=\linewidth
\begin{longfigure}{c}
  \caption{\label{fig:orbits_3sigma_2} \emph{Top panels:} Astrometric orbits of 3$\sigma$ detections projected on the sky. North is up and east is left. The solid red line shows the model orbit and open circles mark the individual HIPPARCOS measurements. \emph{Bottom panels:} O--C residuals for the normal points of the orbital solution (filled blue circles) and of the five-parameter model without companion (open squares).} 
  \endLFfirsthead
	\caption{Continued.}
	\endLFhead
	\vspace{20mm}
\includegraphics[width= 0.3\linewidth]{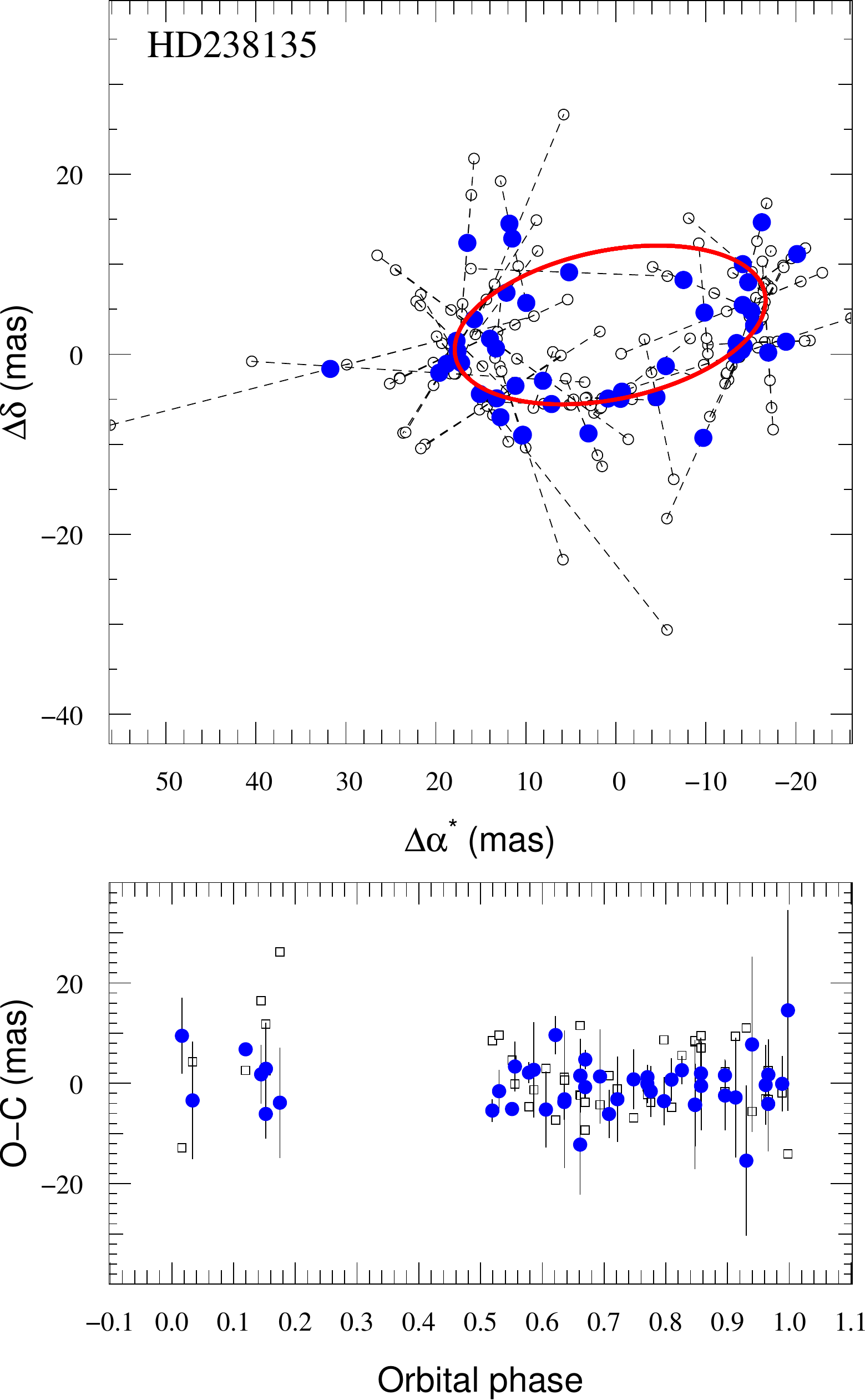} 
\includegraphics[width= 0.3\linewidth]{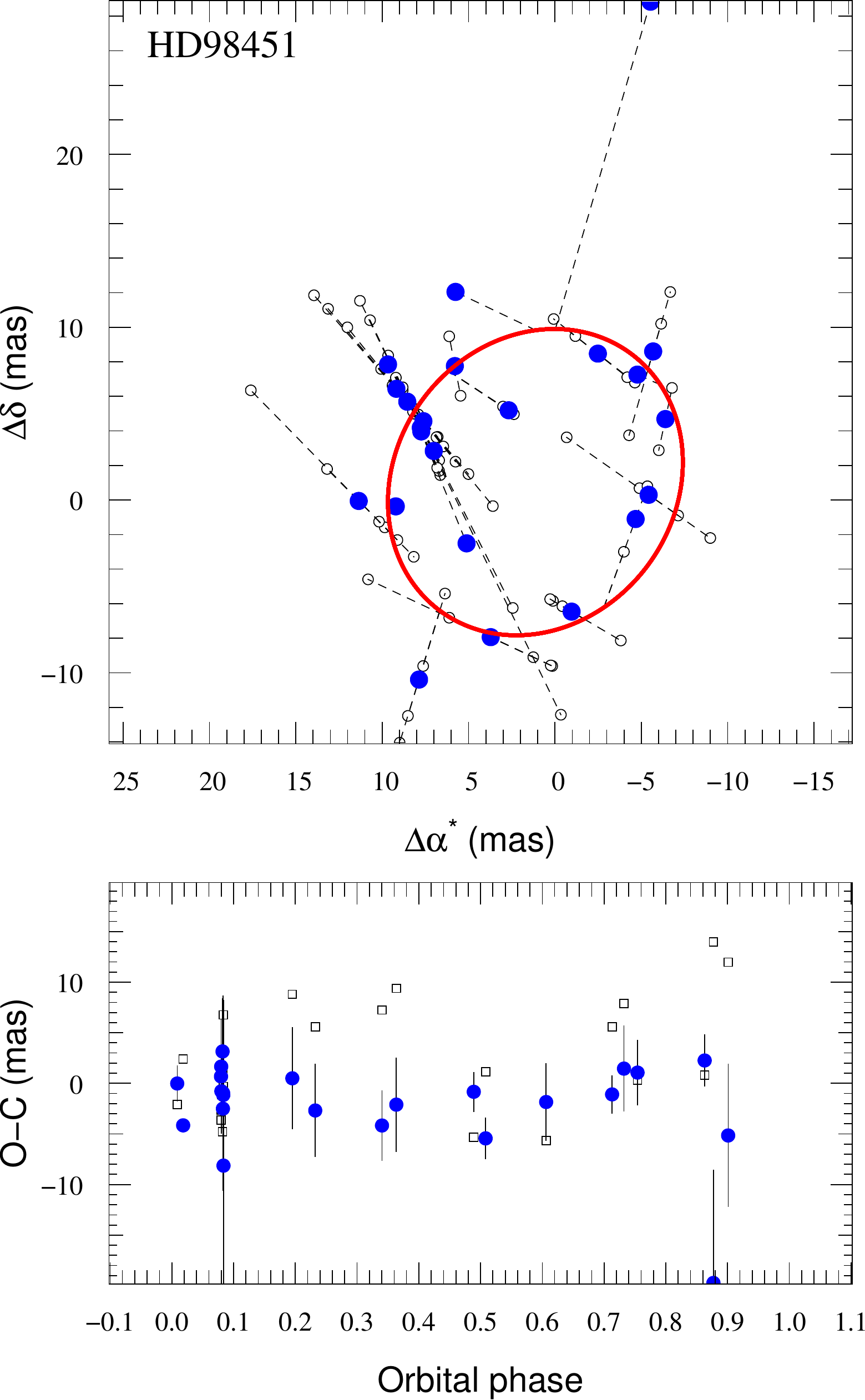}

\includegraphics[width= 0.3\linewidth]{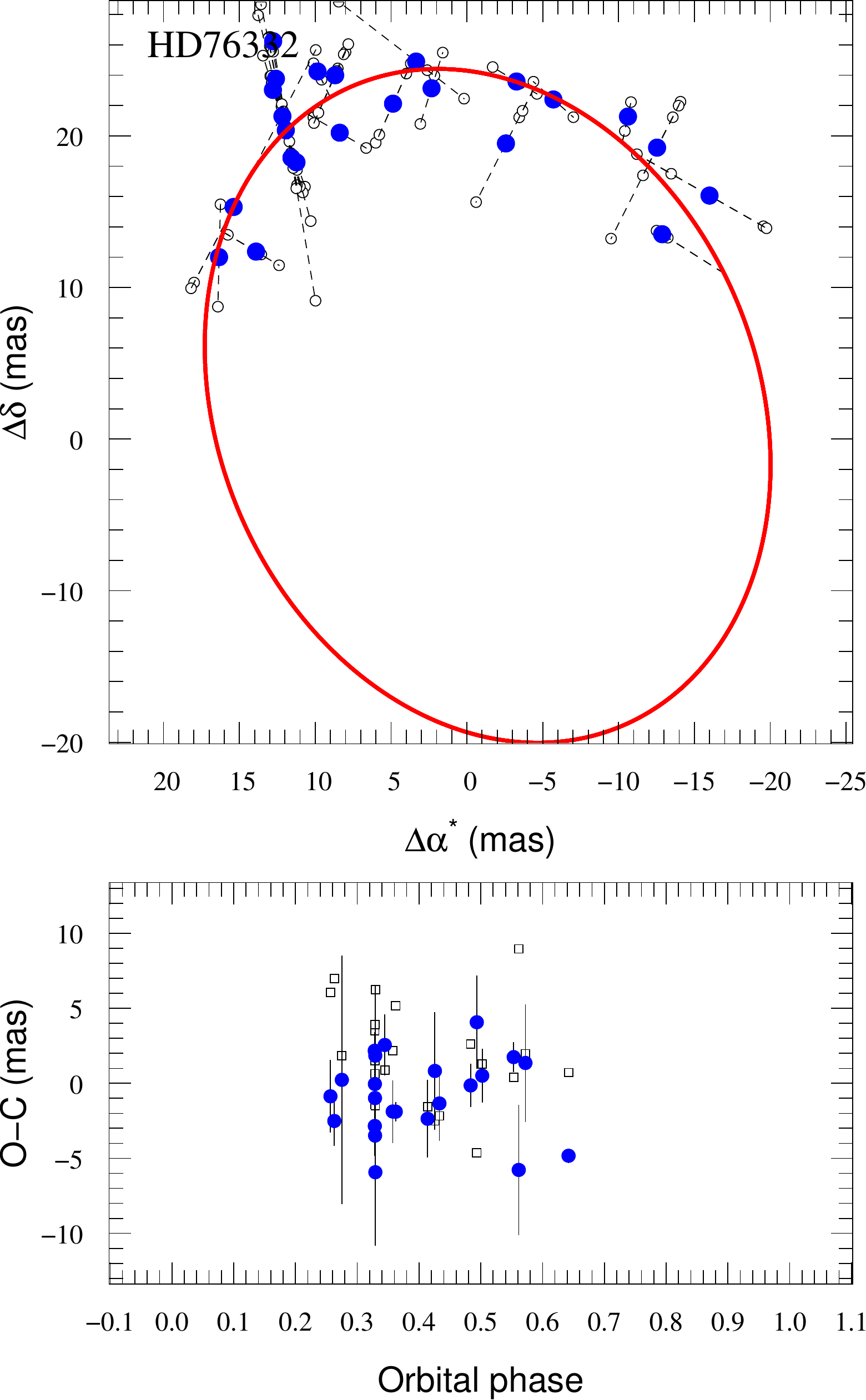} \\
\includegraphics[width= 0.3\linewidth]{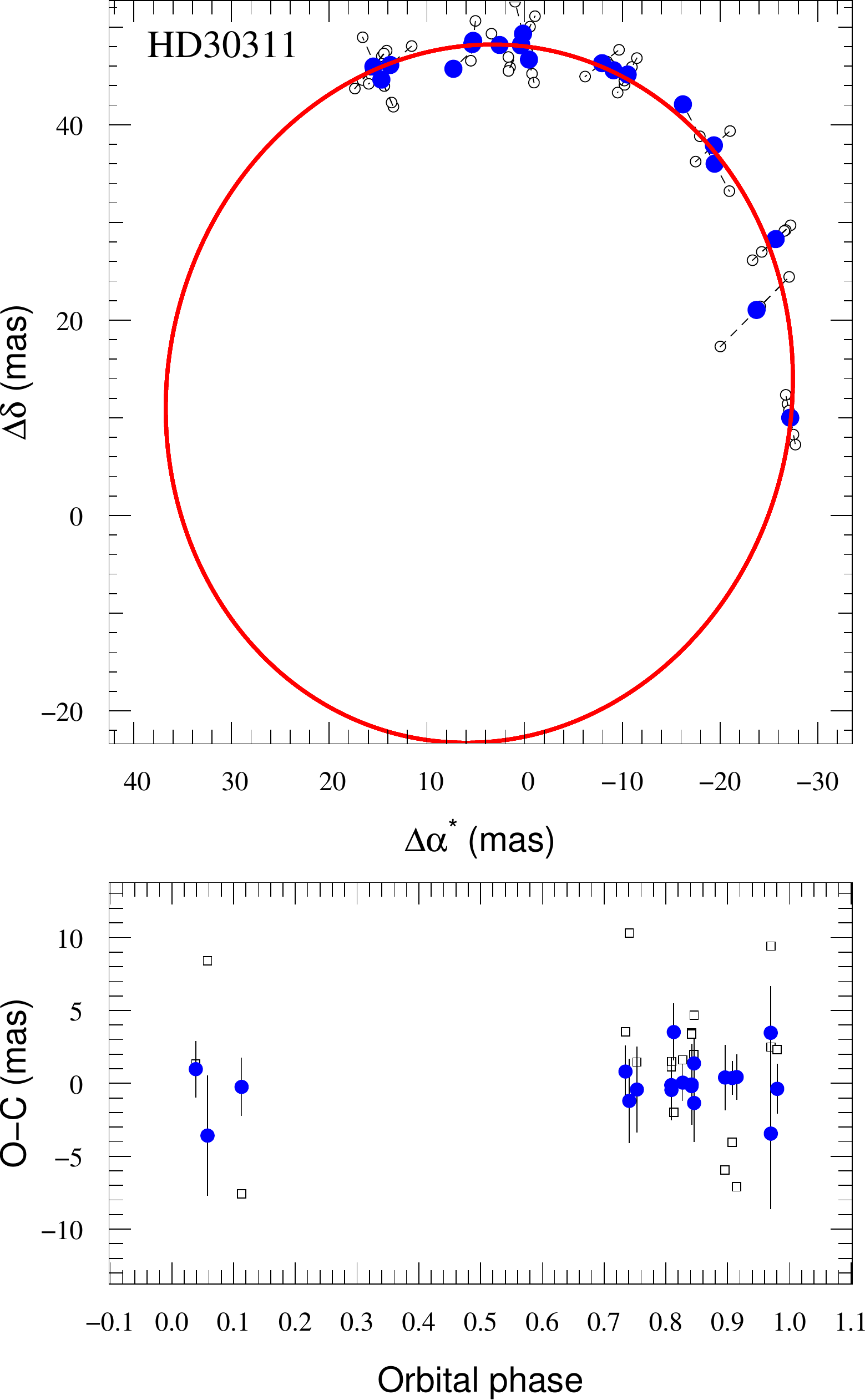} 
\includegraphics[width= 0.3\linewidth]{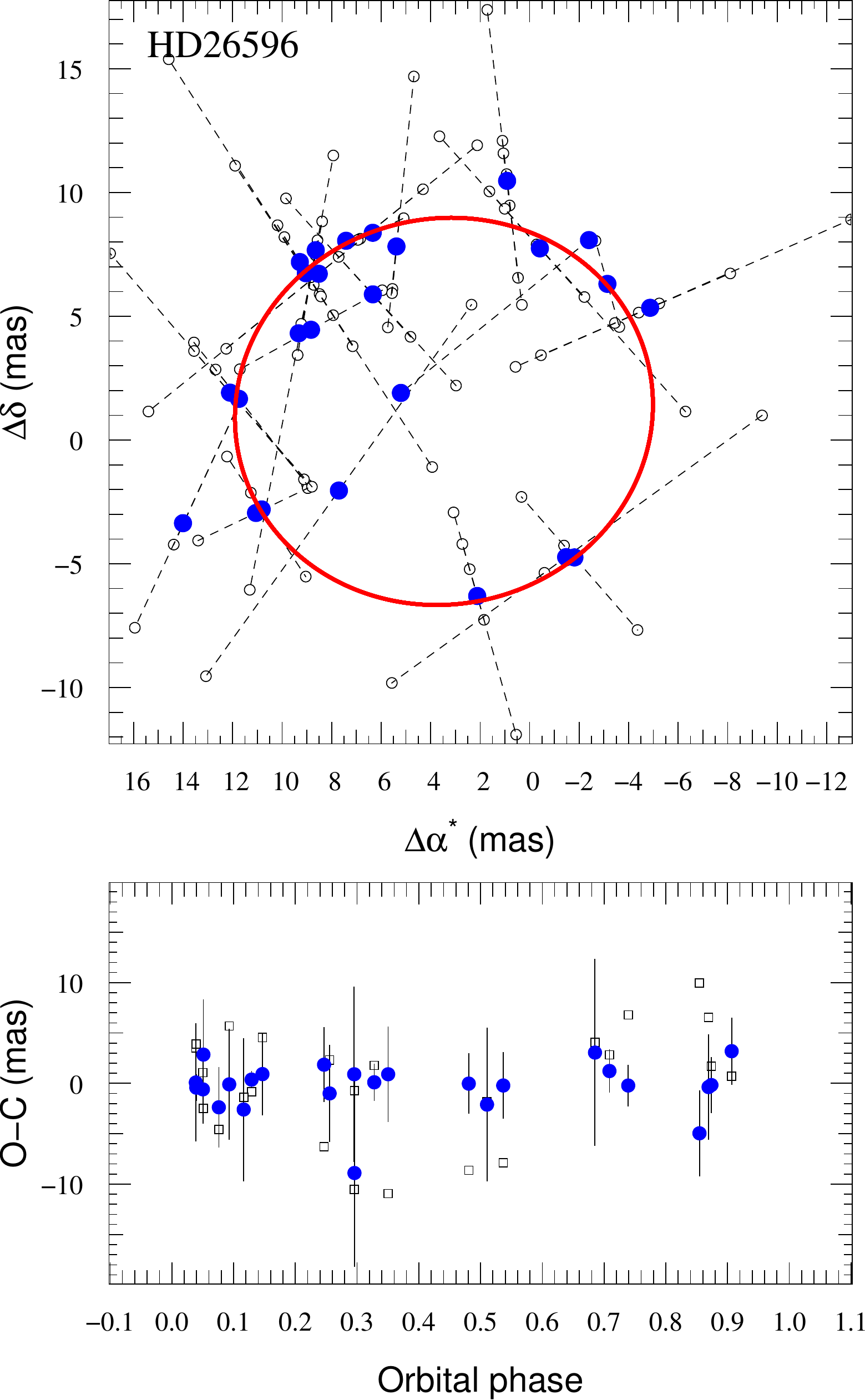} 
\includegraphics[width= 0.3\linewidth]{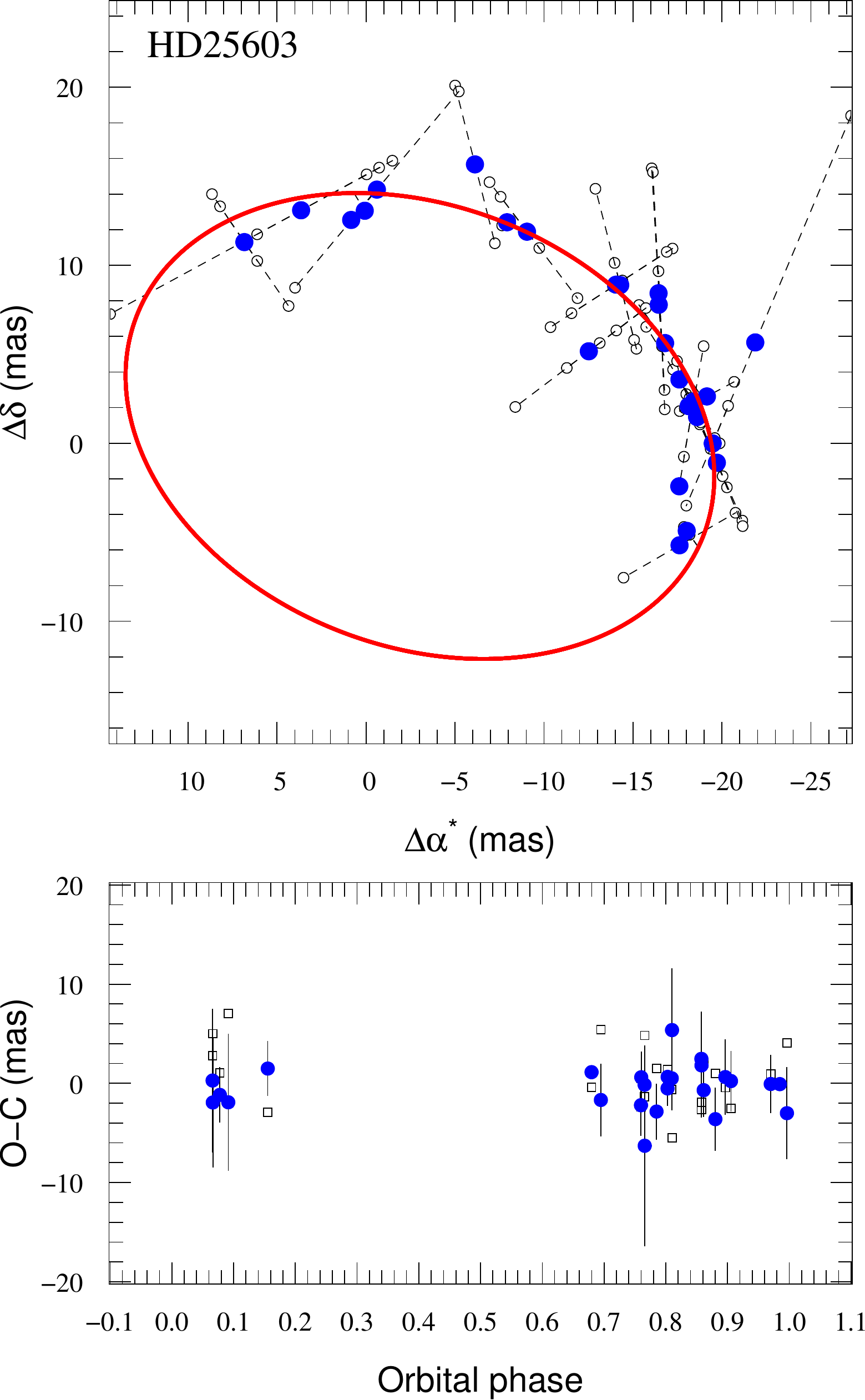}\\
\includegraphics[width= 0.3\linewidth]{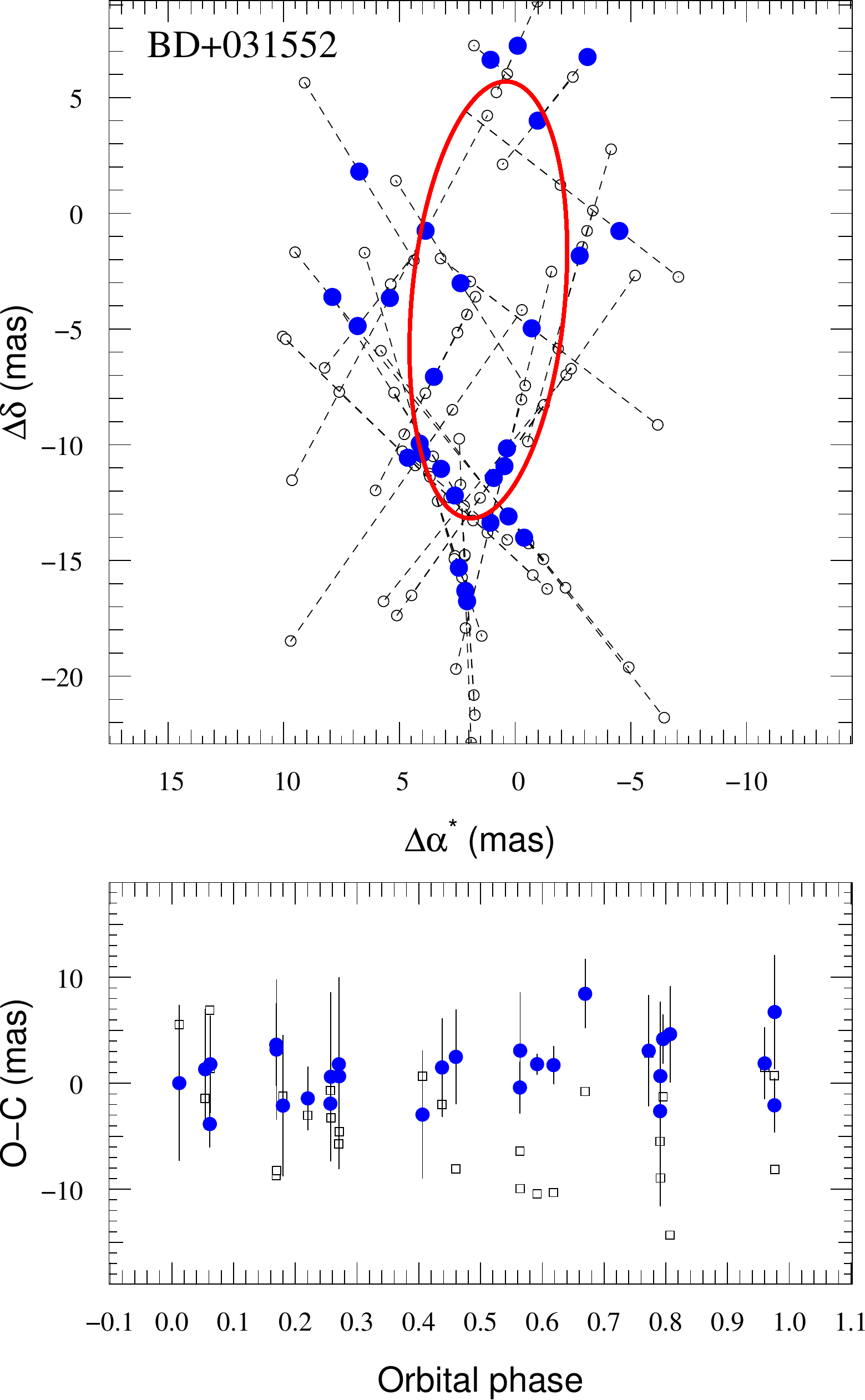} 
\end{longfigure}

\end{appendix}

\end{document}